\documentclass[a4paper,11pt]{article}%
\pdfoutput=1 
\usepackage{jheppub}
\usepackage{amsmath}
\usepackage{amsfonts}
\usepackage{amssymb}
\usepackage{graphicx}%
\usepackage[T1]{fontenc} 

\newcommand{\dhd}{{\textstyle d}
\lower.03ex\hbox{\kern-0.40em$^{\scriptstyle-}$}\kern-0.08em{}}  

\newcommand{\dbar}{{\textstyle \delta}
\lower.03ex\hbox{\kern-0.38em$^{\scriptstyle-}$}\kern-0.05em{}}
\newcommand{\half}{{1\over 2}}
\newcommand{\bu}{{\bullet}}

\newcommand{\bsi}{{\bar \psi}}

\newcommand{\cald}{{\cal D}}  
\newcommand{\calf}{{\cal F}} 
\newcommand{\calg}{{\cal G}} 
\newcommand{\calh}{{\cal H}}     
  
\newcommand{\calo}{{\cal O}}  
\newcommand{\calq}{{\cal Q}}

\newcommand{\calv}{{\cal V}}

\newcommand{\tilcaf}{\tilde{\cal F}} 
\newcommand{\ticalo}{\tilde{\cal O}} 
\newcommand{\ticalf}{\tilde{\cal F}} 
\newcommand{\ticalg}{\tilde{\cal G}} 
\newcommand{\ticalq}{\tilde{\cal Q}}

\newcommand{\tilA}{\tilde{A}} 
\newcommand{\tilD}{\tilde{D}} 
\newcommand{\tiL}{\tilde{L}} 
\newcommand{\tilF}{\tilde{F}} 
 
\newcommand{\tilT}{\tilde{T}} 
\newcommand{\tilU}{\tilde{U}}

\newcommand{\calr}{{\cal R}}

\pdfoutput=1
\abstract{We study how the rapidity evolution of gluon transverse momentum dependent distribution 
changes from nonlinear evolution at small $x\ll 1$ to linear  evolution at moderate $x\sim 1$.}
\arxivnumber{}
\affiliation{$^{*}$ Physics Dept., Old Dominion University, Norfolk VA 23529,USA
and Theory Group, JLAB, 12000 Jefferson Ave, Newport News, VA 23606,USA}
\affiliation{$^{\dag}$ Theory Group, JLAB, 12000 Jefferson Ave, Newport News, VA 23606,USA}
\emailAdd{balitsky@jlab.org}
\emailAdd{atarasov@jlab.org}
\begin{document}

\title{\boldmath Rapidity evolution of gluon TMD from low to moderate $x$}
\author{I. Balitsky$^{*}$ and A. Tarasov$^{\dag}$}
\preprint{JLAB-THY-15-2040}
\maketitle

\flushbottom

\section{Introduction\label{aba:sec1}}

\bigskip
A TMD factorization \cite{cs1, jimayuan,collinsbook} generalizes the usual concept of parton density by allowing PDFs to
depend on intrinsic transverse momenta in addition to the usual longitudinal momentum fraction variable. 
These transverse-momentum dependent parton distributions (also called unintegrated parton distributions) are widely used in the analysis of 
semi-inclusive processes like semi-inclusive deep inelastic scattering (SIDIS) 
or dijet production in hadron-hadron collisions (for a review, see Ref. \cite{collinsbook}). However, the analysis of TMD evolution in these cases is mostly
restricted to the evolution of quark TMDs, whereas at high collider energies the majority of produced particles
will be small-$x$ gluons. In this case one has to understand the transition between non-linear dynamics at small $x$ and
presumably linear evolution of gluon TMDs at intermediate $x$.

The study of the transition between the low-$x$ and moderate-$x$ TMDs is complexified by the fact that there are
  two non-equivalent definitions of gluon TMDs in small-$x$ and ``medium $x$'' communities.
In the small-$x$ literature the Weizsacker-Williams (WW) unintegrated gluon distribution \cite{domarxian} is defined in terms of the matrix element 
\begin{equation}
\sum_X {\rm tr}\langle p| D^iU U^\dagger(z_\perp)|X\rangle \langle X| D_iU U^\dagger(0_\perp) |p\rangle
\label{WW}
\end{equation}
between target states (typically protons). Here tr is a color trace in the fundamental representation, $\sum_X$ denotes the sum over full set of hadronic states and $U_z$ is a Wilson-line operator - infinite gauge link ordered along the light-like line
\begin{equation}
U(z_\perp)~=~[\infty n+z_\perp,-\infty n+z_\perp],~~~~[x,y]~\equiv~{\rm P}e^{ig\int\! du~(x-y)^\mu A_\mu(ux+(1-u)y)}
\label{defu}
\end{equation}
and $D^iU(z_\perp)=\partial^iU(z_\perp)-iA^i(\infty n+z_\perp)U(z_\perp)+iA^i(-\infty n+z_\perp)U(z_\perp)$.
In the spirit of rapidity factorization, Bjorken $x$ enters this expression as a rapidity cutoff for Wilson-line operators. 
Roughly speaking, each gluon emitted by Wilson line has rapidity restricted from above by $\ln x_B$.
 
One can rewrite the above matrix element (up to some trivial factor) in the form
\begin{eqnarray}
&&\hspace{-0mm}
\alpha_s\cald(x_B,z_\perp)~=~-{1\over8\pi^2(p\cdot n)x_B}\!\int\! du \sum_X \langle p|\tilcaf^a_\xi(z_\perp+un)|X\rangle \langle X| \calf^{a\xi}(0)|p\rangle
\label{dgTMD}
\end{eqnarray}
where
\begin{eqnarray}
&&\hspace{-0mm}
\calf^a_\xi(z_\perp+un)~\equiv~[\infty n+z_\perp,u n+z_\perp]^{am}n^\mu gF^m_{\mu \xi}(un+z_\perp)
\nonumber\\
&&\hspace{-0mm}
\tilcaf^a_\xi(z_\perp+un)~\equiv~n^\mu g\tilF^m_{\mu \xi}(un+z_\perp)[un+z_\perp,\infty n+z_\perp]^{ma}
\label{1.4}
\end{eqnarray}
and define the ``WW unintegrated gluon distribution''
\begin{equation}
\cald(x_B,k_\perp)~=~\!\int\!d^2z_\perp~e^{i(k,z)_\perp}\cald(x_B,z_\perp)
\end{equation}
(Here $(k,z)_\perp$ denotes the scalar product in 2-dim transverse Euclidean space.)
It should be noted that since Wilson lines in Eq. (\ref{WW}) are renorm-invariant $\alpha_s\cald(x_B,k_\perp)$ does not depend on the renormalization scale $\mu$.

On the other hand, at moderate $x_B$ the unintegrated gluon distribution is defined as \cite{muldrod}
\begin{eqnarray}
&&\hspace{-0mm}
\cald(x_B,k_\perp,\eta)~=~\!\int\!d^2z_\perp~e^{i(k,z)_\perp}\cald(x_B,z_\perp,\eta),
\label{gTMD}\\
&&\hspace{-0mm}
\alpha_s\cald(x_B,z_\perp,\eta)~=~{-x_B^{-1}\over 8\pi^2(p\cdot n)}\!\int\! du ~e^{-ix_Bu(pn)}\sum_X \langle p|\tilcaf^a_\xi(z_\perp+un)|X\rangle \langle X| \calf^{a\xi}(0)|p\rangle
\nonumber
\end{eqnarray}
where $|p\rangle$ is an unpolarized target with momentum $p$ (typically proton).
There are more involved definitions with Eq. (\ref{gTMD}) multiplied by some Wilson-line factors \cite{collinsbook, echevidsci} 
following from CSS factorization \cite{cs2} but we will discuss the ``primordial'' TMD (\ref{gTMD}).
The Bjorken $x$ is now introduced explicitly in the definition of gluon TMD. However, because light-like Wilson lines exhibit rapidity divergencies, we need a 
separate cutoff $\eta$ (not necessarily equal to $\ln x_B$) for the rapidity of the gluons emitted by Wilson lines. In addition,
 the matrix elements (\ref{gTMD}) may have double-logarithmic contributions 
of the type $(\alpha_s\eta\ln x_B)^n$ while the WW distribution (\ref{dgTMD}) has only single-log terms $(\alpha_s \ln x_B)^n$ 
described by the BK evolution \cite{npb96,yura}.

 In the present paper we study the connection between rapidity evolution of WW TMD  (\ref{dgTMD}) at low $x_B$
 and (\ref{gTMD}) at moderate $x_B\sim 1$.  
We will assume $k_\perp^2\geq $ few GeV$^2$ so that we can use perturbative QCD (but otherwise $k_\perp$ 
is arbitrary and can be of order of $s$ as in the DGLAP evolution). In this kinematic region we will vary Bjorken $x_B$ and look how non-linear evolution at small $x$ transforms
 into linear evolution at moderate $x_B$.  It should be noted that at least at moderate $x_B$ gluon TMDs mix with the quark ones. 
 In this paper we disregard this mixing leaving the calculation of full matrix for future publications. (For the study of 
 quark TMDs in the low-$x$ region see recent preprint \cite{kovsievert}.)
 
 In addition, we will present the evolution equation for the
 fragmentation function
\begin{eqnarray}
&&\hspace{-0mm}
\cald^{\rm f}(\beta_F,k_\perp,\eta)~=~\!\int\!d^2z_\perp~e^{-i(k,z)_\perp}\cald^{\rm f}(\beta_F,z_\perp,\eta),
\label{frTMD}\\
&&\hspace{-0mm}
\alpha_s\cald^{\rm f}(\beta_F,z_\perp,\eta)
~=~{-\beta_F^{-1}\over 8\pi^2(p\cdot n)}\!\int\! du ~e^{i\beta_Fu(pn)}
\sum_X \langle 0|\tilcaf^a_\xi(z_\perp+un)|p+X\rangle \langle p+X| \calf^{a\xi}(0)|0\rangle
\nonumber
\end{eqnarray}
 where $p$ is the momentum of the registered hadron. It turns out to be free of non-linear terms, at least in the leading log 
 approximation.
 
It should be emphasized that we consider gluon TMDs with Wilson links going to $+\infty$ in the longitudinal direction relevant for 
SIDIS \cite{collins1}.  Note that in the leading order SIDIS is determined solely by quark TMDs but beyond that the gluon TMDs 
should be taken into account, especially for the description of various processes at future EIC collider (see e.g. the report \cite{EICase}).

It is worth noting that another gluon TMD with links going to $-\infty$ arises in the study of processes with exclusive particle production (like Drell-Yan or Higgs production), see for example the discussion in Ref. \cite{muxiyu}. We plan to study it in future publications.
 
 The paper is organized as follows. In Sec. 2 we remind the general logic of rapidity factorization and rapidity evolution. 
 In Sec. 3 we derive the evolution equation of gluon TMD in the light-cone (DGLAP) limit.
 In Sec. 4 we calculate the Lipatov vertex of the gluon production by the $\calf^a_i$  operator and the so-called virtual corrections. 
 The final TMD evolution 
 equation for all $x_B$ and  transverse momenta is presented in Sec. 5 and in Sec. 6 we discuss the DGLAP, BK and Sudakov limits of our equation. In Sec. 7 we demonstrate that the linearized evolution equation for unintegrated gluon distribution interpolates between 
 BFKL and DGLAP equations. 
 In Sec. 8 we present the evolution equations for fragmentation TMD and Sec. 9 contains conclusions and outlook. 
 The necessary formulas for propagators near the light cone and in the shock-wave
 background can be found in  Appendices.

\section{Rapidity factorization and evolution \label{sec2}}

In the spirit of high-energy OPE, the rapidity of the gluons is restricted from above by the ``rapidity divide'' $\eta$ separating the impact factor and the matrix element so the proper definition of $U_x$ is 
\footnote
{Alternatively, with the leading-log accuracy one can take the Wilson line slightly off the light cone, see Ref. \cite{collinsbook}. 
To pave the way for future NLO calculation we prefer the ``rigid cutoff'' Eq. (\ref{cutoff}) which was used for the NLO calculations in the 
 low-$x$ case \cite{nlobk}.}
\begin{eqnarray}
&&\hspace{-0mm} 
 U^\eta_x~=~{\rm Pexp}\Big[ig\!\int_{-\infty}^\infty\!\! du ~p_1^\mu A^\eta_\mu(up_1+x_\perp)\Big],
\nonumber\\
&&\hspace{-0mm} 
A^\eta_\mu(x)~=~\int\!{d^4 k\over 16\pi^4} ~\theta(e^\eta-|\alpha|)e^{-ik\cdot x} A_\mu(k)
\label{cutoff}
\end{eqnarray}
where  the  Sudakov variable $\alpha$ is defined as usual,  $k=\alpha p_1+\beta p_2+k_\perp$.
We define the light-like vectors $p_1$ and $p_2$ such that $p_1=n$  and $p_2=p-{m^2\over s}n$, where $p$ is the momentum of the target particle of mass $m$. 
We use metric $g^{\mu\nu}~=~(1,-1,-1,-1)$ so $p\cdot q~=~(\alpha_p\beta_q+\alpha_q\beta_p){s\over 2}-(p,q)_\perp$. For the coordinates we use 
the notations $x_\bu\equiv x_\mu p_1^\mu$ and $x_\ast\equiv x_\mu p_2^\mu$ related to the light-cone coordinates by $x_\ast=\sqrt{s\over 2}x_+$ and $x_\bu=\sqrt{s\over 2}x_-$.
It is convenient to define Fourier transform of the operator $\calf^a_i$
\begin{eqnarray}
&&\hspace{-0mm} 
\calf^{a\eta}_{i}(\beta_B,k_\perp)~=~\!\int\! d^2z_\perp~e^{-i(k,z)_\perp}\calf^{a\eta}_{i}(\beta_B,z_\perp),
\nonumber\\
&&\hspace{-0mm} 
\calf^{a\eta}_{i}(\beta_B,z_\perp)~\equiv~{2\over s}
\!\int\! dz_\ast ~e^{i\beta_B z_\ast} 
\big([\infty,z_\ast]_z^{am}gF^m_{\bu i}(z_\ast,z_\perp))^\eta
\label{kalf}
\end{eqnarray}
where the index $\eta$ denotes the rapidity cutoff (\ref{cutoff}) for all gluon fields in this operator.
Here we introduced the ``Bjorken $\beta_B$''  to have similar formulas for 
 the DIS and annihilation matrix elements ($\beta_B=x_B$ in DIS and $\beta_B=\beta_F={1\over z_F}$
 for fragmentation functions).
Also, hereafter we use the notation 
$[\infty, z_\ast]_z\equiv[\infty_\ast p_1+z_\perp, {2\over s}z_\ast p_1+z_\perp]$  where  
$[x,y]$ stands for the straight-line gauge link connecting points $x$ and $y$ as defined in Eq. (\ref{defu}). 
Our convention is that the Latin Lorentz indices always correspond to transverse coordinates while Greek Lorentz indices are four-dimensional.

Similarly, we define 
\begin{eqnarray}
&&\hspace{-0mm} 
\tilcaf_i^{a\eta}(\beta_B,k_\perp)~=~\!\int\! d^2z_\perp~e^{i(k,z)_\perp}\tilcaf_i^{a\eta}(\beta_B,z_\perp),
\nonumber\\
&&\hspace{-0mm} 
\tilcaf^{a\eta}_i(\beta_B,z_\perp)~\equiv~{2\over s}
\!\int\! dz_\ast ~e^{-i\beta_B z_\ast} 
g\big(\tilF^m_{\bu i}(z_\ast,z_\perp)[z_\ast,\infty]_z^{ma}\big)^\eta
\label{tilkaf}
\end{eqnarray}
in the complex-conjugate part of the amplitude.

In this notations the unintegrated gluon TMD (\ref{gTMD}) can be represented as 
\begin{eqnarray}
&&\hspace{-0mm} 
\langle p|\tilcaf^{a\eta}_i(\beta_B,z_\perp)\calf^{ai\eta}(\beta_B,0_\perp)|p+\xi p_2\rangle 
\equiv\sum_X\langle p|\tilcaf^{a\eta}_i(\beta_B,z_\perp)|X\rangle\langle X|\calf^{ai\eta}(\beta_B,0_\perp)|p+\xi p_2\rangle 
\nonumber\\
&&\hspace{-0mm} 
=~-4\pi^2\delta(\xi)\beta_Bg^2\cald(\beta_B,z_\perp,\eta)
\label{TMD}
\end{eqnarray}
Hereafter we use a short-hand notation
\begin{eqnarray}
&&\hspace{-0mm} 
\langle p|\ticalo_1...\ticalo_m\calo_1...\calo_n |p'\rangle
\equiv~\sum_X\langle p| \tilT\{ \ticalo_1...\ticalo_m\}|X\rangle\langle X|T\{\calo_1...\calo_n\}|p'\rangle
\label{ourop}
\end{eqnarray}
where tilde on the operators in the l.h.s. of this formula stands as a reminder that they should be inverse time ordered
as indicated by inverse-time ordering $\tilT$ in the r.h.s. of the above equation.

As discussed e.g. in Ref. \cite{keld},  such martix element can be represented by a double functional integral
\begin{eqnarray}
&&\hspace{-0mm} 
\langle \ticalo_1...\ticalo_m\calo_1...\calo_n \rangle
\nonumber\\
&&\hspace{-0mm} 
=~\int\! D\tilA D\tilde{\bar\psi}D\tilde{\psi}~e^{-iS_{\rm QCD}(\tilA,\tilde{\psi})}\!\int\! DA D\bar{\psi} D\psi ~e^{iS_{\rm QCD}(A,\psi)} \ticalo_1...\ticalo_m\calo_1...\calo_n
\label{funtegral}
\end{eqnarray}
with the boundary condition $\tilA(\vec{x},t=\infty)=A(\vec{x},t=\infty)$ (and similarly for quark fields) reflecting the sum over all intermediate states $X$. 
Due to this condition, the matrix element (\ref{TMD}) can be made gauge-invariant by connecting the endpoints of Wilson lines at infinity with the gauge link \footnote{Similarly, this gauge link is implied in Eq. (1.1) which is Eq. (\ref{inftylink}) at $\beta_B=0$.}
\begin{eqnarray}
&&\hspace{-0mm} 
\langle p|\tilcaf^a_i(\beta_B,x_\perp)\calf^{ai}(\beta_B',y_\perp)|p'\rangle~
\nonumber\\
&&\hspace{-0mm} 
\rightarrow~\langle p|\tilcaf^a_i(\beta_B,x_\perp)[x_\perp+\infty p_1,y_\perp+\infty p_1]\calf^{ai}(\beta'_B,y_\perp)|p'\rangle
\label{inftylink}
\end{eqnarray}
This gauge link is important if we use the light-like gauge $p_1^\mu A_\mu=0$ for calculations \cite{bejuan}, but in all other gauges it can be neglected.
We will not write it down explicitly but will always assume it in our formulas.

We will study the rapidity evolution of the operator 
\begin{equation}
\tilcaf^{a\eta}_i(\beta_B,x_\perp)\calf^{a\eta}_j(\beta_B,y_\perp)
\label{operator}
\end{equation}
Matrix elements of this operator between unpolarized hadrons can be parametrized as \cite{muldrod}
\begin{eqnarray}
&&\hspace{0mm} 
\int\! d^2z_\perp~e^{i(k,z)_\perp}\langle p|\tilcaf^{a\eta}_i(\beta_B,z_\perp)\calf_j^{a\eta}(\beta_B,0_\perp)|p+\xi p_2\rangle
~=~ 2\pi^2\delta(\xi) \beta_Bg^2\calr_{ij}(\beta_B, k_\perp;\eta)
\nonumber\\
&&\hspace{0mm}
\calr_{ij}(\beta_B, k_\perp;\eta)~=~-g_{ij}\cald(\beta_B,k_\perp,\eta)
+\big({2k_ik_j\over m^2}+g_{ij}{k_\perp^2\over m^2}\Big)\calh(\beta_B,k_\perp,\eta)
\label{mael}
\end{eqnarray}
where $m$ is the mass of the target hadron (typically proton).
The reason we study the evolution of the operator (\ref{operator}) with non-convoluted indices $i$ and $j$ is that, as we shall see below, the rapidity evolution mixes functions $\cald$ and $\calh$. It should be also noted that our final equation for the evolution of the operator
(\ref{operator}) is applicable for polarized targets as well.

We shall also study the  evolution of fragmentation functions defined by ``fragmentation matrix elements'' (\ref{frTMD}) 
of the operator (\ref{operator}). If the polarization of the fragmentation hadron is not registered, this matrix element can be parametrized 
similarly to Eq. (\ref{mael}) (cf. Ref. \cite{muldrod})
\begin{eqnarray}
&&\hspace{0mm} 
\int\! d^2z_\perp~e^{-i(k,z)_\perp}
\sum_X\langle 0|\tilcaf^{a\eta}_i(-\beta_F,z_\perp)|p+X\rangle
\langle p+\xi p_2+X|\calf_j^{a\eta}(-\beta_F,0_\perp)|0\rangle
\nonumber\\
&&\hspace{0mm}
=~ 2\pi^2\delta(\xi) \beta_Fg^2\Big[-g_{ij}\cald^{\rm f}(\beta_F,k_\perp,\eta)
+\Big({2k_ik_j\over m^2}+g_{ij}{k_\perp^2\over m^2}\Big)\calh^{\rm f}(\beta_F,k_\perp,\eta)\Big]
\label{maelfr}
\end{eqnarray}
Note that $\beta_F$ should be greater than 1 in this equation, otherwise the cross section vanishes.  As to matrix element
(\ref{TMD}), it can be defined with either sign of $\beta_B$ but the deep inelastic scattering corresponds to
$\beta_B=x_B>0$. In our calculations we will consider $\beta_B>0$ for simplicity and perform the trivial analytic continuation to negative $\beta_B$ in the final formula (\ref{master1alt}).

In the spirit of rapidity factorization, in order to find the evolution of the operator (\ref{operator})
with respect to rapidity cutoff $\eta$ 
(see Eq. (\ref{cutoff})) one should integrate in the matrix element  (\ref{TMD}) over gluons and quarks with rapidities $\eta>Y>\eta'$ and temporarily ``freeze'' fields with $Y<\eta'$ to be integrated over later. (For a review, see Refs. \cite{mobzor,nlolecture}.) In this case, we obtain functional integral of Eq. (\ref{funtegral}) type over fields with $\eta>Y>\eta'$ in the ``external'' fields with $Y<\eta'$. 
In terms of Sudakov variables we integrate over gluons with $\alpha$ between $\sigma=e^\eta$ and $\sigma'=e^{\eta'}$ and, in the leading order, only
the diagrams with gluon emissions are relevant - the quark diagrams will enter as loops at the next-to-leading (NLO) level.

To make connections with parton model we will have in mind the frame where target's velocity is large and call the small $\alpha$ fields by the name ``fast fields'' and large $\alpha$ fields by 
``slow'' fields.  Of course, ``fast'' {\it vs} ``slow'' depends on frame but we will stick to naming fields as they appear 
in the projectile's frame. (Note that in Ref. \cite{npb96} the terminology is opposite, as appears in the target's frame).
As discussed in Ref. \cite{npb96},  the interaction of ``slow'' gluons of large $\alpha$ with ``fast'' fields of 
small $\alpha$  is described by eikonal  gauge factors and the integration over slow fields  results in  
Feynman diagrams in the background of fast fields which form a thin shock wave due to Lorentz contraction.
 However,  in Ref. \cite{npb96} (as well as in all small-$x$ literature)
it was assumed that the characteristic transverse momenta of fast and slow fields are of the same order of
magnitude. For our present purposes we need to relax this condition and consider cases where the 
transverse momenta of fast and slow fields do differ. In this case, we need to rethink the shock-wave approach.

Let us figure out how the relative longitudinal size of fast and slow fields depends on their transverse momenta.
 The typical longitudinal size of fast 
fields is $\sigma_\ast\sim {\sigma' s\over l_\perp^2}$ where $l_\perp$ is the characteristic 
scale of transverse momenta of fast fields. The typical distances  traveled by slow gluons are 
$\sim{\sigma s\over k_\perp^2}$ where $k_\perp$ is the characteristic 
scale of transverse momenta of slow fields. Effectively, the 
large-$\alpha$ gluons propagate in the external field of the small-$\alpha$ shock wave, except the case
$l_\perp^2\ll k_\perp^2$ which should be treated separately since the ``shock wave'' is not necessarily thin in this case.  
Fortunately, when $l_\perp^2\ll k_\perp^2$ one can use  
the light-cone expansion of slow fields and leave at the leading order only the light-ray operators of the leading twist.
We will use the combination of shock-wave and light-cone expansions and write the interpolating formulas 
which describe the leading-order contributions in both cases.

\section{Evolution kernel in the light-cone limit}

As we discussed above, we will obtain the evolution kernel in two separate cases: the ``shock wave'' case when  the characteristic transverse momenta 
of the background gluon (or quark) fields $l_\perp$ are of the order of typical momentum of emitted gluon $k_\perp$ and  
the ``light cone'' case when $l_\perp^2\ll k_\perp^2$.  It is convenient to start with the light-cone situation and consider 
 the one-loop evolution of the operator $\tilcaf^{a\eta}_i(\beta_B,x_\perp)\calf^{ai\eta}(\beta_B,y_\perp)$ 
 in the case when the background fields are soft so we can use the 
 expansion of propagators in external fields near the light cone \cite{bbr}. 
 
\begin{figure}[htb]
\begin{center}
\includegraphics[width=104mm]{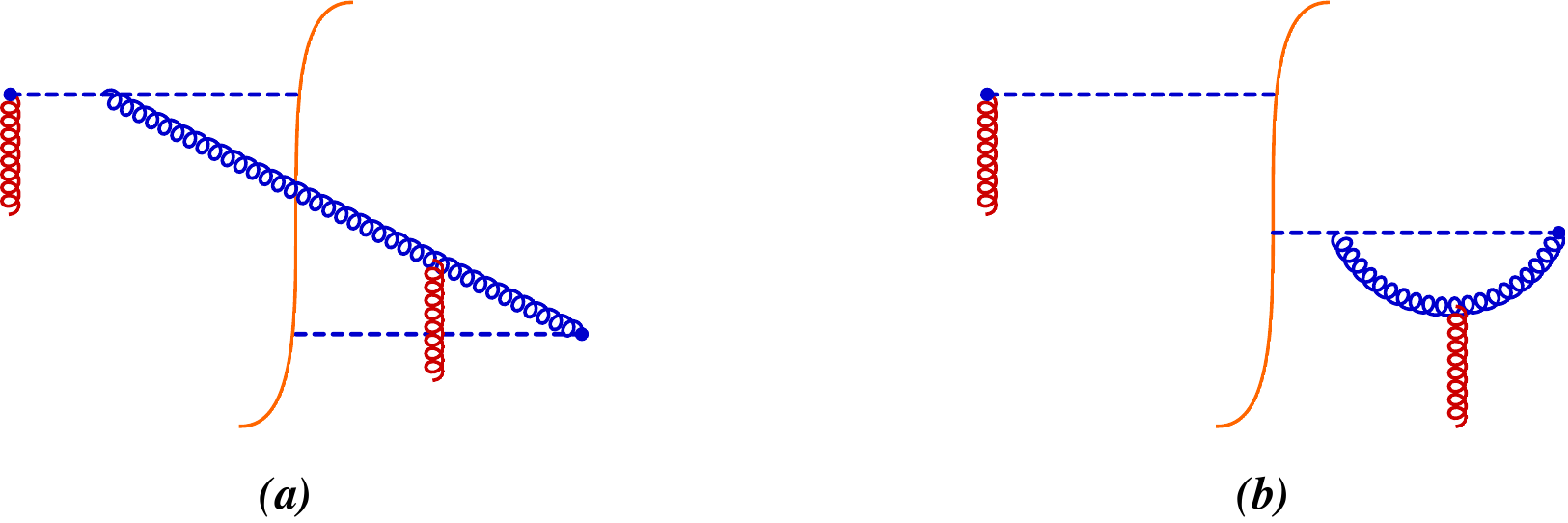}
\end{center}
\caption{Typical diagrams for production (a) and virtual (b) contributions to the evolution kernel. The dashed lines denote gauge links.\label{fig:1}}
\end{figure}

 In the leading order there is only one ``quantum'' gluon and we get the typical diagrams of Fig. \ref{fig:1} type. 
One sees that the evolution kernel consist of two parts: ``real'' part with the emission of a real gluon and a ``virtual'' part without such emission. 
The ``real'' production part of the kernel can be obtained as a square of a Lipatov vertex - the amplitude of the emission of a real gluon by the Wilson-line operator $\calf^a_i$:  
\begin{eqnarray}
&&\hspace{-1mm}
\langle \tilcaf_i^a(\beta_B,x_\perp) \calf_j^a(\beta_B,y_\perp)\rangle^{\ln\sigma}~
\nonumber\\
&&\hspace{-1mm}
=~-\!\int_{\sigma'}^{\sigma}\!{\dhd\alpha\over 2\alpha}\dhd^2k_\perp\big(
\langle \lim_{k^2\rightarrow 0}k^2\tilcaf_i^a(\beta_B,x_\perp)\tilA^m_\rho(k)\rangle
\langle \lim_{k^2\rightarrow 0}k^2A^{m\rho}(k)\calf_j^a(\beta_B,y_\perp)\rangle\big)^{\ln\sigma'}
\label{3.1}
\end{eqnarray}
Hereafter we use the space-saving notation $\dhd^np\equiv {d^np\over(2\pi)^n}$.

 \subsection{Lipatov vertex}
As we mentioned, the production (``real'') part of the kernel
corresponds to square of Lipatov vertex describing the emission of a gluon by the operator $\calf^a_i$.
The Lipatov vertex is defined as
\begin{eqnarray}
&&\hspace{-1mm}
L^{ab}_{\mu i}(k,y_\perp,\beta_B)~=~i\lim_{k^2\rightarrow 0}k^2\langle T\{A^a_\mu(k)\calf^b_i(\beta_B,y_\perp)\}\rangle
\label{3.2}
\end{eqnarray}
(To simplify our notations, we will often omit 
label $\eta$ for the rapidity cutoff (\ref{cutoff}) but it will be always assumed when not displayed).

\begin{figure}[htb]
\begin{center}
\includegraphics[width=104mm]{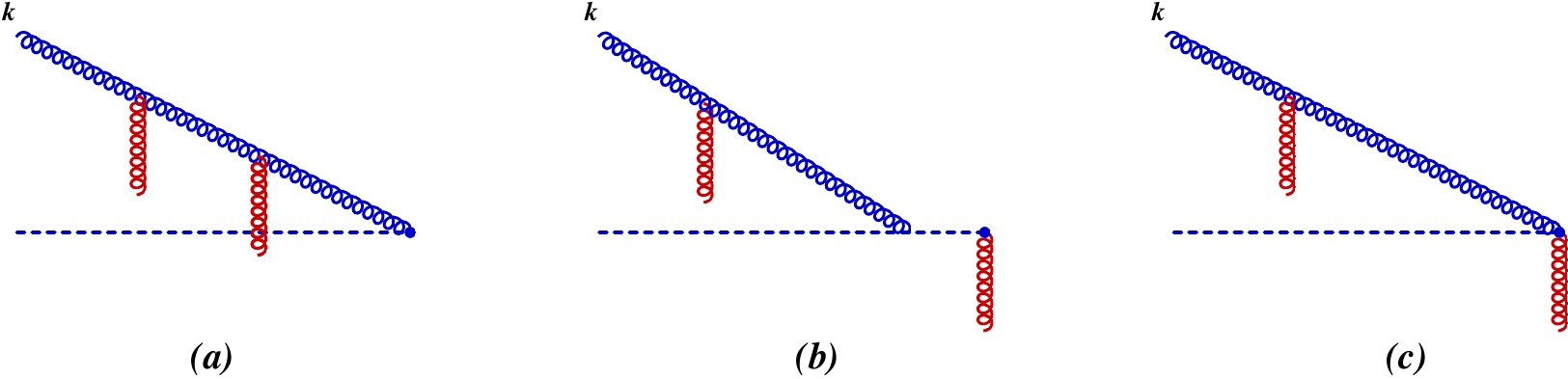}
\end{center}
\caption{Lipatov vertex of gluon emission. \label{fig:2}}
\end{figure}

We will use the background-Feynman gauge. 
The three corresponding diagrams are shown in Fig. \ref{fig:2}. 

\subsubsection{Emission of soft gluon near the light cone}

In accordance with general background-field formalism we separate the gluon field into the ``classical'' background part and
``quantum'' part
$$
A_\mu~\rightarrow~A^{\rm q}_\mu+A^{\rm cl}_\mu
$$
where the ``classical'' fields are fast $(\alpha<\sigma')$ and ``quantum'' fields are slow  $(\alpha>\sigma')$.
It should be emphasized that our ``classical'' field does not satisfy the equation $D^\mu F_{\mu\nu}=0$; rather, 
$(D^\mu F^{\rm cl}_{\mu\nu})^a=-g\bsi\gamma_\nu t^a\psi$, where $\psi$ are the ``classical'' (i.e. fast) quark fields. 
In addition, in this Section it is assumed that the slow fields are hard and the fast fields are soft so one can use the light-cone expansion. 
We will perform calculations in the background-Feynman gauge, where the gluon propagator is $\big({1\over P^2+2iF}\big)_{\mu\nu}$,
see Appendix A.

The first-order term in the expansion of the operator $[\infty,y_\ast]_y^{nm}F^m_{\bu i}(y_\ast,y_\perp)$ in quantum fields has the form
\begin{eqnarray}
&&\hspace{-11mm}
[\infty,y_\ast]_y^{nm} F_{\bu i}^m(y_\ast,y_\perp)~\stackrel{\rm 1st}{=}~
{s\over 2}{\partial\over\partial y_\ast}[\infty,y_\ast]_y^{nm}A_i^{m{\rm q}}(y_\ast,y_\perp)
\label{3.3}\\
&&\hspace{-11mm}
-~[\infty,y_\ast]_y^{nm}\partial_iA_\bu^{m{\rm q}}(y_\ast,y_\perp)
+i\!\int^{\infty}_{y_\ast}\! d{2\over s}z'_\ast ~[\infty,z'_\ast]_yA_\bu^{\rm q}(z'_\ast,y_\perp)[z'_\ast,y_\ast]_y^{nm}F_{\bu i}^m(y_\ast,y_\perp)
\nonumber
\end{eqnarray}
(to save space, we omit the label ${}^{\rm cl}$ from classical fields).  The corresponding vertex of gluon emission is given by
\begin{eqnarray}
&&\hspace{-1mm}
\lim_{k^2\rightarrow 0}k^2\langle A^{a{\rm q}}_\mu(k)\big([\infty,y_\ast]_y^{nm} F_{\bu i}^m(y_\ast,y_\perp)\big)^{\rm 1st}\rangle
\label{3.4}\\
&&\hspace{-1mm}
=~\lim_{k^2\rightarrow 0}k^2\Big[{s\over 2}{\partial\over\partial y_\ast}[\infty,y_\ast]_y^{nm}\langle A^{a{\rm q}}_\mu(k)A_i^{m{\rm q}}(y_\ast,y_\perp)\rangle
-~[\infty,y_\ast]_y^{nm}\langle A^{a{\rm q}}_\mu(k)\partial_iA_\bu^{m{\rm q}}(y_\ast,y_\perp)\rangle
\nonumber\\
&&\hspace{11mm}
+~i{2\over s}\!\int^{\infty}_{y_\ast}\! dz_\ast ~[\infty,z_\ast]_y\langle A^{a{\rm q}}_\mu(k)A_\bu^{\rm q}(z_\ast,y_\perp)\rangle[z_\ast,y_\ast]_y^{nm}F_{\bu i}^m(y_\ast,y_\perp)\Big]
\nonumber
\end{eqnarray}
To calculate the r.h.s. we can use formulas (\ref{A.47})-(\ref{A.48}) from Appendix A.
As we mentioned,  we need contributions to production part of the kernel with the collinear twist up to two.
However, it is easy to see that the light-cone expansion of gluon emission vertex starts with the operators of twist one ($\sim F_{\bu i}$)
since the gauge links in the first term in Eq. (\ref{A.20}) cancel in Eq. (\ref{3.4}) and the remaining background-free emission of
gluon is proportional to $\delta(\beta_B+{k_\perp^2\over\alpha s}\big)$ which vanishes for $\beta_B>0$. 
Thus, to get the contribution to the production part of the kernel of collinear twist up to two it is sufficient to use formula (\ref{A.20}) for 
Feynman amplitude and formula (\ref{A.23}) for complex conjugate amplitude with twist-one (one $F_{\bu i}$) accuracy. 
In this case the quark terms do not contribute and the gluon terms simplify to
\begin{eqnarray}
&&\hspace{-3mm}
\lim_{k^2\rightarrow 0}k^2\langle A_\mu^{a{\rm q}}(k) A^{b{\rm q}}_\nu(y)\rangle~
=~-ie^{i{k_\perp^2\over\alpha s}y_\ast-i(k,y)_\perp}\calo^{ab}_{\mu\nu}(\infty,y_\ast,y_\perp;k),
\label{3.5}
\\
&&\hspace{1mm}
\calo_{\mu\nu}(\infty,y_\ast,y_\perp;k)~=~g_{\mu\nu}[\infty,y_\ast]_y
\nonumber\\
&&\hspace{34mm}
+~g\!\int_{y_\ast}^{\infty}\!\!\!dz_\ast~
\Big(-{4i\over\alpha s^2}k^j(z-y)_\ast g_{\mu\nu}[\infty,z_\ast]_yF_{\bu j}(z_\ast,y_\perp)[z_\ast,y_\ast]_y
\nonumber\\
&&\hspace{34mm}
+~{4\over \alpha s^2}(\delta_\mu^jp_{2\nu}-\delta_\nu^jp_{2\mu})[\infty,z_\ast]_yF_{\bu j}(z_\ast,y_\perp)[z_\ast,y_\ast]_y\Big)
\nonumber
\end{eqnarray}
With the help of this formula Eq. (\ref{3.4}) reduces to
\begin{eqnarray}
&&\hspace{-1mm}
\lim_{k^2\rightarrow 0} k^2\langle A_\mu^{a{\rm q}}(k)\Big([\infty,y_\ast]_y^{nm} F_{\bu i}^m(y_\ast,y_\perp)\big)^{\rm 1st}\rangle~
\label{3.6}\\
&&\hspace{-1mm}
=~\Big\{\big({k_\perp^2\over 2\alpha}g^\perp_{\mu i}-p_{1\mu}k_i\big)
\Big(1-{4igk^j\over\alpha s^2}\!\int_{y_\ast}^\infty\! dz_\ast~(z-y)_\ast ~[\infty,z_\ast]_yF_{\bu j}(z_\ast,y_\perp)[z_\ast,\infty]_y
 \Big)^{an}
\nonumber\\
&&\hspace{5mm}
+~ig\big({2\alpha\over k_\perp^2}p_{1\mu}-{2\over\alpha s}p_{2\mu}\big)\Big([\infty,y_\ast]_yF_{\bu i}(y_\ast,y_\perp)[y_\ast,\infty]_y
\nonumber\\
&&\hspace{44mm}
~-{ik_\perp^2\over\alpha s}\!\int_{y_\ast}^\infty\!\!dz_\ast[\infty,z_\ast]_yF_{\bu i}(z_\ast, y_\perp)[z_\ast,\infty]_y\Big)^{an}
\nonumber\\
&&\hspace{10mm}
+~{2g\over\alpha s}(g_{\mu i}k^j-\delta_\mu^jk_i)\!\int_{y_\ast}^\infty\! dz_\ast~([\infty,z_\ast]_yF_{\bu j}(z_\ast, y_\perp)[z_\ast,\infty]_y)^{an}
\Big\}e^{i{k_\perp^2\over\alpha s}y_\ast -i(k,y)_\perp}
\nonumber
\end{eqnarray}
Note that $k_\mu\times \big({\rm r.h.s.~of ~Eq. ~(\ref{3.6})}\big)^\mu~=~0$ as required by gauge invariance. Integrating the r.h.s. of Eq. (\ref{3.6})
over $y_\ast$ we obtain
\begin{eqnarray}
&&\hspace{-3mm}
L^{ab}_{\mu i}(k,y_\perp,\beta_B)~=~i\lim_{k^2\rightarrow 0} k^2\langle A_\mu^{a{\rm q}}(k)(\calf_i^b(\beta_B,y_\perp)\big)^{\rm 1st}\rangle
~=~{2ge^{-i(k,y)_\perp}\over \alpha\beta_Bs+k_\perp^2}
\label{3.7}\\
&&\hspace{-3mm}
\times~
\Big[{\alpha\beta_Bs\over k_\perp^2}\big({k_\perp^2\over\alpha s}p_{2\mu}-\alpha p_{1\mu}\big)\delta_i^l-\delta_\mu^lk_i
+{\alpha\beta_Bsg_{\mu i}k^l\over k_\perp^2+\alpha\beta_Bs}+{2\alpha k_ik^l\over k_\perp^2+\alpha\beta_Bs} p_{1\mu}\Big]
\calf_l^{ab}(\beta_B+{k_\perp^2\over\alpha s},y_\perp)
\nonumber
\end{eqnarray}
At this point it is convenient to switch to the light-like gauge $p_2^\mu A_\mu=0$. Since $k_\mu\times \big({\rm r.h.s.~of ~Eq. ~(\ref{3.7})}\big)^\mu~=~0$ 
it is sufficient to replace $\alpha p_1^\mu$ in the r.h.s. of Eq. (\ref{3.7}) by $\alpha p_1^\mu-k^\mu~=~-k_\perp^\mu-{k_\perp^2\over\alpha s}p_2^\mu$. One obtains
\begin{eqnarray}
&&\hspace{-3mm}
L^{ab}_{\mu i}(k,y_\perp,\beta_B)^{\rm light-like}~=~2ge^{-i(k,y)_\perp}
\label{3.8}\\
&&\hspace{-3mm}
\times~\Big[
{k^\perp_\mu\delta_i^l\over k_\perp^2}
-{\delta_\mu^lk_i+\delta_i^lk^\perp_\mu-g_{\mu i}k^l\over \alpha\beta_Bs+k_\perp^2}-{k_\perp^2g_{\mu i}k^l+2k^\perp_\mu k_ik^l\over(\alpha\beta_Bs+k_\perp^2)^2}\Big]
\calf_l^{ab}(\beta_B+{k_\perp^2\over\alpha s},y_\perp)~+~O(p_{2\mu})
\nonumber
\end{eqnarray}
We do not write down the terms $\sim p_{2\mu}$ since they do not contribute to the production kernel ($\sim$ square of the expression in the r.h.s. of Eq. (\ref{3.8})).

For the complex conjugate amplitude one obtains from Eq. (\ref{A.49})
\begin{eqnarray}
&&\hspace{-3mm}
-i\!\lim_{k^2\rightarrow 0}k^2\langle \tilA^{a}_\mu(x)\tilA_\nu^{b}(k) \rangle
=e^{-i{k_\perp^2\over\alpha s}x_\ast+i(k,x)_\perp}\ticalo^{ab}_{\mu\nu}(x_\ast,\infty,x_\perp;k),
\nonumber\\
&&\hspace{-3mm}
\ticalo_{\mu\nu}(x_\ast,\infty,x_\perp;k)~=~g_{\mu\nu}[x_\ast,\infty]_x+g\!\int_{x_\ast}^{\infty}\!\!\!dz_\ast~[x_\ast,z_\ast]_x
\Big({4i\over\alpha s^2}(z-x)_\ast g_{\mu\nu}\tilF_{\bu j}(z_\ast,x_\perp)[z_\ast,\infty]_x k^j
\nonumber\\
&&\hspace{31mm}
-~{4\over \alpha s^2}(\delta_\mu^jp_{2\nu}-\delta_\nu^jp_{2\mu})\tilF_{\bu j}(z_\ast,x_\perp)[z_\ast,\infty]_x
\Big)
\label{3.9}
\end{eqnarray}
where $\ticalo_{\mu\nu}$ is obtained from the Eq. (\ref{A.23}) with twist-two accuracy (as we mentioned, quark operators start from twist two and
therefore do not contribute to the production kernel).

Repeating steps which lead us to Eq. (\ref{3.8}) we obtain
\begin{eqnarray}
&&\hspace{-11mm}
\tiL^{ab}_{i\mu}(k,x_\perp,\beta_B)^{\rm light-like}~=~-i\lim_{k^2\rightarrow 0}k^2\langle \big(\ticalf_i^a(\beta_B,x_\perp)\big)^{\rm 1st}\tilA^{b {\rm q}}_\mu(k)\rangle^{\rm light-like}~=~2ge^{i(k,x)_\perp}
\label{3.10}\\
&&\hspace{-11mm}
\times~\Big[{k^\perp_\mu\delta_i^k\over k_\perp^2}
-{\delta_\mu^k k_i+k^\perp_\mu\delta_i^k-g_{\mu i}k^k\over \alpha\beta_Bs+k_\perp^2}
-{k_\perp^2 g_{\mu i}k^k+2k^\perp_\mu k_ik^k\over(\alpha\beta_Bs+k_\perp^2)^2} 
\Big]\ticalf_k^{ab}\big(\beta_B+{k_\perp^2\over\alpha s},x_\perp\big)~+~O(p_{2\mu})
\nonumber
\end{eqnarray}
The product of Lipatov vertices (\ref{3.8}) and (\ref{3.10}) integrated according to Eq. (\ref{3.1}) gives 
the production part of the evolution kernel in the light-cone limit. 
To get the full kernel, we need to add the virtual contribution coming from diagrams of Fig. \ref{fig:1}b type.

\subsection{Virtual part of the kernel \label{sec:3.2}}
To get the virtual part coming from diagrams of Fig. \ref{fig:1}b type we need to expand the operator $\calf$ up to the second order in quantum field
\begin{eqnarray}
&&\hspace{-3mm}
\langle[\infty,y_\ast]_y^{nm}F_{\bu i}^m(y_\ast,y_\perp)\rangle^{\rm 2nd}~=~
\label{3.11}\\
&&\hspace{-3mm}
=~
-~{4g^2\over s^2}\!\int^{\infty}_{y_\ast}dz_\ast\!\int_{y_\ast}^{z_\ast}\! dz'_\ast 
~([\infty,z_\ast]_y\langle A^{\rm q}_\bu(z_\ast,y_\perp)[z_\ast,z'_\ast]_yA^{\rm q}_\bu(z'_\ast,y_\perp)[z'_\ast,y_\ast]_y\rangle)^{nm}F_{\bu i}^m(y_\ast,y_\perp)
\nonumber\\
&&\hspace{41mm}
+~ig{\partial\over\partial y_\ast}\!\int^{\infty}_{y_\ast}\!\! dz_\ast \big([\infty,z_\ast]_y
\langle A^{\rm q}_\bu(z_\ast,y_\perp)[z_\ast,y_\ast]_y\big)^{nm}A_i^{m{\rm q}}(y_\ast,y_\perp)\rangle
\nonumber\\
&&\hspace{41mm}
-~{2ig\over s}\!\int^{\infty}_{y_\ast}\!\! dz_\ast \big([\infty,z_\ast]_y\langle A^{\rm q}_\bu(z_\ast,y_\perp)[z_\ast,y_\ast]_y\big)^{nm}\partial_iA_\bu^{m{\rm q}}(y_\ast,y_\perp)\rangle
\nonumber
\end{eqnarray}
As we mentioned above, we are interested in operators up to (collinear) twist one. Looking at the explicit expressions for propagators 
in Appendix A it is easy to see that the only contribution of  twist one comes from 
$\langle A^{\rm q}_\bu(z_\ast,y_\perp) A_i^{{\rm q}}(y_\ast,y_\perp)\rangle$ 
propagator, which is given by Eq. (\ref{A.45}) with
\begin{eqnarray}
&&\hspace{-3mm}
\calg_{\bu i}(z_\ast,y_\ast;p_\perp)~
=~-{2g\over \alpha s}\!\int_{y_\ast}^{z_\ast}\! dz'_\ast~[z_\ast,z'_\ast]F_{\bu i}(z'_\ast)[z'_\ast,y_\ast]
\label{3.12}
\end{eqnarray}
We obtain
\begin{eqnarray}
&&\hspace{-1mm}
\langle[\infty,y_\ast]_y^{nm}F_{\bu i}^m(y_\ast,y_\perp)\rangle^{\rm 2nd}~
=~ g^2N_c\!\int_0^\infty\!{\dhd\alpha\over \alpha}\Big[
-(y_\perp|{1\over p_\perp^2}|y_\perp)[\infty,y_\ast]^{nm}F_{\bu i}^m(y_\ast,y_\perp)
\nonumber\\
&&\hspace{34mm}
+~{i\over\alpha s}\!\int_{y_\ast}^\infty\! dz_\ast~
(y_\perp|e^{-i{p_\perp^2\over\alpha s}(z_\ast-y_\ast)}|y_\perp)
[\infty,z_\ast]^{nm}F_{\bu i}^m(z_\ast,y_\perp)\Big]
\label{3.13}
\end{eqnarray}
where we used  Schwinger's notations
\begin{equation}
(x_\perp|f(p_\perp)|y_\perp)~\equiv~ \int\! \dhd^2p_\perp~e^{i(p,x-y)_\perp}f(p), ~~~~~(x_\perp|p_\perp)~=~e^{i(p,x)_\perp}
\label{schwingerperp}
\end{equation}
For the operator $\calf(\beta_B, y_\perp)$ the Eq.  (\ref{3.13}) gives
\begin{eqnarray}
&&\hspace{-1mm}
\langle\calf_i^n(\beta_B,y_\perp)\rangle^{\rm 2nd}~
=~ -g^2N_c\!\int_0^\infty\!{\dhd\alpha\over \alpha}
(y_\perp|{\alpha\beta_Bs\over p_\perp^2(\alpha\beta_Bs+p_\perp^2)}|y_\perp)
\calf_i^n(\beta_B,y_\perp)
\label{3.15}
\end{eqnarray}
For the complex conjugate amplitude 
\begin{eqnarray}
&&\hspace{-3mm}
\langle \tilF_{\bu i}^m(x_\ast,x_\perp)[x_\ast,\infty]_x^{mn}\rangle^{\rm 2nd}~=~
\label{3.16}\\
&&\hspace{-3mm}
=~
-~{4g^2\over s^2}\!\int^{\infty}_{x_\ast}dz_\ast\!\int_{z_\ast}^{\infty}\! dz'_\ast 
~\tilF_{\bu i}^m(x_\ast,x_\perp)\big([x_\ast,z_\ast]_x\langle\tilA^{\rm q}_\bu(z_\ast,x_\perp)[z_\ast,z'_\ast]_x \tilA^{\rm q}_\bu(z'_\ast,x_\perp)\rangle[z'_\ast,\infty]_x\big)^{mn}
\nonumber\\
&&\hspace{41mm}
-~ig{\partial\over\partial x_\ast}\!\int^{\infty}_{x_\ast}\!\! dz_\ast \langle \tilA_i^{m{\rm q}}(x_\ast,x_\perp)\big([x_\ast,z_\ast]_x \tilA^{\rm q}_\bu(z_\ast,x_\perp)[z_\ast,\infty]_x\big)^{mn}\rangle
\nonumber\\
&&\hspace{41mm}
+~{2i\over s}g\!\int^{\infty}_{x_\ast}\!\! dz_\ast\langle \partial_i\tilA_\bu^{m{\rm q}}\big([x_\ast,z_\ast]_x\tilA^{\rm q}_\bu(z_\ast,x_\perp)[z_\ast,\infty]_x\big)^{mn}\rangle
\nonumber
\end{eqnarray}
Again, the only contribution of twist one comes from $\langle \tilA_i^{{\rm q}}(x_\ast,x_\perp) \tilA^{{\rm q}}_\bu(z_\ast,x_\perp)\rangle$ 
given by Eq. (\ref{A.46}) with
\begin{eqnarray}
&&\hspace{-3mm}
\ticalg_{i\bu}^{kl}(x_\ast,z_\ast;p_\perp)~=
-~{2g\over \alpha s}\!\int_{x_\ast}^{z_\ast}\! dz'_\ast~([x_\ast,z'_\ast]\tilF_{\bu i}(z'_\ast)[z'_\ast,z_\ast])^{kl}
\label{3.17}
\end{eqnarray}
(see Eq. (\ref{A.23})) so the virtual correction in the complex conjugate amplitude is proportional to
\begin{eqnarray}
&&\hspace{-1mm}
\langle\ticalf_i^n(\beta_B,x_\perp)\rangle^{\rm 2nd}~
=~ -g^2N_c\!\int_0^\infty\!{\dhd\alpha\over \alpha}
(x_\perp|{\alpha\beta_Bs\over p_\perp^2(\alpha\beta_Bs+p_\perp^2)}|x_\perp)
\ticalf_i^n(\beta_B,x_\perp)
\label{3.18}
\end{eqnarray}
The total virtual correction is
\begin{eqnarray}
&&\hspace{-1mm}
\langle \tilcaf_i^a(\beta_B, x_\perp) \calf_j^a(\beta_B,y_\perp)\rangle_{\rm virt}~
\nonumber\\
&&\hspace{-1mm}
=~-2g^2N_c\tilcaf_i^a(\beta_B,x_\perp) \calf_j^a(\beta_B,y_\perp)\!\int_0^\infty\!{\dhd\alpha\over \alpha}
\!\int\! \dhd^2p_\perp {\alpha\beta_Bs\over p_\perp^2(\alpha\beta_Bs+p_\perp^2)}
\label{3.19}
\end{eqnarray}
Note that with our rapidity cutoff in $\alpha$ (Eq. (\ref{cutoff})) the contribution (\ref{3.19}) coming from the diagram 
in Fig. \ref{fig:1}b is UV finite. Indeed, regularizing the IR divergence with a small gluon mass $m^2$ we obtain 
\begin{eqnarray}
&&\hspace{-1mm}
\!\int_0^\sigma\!{d\alpha\over \alpha}
\!\int\! d^2p_\perp {\alpha\beta_Bs\over (p_\perp^2+m^2)(\alpha\beta_Bs+p_\perp^2+m^2)}~\simeq~{\pi\over 2}\ln^2{\sigma\beta_Bs+m^2\over m^2}
\label{add1}
\end{eqnarray}
which is finite without any UV regulator (the IR divergence is canceled with the corresponding term in the real correction, see Eq. (\ref{3.20}) below).
This feature -  simultaneous regularization of UV and rapidity divergence - is a consequence of our specific choice of cutoff in rapidity. For a different 
 rapidity cutoff we may have the UV divergence in the remaining integrals which has to be regulated with suitable UV cutoff (for example, see Refs. \cite{cgkh,bneu}). Let us illustrate this
 using the example of the Fig. \ref{fig:1}b diagram calculated above.
Technically, we calculated the loop integral in this diagram
\begin{eqnarray}
&&\hspace{-1mm}
\!\int\! \dhd\alpha\dhd\beta\dhd\beta'\dhd^2p_\perp {-\beta_Bs\over (\beta-i\epsilon)(\beta'+\beta_B-i\epsilon)(\alpha\beta s-p_\perp^2-m^2+i\epsilon)
(\alpha\beta's-p_\perp^2-m^2+i\epsilon)}
\nonumber\\
\label{add2}
\end{eqnarray}
by taking residues in the integrals over Sudakov variables $\beta$ and $\beta'$ and cutting the obtained integral over $\alpha$ from above by the cutoff (\ref{cutoff}).  
Instead, let us take the residue over $\alpha$:
\begin{eqnarray}
&&\hspace{-1mm}
i\beta_B\!\int\! {\dhd^2p_\perp \over m^2+p_\perp^2}\!\int\! \dhd\beta\dhd\beta'
{\theta(\beta)\theta(-\beta')-\theta(-\beta)\theta(\beta')\over (\beta'+\beta_B-i\epsilon)(\beta-i\epsilon)(\beta'-\beta)}
\label{add3}\\
&&\hspace{-1mm}
=~i\beta_B\!\int\! {\dhd^2p_\perp \over m^2+p_\perp^2}\!\int\! {\dhd\beta\dhd\beta'\over \beta'+\beta_B-i\epsilon}
\Big[{\theta(\beta)\theta(-\beta')-\theta(-\beta)\theta(\beta')\over (\beta-i\epsilon)(\beta'-\beta)}
+{\theta(\beta')\over (\beta-i\epsilon)(\beta'-\beta+i\epsilon)}\Big]
\nonumber\\
&&\hspace{-1mm}
=~i\beta_B\!\int\! {\dhd^2p_\perp \over m^2+p_\perp^2}\!\int\! 
{\dhd\beta\dhd\beta'\over \beta'+\beta_B-i\epsilon}{\theta(\beta)\over (\beta-i\epsilon)(\beta'-\beta+i\epsilon)}
~=~\beta_B\!\int\! {\dhd^2p_\perp \over m^2+p_\perp^2}\!\int_0^\infty\! 
{\dhd\beta\over \beta(\beta+\beta_B)}
\nonumber
\end{eqnarray}
which is integral (\ref{add1}) with the replacement of variable $\beta={p_\perp^2\over\alpha s}$.

A conventional way of rewriting this integral in the framework of collinear factorization approach is 
\begin{eqnarray}
&&\hspace{-1mm}
\beta_B\!\int\! {\dhd^2p_\perp \over m^2+p_\perp^2}\!\int_0^\infty\! 
{d\beta\over \beta(\beta+\beta_B)}~=~\!\int\! {\dhd^2p_\perp \over m^2+p_\perp^2}\!\int_0^1 \! {dz\over 1-z}
\label{add4}
\end{eqnarray}
where   $z={\beta_B\over\beta_B+\beta}$
 is  a fraction of momentum $(\beta_B+\beta)p_2$ of ``incoming gluon'' (described by $\calf_i$ in our formalism) carried by the emitted 
``particle''  with fraction $\beta_B p_2$, see the discussion of the DGLAP kernel in the next Section.  Now, if we cut the rapidity
of the emitted gluon by cutoff in fraction of momentum $z$, we would still have the UV divergent expression which must be regulated by 
a suitable UV cutoff.  

\subsection{Evolution kernel in the light-cone limit \label{Sect3.3}}

Summing the product of Lipatov vertices (\ref{3.8}) and (\ref{3.10}) (integrated according to Eq. (\ref{3.1})) and
the virtual correction (\ref{3.19}) we obtain the one-loop evolution kernel in the light-cone approximation
\begin{eqnarray}
&&\hspace{-1mm}
\big(\ticalf_i^a(\beta_B,x_\perp)\calf_j^a(\beta_B,y_\perp)\big)^{\ln\sigma}~
\label{3.20}\\
&&\hspace{-1mm}
=~2g^2N_c\!\int_{\sigma'}^{\sigma}\!{\dhd\alpha\over \alpha}\dhd^2k_\perp~\Big\{e^{i(k,x-y)_\perp}
\ticalf_k^a\big(\beta_B+{k_\perp^2\over\alpha s},x_\perp\big)\calf_l^a\big(\beta_B+{k_\perp^2\over\alpha s},y_\perp\big)
\nonumber\\
&&\hspace{-1mm}
\times~\Big[{\delta_i^k\delta_j^l\over k_\perp^2}-{2\delta_i^k\delta_j^l\over\alpha\beta_Bs+k_\perp^2}
+~{k_\perp^2\delta_i^k\delta_j^l+\delta_j^kk_ik^l+\delta_i^lk_jk^k-\delta_j^lk_ik^k-\delta_i^kk_jk^l-g^{kl}k_ik_j-g_{ij}k^kk^l
\over (\alpha\beta_Bs+k_\perp^2)^2}
\nonumber\\
&&\hspace{22mm}
+~k_\perp^2{2g_{ij}k^kk^l+\delta_i^kk_jk^l+\delta_j^lk_ik^k-\delta_j^kk_ik^l-\delta_i^lk_jk^k\over (\alpha\beta_Bs+k_\perp^2)^3}
-{k_\perp^4g_{ij}k^kk^l\over  (\alpha\beta_Bs+k_\perp^2)^4}\Big]
\nonumber\\
&&\hspace{44mm}
-~ {\alpha\beta_Bs\over k_\perp^2(\alpha\beta_Bs+k_\perp^2)} 
\ticalf_i^a(\beta_B,x_\perp)\calf_j^a(\beta_B,y_\perp)\Big\}^{\ln\sigma'}
\nonumber
\end{eqnarray}
where rapidities of gluons in the operators in the r.h.s. are restricted from above by $\ln\sigma'$.

Let us write down now the evolution equation for gluon TMDs defined by the matrix element (\ref{mael}). 
If we define $\beta_B$ as a fraction of the momentum $p$ of the original hadron we have $\beta_B<1$. 
Moreover, in the production part of the amplitude  we have a kinematical restriction that the sum of $\beta_B$ and the fraction carried by 
emitted gluon $ {k_\perp^2\over\alpha s}$ should be less than one. This leads to the upper cutoff in the $k_\perp$ integral 
$k_\perp^2\leq \alpha(1-\beta_B)s$ and we get the equation
\begin{eqnarray}
&&\hspace{-1mm}
{d\over d\ln\sigma}\langle p|\ticalf_i^a(\beta_B,x_\perp)\calf_j^a(\beta_B,y_\perp)|p\rangle~
\label{3.21}\\
&&\hspace{-1mm}
=~{g^2N_c\over\pi}\!\int\!\dhd^2k_\perp~\Big\{e^{i(k,x-y)_\perp}
\langle p|\ticalf_k^a\big(\beta_B+{k_\perp^2\over\sigma s},x_\perp\big)\calf_l^a\big(\beta_B+{k_\perp^2\over\sigma s},y_\perp\big)
|p\rangle
\nonumber\\
&&\hspace{-1mm}
\times~\Big[{\delta_i^k\delta_j^l\over k_\perp^2}-{2\delta_i^k\delta_j^l\over\sigma\beta_Bs+k_\perp^2}
+~{k_\perp^2\delta_i^k\delta_j^l+\delta_j^kk_ik^l+\delta_i^lk_jk^k-\delta_j^lk_ik^k-\delta_i^kk_jk^l-g^{kl}k_ik_j-g_{ij}k^kk^l
\over (\sigma\beta_Bs+k_\perp^2)^2}
\nonumber\\
&&\hspace{2mm}
+~k_\perp^2{2g_{ij}k^kk^l+\delta_i^kk_jk^l+\delta_j^lk_ik^k-\delta_j^kk_ik^l-\delta_i^lk_jk^k\over (\sigma\beta_Bs+k_\perp^2)^3}
-{k_\perp^4g_{ij}k^kk^l\over  (\sigma\beta_Bs+k_\perp^2)^4}\Big]
\theta\big(1-\beta_B-{k_\perp^2\over \sigma s}\big)
\nonumber\\
&&\hspace{58mm}
-~ {\sigma\beta_Bs\over k_\perp^2(\sigma\beta_Bs+k_\perp^2)} 
\langle p|\ticalf_i^a(\beta_B,x_\perp)\calf_j^a(\beta_B,y_\perp)|p\rangle\Big\}
\nonumber
\end{eqnarray}
(there is obviously no restriction on $k_\perp$ in the virtual diagram).

If the target hadron is unpolarized one can use the parametrization (\ref{mael}) 
\begin{eqnarray}
&&\hspace{-1mm}
\langle p|\ticalf_i^a(\beta_B,z_\perp)\calf_j^a(\beta_B,0_\perp)|p+\xi p_2\rangle^\eta~
\nonumber\\
&&\hspace{2mm}
=~
 2\pi^2\delta(\xi) \beta_Bg^2\Big[-g_{ij}\cald(\beta_B,z_\perp,\eta)
-{1\over m^2}(2\partial_i\partial_j+g_{ij}\partial_\perp^2)\calh(\beta_B,z_\perp,\eta)\Big]
\nonumber\\
&&\hspace{2mm}
=~
 2\pi^2\delta(\xi) \beta_Bg^2\Big[-g_{ij}\cald(\beta_B,z_\perp,\eta)
-{4\over m^2}(2z_iz_j+g_{ij}z_\perp^2)\calh''(\beta_B,z_\perp,\eta)\Big]
\label{3.22}
\end{eqnarray}
where $\eta\equiv\ln\sigma$, $\calh(\beta_B,z_\perp,\eta)\equiv\int\! \dhd^2k_\perp~e^{-i(k,z)_\perp}\calh(\beta_B,k_\perp,\eta)$ and
\\
 $\calh''(\beta_B,z_\perp,\eta)\equiv\big({\partial\over\partial z^2}\big)^2\calh(\beta_B,z_\perp,\eta)$. 
Rewriting Eq. (\ref{3.21}) in terms of variable $\beta\equiv{k_\perp^2\over\sigma s}$ one obtains 
\begin{eqnarray}
&&\hspace{-1mm}
{d\over d\eta}\Big[g_{ij}\alpha_s\cald(\beta_B,z_\perp,\eta)
+{4\over m^2}(2z_iz_j+g_{ij}z_\perp^2)\alpha_s\calh''(\beta_B,z_\perp,\eta)\Big]
\label{eveq1}\\
&&\hspace{1mm}
=~{\alpha_sN_c\over \pi}\!\int_0^{1-\beta_B}\!d\beta~\Big\{
g_{ij}J_0\big(|z_\perp|\sqrt{\sigma\beta s}\big)\alpha_s\cald(\beta_B+\beta,z_\perp,\eta)
\nonumber\\
&&\hspace{33mm}
\times~\Big[{\beta_B+\beta\over \beta\beta_B}-{2\over\beta_B}
+{3\beta\over \beta_B(\beta_B+\beta)}
-{2\beta^2\over \beta_B(\beta_B+\beta)^2}
+{\beta^3\over  \beta_B(\beta_B+\beta)^3}\Big]
\nonumber\\
&&\hspace{1mm}
+~J_2(|z_\perp|\sqrt{\sigma\beta s})\big(2{z_iz_j\over z_\perp^2}+g_{ij}\big)
\alpha_s\cald(\beta_B+\beta,z_\perp,\eta){\beta\over \beta_B(\beta_B+\beta)}
\nonumber\\
&&\hspace{1mm}
+~
{4\over m^2}J_0\big(|z_\perp|\sqrt{\sigma\beta s}\big)(2z_iz_j+g_{ij}z_\perp^2)
\alpha_s\calh''(\beta_B+\beta,z_\perp,\eta)
\Big[{\beta_B+\beta\over \beta_B\beta}-{2\over\beta_B}
+~{\beta\over \beta_B(\beta_B+\beta)}\Big]
\nonumber\\
&&\hspace{1mm}
+~{4g_{ij}\over m^2}
z_\perp^2J_2\big(|z_\perp|\sqrt{\sigma\beta s}\big)
\alpha_s\calh''(\beta_B+\beta,z_\perp,\eta)
\nonumber\\
&&\hspace{33mm}
\times~\Big[
{\beta\over \beta_B(\beta_B+\beta)}
-{2\beta^2\over \beta_B(\beta_B+\beta)^2}
+{\beta^3\over  \beta_B(\beta_B+\beta)^3}\Big]\Big\}
\nonumber\\
&&\hspace{1mm}
-~ {\alpha_sN_c\over \pi}\!\int_0^\infty\!d\beta{\beta_B\over \beta(\beta_B+\beta)} 
\Big[g_{ij}\alpha_s\cald(\beta_B,z_\perp,\eta)
+{4\over m^2}(2z_iz_j+g_{ij}z_\perp^2)\alpha_s\calh''(\beta_B,z_\perp,\eta)\Big]
\nonumber
\end{eqnarray}
where  we used the formula
\begin{equation}
{1\over 2\pi}\!\int\! d\theta e^{i(k,z)_\perp}[2k_ik_j+k_\perp^2g_{ij}]
~=~-J_2(kz)k^2_\perp\Big({2\over z_\perp^2}z_iz_j+g_{ij}\Big)
\label{formula}
\end{equation}
The evolution equation (\ref{eveq1})
 can be rewritten as a system of evolution equations for $\cald$ and $\calh''$ functions
 ($z'\equiv {\beta_B\over\beta+\beta_B}$):
\begin{eqnarray}
&&\hspace{-1mm}
{d\over d\eta}\alpha_s\cald(\beta_B,z_\perp,\eta)
\label{liconefinal}\\
&&\hspace{-1mm}
=~ {\alpha_sN_c\over \pi}\!\int_{\beta_B}^1\!{dz'\over z'}\Big\{
J_0\Big(|z_\perp|\sqrt{\sigma s\beta_B{1-z'\over z'}}\Big)
\Big[\big({1\over 1-z'}\big)_+ +{1\over z'}-2+z'(1-z')\Big]\alpha_s\cald\big({\beta_B\over z'},z_\perp,\eta\big)   
\nonumber\\
&&\hspace{-1mm}
+~  
{4\over m^2}(1-z')z'
z_\perp^2J_2\Big(|z_\perp|\sqrt{\sigma s\beta_B{1-z'\over z'}}\Big)
\alpha_s\calh''({\beta_B\over z'},z_\perp,\eta)\Big\},
\nonumber\\
&&\hspace{-1mm}
{d\over d\eta}\alpha_s\calh''(\beta_B,z_\perp,\eta)
\nonumber\\
&&\hspace{-1mm}
=~ {\alpha_sN_c\over \pi}\!\int_{\beta_B}^1\!{dz'\over z'}\Big\{
J_0\Big(|z_\perp|\sqrt{\sigma s\beta_B{1-z'\over z'}}\Big)
\Big[\big({1\over 1-z'}\big)_+ -1\Big]\alpha_s\calh''\big({\beta_B\over z'},z_\perp,\eta\big)
\nonumber\\
&&\hspace{-1mm}
+~~{m^2\over 4z_\perp^2}{1-z'\over z'}J_2\Big(|z_\perp|\sqrt{\sigma s\beta_B{1-z'\over z'}}\Big)
\alpha_s\cald\big({\beta_B\over z'},z_\perp,\eta\big)\Big\}
\nonumber
\end{eqnarray}
where $\int^1_x dz f(z)g(z)_+=\int^1_x dz f(z)g(z)-\int^1_0 dz f(1)g(z)$. 
The above equation is our final result for the rapidity evolution of gluon TMDs in the near-light-cone case.

It is instructive to check that the evolution equation (\ref{liconefinal}) agrees with the (one-loop) DGLAP kernel. 
If we take the light-cone limit $x_\perp=y_\perp$ ($\Leftrightarrow z_\perp=0$) we get
\begin{equation}
\hspace{-1mm}
{d\over d\eta}\alpha_s\cald(\beta_B,0_\perp,\eta)~=~{\alpha_s\over\pi}N_c
\!\int_{\beta_B}^1\! {dz'\over z'}~
\Big[\big({1\over 1-z'}\big)_+  +{1\over z'}- 2+ z'(1-z')\Big]
\alpha_s\cald\big({\beta_B\over z'},0_\perp,\eta\big)
\label{3.24}
\end{equation}
One immediately recognizes the expression in the square brackets as gluon-gluon DGLAP kernel
(the term ${11\over 12}\delta(1-z')$ is absent since we consider the gluon light-ray operator multiplied by 
 an extra  $\alpha_s$).  
 It should be mentioned, however,  that Eq. (\ref{3.24}) is not a proper DGLAP equation since the latter is formulated for the gluon parton density 
 on the light cone  defined by
\begin{eqnarray}
&&\hspace{-0mm}
d_g(x_B,\ln\mu^2)~=~-{1\over 8\pi^2\alpha_s(p\cdot n)x_B}
\!\int\! du ~e^{-ix_Bu(pn)} \langle p|\tilcaf^a_i(un)\calf^{ai}(0)|p\rangle^{\mu}
\label{3.25}
\end{eqnarray}
where the light-ray gluon operator $F^a_i(un)[un,0]F^{ai}(0)$ is regularized with counterterms at normalization point $\mu^2$
(recall that on the light ray T-product of operators coincide with the usual product).

Comparing Eqs. (\ref{TMD}) and (\ref{3.25}) we see that $d_g(x_B)=\cald(x_B,z_\perp=0)$ modulo different cutoffs: by 
counterterms for $d_g(x_B,\mu^2)$ and by ``brute force'' rapidity cutoff $Y<\eta$ in $D(x_B,z_\perp=0,\eta=\ln\sigma)$.
However, with the leading-log accuracy subtracting the counterterms is equivalent to imposing a  cutoff in transverse 
momenta of the emitted gluons $k_\perp^2<\mu^2$.  If we would calculate the leading-order renorm-group equation for 
the light-ray operator $\ticalf_i^a(\beta_B,x_\perp)\calf_j^a(\beta_B,x_\perp)$ we would cut the integral over $k_\perp^2$ from above by $\mu^2$ and leave the integration
over rapidity ($\alpha$) unrestricted. Thus, we would obtain
\begin{eqnarray}
&&\hspace{-1mm}
\langle p|\ticalf_i^n(\beta_B,x_\perp)\calf^{ni}(\beta_B,x_\perp)|p\rangle^{\mu}~
\label{3.26}\\
&&\hspace{-1mm}
=~{\alpha_s\over\pi}N_c\!\int_0^\infty\!{d\alpha\over\alpha} \!\int_{{\mu'}^2\over\alpha s}^{\mu^2\over\alpha s}\! d\beta~\Big\{
\theta(1-\beta_B-\beta)\Big[{1\over \beta}
-{2\beta_B\over(\beta_B+\beta)^2}
+{\beta_B^2\over(\beta_B+\beta)^3}-{\beta_B^3\over(\beta_B+\beta)^4}\Big]\nonumber\\
&&\hspace{1mm}
\times~
\langle p|\ticalf_i^{n}(\beta_B+\beta,x_\perp)\calf^{ni}(\beta_B+\beta,x_\perp)|p\rangle^{\mu'}
-{\beta_B\over\beta(\beta_B+\beta)}
\langle p| \ticalf_i^n(\beta_B,x_\perp)\calf^{ni}(\beta_B,x_\perp)|p\rangle^{\mu'}\Big\}
\nonumber\\
&&\hspace{-1mm}
=~{\alpha_s\over\pi}N_c\!\int_0^\infty\!d\beta \!\int_{{\mu'}^2\over\beta s}^{\mu^2\over\beta s}\! {d\alpha\over\alpha}~\Big\{
\theta(1-\beta_B-\beta)\Big[{1\over \beta}
-{2\beta_B\over(\beta_B+\beta)^2}
+{\beta_B^2\over(\beta_B+\beta)^3}-{\beta_B^3\over(\beta_B+\beta)^4}\Big]\nonumber\\
&&\hspace{1mm}
\times~
\langle p|\ticalf_i^{n}(\beta_B+\beta,x_\perp)\calf^{ni}(\beta_B+\beta,x_\perp)|p\rangle^{\mu'}
-{\beta_B\over\beta(\beta_B+\beta)}
\langle p| \ticalf_i^n(\beta_B,x_\perp)\calf^{ni}(\beta_B,x_\perp)|p\rangle^{\mu'}\Big\}
\nonumber
\end{eqnarray}
which should be compared to Eq. (\ref{3.20}) with $x_\perp=y_\perp$
\begin{eqnarray}
&&\hspace{-1mm}
\langle p|\ticalf_i^n(\beta_B,x_\perp)\calf^{in}(\beta_B,x_\perp)|p\rangle^{\ln\sigma}~
\nonumber\\
&&\hspace{-1mm}
=~{\alpha_s\over\pi}N_c\!\int_{\sigma'}^{\sigma}\!{d\alpha\over\alpha} \!\int_0^{\infty}\! d\beta~\Big\{
\theta(1-\beta_B-\beta)\Big[{1\over \beta}
-{2\beta_B\over(\beta_B+\beta)^2}
+{\beta_B^2\over(\beta_B+\beta)^3}-{\beta_B^3\over(\beta_B+\beta)^4}\Big]\nonumber\\
&&\hspace{-1mm}
\times~
\langle p|\ticalf_i^{n}(\beta_B+\beta,x_\perp)\calf^{ni}(\beta_B+\beta,x_\perp)|p\rangle^{\ln\sigma'}
-{\beta_B\beta^{-1}\over \beta_B+\beta}
\langle p| \ticalf_i^n(\beta_B,x_\perp)\calf^{in}(\beta_B,x_\perp)|p\rangle^{\ln\sigma'}\Big\}
\nonumber
\end{eqnarray}
In the leading log approximation $\beta\sim\beta_B=x_B$ so one can replace the cutoff 
 ${\mu^2\over\beta s}$ in Eq. (\ref{3.26}) by the cutoff ${\mu^2\over x_B s}=\sigma$ 
and hence $d_g(x_B,\ln\mu^2)=D(x_B,z_\perp=0,\ln\sigma s)$ with the leading-log accuracy.   
The equation (\ref{3.26}) can be rewritten as an evolution equation 
\begin{eqnarray}
&&\hspace{-11mm}
\mu^2{d\over d\mu^2}\langle p|\ticalf_i^n(\beta_B,x_\perp)\calf^{in}(\beta_B,x_\perp)|p\rangle^{\mu}
\label{3.27}\\
&&\hspace{-11mm}
=~{\alpha_s(\mu)\over\pi}N_c\!\int_0^\infty\! d\beta~\Big\{\theta(1-\beta_B-\beta)
\Big[{1\over \beta}
-{2\beta_B\over(\beta_B+\beta)^2}
+{\beta_B^2\over(\beta_B+\beta)^3}
-~{\beta_B^3\over(\beta_B+\beta)^4}\Big]
\nonumber\\
&&\hspace{-11mm}
\times~
\langle p|\ticalf_i^{n}(\beta_B+\beta,x_\perp)\calf^{ni}(\beta_B+\beta,x_\perp)|p\rangle^\mu
-{\beta_B\over\beta(\beta_B+\beta)}
\langle p| \ticalf_i^n(\beta_B,x_\perp)\calf^{in}(\beta_B,x_\perp)|p\rangle^\mu\Big\}
\nonumber
\end{eqnarray}
which can be transformed to the standard DGLAP form \cite{dglap}
\begin{eqnarray}
&&\hspace{-1mm}
\mu^2{d\over d\mu^2}\alpha_s(\mu)d_g(x_B,\ln\mu^2)
\label{3.28}\\
&&\hspace{-1mm}=~{\alpha_s(\mu)\over\pi}N_c
\!\int_{\beta_B}^1\! {dz'\over z'}~\Big[\big({1\over 1-z'}\big)_+ +{1\over z'}- 2+ z'(1-z')\Big]
\alpha_s(\mu)d_g\big({\beta_B\over z'},\ln\mu^2\big)
\nonumber
\end{eqnarray}
There is a subtle point in comparison of our rapidity evolution of light-ray operators to the 
conventional $\mu^2$ evolution described by renorm-group equations: the self-energy
diagrams  are not regulated by our rapidity cutoff
so the $\delta$-function terms in our version of the DGLAP equations are absent. 
\footnote{For Eq (\ref{3.28}) the absence of these terms is accidental, due to an extra $\alpha_s$ in the definition (\ref{TMD}).}
Indeed, in our analysis we do not change the UV treatment of the theory, we just define
the Wilson-line (or light-ray) operators by the requirement that gluons emitted by those 
operators have rapidity cutoff (\ref{cutoff}). The UV divergences in self-energy and other 
internal loop diagrams appearing in higher-order calculations are absorbed in the usual $Z$-factors. 
So, in a way, we will have two evolution equations for our operators: the trivial $\mu^2$ evolution
described by anomalous dimensions of corresponding gluon (or quark) fields and the 
rapidity evolution. Combined together, the two should describe the $Q^2$ evolution 
of DIS structure functions. Presumably, the argument of coupling constant
in LO equation (\ref{3.24}) (which is $\mu^2$ by default) will be replaced by $\sigma\beta_Bs$ 
in accordance with common lore that this argument is determined by characteristic transverse momenta. 
\footnote{Note that while in the usual renorm-group DGLAP the argument of coupling constant is a part of LO equation, 
with our cutoff this argument can be determined only at the NLO level, same as in the case of NLO BK
equation at low $x$ \cite{nlobk}. This is not surprising since we use the rapidity cutoff borrowed from the NLO BK analysis.}
We plan to return to this point in the future NLO analysis.

\section{Evolution kernel in the general case}

In this section we will find the leading-order rapidity evolution of gluon operator (\ref{operator}) 
$$
\big(\ticalf_i^a(\beta_B,x_\perp)\calf_j^a(\beta_B,y_\perp)\big)^{\ln\sigma}
$$
with the rapidity cutoff $Y<\eta=\ln\sigma$ for all emitted gluons. As we mentioned in the Introduction, 
in order to find the evolution kernel we need to integrate over slow gluons  
with $\sigma>\alpha>\sigma'$ and temporarily freeze fast fields with $\alpha<\sigma'$ to be integrated over later. 
To this end we need the one-loop diagrams in the fast background fields with arbitrary transverse momenta. 
In the previous section we have found the evolution kernel in background fields with  transverse momenta 
$l_\perp\ll p_\perp$ where $p_\perp$ is a characteristic momentum of our quantum slow fields. In this section
at first we will find the Lipatov vertex and virtual correction for the case $l_\perp\sim p_\perp $ and then write down 
general formulas which are correct in the whole region of the transverse momentum. 

The key observation is that for transverse momenta of quantum and background field of the same order we can use the shock-wave approximation developed for small-$x$ physics.  To find the evolution kernel we consider 
the operator (\ref{operator}) in the background of external field $A_\bu(x_\ast,x_\perp)$ (the absence of $x_\bu$ in the argument corresponds to $\alpha=0$). Moreover, we assume that the background field $A_\bu(x_\ast,x_\perp)$ has a narrow
support and vanishes outside the $[-\sigma_\ast,\sigma_\ast]$ interval. This is obviously not the most general form of the external field, 
but it turns out that after obtaining the kernel of the evolution equation it is easy to restore the result for any background field by 
insertion of gauge links at $\pm\infty p_1$, see the discussion after Eq. (\ref{master2alt}).

Since the typical $\beta$'s of the external field are $\beta_{\rm fast}\sim {l_\perp^2\over\alpha_{\rm fast}s}$ the support of the shock wave $\sigma_\ast$ is of order of ${1\over \beta_{\rm fast}}\sim{\sigma's\over l_\perp^2}$. This is to be compared to the typical scale of slow fields ${1\over\beta_{\rm slow}}\sim {\alpha s\over p_\perp^2}\gg \sigma_\ast$ 
so we see that the fast background field can be approximated by a narrow shock wave.
In the ``pure'' low-x case $\beta_B=0$ one can
assume that the support of this shock wave is infinitely narrow. As we shall see below, in our case of arbitrary $\beta_B$ we need to look inside the shock wave so we will separate all integrals over longitudinal distances $z_\ast$ in 
parts ``inside the shock wave''
$|z_\ast|<\sigma_\ast$ and ``outside the shock wave" $|z_\ast|>\sigma_\ast$, calculate them separately and
check that the sum of ``inside'' and ``outside'' contributions does not depend on $\sigma_\ast$ with our accuracy.

\subsection{Production part of the evolution kernel}

\begin{figure}[htb]
\begin{center}
\includegraphics[width=104mm]{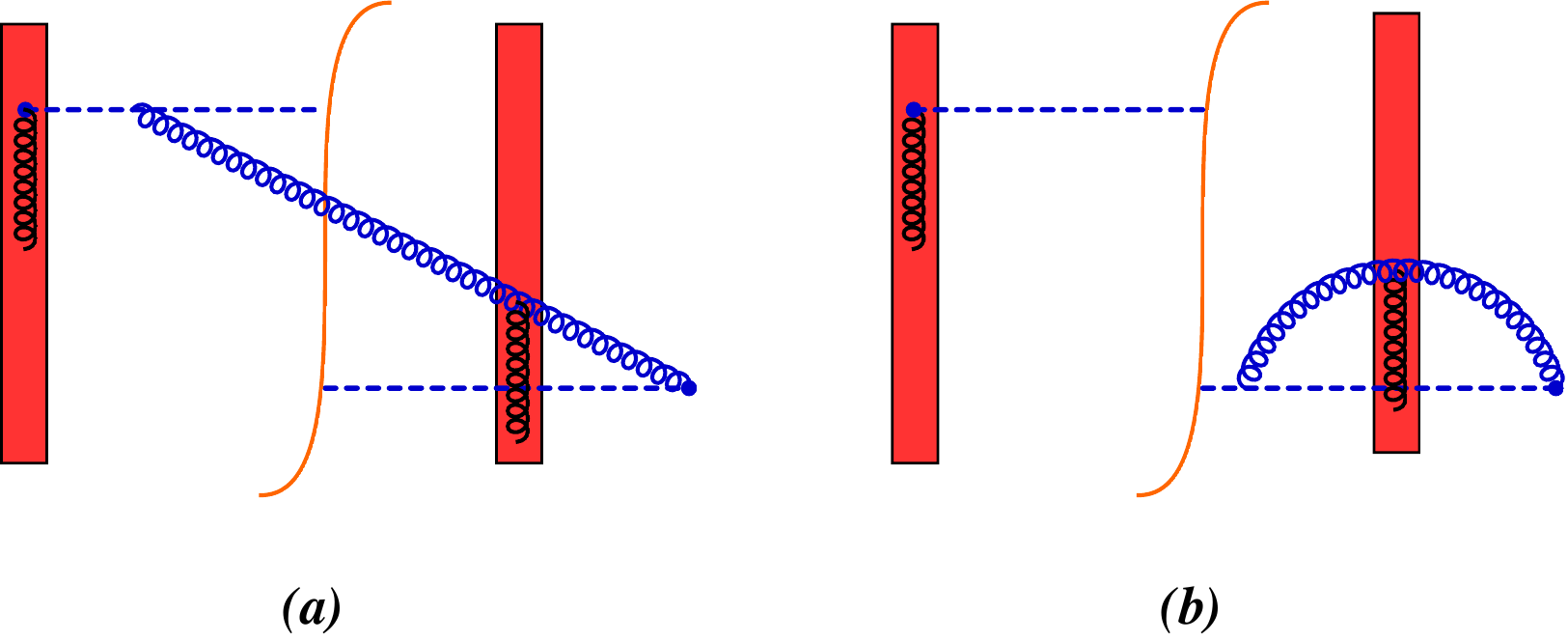}
\end{center}
\caption{Typical diagrams for production (a) and virtual (b) contributions to the evolution kernel. The shaded area denotes shock
wave of background fast fields. \label{fig:3}}
\end{figure}

In the leading order there is only one extra gluon and we get the typical diagrams of Fig. \ref{fig:3} type. 
The production part of the kernel can be obtained as a square of a Lipatov vertex - the amplitude of the emission of a real gluon by the operator $\calf^a_i$ (see Eq. (\ref{3.1}))
\begin{eqnarray}
&&\hspace{-8mm}
\langle \tilcaf^{a}_i(\beta_B, x_\perp) \calf_j^a(\beta_B, y_\perp)\rangle^{\ln\sigma}~
=~-\!\int_{\sigma'}^{\sigma}\!{\dhd\alpha\over 2\alpha}\dhd^2k_\perp
~\big(\tiL^{ba;\mu}_i(k,x_\perp,\beta_B)
L^{ab}_{\mu j}(k,y_\perp,\beta_B)\big)^{\ln\sigma'}
\label{4.1}
\end{eqnarray}
where the Lipatov vertices of gluon emission are defined as  
\begin{eqnarray}
&&\hspace{-1mm}
L^{ab}_{\mu i}(k,y_\perp,\beta_B)~=~i\lim_{k^2\rightarrow 0}k^2\langle A^a_\mu(k)\calf^b_i(\beta_B,y_\perp)\rangle
\nonumber\\
&&\hspace{-1mm}
\tiL^{ba}_{i\mu}(k,x_\perp,\beta_B)
=-i\!\lim_{k^2\rightarrow 0}k^2\langle \tilcaf^{b}_i(\beta_B,x_\perp)\tilA_\mu^{a}(k) \rangle
\label{4.2}
\end{eqnarray}
(cf. Eqs. (\ref{3.2}) and (\ref{3.10})). Hereafter $\langle\calo\rangle$ means the average of operator $\calo$ in the shock-wave background.

\subsection{Lipatov vertex of gluon emission in the shock wave background}

\begin{figure}[htb]
\begin{center}
\includegraphics[width=141mm]{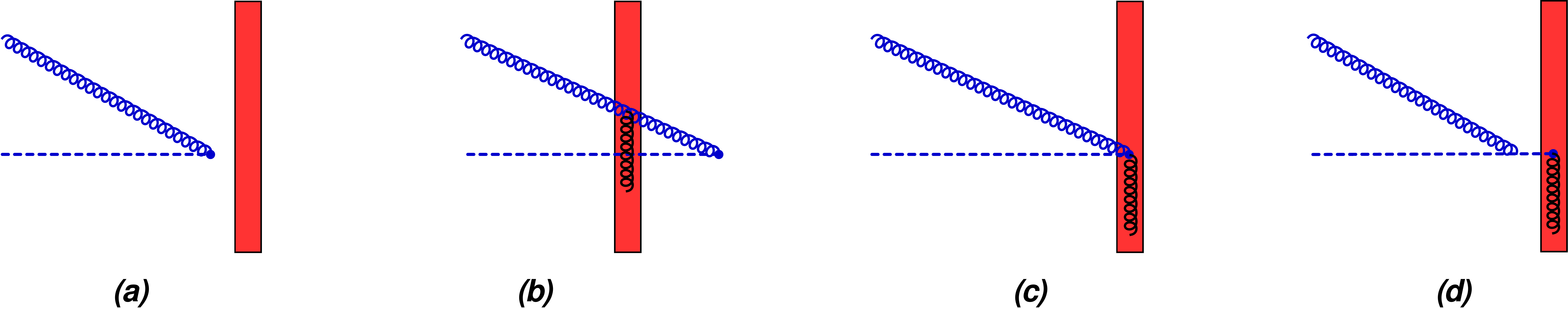}
\end{center}
\caption{Lipatov vertex of gluon emission. \label{fig:4}}
\end{figure}

As we discussed above, we calculate the diagrams in the background of a shock wave of 
width $\sim{\sigma' s\over l_\perp^2}$ 
where $l_\perp$ is the characteristic transverse momentum of the external shock-wave field. Note that the factor in the exponent
in the definition of $\calf(\beta_B)$ is $\sim \beta_B{\sigma' s\over l_\perp^2}$ which is not necessarily small at various $\beta_B$ and
$l_\perp^2$ and therefore we need to take into account the diagram in Fig. \ref{fig:4}c with emission point inside the shock wave. 
We will do this in a following way: we assume that all of the shock wave is contained within $\sigma_\ast>z_\ast>-\sigma_\ast$,
calculate diagrams in  Fig. \ref{fig:4}a-d and check that the dependence on $\sigma_\ast$ cancels in the
final result for the sum of these diagrams. 
 
We start the calculation with the expansion of the gluon fields in $\calf(\beta_B, z_\perp)$ in the first order in slow ``quantum'' field:
\begin{eqnarray}
&&\hspace{-1mm}
\calf_i^n(\beta_B, y_\perp)~\stackrel{1st}{=}~\!g\int\! d{2\over s}y_\ast~e^{i\beta_By_\ast} \Big[
-i{s\over 2}\beta_B[\infty,y_\ast]_y^{nm}A_i^{{\rm q}m}(y_\ast,y_\perp)
\nonumber\\
&&\hspace{11mm}
-~[\infty,y_\ast]_y^{nm}\partial_iA_\bu^{{\rm q}m}(y_\ast,y_\perp)
+{2ig\over s}\!\int^{\infty}_{y_\ast}\! dz_\ast ~[\infty,z_\ast]_yA^{\rm q}_\bu(z_\ast,y_\perp)[z_\ast,y_\ast]_y^{nm}F_{\bu i}^m(y_\ast,y_\perp)
\Big]
\nonumber\\
\label{fla}
\end{eqnarray}
where the gauge links and $F_{\bu i}^m$ are made of fast ``external'' fields. 
The corresponding vertex of gluon emission is given by
\begin{eqnarray}
&&\hspace{-5mm}
\lim_{k^2\rightarrow 0}k^2\langle A^{a{\rm q}}_\mu(k)\calf_i^n(\beta_B, y_\perp)^{\rm 1st}\rangle~
=\lim_{k^2\rightarrow 0}k^2g\!\int\! dy_\ast~e^{i\beta_By_\ast} 
\langle A^{a{\rm q}}_\mu(k)\big\{
-i\beta_B[\infty,y_\ast]_y^{nm}A_i^{m{\rm q}}(y_\ast,y_\perp)
\nonumber\\
&&\hspace{-5mm}
-~{2\over s}[\infty,y_\ast]_y^{nm}\partial_iA_\bu^{m{\rm q}}(y_\ast,y_\perp)
+{4ig\over s^2}\!\int^{\infty}_{y_\ast}\! dz_\ast ~
\big([\infty,z_\ast]A_\bu^{\rm q}(z_\ast,y_\perp)[z_\ast,y_\ast]\big)^{nm}F_{\bu i}^m(y_\ast,y_\perp)
\big\}\rangle
\label{4.4}
\end{eqnarray}
The diagrams in  
Fig. \ref{fig:4}a, \ref{fig:4}b, and \ref{fig:4}(c-d) correspond to different regions of integration over $y_\ast$ in Eq. (\ref{fla}):
$y_\ast>\sigma_\ast$, $-\sigma_\ast >y_\ast$, and $\sigma_\ast >y_\ast>-\sigma_\ast$, respectively.

The trivial calculation of Fig. \ref{fig:4}a contribution yields
\begin{eqnarray}
&&\hspace{-1mm}
i\lim_{k^2\rightarrow 0}k^2\langle A^{a q}_\mu(k)\calf_i^n(\beta_B, y_\perp)\rangle_{\rm Fig.\ref{fig:4}a}~
\label{4.5}\\
&&\hspace{-1mm}
=~
ig\!\int_{\sigma_\ast}^\infty\! dy_\ast~e^{i\beta_By_\ast} [\infty,y_\ast]^{nm}
\lim_{k^2\rightarrow 0}k^2\langle A^a_\mu(k)\big[
-i\beta_BA_i^m(y_\ast,y_\perp)-{2\over s}\partial_iA_\bu^m(y_\ast,y_\perp)\big]\rangle
\nonumber\\
&&\hspace{-1mm}
=~g\delta^{an}{g_{\mu i}\alpha \beta_Bs+2\alpha k_ip_{1\mu}\over\alpha \beta_Bs+k_\perp^2}~
e^{i(\beta_B +{k_\perp^2\over\alpha s})\sigma_\ast-i(k,y)_\perp}
\nonumber
\end{eqnarray}

\subsubsection{Diagram in Fig. \ref{fig:4}b }
Next step is the calculation of Fig.\ \ref{fig:4}b contribution. Using the vertex of gluon emission from the shock wave (\ref{B.31}) 
one obtains
\begin{eqnarray}
&&\hspace{-3mm}
ig\lim_{k^2\rightarrow 0}\!\int_{-\infty}^{-\sigma_\ast}\! dy_\ast~e^{i\beta_By_\ast} 
k^2\langle A^{a{\rm q}}_\mu(k)[\infty,y_\ast]_y^{nm}\big\{
-i\beta_BA_i^{m{\rm q}}(y_\ast,y_\perp)
-~{2\over s}\partial_iA_\bu^{m{\rm q}}(y_\ast,y_\perp)
\big\}\rangle
\nonumber\\
&&\hspace{-3mm}
=~\!g\int_{-\infty}^{-\sigma_\ast}\! dy_\ast~e^{i\beta_By_\ast} \!\int\! d^2z_\perp~e^{-i(k,z)_\perp}
\big\{-i\beta_B\calo_{\mu i}(\infty,y_\ast,z_\perp;k)(z_\perp|e^{i{p_\perp^2\over\alpha s}y_\ast}|y_\perp)
\nonumber\\
&&\hspace{45mm}
-~{2i\over s}\calo_{\mu \bu}(\infty,y_\ast,z_\perp;k)(z_\perp|p_ie^{i{p_\perp^2\over\alpha s}y_\ast}|y_\perp)
\big\}^{am}(U_y^\dagger)^{mn}
\label{4.6}
\end{eqnarray}
where $\calo$ is given by Eqs. (\ref{B.32}):
\begin{eqnarray}
&&\hspace{-5mm}
\calo^{ab}_{\mu\nu}(\infty,y_\ast,z_\perp;k)~\stackrel{y_\ast<-\sigma_\ast}{=}~g_{\mu\nu}U^{ab}_z
\nonumber\\
&&\hspace{-5mm}
+~g\!\int_{-\infty}^{\infty}\!\!\!dz_\ast~\big([\infty,z_\ast]_z\big[-{2iz_\ast\over\alpha s^2} g_{\mu\nu}(2k^j-iD^j)
+{4\over \alpha s^2}(\delta_\mu^jp_{2\nu}-\delta_\nu^jp_{2\mu})\big]F_{\bu j}(z_\ast,z_\perp)[z_\ast,-\infty]_z\big)^{ab}
\nonumber\\
&&\hspace{-5mm}
+~{4g\over\alpha^2 s^3}\!\int_{-\infty}^{\infty}\!\! \!dz_\ast~\big([\infty,z_\ast]_z 
\big\{-ip_{2\mu}p_{2\nu}D^jF_{\bu j}(z_\ast,z_\perp)[z_\ast,-\infty]_z
\nonumber\\
&&\hspace{11mm}
+~g\!\int_{-\infty}^{z_\ast}\!\!dz'_\ast~\big[2i\alpha g_{\mu\nu}z'_\ast-{4\over s}p_{2\mu}p_{2\nu}\big]
~F_{\bu j}(z_\ast,z_\perp) [z_\ast,z'_\ast]_zF_\bu^{~j}(z'_\ast,z_\perp)[z'_\ast,-\infty]_z
\big\}\big)^{ab}
\nonumber\\
&&\hspace{-5mm} 
+~{g^2\over \alpha s^2}
\Big\{\!\int_{-\infty}^{\infty}\! \! dz_\ast\!\int_{-\infty}^{z_\ast}\! \! dz'_\ast
\bsi(z_\ast,z_\perp)[z_\ast,\infty]_zt^aU_zt^b
[-\infty,z'_\ast]_z\gamma^\perp_\mu\!\not\! p_1\gamma^\perp_\nu\psi(z'_\ast,z_\perp)
\nonumber\\
\hspace{-5mm} 
&&\hspace{11mm}-~\!\int_{-\infty}^{\infty}\! \! dz_\ast\!\int_{z_\ast}^{\infty}\! \! dz'_\ast
\bsi(z_\ast,z_\perp)\gamma^\perp_\nu\!\not\! p_1\gamma^\perp_\mu[z_\ast,-\infty]_z t^bU^\dagger_zt^a[\infty,z'_\ast]_z
\psi(z'_\ast,z_\perp)\Big\}
\label{4.7}
\end{eqnarray}
where we replaced $y_\ast$ by $-\infty$ since we assumed that there is no gauge field outside the $[-\sigma_\ast,\sigma_\ast]$ interval. 

Let us compare relative size of terms in the r.h.s. of this equation. The leading $g_{\mu\nu}$ term is $\sim U_z\sim 1$ and 
it is clear that all other $g_{\mu\nu}$ terms are small. Indeed, the first term in the second line 
is $\sim {g\over\alpha s^2}\!\int\! dz_\ast~z_\ast(2k^j-iD^j)F_{\bu j}\sim {\sigma_\ast\over\alpha s} k^j\partial_jU\sim  {k_\perp^2\over\alpha s}\sigma_\ast\ll1$ since 
the width of the shock wave is $\sim {s\sigma'\over k_\perp^2}$ and $\alpha\gg\sigma'$ (recall that in this Section $l_\perp\sim k_\perp$). Similarly, 
the first term in the fourth line is 
$\sim {g^2\over\alpha s^3}\!\int\! dz_\ast dz'_\ast~z'_\ast F_{\bu j}(z_\ast)F_\bu^{~j}(z'_\ast)\sim {\sigma_\ast\over\alpha s}\partial^jU\partial_jU\sim \sigma_\ast {k_\perp^2\over\alpha s}\ll1$. 

Next, let us find out the relative size of quark terms in Eq. (\ref{4.7}).  The ``power counting'' for external
quark fields in comparison to gluon ones is \\
${g^2\over s}\!\int\! dz_\ast \bsi\!\not\! p_1\psi(z_\ast)~\sim~{g\over s}\!\int\! dz_\ast D^iF_{\bu i}(z_\ast)~\sim k_\perp^2U\sim k_\perp^2$
and each extra integration inside the shock wave brings extra $\sigma_\ast$. Thus, the two last terms in Eq. (\ref{4.7})
are $\sim g^\perp_{\mu\nu}{\sigma_\ast\over\alpha s}k_\perp^2\ll 1$
\footnote{Note, however, that the quark term 
$\sim {g\over\alpha^2 s^3}p_{2\mu}p_{2\nu}\!\int\! dz_\ast D^jF_{\bu j}(z_\ast)\sim p_{2\mu}p_{2\nu}{k_\perp^2\over\alpha^2s^2}U$ is of the same order 
of magnitude as the gluon term $\sim  \frac{g^2}{\alpha^2 s^4}p_{2\mu}p_{2\nu}\!\int\! dz_\ast dz'_\ast F_{\bu j}(z_\ast)F_\bu^{~j}(z'_\ast)$.}. 

After omitting small terms the expression (\ref{4.7}) reduces to
\begin{eqnarray}
&&\hspace{-1mm}
\calo_{\mu\nu}(\infty,-\infty,z_\perp;k)=~g_{\mu\nu}U_z
+~{4g\over \alpha s^2}(\delta_\mu^jp_{2\nu}-\delta_\nu^jp_{2\mu})\!\int_{-\infty}^{\infty}\!\!\!dz_\ast~[\infty,z_\ast]_zF_{\bu j}(z_\ast,z_\perp)[z_\ast,-\infty]_z
\nonumber\\
&&\hspace{-1mm}
+~{4g\over\alpha^2 s^3}p_{2\mu}p_{2\nu}\!\int_{-\infty}^{\infty}\!\! \!dz_\ast~[\infty,z_\ast]_z 
\Big\{-iD^jF_{\bu j}(z_\ast,z_\perp)[z_\ast,-\infty]_z
\nonumber\\
&&\hspace{51mm}
-~{4g\over s}\!\int_{-\infty}^{z_\ast}\!dz'_\ast
~F_{\bu j}(z_\ast,z_\perp) [z_\ast,z'_\ast]_zF_\bu^{~j}(z'_\ast,z_\perp)[z'_\ast,-\infty]_z
\Big\}
\nonumber\\
&&\hspace{51mm}
=~g_{\mu\nu}U_z+{2i\over\alpha s}(\delta_\mu^jp_{2\nu}-\delta_\nu^jp_{2\mu})\partial_jU_z-{2p_{2\mu}p_{2\nu}\over\alpha^2s^2}\partial_\perp^2U_z
\label{4.8}
\end{eqnarray}
where we used the formula
\begin{eqnarray}
&&\hspace{-5mm}
\partial_\perp^2U_z~=~\!g\int_{-\infty}^\infty\! dz_\ast~[\infty,z_\ast]_z
\label{4.9}\\
&&\hspace{-5mm}
\times~\Big({2i\over s} D^jF_{\bu j}(z_\ast,z_\perp)[z_\ast,-\infty]_z+{8g\over s^2}\!\int_{-\infty}^{z_\ast}\!dz'_\ast
~F_{\bu j}(z_\ast,z_\perp) [z_\ast,z'_\ast]_zF_\bu^{~j}(z'_\ast,z_\perp)[z'_\ast,-\infty]_z
\Big)
\nonumber
\end{eqnarray}
Using Eq. (\ref{4.8}) one obtains for the r.h.s. of Eq. (\ref{4.6})
\begin{eqnarray}
&&\hspace{-3mm}
i\lim_{k^2\rightarrow 0}k^2\langle A^{a q}_\mu(k)\mathcal{F}_i^n(\beta_B, y_\perp)\rangle_{\rm Fig.\ref{fig:4}b}~
\label{4.10}\\
&&\hspace{-3mm}
=~ \!g\int\! d^2z_\perp~e^{-i(k,z)_\perp}
\big\{-\calo_{\mu i}(\infty,-\infty,z_\perp;k)(z_\perp|{\alpha\beta_Bs\over\alpha\beta_Bs+p_\perp^2}
e^{-i(\beta_B+{p_\perp^2\over\alpha s})\sigma_\ast}|y_\perp)
\nonumber\\
&&\hspace{32mm}
-~{2\over s}\calo_{\mu \bu}(\infty,-\infty,z_\perp;k)(z_\perp|{p_i\alpha s\over\alpha\beta_Bs+p_\perp^2}e^{-i(\beta_B+{p_\perp^2\over\alpha s})\sigma_\ast}|y_\perp)
\big\}^{am}(U_y^\dagger)^{mn}
\nonumber\\
&&\hspace{-3mm}
=~-ge^{-i\beta_B\sigma_\ast}(k_\perp|\big[g_{\mu i}U-{2ip_{2\mu}\over\alpha s}\partial_iU
\nonumber\\
&&\hspace{32mm}
+~{2\over\beta_Bs}\big(p_{1\mu}U+{i\over\alpha}\partial^\perp_\mu U-{p_{2\mu}\over\alpha^2 s}\partial_\perp^2 U\big)p_i\big]{\alpha\beta_Bs\over\alpha\beta_Bs+p_\perp^2}U^\dagger|y_\perp)^{an}
\nonumber
\end{eqnarray}
where we used the fact that ${p_\perp^2\over \alpha s}\sigma_\ast\ll 1$ when all the transverse momenta are of the same order.

\subsubsection{Diagrams in Fig. \ref{fig:4} c,d}
Next step is the calculation of Fig.\ref{fig:4} c,d contributions. Using the vertex of gluon emission from the shock wave (\ref{A.47}) 
and Eqs. (\ref{B.6}), (\ref{B.7}) one obtains
\begin{eqnarray}
&&\hspace{-3mm}
\lim_{k^2\rightarrow 0}k^2ig\!\int_{-\sigma_\ast}^{\sigma_\ast}\! dy_\ast~e^{i\beta_By_\ast} 
\langle A^{a{\rm q}}_\mu(k)\big\{
-i\beta_B[\infty,y_\ast]_y^{nm}A_i^{m{\rm q}}(y_\ast,y_\perp)
\label{4.11}\\
&&\hspace{3mm}
-~{2\over s}[\infty,y_\ast]_y^{nm}\partial_iA_\bu^{m{\rm q}}(y_\ast,y_\perp)
+{4ig\over s^2}\!\int^{\infty}_{y_\ast}\! dz_\ast ~
\big([\infty,z_\ast]_yA_\bu^{\rm q}(z_\ast,y_\perp)[z_\ast,y_\ast]_y\big)^{nm}F_{\bu i}^m(y_\ast,y_\perp)
\big\}\rangle
\nonumber\\
&&\hspace{-3mm}
=~\!g\int_{-\sigma_\ast}^{\sigma_\ast}\! dy_\ast~\big\{e^{i(\beta_B+{k_\perp^2\over\alpha s})y_\ast} 
\big[-i\beta_B\calo_{\mu i}^{am}(\infty,y_\ast,y_\perp;k)
-~{2\over s}\big(ik_i+{\partial\over\partial y^i}\big)\calo_{\mu \bu}^{am}(\infty,y_\ast,y_\perp;k)
\big]
\nonumber\\
&&\hspace{3mm}
+~{4ig\over s^2}
\!\int_{y_\ast}^\infty\! dz_\ast ~e^{i\beta_By_\ast+i{k_\perp^2\over\alpha s}z_\ast} 
\big(\calo_{\mu\bu}(\infty,z_\ast,y_\perp;k)
[z_\ast,y_\ast]_yF_{\bu i}(y_\ast,y_\perp)\big)^{am}\big\}[y_\ast,\infty]_y^{mn}e^{-i(k,y)_\perp}\nonumber
\nonumber\\
&&\hspace{-3mm}
=ge^{-i(k,y)_\perp}~\!\int_{-\sigma_\ast}^{\sigma_\ast}\! dy_\ast~\Big[e^{i\beta_By_\ast} 
\big(-i\{\beta_B\calo_{\mu i}(\infty,y_\ast,y_\perp;k)
+~{2\over s}k_i\calo_{\mu \bu}(\infty,y_\ast,y_\perp;k)\}[y_\ast,\infty]_y
\nonumber\\
&&\hspace{-3mm}
-~{2\over s}{\partial\over\partial y^i}\big(\calo_{\mu \bu}(\infty,y_\ast,y_\perp;k)[y_\ast,\infty]_y\big)
+~{4ig\over s^2}\!\int_{y_\ast}^\infty\! \!dz_\ast ~\calo_{\mu\bu}(\infty,y_\ast,y_\perp;k)[y_\ast,z_\ast]_yF_{\bu i}(z_\ast,y_\perp)
\nonumber\\
&&\hspace{-3mm}
\times~[z_\ast,\infty]_y\big)
~+~{4ig\over s^2}\!\int_{y_\ast}^\infty\! dz_\ast~~e^{i\beta_By_\ast+i{k_\perp^2\over\alpha s}z_\ast}  \calo_{\mu\bu}(\infty,z_\ast,y_\perp;k)
[z_\ast,y_\ast]_yF_{\bu i}(y_\ast,y_\perp)[y_\ast,\infty]_y\Big]^{an}
\nonumber
\end{eqnarray}
where $\calo_{\mu\nu}~=~\calg_{\mu \nu}+\calq_{\mu \nu}+{\bar\calq}_{\mu \nu}$ and $\calg$, $\calq$ and ${\bar\calq}$ are given by Eqs. (\ref{B.6}) and (\ref{B.7}). 
As we mentioned above, the contributions with extra $(z-\sigma)_\ast$ are  small and 
so are the quark terms (except term $\sim D^jF_{\bu j}$). So, we   have $\calq_{\mu i}={\bar\calq}_{\mu i}=0$ and 
\begin{eqnarray}
&&\hspace{-1mm}
\calo_{\mu i}(\infty,y_\ast,y_\perp;k)[y_\ast,\infty]_y~=~g_{\mu i}
-{4gp_{2\mu}\over\alpha s^2}\!\int_{y_\ast}^{\infty}\!\!\!dz_\ast~[\infty,z_\ast]_yF_{\bu i}(z_\ast,y_\perp)[z_\ast,\infty]_y,
\nonumber\\
&&\hspace{-1mm}
\calo_{\mu \bu}(\infty,y_\ast,y_\perp;k)[y_\ast,\infty]_y~=~p_{1\mu}+g\!\int_{y_\ast}^\infty\! dz_\ast [\infty,z_\ast]_y\big\{{2\over\alpha s}F_{\bu\mu}(z_\ast,y_\perp)[z_\ast,\infty]_y
\label{4.12}\\
&&\hspace{-1mm}
-~{2ip_{2\mu}\over\alpha^2 s^2}D^jF_{\bu j}(z_\ast,y_\perp)[z_\ast,\infty]_y -{8gp_{2\mu}\over\alpha^2s^3}\!\int_{y_\ast}^{z_\ast}\! dz'_\ast F_{\bu j}(z_\ast,y_\perp)[z_\ast,z'_\ast]_yF_\bu^{~j}(z'_\ast,y_\perp)[z'_\ast,\infty]_y\big\}
\nonumber
\end{eqnarray}
After some algebra the r.h.s. of Eq. (\ref{4.11}) reduces to
\begin{eqnarray}
&&\hspace{-1mm}
i\lim_{k^2\rightarrow 0}k^2\langle A^{a q}_\mu(k)\mathcal{F}_i^n(\beta_B, y_\perp)\rangle_{\rm Fig.\ref{fig:4}c+d}~
\nonumber\\
&&\hspace{-1mm}
=-~{g_{\mu i}\alpha\beta_Bs+2\alpha k_ip_{1\mu}\over \alpha\beta_Bs+k_\perp^2}
\big[e^{i(\beta_B+{k_\perp^2\over\alpha s})\sigma_\ast}-e^{-i(\beta_B+{k_\perp^2\over\alpha s})\sigma_\ast}\big]e^{-i(k,y)_\perp}g\delta^{an}
\nonumber\\
&&\hspace{-1mm}
+~ge^{-i(k,y)_\perp}\Big\{{2\over\alpha\beta_Bs}\big(\alpha p_{1\mu}\delta_i^j
+\beta_Bp_{2\mu}\delta_i^j-k_i\delta_\mu^j\big)\big[\calf_j(\beta_B, y_\perp)-i\partial_jU_yU_y^\dagger e^{-i\beta_B\sigma_\ast} \big]
\nonumber\\
&&\hspace{-1mm}
-~{2\alpha p_{1\mu}\over k_\perp^2}\calf_i(\beta_B, y_\perp)+{2p_{2\mu}k_i\over\alpha^2 s^2\beta_B}[{\cal V}(\beta_B,y_\perp)
-e^{-i\beta_B\sigma_\ast}\partial_\perp^2 U_yU_y^\dagger]
\nonumber\\
&&\hspace{-1mm}
+~{2i\over\alpha\beta_Bs}\partial^y_i
\big\{\calf_\mu(y_\perp,\beta_B)-i\partial_\mu U_yU_y^\dagger e^{-i\beta_B\sigma_\ast}
-{p_{2\mu}\over\alpha s}[{\cal V}(\beta_B,y_\perp)-e^{-i\beta_B\sigma_\ast}\partial_\perp^2 U_yU_y^\dagger]\big\}
\nonumber\\
&&\hspace{-1mm}
+~~{4g\over \beta_Bs^2}
\!\int\! dz_\ast dz'_\ast~e^{i\beta_B {\rm min}(z_\ast,z'_\ast)}
\nonumber\\
&&\hspace{-1mm}
\times~[\infty,z_\ast]\big\{{2\over\alpha s}F_{\bu\mu}(z_\ast,y_\perp)[z_\ast,z'_\ast]_y
-~{2ip_{2\mu}\over\alpha^2 s^2}D^jF_{\bu j}(z_\ast,y_\perp)[z_\ast,z'_\ast]_y\big\}
F_{\bu i}(z'_\ast,y_\perp)[z'_\ast,\infty]_y
\nonumber\\
&&\hspace{-1mm}
-~{32g^2p_{2\mu}\over\alpha^2s^5\beta_B}\!\int\! dz_\ast dz'_\ast dz''_\ast
~\theta(z_\ast-z''_\ast)e^{i\beta_B{\rm min}(z'_\ast,z''_\ast)}
[\infty,z_\ast]_yF_{\bu j}(z_\ast,y_\perp)[z_\ast,z''_\ast]_yF_\bu^{~j}(z''_\ast,y_\perp)
\nonumber\\
&&\hspace{-1mm}
\times~
[z''_\ast,z'_\ast]_y
F_{\bu i}(z'_\ast,y_\perp)[z'_\ast,\infty]_y
~-~{2\over\alpha\beta_Bs}e^{-i\beta_B\sigma_\ast}
(\partial_\mu U_y\partial_iU_y^\dagger+{ip_{2\mu}\over\alpha s}\partial_\perp^2U_y\partial_i U_y^\dagger)\Big\}^{an}
\label{4.13}
\end{eqnarray}
where
\begin{eqnarray}
&&\hspace{-1mm}
{\cal V}(\beta_B,y_\perp)~\equiv~g\!\int_{-\infty}^\infty\! dz_\ast~e^{i\beta_Bz_\ast}
\Big({2i\over s}[\infty,z_\ast]_yD^jF_{\bu j}(z_\ast,y_\perp)
\nonumber\\
&&\hspace{18mm}
+~{8g\over s^2}\!\int_{z_\ast}^\infty\!dz'_\ast~
[\infty,z'_\ast]_yF_{\bu j}(z'_\ast,y_\perp) [z'_\ast,z_\ast]_yF_\bu^{~j}(z_\ast,y_\perp)\Big)[z_\ast,\infty]_y
\label{4.14}
\end{eqnarray}

\subsubsection{Lipatov vertex}
The sum of Eqs. (\ref{4.5}), (\ref{4.10}), and (\ref{4.13}) gives the Lipatov vertex of gluon emission in the form
\begin{eqnarray}
&&\hspace{-1mm}
L^{ab}_{\mu i}(k,y_\perp,\beta_B)~=~
g\delta^{ab}{g_{\mu i}\alpha \beta_Bs+2\alpha k_ip_{1\mu}\over\alpha \beta_Bs+k_\perp^2}~
e^{-i\beta_B \sigma_\ast-i(k,y)_\perp}
\nonumber\\
&&\hspace{-3mm}
-ge^{-i\beta_B\sigma_\ast}(k_\perp|\big[g_{\mu i}U-{2ip_{2\mu}\over\alpha s}\partial_iU
+\big(p_{1\mu}U+{i\over\alpha}\partial^\perp_\mu U-{p_{2\mu}\over\alpha^2 s}\partial_\perp^2 U\big)
{2p_i\over\beta_Bs}\big]{\alpha\beta_Bs\over\alpha\beta_Bs+p_\perp^2}U^\dagger|y_\perp)^{ab}
\nonumber\\
&&\hspace{-1mm}
+~ge^{-i(k,y)_\perp}\Big\{{2\over\alpha\beta_Bs}\big(\alpha p_{1\mu}\delta_i^j+\beta_Bp_{2\mu}\delta_i^j-k_i\delta_\mu^j\big)\big[\calf_j(\beta_B, y_\perp)-i\partial_jU_yU_y^\dagger e^{-i\beta_B\sigma_\ast} \big]
\nonumber\\
&&\hspace{-1mm}
-~{2\alpha p_{1\mu}\over k_\perp^2}\calf_i(\beta_B, y_\perp)+{2p_{2\mu}k_i\over\alpha^2 s^2\beta_B}[{\cal V}(\beta_B,y_\perp)-\partial_\perp^2 U_yU_y^\dagger e^{-i\beta_B\sigma_\ast}]
\nonumber\\
&&\hspace{-1mm}
+~{2i\over\alpha\beta_Bs}\partial^y_i
\big\{\calf_\mu(\beta_B, y_\perp)-i\partial_\mu U_yU_y^\dagger e^{-i\beta_B\sigma_\ast}
-{p_{2\mu}\over\alpha s}[{\cal V}(\beta_B,y_\perp)-e^{-i\beta_B\sigma_\ast}\partial_\perp^2 U_yU_y^\dagger]\big\}
\nonumber\\
&&\hspace{-1mm}
+~~{4g\over \beta_Bs^2}
\!\int\! dz_\ast dz'_\ast~e^{i\beta_B {\rm min}(z_\ast,z'_\ast)}
\nonumber\\
&&\hspace{-1mm}
\times~[\infty,z_\ast]\big\{{2\over\alpha s}F_{\bu\mu}(z_\ast,y_\perp)[z_\ast,z'_\ast]_y
-~{2ip_{2\mu}\over\alpha^2 s^2}D^jF_{\bu j}(z_\ast,y_\perp)[z_\ast,z'_\ast]_y\big\}
F_{\bu i}(z'_\ast,y_\perp)[z'_\ast,\infty]_y
\nonumber\\
&&\hspace{-1mm}
-~{32g^2p_{2\mu}\over\alpha^2s^5\beta_B}\!\int\! dz_\ast dz'_\ast dz''_\ast
~\theta(z_\ast-z''_\ast)e^{i\beta_B{\rm min}(z'_\ast,z''_\ast)}
[\infty,z_\ast]_yF_{\bu j}(z_\ast,y_\perp)[z_\ast,z''_\ast]_yF_\bu^{~j}(z''_\ast,y_\perp)
\nonumber\\
&&\hspace{-1mm}
\times~
[z''_\ast,z'_\ast]_y
F_{\bu i}(z'_\ast,y_\perp)[z'_\ast,\infty]_y
~-~{2\over\alpha\beta_Bs}e^{-i\beta_B\sigma_\ast}
(\partial_\mu U_y\partial_iU_y^\dagger+{ip_{2\mu}\over\alpha s}\partial_\perp^2U_y\partial_i U_y^\dagger)\Big\}^{ab}
\label{4.15}
\end{eqnarray}
This expression explicitly depends on the cutoff $\sigma_\ast$. However, we can set $\sigma_\ast=0$ in the r.h.s. of Eq. (\ref{4.15}) (and eliminate few terms as well). 
To demonstrate this, let us consider two cases: $\beta_B\ll {1\over\sigma_\ast}$ and $\beta_B\geq {1\over\sigma_\ast}$. In the first case
\begin{eqnarray}
&&\hspace{-1mm}
L^{ab}_{\mu i}(k,y_\perp,\beta_B)\stackrel{\beta_B\sigma_\ast\ll 1}{=}
-ge^{-i(k,y)_\perp}{2\alpha p_{1\mu}\over k_\perp^2}\calf_i^{ab}(\beta_B, y_\perp)+
g(k_\perp|{g_{\mu i}\alpha \beta_Bs+2\alpha k_ip_{1\mu}\over\alpha \beta_Bs+k_\perp^2}~
\label{4.16}\\
&&\hspace{-1mm}
-\big[~g_{\mu i}U-{2ip_{2\mu}\over\alpha s}\partial_iU
+{2\over\beta_Bs}\big(p_{1\mu}U+{i\over\alpha}\partial^\perp_\mu U-{p_{2\mu}\over\alpha^2 s}\partial_\perp^2 U\big)p_i\big]
{\alpha\beta_Bs\over\alpha\beta_Bs+p_\perp^2}U^\dagger|y_\perp)^{ab}
\nonumber\\
&&\hspace{-1mm}
=~
g(k_\perp|g_{\mu i}\big({\alpha \beta_Bs\over\alpha \beta_Bs+p_\perp^2}-U{\alpha \beta_Bs\over\alpha \beta_Bs+p_\perp^2}U^\dagger\big)
+2\alpha p_{1\mu}\big({p_i\over \alpha\beta_Bs+p_\perp^2}-U{p_i\over \alpha\beta_Bs+p_\perp^2}U^\dagger\big)
\nonumber\\
&&\hspace{-1mm}
+~\big[{2ip_{2\mu}\over\alpha s}\partial_iU-{2i\over\alpha\beta_Bs}\partial^\perp_\mu Up_i+{2p_{2\mu}\over\alpha^2s^2\beta_B}\partial_\perp^2 Up_i
\big]{\alpha\beta_Bs\over\alpha\beta_Bs+p_\perp^2}U^\dagger-{2\alpha p_{1\mu}\over p_\perp^2}i(\partial_iU)U^\dagger|y_\perp)^{ab}
\nonumber
\end{eqnarray}
and all other terms are small since they contain extra factors $e^{i\beta_B z_\ast}-e^{-i\beta_B \sigma_\ast}\simeq (z-\sigma)_\ast$ (or $z'_\ast -\sigma_\ast$ or $z''_\ast-\sigma_\ast$) in the integrand.

In the second case $\sigma'\beta_Bs\geq p_\perp^2$ so $\alpha\beta_Bs\gg p_\perp^2$ and we get
\begin{eqnarray}
&&\hspace{-3mm}
L^{ab}_{\mu i}(k,y_\perp,\beta_B)~
\nonumber\\
&&\hspace{-3mm}
=~e^{-i(k,y)_\perp}\Big(2\big({p_{2\mu}\over\alpha s}-{\alpha p_{1\mu}\over k_\perp^2}\big)\calf_i(\beta_B,y_\perp)-{2k_i\over\alpha\beta_Bs}\calf_\mu(\beta_B,y_\perp)+~{2i\over\alpha\beta_Bs}\partial^y_i\calf_\mu(\beta_B,y_\perp)
\nonumber\\
&&\hspace{-3mm}  
+~{8g^2\over\alpha \beta_Bs^3}
\!\int\! dz_\ast dz'_\ast~e^{i\beta_B {\rm min}(z_\ast,z'_\ast)}~[\infty,z_\ast]F_{\bu\mu}(z_\ast,y_\perp)[z_\ast,z'_\ast]_yF_{\bu i}(z'_\ast,y_\perp)[z'_\ast,\infty]_y\Big)^{ab}
\label{4.17}
\end{eqnarray}
where we used the formula
\begin{eqnarray}
&&\hspace{-3mm}
(k_\perp|g_{\mu i}\big({\alpha \beta_Bs\over\alpha \beta_Bs+p_\perp^2}-U{\alpha \beta_Bs\over\alpha \beta_Bs+p_\perp^2}U^\dagger\big)
+2\alpha p_{1\mu}\big({p_i\over \alpha\beta_Bs+p_\perp^2}-U{p_i\over \alpha\beta_Bs+p_\perp^2}U^\dagger\big)|y_\perp)
\nonumber\\
&&\hspace{-3mm}
=~(k_\perp|{\alpha \beta_Bsg_{\mu i}+2\alpha p_{1\mu}k_i\over\alpha\beta_Bs+k_\perp^2}(2ik^j\partial_jU-\partial_\perp^2 U){1\over\alpha \beta_Bs+p_\perp^2}U^\dagger
+2i\alpha p_{1\mu}\partial_iU{1\over \alpha\beta_Bs+p_\perp^2}U^\dagger|y_\perp)
\nonumber\\
\label{4.18}
\end{eqnarray}
Let us now compare the contributions of various terms in the r.h.s. of Eq. (\ref{4.15}) to the production part of the evolution kernel 
defined by the  square of Lipatov vertices (\ref{4.15}). It is clear that the square of the first term $\sim\big({p_{2\mu}\over\alpha s}-{\alpha p_{1\mu}\over k_\perp^2}\big)\calf_i$
is proportional to $\tilcaf^{i}{1\over k_\perp^2}\calf_{i}$ and contributions of all other terms are down by at least one power of ${p_\perp^2\over\alpha\beta_Bs}$. 
Thus, with our accuracy
\begin{eqnarray}
&&\hspace{-3mm}
L^{ab}_{\mu i}(k,y_\perp,\beta_B)~
\stackrel{\alpha\beta_Bs\gg p_\perp^2}{=}~2e^{-i(k,y)_\perp}\big({p_{2\mu}\over\alpha s}-{\alpha p_{1\mu}\over k_\perp^2}\big)\calf_i^{ab}(\beta_B,y_\perp)
\label{4.19}
\end{eqnarray}
We see that in both cases (\ref{4.16}) and (\ref{4.19}) one can replace $\sigma_\ast$ by 0. Moreover, with our accuracy the Lipatov
vertex (\ref{4.15}) can be reduced to the ``direct sum'' of Eqs. (\ref{4.16}) and (\ref{4.19}):
\begin{eqnarray}
&&\hspace{-1mm}
L^{ab}_{\mu i}(k,y_\perp,\beta_B)~=~2ge^{-i(k,y)_\perp}
\big({p_{2\mu}\over\alpha s}-{\alpha p_{1\mu}\over k_\perp^2}\big)
[\calf_i(\beta_B, y_\perp)-U_i(y_\perp)]^{ab}
\label{4.20}\\
&&\hspace{-1mm}
+~
g(k_\perp|g_{\mu i}\big({\alpha \beta_Bs\over\alpha \beta_Bs+p_\perp^2}
-U{\alpha \beta_Bs\over\alpha \beta_Bs+p_\perp^2}U^\dagger\big)
+2\alpha p_{1\mu}\big({p_i\over \alpha\beta_Bs+p_\perp^2}-U{p_i\over \alpha\beta_Bs+p_\perp^2}U^\dagger\big)
\nonumber\\
&&\hspace{-1mm}
+~\big[2i\beta_Bp_{2\mu}\partial_iU-2i\partial^\perp_\mu Up_i+{2p_{2\mu}\over\alpha s}\partial_\perp^2 Up_i
\big]{1\over\alpha\beta_Bs+p_\perp^2}U^\dagger-{2\alpha p_{1\mu}\over p_\perp^2}U_i|y_\perp)^{ab}
\nonumber
\end{eqnarray}
where we introduced the notation $U_i\equiv \calf_i(0)=i(\partial_iU)U^\dagger$.
It is clear that at $\beta_B\sigma_\ast\ll 1$ the first term in the r.h.s. of this equation disappears and we get the  r.h.s. of Eq. (\ref{4.16}). On the other hand, as we saw above,
at $\beta_B\sigma_\ast\geq 1$ all terms in the last two lines in the r.h.s. of Eq. (\ref{4.20}) are small except
 $(k_\perp|{2p_{2\mu}\over\alpha s}U_i -{2\alpha p_{1\mu}\over p_\perp^2}U_i|y_\perp)^{ab}$ which cancels
 the second term in the first line of Eq. (\ref{4.20}) so we get the r.h.s. of Eq. (\ref{4.19}).
 It is worth noting that at $\beta_B=0$ Eq. (\ref{4.20})  agrees with the Lipatov vertex obtained in Ref. \cite{bal04}.

It is instructive to check the Lipatov vertex property $k^\mu L^{ab}_{\mu i}(k,y_\perp,\beta_B)~=~0$. One obtains
\begin{eqnarray}
&&\hspace{-1mm}
k^\mu\times ({\rm r.h.s.~of~Eq.~(\ref{4.20}}))_\mu~
\label{4.21}\\
&&\hspace{-1mm}
=~
g(k_\perp|k_i\big({\alpha \beta_Bs\over\alpha \beta_Bs+k_\perp^2}-U{\alpha \beta_Bs\over\alpha \beta_Bs+p_\perp^2}U^\dagger\big)
+k_\perp^2\big({k_i\over \alpha\beta_Bs+k_\perp^2}-U{p_i\over \alpha\beta_Bs+p_\perp^2}U^\dagger\big)
\nonumber\\
&&\hspace{-1mm}
+~(\alpha\beta_B s[p_i,U]+[p_\perp^2+\alpha\beta_B s,U]p_i)
{1\over\alpha\beta_Bs+p_\perp^2}U^\dagger-U_i|y_\perp)~=~0
\nonumber
\end{eqnarray}

\subsection{Lipatov vertex for arbitrary transverse momenta}

Let us demonstrate that for arbitrary transverse momenta the Lipatov vertex of gluon emission is given by 
the following ``interpolating formula''
\begin{eqnarray}
&&\hspace{-1mm}
L^{ab}_{\mu i}(k,y_\perp,\beta_B)
\nonumber\\
&&\hspace{-1mm}
=~g(k_\perp|{\alpha \beta_Bsg_{\mu i}+2\alpha p_{1\mu}k_i\over\alpha\beta_Bs+k_\perp^2}(2ik^j\partial_jU-\partial_\perp^2 U){1\over\alpha \beta_Bs+p_\perp^2}U^\dagger
+2i\alpha p_{1\mu}\partial_iU{1\over \alpha\beta_Bs+p_\perp^2}U^\dagger
\nonumber\\
&&\hspace{-1mm}
+~{2i\over \alpha s}p_{2\mu}\partial_i U{\alpha\beta_Bs \over \alpha\beta_Bs+p_\perp^2}U^\dagger
-\big[2i\partial_\mu U
-{2p_{2\mu}\over \alpha s}\partial_\perp^2U\big]
{p_i\over\alpha\beta_Bs+p_\perp^2}U^\dagger-{2\alpha p_{1\mu}\over p_\perp^2}i(\partial_iU)U^\dagger |y_\perp)^{ab}
\nonumber\\
&&\hspace{-1mm}
+~{2ge^{-i(k,y)_\perp}\over\alpha\beta_B s+k_\perp^2}\big[-\delta_\mu^jk_i
+{2\alpha k_ik^jp_{1\mu}\over\alpha\beta_Bs+k_\perp^2}+{\alpha\beta_B sg_{\mu i}k^j\over\alpha\beta_Bs+k_\perp^2}
+\beta_Bp_{2\mu}\delta_i^j-\alpha p_{1\mu}{\alpha\beta_Bs\over k_\perp^2}\delta_i^j\big]
\nonumber\\
&&\hspace{77mm}
\times~\big[\calf_j\big(\beta_B+{k_\perp^2\over\alpha s},y_\perp\big)-U_j(y_\perp)\big]^{ab}
\label{lvresult}
\end{eqnarray}

Let us consider at first the light-cone limit corresponding to the case when the characteristic transverse momenta of the external ``fast'' gluon fields are small in comparison to 
the momenta of ``slow'' gluons which we integrated over.  As we discussed above, the higher-twist terms $\sim D_j F_{\bu k}$ or $\sim F_{\bu j}F_{\bu k}$ exceed our accuracy so 
we can eliminate terms $\sim\partial_\perp^2U$ and commute operators $\partial_jU$ with ${1\over p_\perp^2+\alpha\beta_Bs}$ resulting in
\begin{eqnarray}
&&\hspace{-1mm}
{\rm r.h.s.~of~Eq.~(\ref{lvresult})}
\stackrel{\rm light-cone}{=}2g(k_\perp|k^j{\alpha \beta_Bsg_{\mu i}+2\alpha p_{1\mu}k_i\over(\alpha\beta_Bs+k_\perp^2)^2}U_j
+{\alpha p_{1\mu}\over\alpha\beta_Bs+k_\perp^2}U_i
\nonumber\\
&&\hspace{-1mm}
+~{p_{2\mu}\beta_B \over \alpha\beta_Bs+k_\perp^2}U_i
-{k_i\over\alpha\beta_Bs+k_\perp^2}U_{\mu_\perp}-{\alpha p_{1\mu}\over k_\perp^2}U_i |y_\perp)^{ab}
\nonumber\\
&&\hspace{-1mm}
+~{2ge^{-i(k,y)_\perp}\over\alpha\beta_B s+k_\perp^2}\big[-\delta_\mu^jk_i
+{2\alpha k_ik^jp_{1\mu}\over\alpha\beta_Bs+k_\perp^2}+{\alpha\beta_B sg_{\mu i}k^j\over\alpha\beta_Bs+k_\perp^2}
+\beta_Bp_{2\mu}\delta_i^j-\alpha p_{1\mu}{\alpha\beta_Bs\over k_\perp^2}\delta_i^j\big]
\nonumber\\
&&\hspace{70mm}
\times~\big[\calf_j\big(\beta_B+{k_\perp^2\over\alpha s},y_\perp\big)-U_j(y_\perp)\big]^{ab}
\end{eqnarray}
It is clear now that the first two lines in the r.h.s. cancel the last term in the square brackets in the last line so
we recover the light-cone result (\ref{3.8}).

Next we consider the case when the transverse momenta of fast and slow fields are comparable so the Lipatov vertex is given by Eq. (\ref{4.20}) above.
The difference between the r.h.s.'s of Eq. (\ref{lvresult}) and Eq. (\ref{4.20}) is
\begin{equation}
\hspace{-1mm}
{2ge^{-i(k,y)_\perp}\over\alpha\beta_B s+k_\perp^2}\big[-\delta_\mu^jk_i
+{2\alpha k_ik^jp_{1\mu}\over\alpha\beta_Bs+k_\perp^2}+{\alpha\beta_B sg_{\mu i}k^j\over\alpha\beta_Bs+k_\perp^2}\big]
\big[\calf_j\big(\beta_B+{k_\perp^2\over\alpha s},y_\perp\big)-U_j(y_\perp)\big]^{ab}
\label{4.22}
\end{equation}
where we used Eq. (\ref{4.18}). It is easy to see that the expression (\ref{4.22}) is small in both $\beta_B\ll{1\over\sigma_\ast}$ and $\beta_B\geq{1\over\sigma_\ast}$
cases. Indeed, when  $\beta_B\ll{1\over\sigma_\ast}$
 the integral representing $\calf_j\big(\beta_B+{k_\perp^2\over\alpha s},y_\perp\big)-U_j(y_\perp)$ 
 contains an extra  factor 
 $e^{i(\beta_B+{k_\perp^2\over\alpha s})z_\ast}-1\sim (\beta_B+{k_\perp^2\over\alpha s}\big)\sigma_\ast\ll1$ in the integrand  and in the 
 $\beta_B\geq{1\over\sigma_\ast}$ case the Eq. (\ref{4.22}) is $\sim {k_\perp^2\over\alpha\beta_Bs}\ll 1$ in comparison to the leading term 
 in this limit (\ref{4.19}).
 
 As in the light-cone case, for calculation of the evolution kernel it is convenient to go to the light-like gauge $p_2^\mu A_\mu=0$. 
 Since $k_\mu\times \big({\rm r.h.s.~of ~Eq. ~(\ref{lvresult})}\big)^\mu~=~0$ (see Eq. (\ref{4.21}))
it is sufficient to replace $\alpha p_1^\mu$ in the r.h.s. of Eq. (\ref{lvresult}) by $\alpha p_1^\mu-k^\mu~=~-k_\perp^\mu-{k_\perp^2\over\alpha s}p_2^\mu$. One obtains
\begin{eqnarray}
&&\hspace{-1mm}
L^{ab}_{\mu i}(k,y_\perp,\beta_B)^{\rm light-like}
\label{4.23}\\
&&\hspace{-1mm}
=~g(k_\perp|{\alpha \beta_Bsg_{\mu i}-2k^\perp_{\mu}k_i\over\alpha\beta_Bs+k_\perp^2}(2ik^j\partial_jU-\partial_\perp^2 U){1\over\alpha \beta_Bs+p_\perp^2}U^\dagger
-2ik^\perp_\mu \partial_iU{1\over \alpha\beta_Bs+p_\perp^2}U^\dagger
\nonumber\\
&&\hspace{55mm}
-~2i\partial_\mu U{p_i\over\alpha\beta_Bs+p_\perp^2}U^\dagger+{2k^\perp_\mu\over k_\perp^2}U_i |y_\perp)^{ab}
\nonumber\\
&&\hspace{-1mm}
+~2ge^{-i(k,y)_\perp}\Big[
{k^\perp_\mu\delta_i^j\over k_\perp^2}
-{\delta_\mu^jk_i+\delta_i^jk^\perp_\mu-g_{\mu i}k^j\over \alpha\beta_Bs+k_\perp^2}
-{k_\perp^2g_{\mu i}k^j+2k^\perp_\mu k_ik^j\over(\alpha\beta_Bs+k_\perp^2)^2}\Big]
\nonumber\\
&&\hspace{55mm}
\times~\big[\calf_j\big(\beta_B+{k_\perp^2\over\alpha s},y_\perp\big)-U_j(y_\perp)\big]^{ab}~+~O(p_{2\mu})
\nonumber
\end{eqnarray}
As usual, we do not display the term $\sim p_{2\mu}$ since it does not contribute to the evolution kernel. Using $[p_\perp^2,U]~=~-2ip^j\partial_jU+\partial_\perp^2U$ one can rewrite this vertex as
\begin{eqnarray}
&&\hspace{-1mm}
L^{ab}_{\mu i}(k,y_\perp,\beta_B)^{\rm light-like}~=~i\lim_{k^2\rightarrow 0} k^2\langle A_\mu^{a{\rm q}}(k)\calf_i^b(\beta_B,y_\perp)\rangle^{\rm light-like}
\label{lvertax}\\
&&\hspace{-1mm}
=~g(k_\perp|U{p_\perp^2g_{\mu i}+2p^\perp_{\mu}p_i\over\alpha\beta_Bs+p_\perp^2}U^\dagger-{k_\perp^2g_{\mu i}+2k^\perp_{\mu}k_i\over\alpha\beta_Bs+k_\perp^2} |y_\perp)^{ab}
+{2gk^\perp_\mu\over k_\perp^2}e^{-i(k,y)_\perp}\calf_i^{ab}\big(\beta_B+{k_\perp^2\over\alpha s},y_\perp\big)
\nonumber\\
&&\hspace{-1mm}
-~2ge^{-i(k,y)_\perp}\Big[
{\delta_\mu^jk_i+\delta_i^jk^\perp_\mu-g_{\mu i}k^j\over \alpha\beta_Bs+k_\perp^2}
+{g_{\mu i}k_\perp^2k^j+2k^\perp_\mu k_ik^j\over(\alpha\beta_Bs+k_\perp^2)^2}\Big]
{\breve\calf}^{ab}_j\big(\beta_B+{k_\perp^2\over\alpha s},y_\perp\big)
~+~O(p_{2\mu})
\nonumber
\end{eqnarray}
where we introduced the notation
\begin{equation}
{\breve\calf}_j\big(\beta_B+{k_\perp^2\over\alpha s},y_\perp\big)
~\equiv~\calf_j\big(\beta_B+{k_\perp^2\over\alpha s},y_\perp\big)-\calf_j(0,y_\perp)
\label{4.25}
\end{equation}
(recall that $\calf_j(0,y_\perp)=U_j(y_\perp)\equiv i\partial_jU_yU_y^\dagger$).

It should be emphasized that while we constructed the Lipatov vertex (\ref{lvresult}) as a formula which interpolates between the light-cone result (\ref{3.8}) for small transverse momenta
of background fields and shock-wave result (\ref{4.20}) for comparable transverse momenta, we have just demonstrated that 
with our leading-log accuracy our final expression (\ref{lvresult}) is correct in the whole range of the transverse momenta.

It is convenient to rewrite the  Lipatov vertex (\ref{lvertax})  in a different form without explicit subtraction (\ref{4.25}). Starting from Eq. (\ref{4.23}) 
we get
\begin{eqnarray}
&&\hspace{-1mm}
L^{ab}_{\mu i}(k,y_\perp,\beta_B)^{\rm light-like}   
\label{lvertalt}\\
&&\hspace{5mm}
=~g(k_\perp|\calf^j\big(\beta_B+{k_\perp^2\over\alpha s}\big)\Big\{{\alpha \beta_Bsg_{\mu i}-2k^\perp_{\mu}k_i\over\alpha\beta_Bs+k_\perp^2}(k_jU+Up_j){1\over\alpha \beta_Bs+p_\perp^2}U^\dagger
\nonumber\\
&&\hspace{11mm}
-~2k^\perp_\mu U{ g_{ij}\over \alpha\beta_Bs+p_\perp^2}U^\dagger-~2g_{\mu j} U{p_i\over\alpha\beta_Bs+p_\perp^2}U^\dagger+{2k^\perp_\mu\over k_\perp^2}g_{ij}\Big\}|y_\perp)^{ab}   
~+~O(p_{2\mu})
\nonumber
\end{eqnarray}
where
the operator
$\calf_i(\beta)$
is defined as usual
\begin{eqnarray}
&&\hspace{-1mm}
(k_\perp|\calf_i(\beta)|y_\perp)
~\equiv~{2\over s}\int\! dy_\ast ~e^{i\beta y_\ast-i(k,y)_\perp}\calf_i(y_\ast, y_\perp)
\label{defoperf}
\end{eqnarray}

Let us prove that Eq. (\ref{lvertalt}) coincides with Eq. (\ref{lvertax}) with our accuracy. First, as we discussed above, in the light-cone case ($l_\perp^2\ll p_\perp^2$) 
we can drop higher-twist terms  and commute operators $U$ with $p_i$ and ${1\over p_\perp^2+\alpha\beta_Bs}$ which gives us Eq. (\ref{3.8}). Second, consider
the ``shock-wave'' case  $l_\perp^2\sim p_\perp^2$. When  $\beta_B\ll{1\over\sigma_\ast}$
 the integral representing $\calf_j\big(\beta_B+{k_\perp^2\over\alpha s}\big)$ 
 contains an exponential  factor 
 $e^{i(\beta_B+{k_\perp^2\over\alpha s})z_\ast}\sim e^{i(\beta_B+{k_\perp^2\over\alpha s})\sigma_\ast}$. This factor can be approximated by one, since  
 ${k_\perp^2\over\alpha s}\sigma_\ast\ll 1$ in the shock-wave case (see the discussion above), so we can replace $\calf_j\big(\beta_B+{k_\perp^2\over\alpha s}\big)$
 by $U_j$ and get
\begin{eqnarray}
&&\hspace{-1mm}
L^{ab}_{\mu i}(k,y_\perp,\beta_B)^{\rm light-like}   
\label{4.29}\\
&&\hspace{-1mm}
\simeq~g(k_\perp|\Big\{{\alpha \beta_Bsg_{\mu i}-2k^\perp_{\mu}k_i\over\alpha\beta_Bs+k_\perp^2}(k_ji\partial^jU+i\partial^jUp_j){1\over\alpha \beta_Bs+p_\perp^2}U^\dagger
-2k^\perp_\mu i\partial_iU{1\over \alpha\beta_Bs+p_\perp^2}U^\dagger
\nonumber\\
&&\hspace{55mm}
-~2i\partial_\mu U{p_i\over\alpha\beta_Bs+p_\perp^2}U^\dagger+{2k^\perp_\mu\over k_\perp^2}U_i\Big\}|y_\perp)^{ab}   
~+~O(p_{2\mu})
\nonumber
\end{eqnarray}
which gives the first two lines in the r.h.s. of Eq. (\ref{4.23}). As it was shown above, the last two lines in the r.h.s. of Eq. (\ref{4.23}) are small at  $\beta_B\ll{1\over\sigma_\ast}$
so Eq. (\ref{lvertalt}) coincides with Eq. (\ref{4.23}) at $\beta_B\ll{1\over\sigma_\ast}$ with our accuracy. Finally, in  the 
 $\beta_B\geq{1\over\sigma_\ast}$ case $\alpha\beta_Bs\gg p_\perp^2$ and therefore the Eq. (\ref{lvertalt}) reduces to
\begin{eqnarray}
&&\hspace{-1mm}
L^{ab}_{\mu i}(k,y_\perp,\beta_B)^{\rm light-like}   ~
\simeq~{2k^\perp_\mu\over k_\perp^2}e^{-i(k,y)_\perp}g\calf_i^{ab} \big(\beta_B+{k_\perp^2\over\alpha s},y_\perp\big)  
~+~O(p_{2\mu})
\nonumber
\end{eqnarray}
which is the same as Eq. (\ref{4.23}) in this limit.

Similar calculation for complex-conjugate amplitude gives
\begin{eqnarray}
&&\hspace{-1mm}
\tiL_{i\mu}^{ba}(k,x_\perp,\beta_B)^{\rm light-like}
~=~-i\lim_{k^2\rightarrow 0} k^2\langle \tilcaf_i^b(\beta_B,x_\perp)\tilA^{a q}_\mu(k)\rangle^{\rm light-like}
\label{lvertaxcc}\\
&&\hspace{-1mm}
=~g(x_\perp|\tilU{p_\perp^2g_{\mu i}+2p^\perp_\mu p_i \over \alpha\beta_Bs+p_\perp^2}\tilU^\dagger
-{p_\perp^2g_{\mu i}+2p^\perp_\mu p_i \over \alpha\beta_Bs+p_\perp^2}|k_\perp)^{ba}
+2ge^{i(k,x)_\perp}{k^\perp_{\mu}\over k_\perp^2}\tilcaf_i^{ba}\big(\beta_B+{k_\perp^2\over\alpha s},x_\perp\big)
\nonumber\\
&&\hspace{-1mm}
-~2ge^{i(k,x)_\perp}\Big[{\delta_\mu^jk_i+k^\perp_{\mu}\delta_i^j-g_{\mu i}k^j\over\alpha\beta_B s+k_\perp^2}
+{g_{\mu i}k_\perp^2k^j+2k_ik^jk^\perp_{\mu}\over(\alpha\beta_Bs+k_\perp^2)^2}
\Big]{\breve\ticalf}_j^{ba}\big(\beta_B+{k_\perp^2\over\alpha s},x_\perp\big)~+~O(p_{2\mu})
\nonumber
\end{eqnarray}
where
\begin{equation}
{\breve\ticalf}_j\big(\beta_B+{k_\perp^2\over\alpha s},x_\perp\big)
~\equiv~\ticalf_j\big(\beta_B+{k_\perp^2\over\alpha s},x_\perp\big)
-\ticalf_j(0,x_\perp)
\label{4.27}
\end{equation}
Similarly to Eq. (\ref{lvertalt}) we can rewrite the above expression in the form without subtractions 
\begin{eqnarray}
&&\hspace{-1mm}
\tiL^{ba}_{i\mu}(k,x_\perp,\beta_B)^{\rm light-like}   
\label{lvertaltcc}\\
&&\hspace{-1mm}
=~g(x_\perp|\Big\{ \tilU{1\over\alpha\beta_Bs+p_\perp^2}
(\tilU^\dagger k_j+p_j\tilU^\dagger){\alpha \beta_Bsg_{\mu i}-2k^\perp_{\mu}k_i\over\alpha \beta_Bs+k_\perp^2}
-2k^\perp_\mu g_{ij}\tilU{1\over \alpha\beta_Bs+p_\perp^2}\tilU^\dagger
\nonumber\\
&&\hspace{33mm}
-~2g_{\mu j} \tilU{p_i\over\alpha\beta_Bs+p_\perp^2}\tilU^\dagger+{2k^\perp_\mu\over k_\perp^2}g_{ij}\Big\}
\ticalf^j\big(\beta_B+{k_\perp^2\over\alpha s}\big)|k_\perp)^{ba}   ~+~O(p_{2\mu})
\nonumber
\end{eqnarray}
The production part of the evolution kernel is proportional to the cross section of gluon emission given by the product 
of Eqs. (\ref{lvertax}) and (\ref{lvertaxcc}) integrated according to Eq. (\ref{3.1}). To find the full kernel we should
calculate the virtual part.

\subsection{Virtual correction}

\begin{figure}[htb]
\begin{center}
\includegraphics[width=121mm]{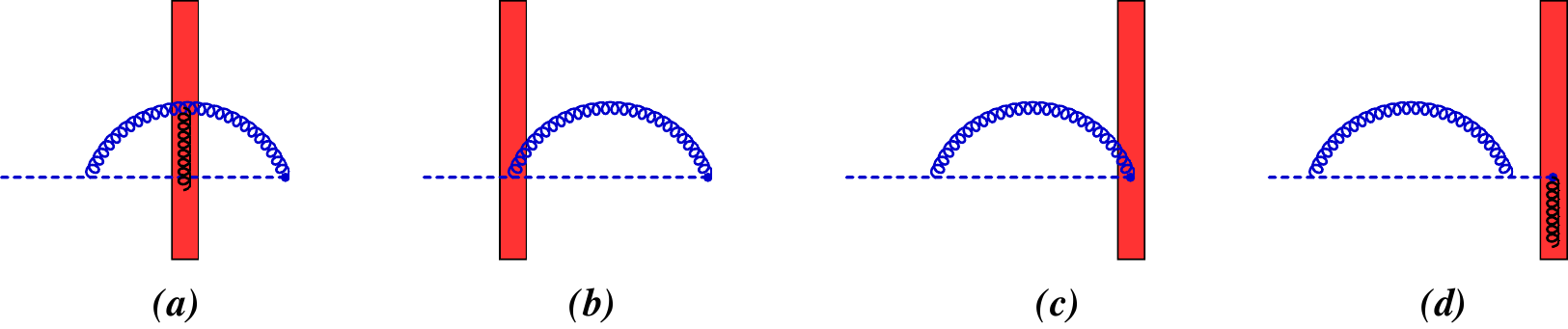}
\end{center}
\caption{Virtual part of the evolution kernel. \label{fig:5}}
\end{figure}

To get the virtual correction shown in Fig. \ref{fig:5} we should use the expansion (\ref{3.11})  of the operator $\calf$ up to the second order in quantum field. From Eq. (\ref{3.11}) one gets
\begin{eqnarray}
&&\hspace{-1mm}
\langle\calf_i^n(\beta_B, y_\perp)\rangle^{\rm 2nd}
\label{5.1}\\
&&\hspace{-1mm}=~{2g^2\over s}\!\int\! dy_\ast~e^{i\beta_B y_\ast}
\Big[\beta_B\!\int^{\infty}_{y_\ast}\!\! dz_\ast \big([\infty,z_\ast]_y
\langle A^{\rm q}_\bu(z_\ast,y_\perp)[z_\ast,y_\ast]_y\big)^{nm}A_i^{m{\rm q}}(y_\ast,y_\perp)\rangle
\nonumber\\
&&\hspace{-1mm}
-~{2i\over s}\!\int^{\infty}_{y_\ast}\!\! dz_\ast \big([\infty,z_\ast]_y\langle A^{\rm q}_\bu(z_\ast,y_\perp)[z_\ast,y_\ast]_y\big)^{nm}\partial_iA_\bu^{m{\rm q}}(y_\ast,y_\perp)\rangle
\nonumber\\
&&\hspace{-1mm}
-~{4g\over s^2}\!\int^{\infty}_{y_\ast}dz_\ast\!\int_{y_\ast}^{z_\ast}\! dz'_\ast 
~([\infty,z_\ast]_y\langle A^{\rm q}_\bu(z_\ast,y_\perp)[z_\ast,z'_\ast]_yA^{\rm q}_\bu(z'_\ast,y_\perp)[z'_\ast,y_\ast]_y\rangle)^{nm}F_{\bu i}^m(y_\ast,y_\perp)\Big]
\nonumber
\end{eqnarray}
As in the case of production
kernel we will calculate the diagrams in Fig. \ref{fig:5}a, \ref{fig:5}b, and \ref{fig:5}c separately and then check that the final result does not depend on the size of the shock wave $\sigma_\ast$
(it is easy to see that the diagram in Fig. \ref{fig:5}d vanishes in Feynman gauge).

\subsubsection{Diagram in Fig. \ref{fig:5}a}
Let us start with the diagram in Fig. \ref{fig:5}a. Using Eq. (\ref{3.11}) and (\ref{B.28}) we get
\begin{eqnarray}
&&\hspace{-3mm}
{2\over s}\!\int_{-\infty}^{-\sigma_\ast}\! dy_\ast~e^{i\beta_By_\ast} \langle[\infty,y_\ast]_y^{nm}
gF_{\bu i}^m(y_\ast,y_\perp)\rangle^{\rm Fig. ~\ref{fig:5}a}~=~
\label{5.2}\\
&&\hspace{3mm}
=~
{2g^2\over s}\!\int_{-\infty}^{-\sigma_\ast}\! dy_\ast~e^{i\beta_By_\ast}\Big[\beta_B\!\int^{\infty}_{\sigma_\ast}\!\! dz_\ast \big(
\langle A^{\rm q}_\bu(z_\ast,y_\perp)U_y\big)^{nm}A_i^{m{\rm q}}(y_\ast,y_\perp)\rangle
\nonumber\\
&&\hspace{46mm}
-~{2i\over s}\!\int^{\infty}_{\sigma_\ast}\!\! dz_\ast 
\big(\langle A^{\rm q}_\bu(z_\ast,y_\perp)U_y\big)^{nm}\partial_iA_\bu^{m{\rm q}}(y_\ast,y_\perp)\rangle\Big]
\nonumber\\
&&\hspace{3mm}
=~-{ig^2\over s}f^{nkl}\!\int_0^\infty\!{\dhd\alpha\over\alpha}\!\int_{-\infty}^{-\sigma_\ast}\! dy_\ast\!\int^{\infty}_{\sigma_\ast}\!\! dz_\ast
(y_\perp|e^{-i{p_\perp^2\over\alpha s}z_\ast}\big\{\beta_B \calg_{\bu i}(\infty,-\infty;p_\perp)
\nonumber\\
&&\hspace{23mm}
+~{2\over s}[\calg_{\bu\bu}(\infty,-\infty;p_\perp)+\calq_{\bu\bu}(\infty,-\infty;p_\perp)]p_i\big\}
e^{i(\beta_B+{p_\perp^2\over\alpha s})y_\ast}U^\dagger|y_\perp)^{kl}
\nonumber\\
&&\hspace{3mm}
=~ig^2f^{nkl}\!\int_0^\infty\!\dhd\alpha(y_\perp|{1\over p_\perp^2}
e^{-i{p_\perp^2\over\alpha s}\sigma_\ast}\big\{\alpha\beta_Bs \calg_{\bu i}(\infty,-\infty;p_\perp)
\nonumber\\
&&\hspace{-1mm}
+~2\alpha[\calg_{\bu\bu}(\infty,-\infty;p_\perp)+\calq_{\bu\bu}(\infty,-\infty;p_\perp)]p_i\big\}
{1\over \alpha\beta_Bs+p_\perp^2}e^{-i(\beta_B+{p_\perp^2\over\alpha s})\sigma_\ast}U^\dagger|y_\perp)^{kl}
\nonumber
\end{eqnarray}
(as usual we assume that there are no external fields outside $[\sigma_\ast,-\sigma_\ast]$ interval). 
Moreover, from Eq. (\ref{B.26}) we see that $\calg_{\bu i}(\infty,-\infty;p_\perp)~=~-{i\over\alpha}\partial_iU$ and from Eqs. (\ref{B.26}), (\ref{B.7}) and
(\ref{4.9}) that $\calg_{\bu \bu}(\infty,-\infty;p_\perp)+\calq_{\bu\bu}(\infty,-\infty;p_\perp)~=~-{1\over 2\alpha^2}\partial_\perp^2U$ so we obtain
\begin{eqnarray}
&&\hspace{-3mm}
{2\over s}g\!\int_{-\infty}^{-\sigma_\ast}\! dy_\ast e^{i\beta_B y_\ast}\langle[\infty,y_\ast]_y^{nm}F_{\bu i}^m(y_\ast,y_\perp)\rangle^{\rm Fig. ~\ref{fig:5}a}
\label{5.3}\\
&&\hspace{3mm}
=~g^2f^{nkl}\!\int_0^\infty\!{\dhd\alpha\over\alpha}(y_\perp|{1\over p_\perp^2}e^{-i{p_\perp^2\over\alpha s}\sigma_\ast}[\alpha \beta_Bs \partial_iU-i\partial_\perp^2Up_i]
{1\over \alpha\beta_Bs+p_\perp^2}e^{-i(\beta_B+{p_\perp^2\over\alpha s})\sigma_\ast}U^\dagger|y_\perp)^{kl}
\nonumber\\
&&\hspace{3mm}
=~g^2f^{nkl}e^{-i\beta_B\sigma_\ast}\!\int_0^\infty\!{\dhd\alpha\over\alpha}(y_\perp|{1\over p_\perp^2}[\alpha \beta_Bs \partial_iU-i\partial_\perp^2Up_i]
{1\over \alpha\beta_Bs+p_\perp^2}U^\dagger|y_\perp)^{kl}
\nonumber
\end{eqnarray}
(recall that ${p_\perp^2\over\alpha s}\sigma_\ast\ll 1$ if the transverse momenta in the loop are of order of transverse momenta of external fields). 

\subsubsection{Diagram in Fig. \ref{fig:5}b}
To get the contribution of the diagram in Fig. \ref{fig:5}b we need the gluon propagator with one point in the shock wave (\ref{B.8}), which we will rewrite as follows
\begin{eqnarray}
&&\hspace{-1mm}
\langle A_\mu^a(z_\ast,z_\perp)A_\nu^b(y_\ast,y_\perp)\rangle~
=~
\!\int_{-\infty}^0\!{\dhd\alpha\over 2\alpha}
\big\{(y_\perp|e^{-i{p_\perp^2\over\alpha s}(y-z)_\ast}\label{5.4}
\\
&&\hspace{-1mm}
\times~\big[\calg^{ba}_{\nu\mu}(y_\ast,z_\ast;p_\perp)+\calq^{ba}_{\nu\mu}(y_\ast,z_\ast;p_\perp)\big]|z_\perp)
+(z_\perp|\bar\calq^{ba}_{\nu\mu}(y_\ast,z_\ast;p_\perp)e^{-i{p_\perp^2\over\alpha s}(y-z)_\ast}|y_\perp)\big\}
\nonumber
\end{eqnarray}
with $\calg$ and $\calq$ given by Eqs. (\ref{B.6}) and (\ref{B.7})
\begin{eqnarray}
&&\hspace{-3mm}
\calg_{i\bu}^{ba}(y_\ast,z_\ast;p_\perp)
~=~{2g\over\alpha s}\!\int_{y_\ast}^{z_\ast}\!\!\!dz'_\ast([z_\ast,z'_\ast]F_{\bu i}(z'_\ast)[z'_\ast,y_\ast])^{ab}\!\!,~~\calq_{i\bu}(y_\ast,z_\ast;p_\perp)=\bar\calq_{i\bu}(y_\ast,z_\ast;p_\perp)=0,
\nonumber\\
&&\hspace{-3mm}
\calg_{\bu\bu}^{ba}(y_\ast,z_\ast;p_\perp)~
=~-{4g^2\over\alpha^2 s^2}\!\int_{y_\ast}^{z_\ast}\!\! \!dz'_\ast\!\int_{y_\ast}^{z'_\ast}\!\!\!dz''_\ast
([z_\ast,z'_\ast] F_{\bu j}(z'_\ast) [z'_\ast,z''_\ast]F_\bu^{~j}(z''_\ast)[z''_\ast,y_\ast])^{ab}
\label{5.5}\\
&&\hspace{-3mm}
\calq_{\bu\bu}^{ba}(y_\ast,z_\ast;p_\perp)
~=~-{ig\over \alpha^2s}\!\!\int_{y_\ast}^{z_\ast}\! \! dz'_\ast 
([z_\ast,z'_\ast]D^jF_{\bu j}(z'_\ast)[z'_\ast,y_\ast])^{ab},~~~~~{\bar\calq}_{\bu\bu}(y_\ast,z_\ast;p_\perp)
~=~0
\nonumber
\end{eqnarray}
and therefore  from Eq. (\ref{5.1}) we get
\begin{eqnarray}
&&\hspace{-3mm}
{2\over s}\!\int_{-\infty}^{-\sigma_\ast}\! dy_\ast e^{i\beta_B y_\ast}\langle[\infty,y_\ast]_y^{nm}gF_{\bu i}^m(y_\ast,y_\perp)\rangle^{\rm Fig. ~\ref{fig:5}b}~=~
\label{5.6}\\
&&\hspace{3mm}
=~
{2g^2\over s}\!\int_{-\infty}^{-\sigma_\ast}\! dy_\ast e^{i\beta_B y_\ast}\Big[\beta_B\!\int_{-\sigma_\ast}^{\sigma_\ast}\!\! dz_\ast \big([\infty,z_\ast]_y
\langle A^{\rm q}_\bu(z_\ast,y_\perp)[z_\ast,y_\ast]_y\big)^{nm}A_i^{m{\rm q}}(y_\ast,y_\perp)\rangle
\nonumber\\
&&\hspace{17mm}
-~{2i\over s}\!\int_{-\sigma_\ast}^{\sigma_\ast}\!\! dz_\ast \big([\infty,z_\ast]_y\langle A^{\rm q}_\bu(z_\ast,y_\perp)[z_\ast,y_\ast]_y\big)^{nm}\partial_iA_\bu^{m{\rm q}}(y_\ast,y_\perp)\rangle\Big]
\nonumber
\end{eqnarray}
First, let us show that the second term in the r.h.s. of this equation vanishes. From Eq. (\ref{5.4}) we see that
\begin{eqnarray}
&&\hspace{-1mm}
\langle A^{a{\rm q}}_\bu(z_\ast,y_\perp)\partial_iA_\bu^{b{\rm q}}(y_\ast,y_\perp)\rangle
\label{5.7}\\
&&\hspace{-1mm}
=~-i\!\int_{-\infty}^0\!{\dhd\alpha\over 2\alpha}
(y_\perp|p_ie^{-i{p_\perp^2\over\alpha s}(y-z)_\ast}\big[\calg^{ba}_{\bullet\bullet}(y_\ast,z_\ast;p_\perp)+\calq^{ba}_{\bullet\bullet}(y_\ast,z_\ast;p_\perp)\big]|y_\perp)
~=~0
\nonumber
\end{eqnarray}
because operators in Eq. (\ref{5.4}) do not contain $p$ and $(y_\perp|p_ie^{-i{p_\perp^2\over\alpha s}(y-z)_\ast}|y_\perp)~=~0$.

Now consider the first term in the r.h.s. of Eq. (\ref{5.6}).   From Eq. (\ref{5.4}) we get
\begin{equation}
\hspace{-1mm}
\langle A_\bu^a(z_\ast,y_\perp)A_i^b(y_\ast,y_\perp)\rangle~
=~
{g\over s}\!\int_0^{\infty}\!{\dhd\alpha\over \alpha^2}
(y_\perp|e^{-i{p_\perp^2\over\alpha s}(z-y)_\ast}\!\int_{y_\ast}^{z_\ast}\!\!\!dz'_\ast~[z_\ast,z'_\ast]F_{\bu i}(z'_\ast)[z'_\ast,y_\ast]|y_\perp)^{ab}
\label{5.8}
\end{equation}
and therefore
\begin{eqnarray}
&&\hspace{-3mm}
{2g\over s}\!\int_{-\infty}^{-\sigma_\ast}\! dy_\ast e^{i\beta_By_\ast}\langle[\infty,y_\ast]_y^{nm}F_{\bu i}^m(y_\ast,y_\perp)\rangle^{\rm Fig. ~\ref{fig:5}b}
~=~2{g^2\over s}if^{nkl}\!\int_{-\sigma_\ast}^{\sigma_\ast}\!\! dz'_\ast~\calf_i^{kl}(z'_\ast,y_\perp)
\nonumber\\
&&\hspace{11mm}
\times~\!\int_0^\infty\!{\dhd\alpha\over\alpha}
(y_\perp|\big[e^{-i{p_\perp^2\over\alpha s}\sigma_\ast}-e^{-i{p_\perp^2\over\alpha s}z'_\ast}\big]
{\alpha\beta_B s\over p_\perp^2(\alpha\beta_B s+p_\perp^2)}e^{-i(\beta_B+{p_\perp^2\over\alpha s})\sigma_\ast}|y_\perp)
\label{5.9}
\end{eqnarray}
where $\calf_i^{kl}(z'_\ast,y_\perp)\equiv g([\infty,z'_\ast]_yF_{\bu i}(z'_\ast,y_\perp)[z'_\ast,\infty]_y)^{kl}$, see Eq. (\ref{1.4}).
Since ${p_\perp^2\over\alpha s}\sigma_\ast\ll 1$ the r.h.s. of this equation
can be simplified to
\begin{eqnarray}
&&\hspace{-11mm}
{2g\over s}\!\int_{-\infty}^{-\sigma_\ast}\! dy_\ast e^{i\beta_B y_\ast} \langle[\infty,y_\ast]_y^{nm}F_{\bu i}^m(y_\ast,y_\perp)\rangle^{\rm Fig. ~\ref{fig:5}b}
~=~\label{5.10}
\\
&&\hspace{-5mm}
=~-2{g^2\over s}f^{nkl}e^{-i\beta_B\sigma_\ast}\!\!\int_0^\infty\!{\dhd\alpha\over\alpha}
(y_\perp|{\beta_B\over \alpha\beta_B s+p_\perp^2}|y_\perp)
\!\int_{-\sigma_\ast}^{\sigma_\ast}\!\! dz'_\ast~(z'-\sigma)_\ast\calf_i^{kl}(z'_\ast,y_\perp)~\simeq~0
\nonumber
\end{eqnarray}
because it is $O({p_\perp^2\over\alpha s}\sigma_\ast)$ in comparison to Eq. (\ref{5.3}).

\subsubsection{Diagram in Fig. \ref{fig:5}c}
As in previous Sections, we start from rewriting Eq. (\ref{3.11})
\begin{eqnarray}
&&\hspace{-1mm}
{2g\over s}\!\int_{-\sigma_\ast}^{\sigma_\ast}\! dy_\ast~e^{i\beta_By_\ast}\langle[\infty,y_\ast]_y^{nm}F_{\bu i}^m(y_\ast,y_\perp)\rangle^{\rm Fig.\ref{fig:5}c}~
\label{5.11}\\
&&\hspace{-1mm}
=~
{2g^2\over s}\!\int_{-\sigma_\ast}^{\sigma_\ast}\! dy_\ast~e^{i\beta_By_\ast}\Big[-{4g\over s^2}\!\int^{\infty}_{y_\ast}dz_\ast\!\int_{y_\ast}^{z_\ast}\! dz'_\ast 
~([\infty,z_\ast]_y\langle A^{\rm q}_\bu(z_\ast,y_\perp)[z_\ast,z'_\ast]_yA^{\rm q}_\bu(z'_\ast,y_\perp)\rangle
\nonumber\\
&&\hspace{-1mm}
\times~[z'_\ast,y_\ast]_y)^{nm}F_{\bu i}^m(y_\ast,y_\perp)+\beta_B\!\int^{\infty}_{y_\ast}\!\! dz_\ast \big([\infty,z_\ast]_y
\langle A^{\rm q}_\bu(z_\ast,y_\perp)[z_\ast,y_\ast]_y\big)^{nm}A_i^{m{\rm q}}(y_\ast,y_\perp)\rangle
\nonumber\\
&&\hspace{41mm}
-~{2i\over s}\!\int^{\infty}_{y_\ast}\!\! dz_\ast \big([\infty,z_\ast]_y\langle A^{\rm q}_\bu(z_\ast,y_\perp)[z_\ast,y_\ast]_y\big)^{nm}\partial_iA_\bu^{m{\rm q}}(y_\ast,y_\perp)\rangle\Big]
\nonumber
\end{eqnarray}
Using the propagator (\ref{B.8}) with point $y$ inside the shock wave (and point $z$ anywhere) 
\footnote{Strictly speaking, one should depict Eq. (\ref{5.11}) as several diagrams with points $z$ (and $z'$) inside and outside the shock wave.} we obtain (hereafter $\partial_i(\mathcal{C})\equiv-i[p_i, \mathcal{C}]$)
\begin{eqnarray}
&&\hspace{-1mm}
{2g\over s}\!\int_{-\sigma_\ast}^{\sigma_\ast}\! dy_\ast~e^{i\beta_By_\ast}\langle[\infty,y_\ast]_y^{nm}F_{\bu i}^m(y_\ast,y_\perp)\rangle^{\rm Fig.\ref{fig:5}c}~
\label{5.12}\\
&&\hspace{-1mm}
=~{g^2\over s}if^{nkl}\!\int_{-\sigma_\ast}^{\sigma_\ast}\!dy_\ast~e^{i\beta_By_\ast} \!\int_{y_\ast}^\infty\! dz_\ast
\!\int_0^\infty\!{\dhd\alpha\over\alpha}\Big\{(y_\perp|e^{-i{p_\perp^2\over\alpha s}(z-y)_\ast}
\big[-\beta_B 
([\infty,z_\ast]
\nonumber\\
&&\hspace{11mm}
\times~\calg_{\bu i}(z_\ast,y_\ast;p_\perp)[y_\ast,\infty])^{kl}~+~{2i\over s}
\big\{\partial_i\big([\infty,z_\ast]\calo_{\bu\bu}(z_\ast,y_\ast;p_\perp)[y_\ast,\infty]\big)
\nonumber\\
&&\hspace{11mm}
-~\partial_i([\infty,z_\ast])\calo_{\bu\bu}(z_\ast,y_\ast;p_\perp)[y_\ast,\infty]
-[\infty,z_\ast]\calo_{\bu\bu}(z_\ast,y_\ast;p_\perp)\partial_i([y_\ast,\infty])\big\}^{kl}\big]|y_\perp)
\nonumber\\
&&\hspace{-1mm}
+~{4g\over s^2}\!\int_{y_\ast}^{z_\ast}\! dz'_\ast 
~(y_\perp|e^{-i{p_\perp^2\over\alpha s}(z-z')_\ast}\big([\infty,z_\ast]\calo_{\bu\bu}(z_\ast,z'_\ast;p_\perp)[z'_\ast,y_\ast]F_{\bu i}(y_\ast)[y_\ast,\infty]\big)^{kl}|y_\perp)\Big\}
\nonumber
\end{eqnarray}
where $\calg_{\bu i}$ is given by Eq. (\ref{B.6}) 
\begin{eqnarray}
&&\hspace{-3mm}
\calg_{\bu i}(x_\ast,y_\ast;p_\perp)~=-~{2g\over \alpha s}\int^{x_\ast}_{y_\ast} dz_\ast[x_\ast,z_\ast]F_{\bu i}(z_\ast)[z_\ast,y_\ast]
\end{eqnarray}
and $\calo_{\bu\bu}~\equiv~\calg_{\bu\bu}+\calq_{\bu\bu}$ by Eqs. (\ref{B.6}), (\ref{B.7})
\begin{eqnarray}
&&\hspace{-3mm}
\calo_{\bu\bu}(x_\ast,y_\ast;p_\perp)~=~{g\over\alpha^2 s}\!\int_{y_\ast}^{x_\ast}\!\! \!dz_\ast~[x_\ast,z_\ast]
\big\{-iD^jF_{\bu j}(z_\ast)[z_\ast,y_\ast]
\nonumber\\
&&\hspace{35mm}
-~{4g\over s}\!\int_{y_\ast}^{z_\ast}\!\!dz'_\ast~
~F_{\bu j}(z_\ast) [z_\ast,z'_\ast]F_\bu^{~j}(z'_\ast)[z'_\ast,y_\ast]
\big\}
\label{5.14}
\end{eqnarray}
Using these expressions, one obtains after some algebra
\begin{eqnarray}
&&\hspace{-1mm}
{2g\over s}\!\int_{-\sigma_\ast}^{\sigma_\ast}\! dy_\ast~e^{i\beta_By_\ast}\langle[\infty,y_\ast]_y^{nm}F_{\bu i}^m(y_\ast,y_\perp)\rangle^{\rm Fig.\ref{fig:5}c}~
\label{5.15}\\
&&\hspace{-1mm}
=~g^2f^{nkl}\!\int_0^\infty\!{\dhd\alpha\over\alpha}(y_\perp|{1\over p_\perp^2(\alpha\beta_Bs+p_\perp^2)}|y_\perp)
\Big\{\!\int^{\sigma_\ast}_{-\sigma_\ast}\!\! dw_\ast ~\big(e^{i\beta_Bw_\ast}-e^{-i\beta_B\sigma_\ast}\big)
\nonumber\\
&&\hspace{-1mm}
\times~\Big(-2i\alpha\beta_B\calf_i(w_\ast,y_\perp)-{1\over s}\partial^y_i\calv(w_\ast,y_\perp)
+~{2i\over s^2}\calv(w_\ast,y_\perp)\!\int_{w_\ast}^{\sigma_\ast}\! dz'_\ast
\calf_i(z'_\ast,y_\perp)\Big)
\nonumber\\
&&\hspace{-1mm}
+~{2ig\over s}\!\int_{-\sigma_\ast}^{\sigma_\ast}\!\!  dz'_\ast\!\int_{z'_\ast}^{\sigma_\ast}\! \!dw_\ast~
\Big({2ig\over s}[\infty,w_\ast]_yD^jF_{\bu j}(w_\ast,y_\perp)[w_\ast,\infty]_y
\big[e^{i\beta_Bz'_\ast}-e^{-i\beta_B\sigma_\ast}\big]
\nonumber\\
&&\hspace{25mm}
+{8g^2\over s^2}\!\int_{z'}^{w_\ast}\!\! \!  dw'_\ast 
[\infty,w_\ast]_yF_{\bu j}(w_\ast,y_\perp)[w_\ast,w'_\ast]_yF_\bu^{~j}(w'_\ast)[w'_\ast,\infty]_y
\nonumber\\
&&\hspace{55mm}
\times~\big[e^{i\beta_Bz'}-e^{-i\beta_B\sigma_\ast}\big]
\Big)\calf_i(z'_\ast,y_\perp)
\Big\}^{kl}
\nonumber
\end{eqnarray}
where, as usual, $\calf_i^{kl}=g([\infty,w_\ast]_yF_{\bu i}(w_\ast,y_\perp)[w_\ast,\infty]_y)^{kl}$ and  
\begin{eqnarray}
&&\hspace{-1mm}
{\cal V}(w_\ast,y_\perp)~\equiv~
2ig[\infty,w_\ast]_yD^jF_{\bu j}(w_\ast,y_\perp)[w_\ast,\infty]_y
\nonumber\\
&&\hspace{18mm}
+~{8g^2\over s}\!\int_{w_\ast}^\infty\!dw'_\ast~
[\infty,w'_\ast]_yF_{\bu j}(w'_\ast,y_\perp) [w'_\ast,w_\ast]_yF_\bu^{~j}(w_\ast,y_\perp)[w_\ast,\infty]_y
\label{5.16}
\end{eqnarray}
(cf. Eq. (\ref{4.14})). 

\subsubsection{The sum of diagrams in Fig. \ref{fig:5} \label{sect4.4.4}}

The total virtual correction coming from Fig. \ref{fig:5} is given by the sum of Eqs. (\ref{5.3}) and (\ref{5.15}) 
\begin{eqnarray}
&&\hspace{-3mm}
{2\over s}g\!\int\! dy_\ast e^{i\beta_B y_\ast}\langle[\infty,y_\ast]_y^{nm}F_{\bu i}^m(y_\ast,y_\perp)\rangle^{\rm Fig. ~\ref{fig:5}}
\label{5.17}\\
&&\hspace{3mm}
=~g^2f^{nkl}e^{-i\beta_B\sigma_\ast}\!\int_0^\infty\!{\dhd\alpha\over\alpha}(y_\perp|{1\over p_\perp^2}[\alpha \beta_Bs \partial_iU-i\partial_\perp^2Up_i]
{1\over \alpha\beta_Bs+p_\perp^2}U^\dagger|y_\perp)^{kl}
\nonumber\\
&&\hspace{-1mm}
+~g^2f^{nkl}\!\int_0^\infty\!{\dhd\alpha\over\alpha}(y_\perp|{1\over p_\perp^2(\alpha\beta_Bs+p_\perp^2)}|y_\perp)
\Big\{\!\int^{\sigma_\ast}_{-\sigma_\ast}\!\! dw_\ast ~\big(e^{i\beta_Bw_\ast}-e^{-i\beta_B\sigma_\ast}\big)
\nonumber\\
&&\hspace{-1mm}
\times~\Big(-2i\alpha\beta_B\calf_i(w_\ast, y_\perp)-{1\over s}\partial^y_i\calv(w_\ast,y_\perp)
+~{2i\over s^2}\calv(w_\ast,y_\perp)\!\int_{w_\ast}^{\sigma_\ast}\! dz'_\ast
\calf_i(z'_\ast,y_\perp)\Big)
\nonumber\\
&&\hspace{-1mm}
+~{2ig\over s}\!\int_{-\sigma_\ast}^{\sigma_\ast}\!\!  dz'_\ast\!\int_{z'_\ast}^{\sigma_\ast}\! \!dw_\ast~
\Big({2ig\over s}[\infty,w_\ast]_yD^jF_{\bu j}(w_\ast,y_\perp)[w_\ast,\infty]_y
\big[e^{i\beta_Bz'_\ast}-e^{-i\beta_B\sigma_\ast}\big]
\nonumber\\
&&\hspace{25mm}
+~{8g^2\over s^2}\!\int_{z'}^{w_\ast}\!\! \!  dw'_\ast 
[\infty,w_\ast]_yF_{\bu j}(w_\ast,y_\perp)[w_\ast,w'_\ast]_yF_\bu^{~j}(w'_\ast)[w'_\ast,\infty]_y
\nonumber\\
&&\hspace{55mm}
\times~\big[e^{i\beta_Bz'_\ast}-e^{-i\beta_B\sigma_\ast}\big]
\Big)\calf_i(z'_\ast,y_\perp)
\Big\}^{kl}
\nonumber
\end{eqnarray}
Let us prove that with our accuracy it can be approximated as
\begin{eqnarray}
&&\hspace{-1mm}
{2\over s}g\!\int\! dy_\ast e^{i\beta_B y_\ast}\langle[\infty,y_\ast]_y^{nm}F_{\bu i}^m(y_\ast,y_\perp)\rangle^{\rm Fig. ~\ref{fig:5}}
\label{5.18}\\
&&\hspace{11mm}
=~g^2f^{nkl}\!\int_0^\infty\!{\dhd\alpha\over\alpha}
(y_\perp|{1\over p_\perp^2}[\alpha \beta_Bs \partial_iU-i\partial_\perp^2Up_i]
{1\over \alpha\beta_Bs+p_\perp^2}U^\dagger|y_\perp)^{kl}
\nonumber\\
&&\hspace{25mm}
-~ig^2f^{nkl}\!\int_0^\infty\!{\dhd\alpha\over\alpha}(y_\perp|{\alpha\beta_Bs\over p_\perp^2(\alpha\beta_Bs+p_\perp^2)}|y_\perp)
[\calf_i(\beta_B,y_\perp)-i\partial_iU_yU_y^\dagger]^{kl}
\nonumber
\end{eqnarray}
To this end we compare the size of different terms in the r.h.s. of equations (\ref{5.17}) and (\ref{5.18}) 
at $\beta_B\sigma_\ast\ll 1$ and $\beta_B\sigma_\ast\geq 1$.  
In the first case (at $\beta_B\sigma_\ast\ll 1$) the only surviving terms in the r.h.s.'s of these 
equations are the first terms and they are obviously equal.

In the second case let us start from Eq. (\ref{5.18}). Since $\beta_B\sigma_\ast\sim\beta_B{\sigma's\over p_\perp^2}\geq 1$ we have $\alpha\beta_Bs\gg p_\perp^2$ 
so 
\begin{eqnarray}
&&\hspace{-1mm}
{\rm r.h.s.~of~Eq.~(\ref{5.18})}~=~-~ig^2f^{nkl}\!\int_0^\infty\!{\dhd\alpha\over\alpha}(y_\perp|{1\over p_\perp^2}|y_\perp)\calf^{kl}_i(\beta_B,y_\perp)
\label{5.19}
\end{eqnarray}
Let us now compare the size of different terms in the r.h.s. of Eq. (\ref{5.17}). Since
${1\over s}\int\! dw_\ast\calv(w_\ast)$ $\sim \partial_\perp^2UU^\dagger$  the first term in the fourth line
$\sim\!\int\! dw_\ast \alpha\beta_BF_{\bu i}(w_\ast, y_\perp)\sim \alpha\beta_Bs\partial_iU_yU^\dagger_y$ is much greater than the second term
$\sim{1\over s}\!\int\! dw_\ast \partial_i\calv(w_\ast, y_\perp)\sim \partial_i\partial_\perp^2U_yU^\dagger_y$ 
or the third term 
$\sim{1\over s^2}\!\int\! dw_\ast \partial_i\calv(w_\ast, y_\perp)
\!\int\! dw'_\ast F_{\bu i}(w'_\ast)\sim \partial_\perp^2U_y \partial_iU^\dagger_y$.  Moreover, it is easy to
see that the terms in the last three lines in Eq. (\ref{5.17}) are of the same order as the terms $\sim \calv$ in the fourth line so they are again small in comparison
to the term $\sim \calf_i$. Thus, we get
\begin{eqnarray}
&&\hspace{-1mm}
{\rm r.h.s.~of~Eq.~(\ref{5.17})}~=~g^2f^{nkl}e^{-i\beta_B\sigma_\ast}\!\int_0^\infty\!{\dhd\alpha\over\alpha}(y_\perp|{1\over p_\perp^2}
\big[\partial_iU-{i\over\alpha\beta_Bs}\partial_\perp^2Up_i\big]U^\dagger|y_\perp)^{kl}
\nonumber\\
&&\hspace{-1mm}
-~ig^2f^{nkl}\!\int_0^\infty\!{\dhd\alpha\over\alpha}(y_\perp|{1\over p_\perp^2}|y_\perp)
{2\over s}\!\int^{\sigma_\ast}_{-\sigma_\ast}\!\! dw_\ast ~\big(e^{i\beta_Bw_\ast}-e^{-i\beta_B\sigma_\ast}\big)\calf^{kl}_i(w_\ast, y_\perp)
\label{5.20}
\end{eqnarray}
which coincides with the r.h.s of Eq. (\ref{5.19}).

 Last but not least, let us prove that one can use the formula (\ref{5.18}) in the light-cone limit $l_\perp^2\ll p_\perp^2$ 
 where it coincides with Eq. (\ref{3.15}).
First we notice that the term $\sim\partial_\perp^2U$ has twist two and so exceeds our twist-one light-cone accuracy.
 Next, since the commutator $[p_\perp^2,\partial_iU]$ consists of operators of collinear twist two (or higher), 
 one can rewrite the first term in the r.h.s of  Eq. (\ref{5.18}) in the form
\begin{eqnarray}
&&\hspace{-1mm}
(y_\perp|{1\over p_\perp^2}\alpha \beta_Bs \partial_iU{1\over \alpha\beta_Bs+p_\perp^2}U^\dagger|y_\perp)~\simeq~
 (y_\perp|{\alpha \beta_Bs \over p_\perp^2(\alpha\beta_Bs+p_\perp^2)}|y_\perp)\partial_iU_yU_y^\dagger
 \end{eqnarray}
so it cancels with last term in the r.h.s of  Eq. (\ref{5.18}) and we obtain
\begin{eqnarray}
&&\hspace{-1mm}
{2\over s}g\!\int\! dy_\ast e^{i\beta_B y_\ast}\langle[\infty,y_\ast]_y^{nm}F_{\bu i}^m(y_\ast,y_\perp)\rangle^{\rm Fig. ~\ref{fig:5}}
\label{5.22}\\
&&\hspace{-1mm}
=~
-~ig^2f^{nkl}\!\int_0^\infty\!{\dhd\alpha\over\alpha}(y_\perp|{\alpha\beta_Bs\over p_\perp^2(\alpha\beta_Bs+p_\perp^2)}|y_\perp)
\calf_i^{kl}(\beta_B,y_\perp)
\nonumber
\end{eqnarray}
which is the light-cone result Eq. (\ref{3.15}). 

Thus, the final result for the sum of diagrams in Fig. \ref{fig:5} is Eq. (\ref{5.18}) 
\begin{eqnarray}
&&\hspace{-1mm}
\langle\calf_i^n(\beta_B, y_\perp)\rangle^{\rm Fig. ~\ref{fig:5}}
=~-ig^2f^{nkl}\!\int_{\sigma'}^{\sigma}\!{\dhd\alpha\over\alpha}\Big\{
(y_\perp|{1\over p_\perp^2}[\alpha \beta_Bs i\partial_iU+\partial_\perp^2Up_i]
{1\over \alpha\beta_Bs+p_\perp^2}U^\dagger|y_\perp)^{kl}
\nonumber\\
&&\hspace{32mm}
+~(y_\perp|{\alpha\beta_Bs\over p_\perp^2(\alpha\beta_Bs+p_\perp^2)}|y_\perp)
[\calf_i(\beta_B,y_\perp)-U_i(y_\perp)]^{kl}\Big\}
\label{5.23}\\
&&\hspace{-2mm}
=~-ig^2f^{nkl}\!\int_{\sigma'}^{\sigma}\!{\dhd\alpha\over\alpha}
(y_\perp|{p^j\over p_\perp^2}(2\partial_i\partial_jU+g_{ij}\partial_\perp^2U){1\over\alpha\beta_B s+p_\perp^2}U^\dagger
+{\alpha\beta_Bsp_\perp^{-2}\over \alpha\beta_Bs+p_\perp^2}\calf_i(\beta_B)|y_\perp)^{kl}
\nonumber
\end{eqnarray}
where we imposed our cutoff $\sigma>\alpha>\sigma'$. Again, let us note that the above expression 
is valid with our accuracy in the whole range of transverse momenta.

Similarly to Eq. (\ref{lvertalt}) we can rewrite this formula in the form without subtractions
\begin{eqnarray}
&&\hspace{-1mm}
\langle\calf_i^n(\beta_B, y_\perp)\rangle^{\rm Fig. ~\ref{fig:5}}~=~-ig^2f^{nkl}\!\int_{\sigma'}^{\sigma}\!{\dhd\alpha\over\alpha}
(y_\perp|-{p^j\over p_\perp^2}\calf_k(\beta_B)(i\!\stackrel{\leftarrow}{\partial}_l+U_l)\label{virtalt}\\
&&\hspace{31mm}
\times~(2\delta_j^k\delta_i^l-g_{ij}g^{kl})
U{1\over \alpha\beta_Bs+p_\perp^2}U^\dagger
+\calf_i(\beta_B){\alpha\beta_Bs\over p_\perp^2(\alpha\beta_Bs+p_\perp^2)}|y_\perp)^{kl}
\nonumber
\end{eqnarray}
where $\calf_k\stackrel{\leftarrow}{\partial}_l~\equiv~\partial_l\calf_k~=-i[p_l,\calf_k]$. Indeed, in the light-cone case $l_\perp^2\ll p_\perp^2$ 
one can neglect the operators with high collinear twist so both equations (\ref{5.23}) and (\ref{virtalt}) reduce to the last terms in the r.h.s's which are the same. 
Also, as we discussed above, in the shock-wave case ($l_\perp^2\sim p_\perp^2$) and $\beta_B$ small one can replace 
$\calf_i(\beta_B)$ by $U_i$ so the r.h.s's of Eq. (\ref{5.23}) and Eq. (\ref{virtalt})  coincide after some trivial algebra. Finally, if $l_\perp^2\sim p_\perp^2$ 
and $\beta_B\geq {1\over\sigma_\ast}$ we have $\alpha\beta_Bs\gg p_\perp^2$ so again the equations (\ref{5.23}) and (\ref{virtalt}) reduce to the last terms in the r.h.s's. 

\subsubsection{Virtual correction for the complex conjugate amplitude}
The calculation of the virtual correction in the complex conjugate amplitude is very similar so we will only outline it.
As in the previous Section, we start with the formula (\ref{3.16}) which can be rewritten as
\begin{eqnarray}
&&\hspace{-3mm}
{2\over s}\!\int\! dx_\ast~e^{-i\beta_Bx_\ast}\langle \tilF_{\bu i}^m(x_\ast,x_\perp)[x_\ast,\infty]_x^{mn}\rangle^{\rm 2nd}~=~
\label{5.24}\\
&&\hspace{-3mm}
=~
{2\over s}\!\int\! dx_\ast~e^{-i\beta_Bx_\ast}\Big\{
\beta_B\!\int^{\infty}_{x_\ast}\!\! dz_\ast \tilA_i^{m{\rm q}}(x_\ast,x_\perp)\big([x_\ast,z_\ast]_x \tilA^{\rm q}_\bu(z_\ast,x_\perp)[z_\ast,\infty]_x\big)^{mn}\rangle
\nonumber\\
&&\hspace{41mm}
+~{2i\over s}\!\int^{\infty}_{x_\ast}\!\! dz_\ast\langle \partial_i\tilA_\bu^{m{\rm q}}\big([x_\ast,z_\ast]_x\tilA^{\rm q}_\bu(z_\ast,x_\perp)[z_\ast,\infty]_x\big)^{mn}\rangle
\nonumber\\
&&\hspace{1mm}
-~{4\over s^2}\!\int^{\infty}_{x_\ast}dz_\ast\!\int_{z_\ast}^{\infty}\! dz'_\ast 
~\tilF_{\bu i}^m(x_\ast,x_\perp)\big([x_\ast,z_\ast]_x\tilA^{\rm q}_\bu(z_\ast,x_\perp)[z_\ast,z'_\ast]_x \tilA^{\rm q}_\bu(z'_\ast,x_\perp)[z'_\ast,\infty]_x\big)^{mn}
\Big\}
\nonumber
\end{eqnarray}
Using Eq. (\ref{B.29}) we get 
\begin{eqnarray}
&&\hspace{-3mm}
{2\over s}\!\int\! dx_\ast~e^{-i\beta_Bx_\ast} \langle \tilF_{\bu i}^m(x_\ast,x_\perp)[x_\ast,\infty]_x^{nm}\rangle^{\rm Fig. ~\ref{fig:5}}~=~
\label{5.25}\\
&&\hspace{3mm}
=~-{ig^2\over s}f^{nkl}\!\int_0^\infty\!{\dhd\alpha\over\alpha}\!\int_{-\infty}^{-\sigma_\ast}\! dx_\ast\!\int^{\infty}_{\sigma_\ast}\!\! dz_\ast
(x_\perp|\tilU e^{-i(\beta_B+{p_\perp^2\over\alpha s})x_\ast}\big\{\beta_B \ticalg_{i\bu}(-\infty,\infty;p_\perp)
\nonumber\\
&&\hspace{44mm}
+~{2\over s}p_i[\ticalg_{\bu\bu}(-\infty,\infty;p_\perp)+\ticalq_{\bu\bu}(-\infty,\infty;p_\perp)]\big\}
e^{i{p_\perp^2\over\alpha s}z_\ast}|x_\perp)^{kl}
\nonumber\\
&&\hspace{-1mm}
+~{g^2\over s}if^{nkl}\!\int_{-\sigma_\ast}^{\sigma_\ast}\!dx_\ast~e^{-i\beta_Bx_\ast} \!\int_{x_\ast}^\infty\! dz_\ast
\!\int_0^\infty\!{\dhd\alpha\over\alpha}\Big\{(x_\perp|
\big[-\beta_B 
([\infty,x_\ast]\ticalg_{i \bu}(x_\ast,z_\ast;p_\perp)[z_\ast,\infty])^{kl}
\nonumber\\
&&\hspace{-1mm}
-~{2i\over s}
\big\{\partial_i\big([\infty,x_\ast]\ticalo_{\bu\bu}(x_\ast,z_\ast;p_\perp)[z_\ast,\infty]\big)
-~\partial_i([\infty,x_\ast])\ticalo_{\bu\bu}(x_\ast,z_\ast;p_\perp)[z_\ast,\infty]
\nonumber\\
&&\hspace{44mm}
-~[\infty,x_\ast]\ticalo_{\bu\bu}(x_\ast,z_\ast;p_\perp)\partial_i([z_\ast,\infty])\big\}^{kl}\big]e^{-i{p_\perp^2\over\alpha s}(x-z)_\ast}|x_\perp)
\nonumber\\
&&\hspace{-1mm}
+~{4\over s^2}\!\int_{z_\ast}^{\infty}\! dz'_\ast 
~(x_\perp|\big([\infty,x_\ast]\tilF_{\bu i}(x_\ast)[x_\ast,z_\ast]\ticalo_{\bu\bu}(z_\ast,z'_\ast;p_\perp)[z'_\ast,\infty]\big)^{kl}
e^{-i{p_\perp^2\over\alpha s}(z-z')_\ast}|x_\perp)\Big\}
\nonumber
\end{eqnarray}
Similarly to Eq. (\ref{5.17}) it is possible to demonstrate that the last three lines in the r.h.s. of this equation
exceed our accuracy, and moreover, one can neglect factors $e^{-i\beta_B\sigma_\ast}$.  Using formulas (\ref{B.30}) 
for $\ticalg_{\mu\nu}$ and (\ref{B.10})  for $\ticalq_{\mu\nu}$
we obtain the virtual correction in the complex conjugate amplitude in the form
\begin{eqnarray}
&&\hspace{-1mm}
\langle \ticalf_i^n(\beta_B, x_\perp)\rangle^{\sigma} ~=~{2g\over s}\!\int\! dx_\ast~e^{-i\beta_Bx_\ast} \langle \tilF_{\bu i}^m(x_\ast,x_\perp)[x_\ast,\infty]_x^{mn}\rangle^{\sigma}~
\nonumber\\
&&\hspace{11mm}
=~-ig^2f^{nkl}\!\int_{\sigma'}^{\sigma}{\!\dhd\alpha\over \alpha}\Big\{(x_\perp|\tilU {1\over\alpha\beta_Bs+p_\perp^2}
\big\{-i\alpha\beta_Bs \partial_i\tilU^\dagger
+p_i\partial_\perp^2\tilde{U}^\dagger\big\}{1\over p_\perp^2}|x_\perp)^{kl}
\nonumber\\
&&\hspace{33mm}
+~\big(\tilcaf_i(\beta_B, x_\perp)-i\partial_i\tilU_x\tilU^\dagger_x\big)^{kl}
(x_\perp|{\alpha\beta_Bs\over p_\perp^2(\alpha\beta_Bs+p_\perp^2)}|x_\perp)\Big\}
\label{5.26}
\end{eqnarray}
where we have imposed our cutoffs in $\alpha$ and used the formula
\begin{eqnarray}
&&\hspace{-5mm}
\partial_\perp^2U^\dag_z~=~\!g\int_{-\infty}^\infty\! dz_\ast~[-\infty,z_\ast]_z
\nonumber\\
&&\hspace{-5mm}
\times~\Big(-{2i\over s} D^jF_{\bu j}(z_\ast,z_\perp)[z_\ast,\infty]_z+{8g\over s^2}\!\int^{\infty}_{z_\ast}\!dz'_\ast
~F_{\bu j}(z_\ast,z_\perp) [z_\ast,z'_\ast]_zF_\bu^{~j}(z'_\ast,z_\perp)[z'_\ast,\infty]_z
\Big)
\nonumber
\end{eqnarray}
Similarly to Eq. (\ref{5.18})  this expression is also valid in the light-cone case $l_\perp^2\ll p_\perp^2$ where it coincides with Eq. (\ref{3.18}).

Alternatively, one can use the expression without subtractions (cf. Eq. (\ref{virtalt}))
\begin{eqnarray}
&&\hspace{-1mm}
\langle \ticalf_i^n(\beta_B, x_\perp)\rangle^{\sigma} ~=~-ig^2f^{nkl}\!\int_{\sigma'}^{\sigma}{\!\dhd\alpha\over \alpha}
(x_\perp|
\tilU{1\over \alpha \beta_Bs+p_\perp^2}\tilU^\dagger
\label{virtaltcc}\\
&&\hspace{25mm}
\times~(2\delta_i^k\delta_j^l-g_{ij}g^{kl} )(i\partial_k-\tilU_k)\tilde{\calf}_l(\beta_B)
{p^j\over p_\perp^2}
+~\tilcaf_i(\beta_B)
{\alpha\beta_Bs\over p_\perp^2(\alpha\beta_Bs+p_\perp^2)}|x_\perp)^{kl}
\nonumber
\end{eqnarray}
%

\section{Evolution equation for gluon TMD}

 Now we are in a position to assemble all leading-order contributions to the rapidity evolution of gluon TMD. Adding the production part 
 (\ref{3.1}) with Lipatov vertices (\ref{lvertalt}) and (\ref{lvertaltcc}) and the virtual parts from previous Section (\ref{virtalt}) and (\ref{virtaltcc}) we obtain
\begin{eqnarray}
&&\hspace{-1mm}
\big( \tilcaf_i^a(\beta_B,x_\perp) \calf_j^a(\beta_B, y_\perp)\big)^{\ln\sigma}~
\label{6.1}\\
&&\hspace{-1mm}
=~-g^2\!\int_{\sigma'}^{\sigma}\!{\dhd\alpha\over 2\alpha}\!\int\!\dhd^2k_\perp
~{\rm Tr}\{\tiL_i^{~\mu}(k,x_\perp,\beta_B)^{\rm light-like}
L_{\mu j}(k,y_\perp,\beta_B)^{\rm light-like}\}
\nonumber\\
&&\hspace{-1mm}
-~g^2\!\int_{\sigma'}^{\sigma}\!{\dhd\alpha\over \alpha}{\rm Tr}\Big\{\ticalf_i(\beta_B,x_\perp)
(y_\perp|-{p^m\over p_\perp^2}\calf_k(\beta_B)(i\!\stackrel{\leftarrow}{\partial}_l+U_l)(2\delta_m^k\delta_j^l-g_{jm}g^{kl})
U{1\over \alpha\beta_Bs+p_\perp^2}U^\dagger
\nonumber\\
&&\hspace{33mm}
+~\calf_j(\beta_B){\alpha\beta_Bs\over p^2_\perp(\alpha\beta_Bs+p_\perp^2)}|y_\perp)
\nonumber\\
&&\hspace{-1mm}
+~(x_\perp|
\tilU{1\over \alpha \beta_Bs+p_\perp^2}\tilU^\dagger(2\delta_i^k\delta_m^l-g_{im}g^{kl} )(i\partial_k-\tilU_k)\ticalf_l(\beta_B)
{p^m\over p_\perp^2}
\nonumber\\
&&\hspace{33mm}
+~\tilcaf_i(\beta_B)
{\alpha\beta_Bs\over p_\perp^2(\alpha\beta_Bs+p_\perp^2)}|x_\perp)\calf_j\big(\beta_B, y_\perp\big)\Big\}~+~O(\alpha_s^2)
\nonumber
\end{eqnarray}
where Tr is a trace in the adjoint representation.
In the explicit form the evolution equation reads
\begin{eqnarray}
&&\hspace{-1mm}
{d\over d\ln\sigma}\tilcaf_i^a(\beta_B, x_\perp) \calf_j^a(\beta_B, y_\perp)~
\label{master1alt}\\
&&\hspace{-1mm}
=~-\alpha_s{\rm Tr}\Big\{\!\int\!\dhd^2k_\perp
(x_\perp|\Big\{\tilU{1\over\sigma\beta_Bs+p_\perp^2}
(\tilU^\dagger k_k+p_k\tilU^\dagger){\sigma \beta_Bsg_{\mu i}-2k^\perp_{\mu}k_i\over\sigma \beta_Bs+k_\perp^2}
\nonumber\\
&&\hspace{-1mm}
-~2k^\perp_\mu g_{ik}\tilU{1\over \sigma\beta_Bs+p_\perp^2}\tilU^\dagger -2g_{\mu k} \tilU{p_i\over\sigma\beta_Bs+p_\perp^2}
\tilU^\dagger+{2k^\perp_\mu\over k_\perp^2}g_{ik}\Big\}
\ticalf^k\big(\beta_B+{k_\perp^2\over\sigma s}\big)|k_\perp)
\nonumber\\
&&\hspace{5mm}
\times~(k_\perp|\calf^l\big(\beta_B+{k_\perp^2\over\sigma s}\big)
\Big\{{\sigma \beta_Bs\delta^\mu_j-2k_\perp^{\mu}k_j\over\sigma\beta_Bs+k_\perp^2}(k_lU+Up_l){1\over\sigma \beta_Bs+p_\perp^2}U^\dagger
\nonumber\\
&&\hspace{22mm}
-2k_\perp^\mu g_{jl}U{1\over \sigma\beta_Bs+p_\perp^2}U^\dagger-~2\delta_l^\mu U{p_j\over\sigma\beta_Bs+p_\perp^2}U^\dagger
+2g_{jl}{k_\perp^\mu\over k_\perp^2}\Big\}|y_\perp)
\nonumber\\
&&\hspace{5mm}
+~2\ticalf_i(\beta_B, x_\perp)
(y_\perp|-{p^m\over p_\perp^2}\calf_k(\beta_B)(i\!\stackrel{\leftarrow}{\partial}_l+U_l)(2\delta_m^k\delta_j^l-g_{jm}g^{kl})
U{1\over \sigma\beta_Bs+p_\perp^2}U^\dagger
\nonumber\\
&&\hspace{77mm}
+\calf_j(\beta_B){\sigma\beta_Bs\over p^2_\perp(\sigma\beta_Bs+p_\perp^2)}|y_\perp)
\nonumber\\
&&\hspace{5mm}
+~2(x_\perp|
\tilU{1\over \sigma \beta_Bs+p_\perp^2}\tilU^\dagger(2\delta_i^k\delta_m^l-g_{im}g^{kl} )(i\partial_k-\tilU_k)\ticalf_l(\beta_B)
{p^m\over p_\perp^2}
\nonumber\\
&&\hspace{44mm}
+~\tilcaf_i(\beta_B)
{\sigma\beta_Bs\over p_\perp^2(\sigma\beta_Bs+p_\perp^2)}|x_\perp)
\calf_j(\beta_B, y_\perp)\Big\}~+~O(\sigma_s^2)
\nonumber
\end{eqnarray}
The operators $\tilcaf_j(\beta)$ and 
$\calf_i(\beta)$
are defined as usual, see Eq. (\ref{defoperf})
\begin{eqnarray}
&&\hspace{-1mm}
(x_\perp|\tilcaf_i(\beta)|k_\perp)
~=~{2\over s}\int\! dx_\ast ~\ticalf_i(x_\ast, x_\perp)e^{-i\beta x_\ast+i(k,x)_\perp}
\nonumber\\
&&\hspace{-1mm}
(k_\perp|\calf_i(\beta)|y_\perp)
~=~{2\over s}\int\! dy_\ast ~e^{i\beta y_\ast-i(k,y)_\perp}\calf_i(y_\ast, y_\perp)
\label{6.4}
\end{eqnarray}

The evolution equation (\ref{master1alt}) can be rewritten in the form where cancellation of IR and UV divergencies is evident
\begin{eqnarray}
&&\hspace{-1mm}
{d\over d\ln\sigma}\tilcaf_i^a(\beta_B,x_\perp) \calf_j^a(\beta_B,y_\perp)~
\label{master2alt}\\
&&\hspace{-1mm}
=~-\alpha_s{\rm Tr}\Big\{\!\int\!\dhd^2k_\perp
(x_\perp|\Big\{\tilU{1\over\sigma\beta_Bs+p_\perp^2}
(\tilU^\dagger k_k+p_k\tilU^\dagger){\sigma \beta_Bsg_{\mu i}-2k^\perp_{\mu}k_i\over\sigma \beta_Bs+k_\perp^2}
\nonumber\\
&&\hspace{-1mm}
-~2k^\perp_\mu g_{ik}\tilU{1\over \sigma\beta_Bs+p_\perp^2}\tilU^\dagger
-2g_{\mu k} \tilU{p_i\over\sigma\beta_Bs+p_\perp^2}\tilU^\dagger\Big\}
\ticalf^k\big(\beta_B+{k_\perp^2\over\sigma s}\big)|k_\perp)
\nonumber\\
&&\hspace{-1mm}
\times~(k_\perp|\calf^l\big(\beta_B+{k_\perp^2\over\sigma s}\big)
\Big\{{\sigma \beta_Bs\delta^\mu_j-2k_\perp^{\mu}k_j\over\sigma\beta_Bs+k_\perp^2}(k_lU+Up_l){1\over\sigma \beta_Bs+p_\perp^2}U^\dagger
\nonumber\\
&&\hspace{-1mm}-2k_\perp^\mu g_{jl}U{1\over \sigma\beta_Bs+p_\perp^2}U^\dagger
-~2\delta_l^\mu U{p_j\over\sigma\beta_Bs+p_\perp^2}U^\dagger\Big\}|y_\perp)+~2\int\dhd^2k_\perp(x_\perp|\ticalf_i\big(\beta_B+{k_\perp^2\over\sigma s}\big)|k_\perp)
\nonumber\\
&&\hspace{-1mm}
\times(k_\perp|\calf^l\big(\beta_B+{k_\perp^2\over\sigma s}\big)
\Big\{{k_j\over k_\perp^2}{\sigma \beta_Bs+2k_\perp^2\over\sigma\beta_Bs+k_\perp^2}(k_lU+Up_l)
{1\over\sigma \beta_Bs+p_\perp^2}U^\dagger
\nonumber\\
&&\hspace{44mm}
+~2U{g_{jl}\over \sigma\beta_Bs+p_\perp^2}U^\dagger-2{k_l\over k_\perp^2}U{p_j\over \sigma\beta_Bs+p_\perp^2}U^\dagger
\Big\}
|y_\perp)
\nonumber\\
&&\hspace{-1mm}
+~2\int\dhd^2k_\perp(x_\perp|\Big\{\tilU{1\over\sigma\beta_Bs+p_\perp^2}
(\tilU^\dagger k_k+p_k\tilU^\dagger){k_i\over k_\perp^2}
{\sigma \beta_Bs+2k_\perp^2\over\sigma \beta_Bs+k_\perp^2}
+2\tilU{g_{ik}\over \sigma\beta_Bs+p_\perp^2}\tilU^\dagger
\nonumber\\
&&\hspace{-1mm}
-~2\tilU{p_i\over\sigma\beta_Bs+p_\perp^2}\tilU^\dagger{k_k\over k_\perp^2}\Big\}
\ticalf^k\big(\beta_B+{k_\perp^2\over\sigma s}\big)|k_\perp)(k_\perp|\calf_j\big(\beta_B+{k_\perp^2\over\sigma s}\big)|y_\perp)
\nonumber\\
&&\hspace{-1mm}
+~2\ticalf_i(\beta_B, x_\perp)
(y_\perp|-{p^m\over p_\perp^2}\calf_k(\beta_B)(i\!\stackrel{\leftarrow}{\partial}_l+U_l)(2\delta_m^k\delta_j^l-g_{jm}g^{kl})
U{1\over \sigma\beta_Bs+p_\perp^2}U^\dagger|y_\perp)
\nonumber\\
&&\hspace{-1mm}
+~2(x_\perp|
\tilU{1\over \sigma \beta_Bs+p_\perp^2}\tilU^\dagger(2\delta_i^k\delta_m^l-g_{im}g^{kl} )(i\partial_k-\tilU_k)\tilde{\calf}_l(\beta_B)
{p^m\over p_\perp^2}|x_\perp)\calf_j(\beta_B, y_\perp)
\nonumber\\
&&\hspace{-1mm}
-~4\!\int\!{\dhd^2k_\perp\over k_\perp^2}\Big[\ticalf_i\big(\beta_B+{k_\perp^2\over\sigma s}, x_\perp\big)
\calf_j\big(\beta_B+{k_\perp^2\over\sigma s}, y_\perp\big)e^{i(k,x-y)_\perp}
\nonumber\\
&&\hspace{55mm}
~-{\sigma\beta_Bs\over \sigma\beta_Bs+k_\perp^2}\ticalf_i(\beta_B, x_\perp)\calf_j(\beta_B, y_\perp)\Big]\Big\}
~+~O(\alpha_s^2).
\nonumber
\end{eqnarray}
The evolution equation (\ref{master2alt}) is one of the main results of this paper. 
It describes the rapidity evolution of the operator (\ref{operator})  at any Bjorken $x_B\equiv\beta_B$ and any transverse momenta. 

Let us discuss the gauge invariance of this equation. The l.h.s. is gauge invariant after taking into account gauge link at $+\infty$ as shown
in Eq. (\ref{inftylink}). As to the right side, it was obtained by calculation in the background field and promoting the background fields 
to operators in a usual way. However, we performed our calculations in a specific background field $A_\bu(x_\ast, x_\perp)$ with a finite support in $x_\perp$ and we need to address the question how can we restore the r.h.s. of Eq. (\ref{master2alt}) in a generic field $A_\mu$. 
It is easy to see how one can restore the gauge-invariant form: just add gauge link at $+\infty p_1$ or $-\infty p_1$ appropriately. 
For example, the terms $U_z(z|{1\over\sigma\beta s+p_\perp^2}|z')U^\dagger_{z'}$ in r.h.s. of should be replaced by 
$U_z[z_\perp-\infty p_1,z'_\perp-\infty p_1](z|{1\over\sigma\beta s+p_\perp^2}|z')U^\dagger_{z'}$. After performing these insertions we will have the result which is (i) gauge invariant and (ii) coincides with Eq. (\ref{master2alt}) for our choice of background field. At this step, the  background 
fields in the r.h.s. of Eq. (\ref{master2alt}) can be promoted to operators. However, the explicit display of these gauge links at $\pm \infty$  will
make the evolution equation much less readable so we will assume they are always in place rather than written explicitly.

When we consider the evolution of gluon TMD (\ref{gTMD}) given by the matrix element (\ref{TMD}) of the operator (\ref{operator}) we need 
to take into account the kinematical constraint $k_\perp^2\leq \alpha(1-\beta_B)s$ in the production part of the amplitude. 
Indeed, as we discussed in Sect. \ref{Sect3.3}, the initial hadron's momentum is $p\simeq p_2$ so the sum of 
the fraction $\beta_Bp_2$  and the fraction ${k_\perp^2\over\alpha s}p_2$ carried by the emitted gluon should be smaller than $p_2$.
We obtain ($\eta\equiv\ln\sigma$)
\footnote{
Strictly speaking, we need to consider matrix element $\langle p|\tilcaf^{ai}(\beta_B, x_\perp) \calf_i^a(\beta_B, y_\perp)|p+\xi p_2\rangle$ 
proportional to $\delta(\xi)$, see Eq. (\ref{TMD})
} 
\begin{eqnarray}
&&\hspace{-1mm}
{d\over d\eta}\langle p|\tilcaf_i^a(\beta_B, x_\perp) \calf_j^a(\beta_B, y_\perp)|p\rangle^\eta~
\label{masterdis}\\
&&\hspace{-1mm}
=~-\alpha_s\langle p|{\rm Tr}\Big\{\!\int\!\dhd^2k_\perp\theta\big(1-\beta_B-{k_\perp^2\over\sigma s}\big)\Big[
(x_\perp|\Big( U{1\over\sigma\beta_Bs+p_\perp^2}
( U^\dagger k_k+p_k U^\dagger)
\nonumber\\
&&\hspace{-1mm}
\times~{\sigma \beta_Bsg_{\mu i}-2k^\perp_{\mu}k_i\over\sigma \beta_Bs+k_\perp^2}-~2k^\perp_\mu g_{ik} U{1\over \sigma\beta_Bs+p_\perp^2} U^\dagger
-2g_{\mu k}  U{p_i\over\sigma\beta_Bs+p_\perp^2} U^\dagger\Big)
\ticalf^k\big(\beta_B+{k_\perp^2\over\sigma s}\big)|k_\perp)
\nonumber\\
&&\hspace{-1mm}
\times~(k_\perp|\calf^l\big(\beta_B+{k_\perp^2\over\sigma s}\big)
\Big({\sigma \beta_Bs\delta^\mu_j-2k_\perp^{\mu}k_j\over\sigma\beta_Bs+k_\perp^2}(k_lU+Up_l){1\over\sigma \beta_Bs+p_\perp^2}U^\dagger
\nonumber\\
&&\hspace{-1mm}-2k_\perp^\mu g_{jl}U{1\over \sigma\beta_Bs+p_\perp^2}U^\dagger
-~2\delta_l^\mu U{p_j\over\sigma\beta_Bs+p_\perp^2}U^\dagger\Big)|y_\perp)
\nonumber\\
&&\hspace{-1mm}
+~2(x_\perp|\ticalf_i\big(\beta_B+{k_\perp^2\over\sigma s}\big)|k_\perp)
(k_\perp|\calf^l\big(\beta_B+{k_\perp^2\over\sigma s}\big)
\Big({k_j\over k_\perp^2}{\sigma \beta_Bs+2k_\perp^2\over\sigma\beta_Bs+k_\perp^2}(k_lU+Up_l)
{1\over\sigma \beta_Bs+p_\perp^2}U^\dagger
\nonumber\\
&&\hspace{-1mm}
+~2U{g_{jl}\over \sigma\beta_Bs+p_\perp^2}U^\dagger-2{k_l\over k_\perp^2}U{p_j\over \sigma\beta_Bs+p_\perp^2}U^\dagger
\Big)
|y_\perp)
\nonumber\\
&&\hspace{-1mm}
+~2(x_\perp|\Big( U{1\over\sigma\beta_Bs+p_\perp^2}
( U^\dagger k_k+p_k U^\dagger){k_i\over k_\perp^2}
{\sigma \beta_Bs+2k_\perp^2\over\sigma \beta_Bs+k_\perp^2}
+2 U{g_{ik}\over \sigma\beta_Bs+p_\perp^2} U^\dagger
\nonumber\\
&&\hspace{-1mm}
-~2 U{p_i\over\sigma\beta_Bs+p_\perp^2} U^\dagger{k_k\over k_\perp^2}\Big)
\ticalf^k\big(\beta_B+{k_\perp^2\over\sigma s}\big)|k_\perp)(k_\perp|\calf_j\big(\beta_B+{k_\perp^2\over\sigma s}\big)|y_\perp)\Big]
\nonumber\\
&&\hspace{-1mm}
+~2\ticalf_i(\beta_B, x_\perp)
(y_\perp|-{p^m\over p_\perp^2}\calf_k(\beta_B)(i\!\stackrel{\leftarrow}{\partial}_l+U_l)(2\delta_m^k\delta_j^l-g_{jm}g^{kl})
U{1\over \sigma\beta_Bs+p_\perp^2}U^\dagger|y_\perp)
\nonumber\\
&&\hspace{-1mm}
+~2(x_\perp|
 U{1\over \sigma \beta_Bs+p_\perp^2} U^\dagger(2\delta_i^k\delta_m^l-g_{im}g^{kl} )(i\partial_k- U_k)\tilde{\calf}_l(\beta_B)
{p^m\over p_\perp^2}|x_\perp)\calf_j(\beta_B, y_\perp)
\nonumber\\
&&\hspace{-1mm}
-~4\!\int\!{\dhd^2k_\perp\over k_\perp^2}\Big[\theta\big(1-\beta_B-{k_\perp^2\over\sigma s}\big)
\ticalf_i\big(\beta_B+{k_\perp^2\over\sigma s}, x_\perp\big)
\calf_j\big(\beta_B+{k_\perp^2\over\sigma s}, y_\perp\big)e^{i(k,x-y)_\perp}
\nonumber\\
&&\hspace{-1mm}
-~{\sigma\beta_Bs\over \sigma\beta_Bs+k_\perp^2}\ticalf_i(\beta_B, x_\perp)\calf_j(\beta_B, y_\perp)\Big]\Big\}|p\rangle^\eta
~+~O(\alpha_s^2)
\nonumber
\end{eqnarray}
Note that we erased tilde from Wilson lines since we have a sum over full set of states and gluon operators at space-like (or light-like) intervals commute with each other. \footnote{We have left $\tilF$ as a reminder of different signs in the exponents of Fourier transforms 
in the definitions (\ref{kalf}) and (\ref{tilkaf}).}
 This equation describes the rapidity evolution of gluon TMD (\ref{gTMD}) with rapidity cutoff (\ref{cutoff}) in the whole
range of $\beta_B=x_B$ and $k_\perp$ ($\sim|x-y|_\perp^{-1}$). In the next section we will consider some specific cases.

\section{BK, DGLAP, and Sudakov limits of TMD evolution equation}

\subsection{Small-x case: BK evolution of the Weizsacker-Williams distribution}

First, let us consider the evolution of  Weizsacker-Williams (WW) unintegrated gluon distribution (\ref{WW}) which can be obtained 
from Eq. (\ref{masterdis}) by setting $\beta_B=0$. Moreover, in the small-$x$ regime 
it is assumed that the energy is much higher than anything else so the characteristic transverse momenta 
$p_\perp^2\sim(x-y)_\perp^{-2}\ll s$ and in the whole range of evolution ($1\gg\sigma\gg {(x-y)_\perp^{-2}\over s}$) we have 
${p_\perp^2\over\sigma s}\ll 1$, hence the kinematical constraint $\theta\big(1-\beta_B-{k_\perp^2\over\alpha s}\big)$ in Eq. (\ref{masterdis}) can be omitted. 
Under these assumptions, all $\calf_i\big(\beta_B+{p_\perp^2\over\sigma s}\big)$ and $\calf_i(\beta_B)$ 
can be replaced by $i\partial_iUU^\dagger$  and similarly for the complex conjugate amplitude. To simplify algebra, it is convenient to take the production part 
of the kernel in the form of product of Lipatov vertices
(\ref{lvertax}) and (\ref{lvertaxcc}) noting that the ``subtraction terms'' ${\breve{\tilcaf}}_i$ and ${\breve\calf}_j$ vanish in this limit. One obtains
the rapidity evolution of the WW distribution in the form
\begin{eqnarray}
&&\hspace{-1mm}
{d\over d\ln\sigma}\tilU^a_i(x_\perp)U^a_j(y_\perp)~
=~-4\alpha_s{\rm Tr}\Big\{\big(x_\perp\big|
\tilU p_i\tilU^\dagger\big(\tilU{p^k\over p_\perp^2}\tilU^\dagger 
-{p^k\over p_\perp^2}\big)\big(U{p_k\over p_\perp^2}U^\dagger 
-{p_k\over p_\perp^2}\big)Up_j U^\dagger|y_\perp)
\nonumber\\
&&\hspace{37mm}
-~\Big[
(x_\perp|\tilU{p_ip^k\over p_\perp^2}\tilU^\dagger{p_k\over p_\perp^2}|x_\perp)
-\half(x_\perp|{1\over p_\perp^2}|x_\perp)\tilU_i(x_\perp)\Big]
U_j(y_\perp)
\nonumber\\
&&\hspace{44mm}
-~\tilU_i(x_\perp)\Big[
(y_\perp|{p^k\over p_\perp^2}U{p_jp_k\over p_\perp^2}U^\dagger|y_\perp)
-\half(y_\perp|{1\over p_\perp^2}|y_\perp)U_j
(y_\perp)\Big]\Big\}
\label{7.1}
\end{eqnarray}
where we used the formula
\begin{eqnarray}
&&\hspace{-1mm}
-(y_\perp|{p^k\over p_\perp^2}(2\partial_j\partial_kU+g_{jk}\partial_\perp^2U){1\over\sigma\beta s+p_\perp^2}U^\dagger|y_\perp)
\nonumber\\
&&\hspace{37mm}
=~(y_\perp|{p^k\over p_\perp^2}U{p_\perp^2g_{jk}+2p_jp_k\over \sigma \beta_Bs+p_\perp^2}U^\dagger
-{1\over \sigma \beta_Bs+p_\perp^2}U_j|y_\perp)
\label{7.2}
\end{eqnarray}
In this form Eq. (\ref{7.1})  agrees with the results of Ref. \cite{mobzor}.  To see the relation
to the BK equation it
 is convenient to rewrite Eq. (\ref{7.1})  as follows \cite{proceedings} (cf. Ref. \cite{domumuxi}):
\begin{eqnarray}
&&\hspace{-1mm} 
{d\over d\eta}\tilU^a_i(z_1) U^a_j(z_2)
\label{7.3}\\
&&\hspace{-1mm}
=~-{g^2\over 8\pi^3}{\rm Tr}\big\{
(-i\partial^{z_1}_i+\tilU^{z_1}_i)\big[\!\int\! d^2z_3(\tilU_{z_1}\tilU^\dagger_{z_3}-1)
{z_{12}^2\over z_{13}^2z_{23}^2}(U_{z_3}U^\dagger_{z_2}-1)\big]
(i\stackrel{\leftarrow}{\partial^{z_2}_j}+U^{z_2}_j)\big\}
\nonumber 
\end{eqnarray}
where $\eta\equiv\ln\sigma$ as usual. In this equation all indices are 2-dimensional and Tr stands for the trace in the adjoint representation. 
It is easy to see that the expression in the square brackets is actually the BK kernel for the double-functional integral 
for cross sections \cite{mobzor,difope}. 
Hereafter, to ensure gauge invariance, $U_i(z_\perp)$ must be understood as \\
$U_i(z_\perp)~\equiv~\calf_i(0,z_\perp)~=~{2\over s}\!\int\! dz_\ast~[\infty, z_\ast]_zF_{\bu i}(z_\ast,z_\perp)[z_\ast,\infty]$ 
and gauge links at $\infty p_1$ 
must be inserted as discussed after Eq. (\ref{master2alt}).

It is worth noting that Eq. (\ref{7.3}) holds true also at small $\beta_B$ up to $\beta_B\sim {(x-y)_\perp^{-2}\over s}$ since in the whole range of evolution 
$1\gg\sigma\gg{(x-y)_\perp^{-2}\over s}$ one can neglect $\sigma\beta_Bs$ in comparison to $p_\perp^2$ in Eq. (\ref{masterdis}).
This effectively reduces $\beta_B$ to 0 so one reproduces Eq. (\ref{7.3}).

\subsection{Large transverse momenta and the light-cone limit \label{sec:6.2}}

Now let us discuss the case when $\beta_B=x_B\sim 1$ and $(x-y)_\perp^{-2}\sim s$.  At the start of the evolution
(at $\sigma\sim 1$) the cutoff in $p_\perp^2$ in the integrals of Eq. (\ref{master2alt}) is $\sim(x-y)_\perp^{-2}$. However, as the evolution
in rapidity ($\sim\ln\sigma$) progresses the characteristic $p_\perp^2$ becomes smaller due to the kinematical constraint 
$p_\perp^2<\sigma (1-\beta_B)s$. Due to this kinematical constraint evolution in $\sigma$ is correlated with the evolution in $p_\perp^2$:
if $\sigma\gg \sigma'$ the corresponding transverse momenta of background fields ${p'_\perp}^2$ are much smaller than $p_\perp^2$ in quantum loops.
This means that during the evolution we are always in the light-cone case considered in Sect. 3 and therefore the evolution equation for
$\beta_B=x_B\sim 1$ and $(x-y)_\perp^{-2}\sim s$ is Eq. (\ref{3.21}) which reduces to the system of 
evolution equations for  gluon TMDs $\cald(\beta_B,|z_\perp|,\ln\sigma)$ and $\calh(\beta_B,|z_\perp|,\ln\sigma)$ in the case of unpolarized hadron.

\subsection{Sudakov logarithms}

Finally, let us consider the evolution of $\cald(x_B,k_\perp,\eta=\ln\sigma)$ in the region 
where  $x_B\equiv\beta_B\sim 1$ and $k_\perp^2\sim(x-y)_\perp^{-2}\sim$ few $GeV^2$. 
In this case the integrals over $p_\perp^2$ in the production part of the kernel (\ref{masterdis}) are $\sim(x-y)_\perp^{-2}\sim k_\perp^2$   so that 
$p_\perp^2\ll\sigma\beta_Bs$ for the whole range of evolution $1>\sigma>{k_\perp^2\over s}$. 
For the same reason, the kinematical constraint $\theta\big(1-\beta_B-{p_\perp^2\over\sigma s}\big)$ 
in the last line of Eq. (\ref{masterdis}) can
be omitted and we get
\begin{eqnarray}
&&\hspace{-1mm}
{d\over d\ln\sigma}\langle p|\tilcaf^a_i(\beta_B, x_\perp) \calf_j^a(\beta_B, y_\perp)|p\rangle^{\rm real}~
\label{sudreal}\\
&&\hspace{-1mm}
=~4\alpha_sN_c\!\int\!{\dhd^2 p_\perp\over p_\perp^2}e^{i(p,x-y)_\perp} 
\langle p|\tilcaf^a_i\big(\beta_B+{p_\perp^2\over\sigma s},x_\perp\big)
\calf^a_j\big(\beta_B+{p_\perp^2\over\sigma s},y_\perp\big)|p\rangle
\nonumber
\end{eqnarray}
As to the virtual part
\begin{eqnarray}
&&\hspace{-1mm}
{d\over d\ln\sigma}\langle p|\tilcaf^a_i(\beta_B, x_\perp) \calf_j^a(\beta_B, y_\perp)|p\rangle^{\rm virtual}~
\label{sudvirtual}\\
&&\hspace{-1mm}
=~4\alpha_sN_c\!\int\!{\dhd^2 p_\perp\over p_\perp^2}\Big[-{\sigma \beta_Bs\over\sigma \beta_Bs+p_\perp^2}\langle p|\tilcaf^a_i(\beta_B, x_\perp)\calf^a_j(\beta_B, y_\perp)|p\rangle
\Big]
\nonumber\\
&&\hspace{-1mm}
-~2\alpha_s{\rm Tr}\langle p|(x_\perp|\tilU{1\over \sigma \beta_Bs+p_\perp^2}\tilU^\dagger
(2\delta_i^k\delta_m^l-g_{im}g^{kl} )(i\partial_k-\tilU_k)\ticalf_l(\beta_B){p^m\over p_\perp^2}
|x_\perp)\calf_j(\beta_B, y_\perp)
\nonumber\\
&&\hspace{-1mm}
-~\ticalf_i(\beta_B, x_\perp)
(y_\perp|{p^m\over p_\perp^2}\calf_k(\beta_B)(i\!\stackrel{\leftarrow}{\partial}_l+U_l)(2\delta_m^k\delta_j^l-g_{jm}g^{kl})
U{1\over \sigma\beta_Bs+p_\perp^2}U^\dagger|y_\perp)
|p\rangle,
\nonumber
\end{eqnarray}
the two last lines can be omitted. Indeed, as we saw in the end of Sect. \ref{sect4.4.4}, these terms are 
non-vanishing only for the region of large $p_\perp^2\sim \sigma\beta_Bs$. In this region one can expand 
the operator $\calo~\equiv~\calf_k(\beta_B)(i\!\stackrel{\leftarrow}{\partial}_l+U_l)(2\delta_m^k\delta_j^l-g_{jm}g^{kl})U$  
as $\calo(z_\perp)~=~\calo(y_\perp)+(y-z)_i\partial_i\calo(y_\perp)+...$ and get
\begin{eqnarray}
&&\hspace{-1mm} 
(y_\perp|{p_m\over p_\perp^2}\calo{1\over \sigma\beta_Bs+p_\perp^2}|y_\perp)~
=~
\calo_y(y_\perp|{p_m\over p_\perp^2(\sigma\beta_Bs+p_\perp^2)}|y_\perp)
+{i\partial_m\calo_y\over 4\pi\sigma\beta_Bs}+...
\nonumber
\end{eqnarray}
The first term in the r.h.s of this equation is obviously zero while the second is $O\big({m_N^2\over\sigma\beta_Bs}\big)$ 
in comparison to the leading first term in the r.h.s. of Eq. (\ref{sudvirtual}) (the transverse momenta
inside the hadron target are $\sim m_N\sim 1$GeV). 

\begin{figure}[htb]
\begin{center}
\includegraphics[width=38mm]{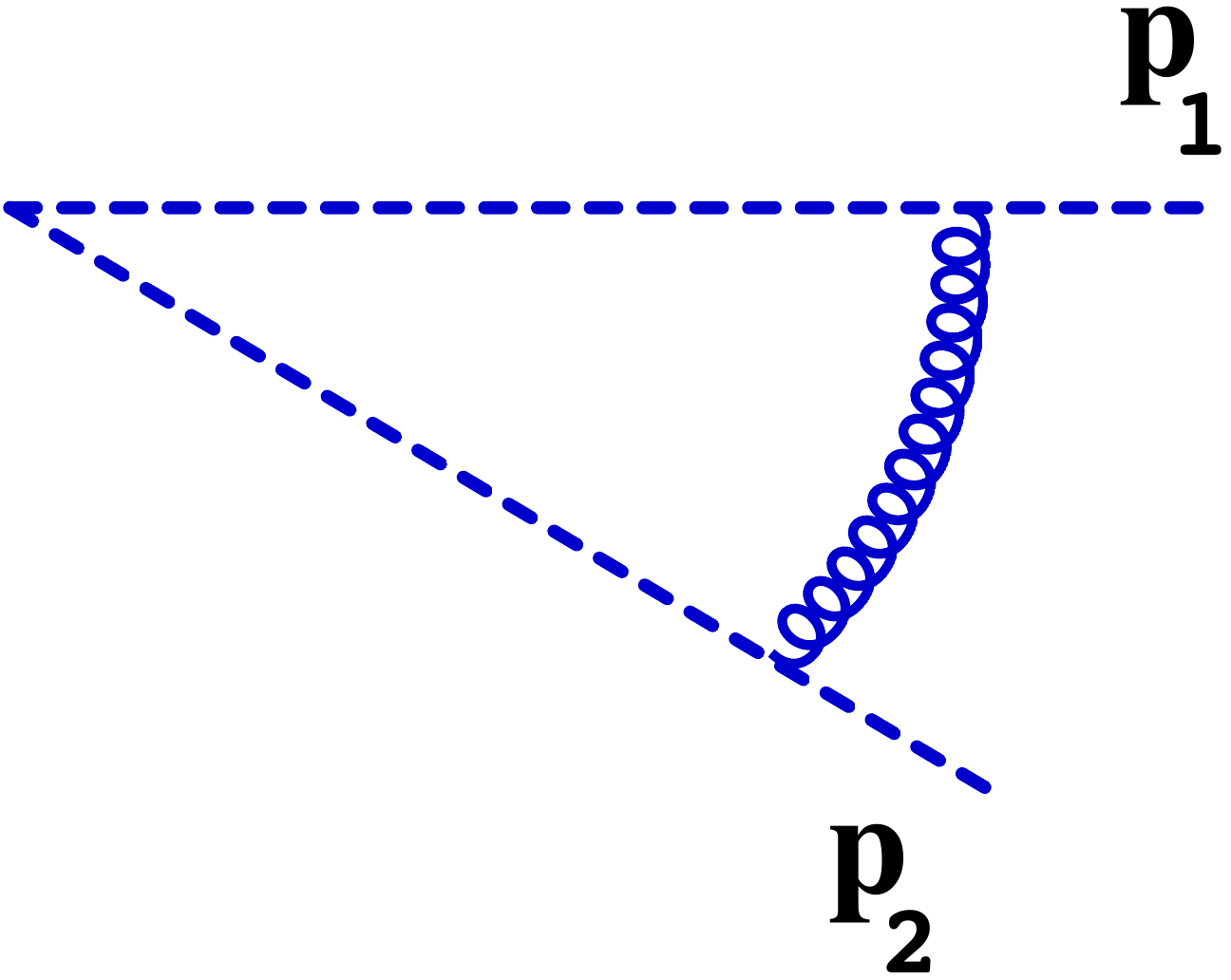}
\end{center}
\caption{Cusp anomalous dimension in the leading order. \label{fig:6}}
\end{figure}

Thus, we obtain the following rapidity evolution equation in the Sudakov region:
\begin{eqnarray}
&&\hspace{-1mm}
{d\over d\ln\sigma}\langle p|\tilcaf^a_i(\beta_B, x_\perp) \calf_j^a(\beta_B, y_\perp)|p\rangle~
\label{7.6}\\
&&\hspace{-1mm}
=~4\alpha_sN_c\!\int\!{\dhd^2 p_\perp\over p_\perp^2}\Big[e^{i(p,x-y)_\perp} 
\langle p|\tilcaf^a_i\big(\beta_B+{p_\perp^2\over\sigma s},x_\perp\big)
\calf^a_j\big(\beta_B+{p_\perp^2\over\sigma s},y_\perp\big)|p\rangle
\nonumber\\
&&\hspace{44mm}
-~{\sigma \beta_Bs\over\sigma \beta_Bs+p_\perp^2}\langle p|\tilcaf^a_i(\beta_B,x_\perp)\calf^a_j(\beta_B,y_\perp)|p\rangle
\Big]
\nonumber
\end{eqnarray}
As we mentioned above, the integrals over $p_\perp^2$ in the production part of the kernel (\ref{7.6}) are $ k_\perp^2$ 
whereas in the virtual part  the logarithmic integrals over $p_\perp^2$ are restricted from 
above by an extra ${1\over p_\perp^2+\sigma\beta_Bs}$ leading to the double-log region where $1\gg\sigma\gg{(x-y)_\perp^{-2}\over s}$ 
and $\sigma\beta_Bs\gg p_\perp^2\gg (x-y)_\perp^{-2}$. In that region only the first term in the r.h.s. of 
Eq. (\ref{7.6}) survives so the evolution equation reduces to
\begin{eqnarray}
&&\hspace{-1mm} 
{d\over d\ln\sigma}\langle p|\tilcaf^a_i(\beta_B, x_\perp)\calf^a_j(\beta_B, y_\perp)|p\rangle^{\eta=\ln\sigma}
\label{7.7}\\
&&\hspace{33mm} 
=~-{g^2N_c\over \pi}
\!\int\!{\dhd^2 p_\perp\over p^2_\perp}\big[1-e^{i(p,x-y)_\perp}\big]
\langle p|\tilcaf^a_i(\beta_B, x_\perp)\calf^a_j(\beta_B, y_\perp)|p\rangle^\eta
\nonumber
\end{eqnarray}
which can be rewritten for the TMD (\ref{gTMD}) as
\begin{eqnarray}
&&\hspace{-1mm} 
{d\over d\ln\sigma}\cald(x_B,z_\perp,\ln\sigma)
~=~-{\alpha_sN_c\over \pi^2}\cald(x_B,z_\perp,\ln\sigma)
\!\int\!{d^2 p_\perp\over p^2_\perp}\big[1-e^{i(p,z)_\perp}\big]
\label{7.8}
\end{eqnarray}
We see that the IR divergence at $p_\perp^2\rightarrow 0$ cancels while the UV divergence in the virtual correction
should be cut from above by the condition $p_\perp^2\leq \sigma s$ following from Eq. (\ref{7.6}). With the double-log accuracy one obtains
\begin{eqnarray}
&&\hspace{-1mm} 
{d\over d\ln\sigma}\cald(x_B,z_\perp,\ln\sigma)
~=~-{\alpha_sN_c\over \pi}\cald(x_B,z_\perp,\ln\sigma)
\ln\sigma s z_\perp^2~+~...
\label{7.9}
\end{eqnarray}
where dots stand for the non-logarithmic contributions. This equation leads to the usual Sudakov double-log result
\begin{eqnarray}
&&\hspace{-1mm} 
\cald(x_B,k_\perp,\ln\sigma)
~\sim~\exp\big\{-{\alpha_sN_c\over 2\pi}\ln^2{\sigma s\over k_\perp^2} \big\}\cald(x_B,k_\perp,\ln{k_\perp^2\over s})
\label{sudakov}
\end{eqnarray}
It is worth noting that the coefficient in front of $\ln^2{\sigma s\over k_\perp^2}$ is determined by the
cusp anomalous dimension of two light-like Wilson lines 
going from point $y$ to $\infty p_1$ and $\infty p_2$ directions (with our cutoff $\alpha<\sigma$). Indeed, if one calculates the 
contribution of the diagram in Fig. \ref{fig:6} for Wilson lines in the adjoint representation, one gets
%
\begin{eqnarray}
&&\hspace{0mm}
\langle[\infty p_1,0][0,\infty p_2]\rangle~=~-ig^2N_c\!\int\!\dhd\alpha\dhd\beta \dhd^2p_\perp~
{\theta(\sigma>|\alpha|>\sigma')
\over\alpha(\beta-i\epsilon)(\alpha\beta s-p_\perp^2+i\epsilon)}~
\nonumber\\
&&\hspace{0mm}
=~-g^2N_c\!\int_{\sigma'}^\sigma\!{\dhd\alpha\over\alpha}\!\int\!{\dhd^2 p_\perp\over p_\perp^2}
~=~-{g^2\over 8\pi^2}N_c\ln{\sigma\over\sigma'}\!\int\!{dp_\perp^2\over p_\perp^2}
\label{7.11}
\end{eqnarray}
which coincides with the coefficient in Eq. (\ref{7.8}), cf Ref. \cite{cusp}.

\section{Rapidity evolution of unintegrated gluon distribution in linear approximation}

 It is instructive to present the evolution kernel (\ref{masterdis}) in the linear (two-gluon) approximation. Since in
  the r.h.s. of Eq. (\ref{masterdis}) we already have $\ticalf_k$ and $\calf_l$ (and each of them has at least one gluon) 
 all factors $U$ and $\tilU$ in the r.h.s. of Eq. (\ref{masterdis}) can be omitted and we get
\begin{eqnarray}
&&\hspace{-3mm}
{d\over d\ln\sigma}\langle p|\tilcaf_i^a(\beta_B, p_\perp) \calf_j^a(\beta_B, p'_\perp)|p\rangle~
\label{masterlin}\\
&&\hspace{-3mm}
=~-\alpha_sN_c\!\int\!\dhd^2k_\perp
\Big\{\theta\big(1-\beta_B-{k_\perp^2\over\sigma s}\big)
\Big[\Big({(p+k)_k\over\sigma\beta_Bs+p_\perp^2}
{\sigma \beta_Bsg_{\mu i}-2k^\perp_{\mu}k_i\over\sigma \beta_Bs+k_\perp^2}
-~2{k^\perp_\mu g_{ik}+p_ig_{\mu k} \over \sigma\beta_Bs+p_\perp^2}\Big)
\nonumber\\
&&\hspace{55mm}
\times~
\Big({\sigma \beta_Bs\delta^\mu_j-2k_\perp^{\mu}k_j\over\sigma\beta_Bs+k_\perp^2}{(p'+k)_l\over\sigma \beta_Bs+{p'}_\perp^2}
-2{k_\perp^\mu g_{jl}+\delta_l^\mu p'_j\over \sigma\beta_Bs+{p'}_\perp^2}\Big)
\nonumber\\
&&\hspace{5mm}
+~2g_{ik}\Big({k_j\over k_\perp^2}{\sigma \beta_Bs+2k_\perp^2\over\sigma\beta_Bs+k_\perp^2}
{(p'+k)_l\over\sigma \beta_Bs+{p'}_\perp^2}
+~{2g_{jl}\over \sigma\beta_Bs+{p'}_\perp^2}-{2p'_jk_l\over k_\perp^2(\sigma\beta_Bs+{p'}_\perp^2)}
\Big)
\nonumber\\
&&\hspace{5mm}
+~2g_{lj}\Big({(p+k)_k\over\sigma\beta_Bs+p_\perp^2}
{k_i\over k_\perp^2}
{\sigma \beta_Bs+2k_\perp^2\over\sigma \beta_Bs+k_\perp^2}
+{2g_{ik}\over \sigma\beta_Bs+p_\perp^2}
-~{2p_ik_k \over k_\perp^2(\sigma\beta_Bs+p_\perp^2)}\Big)\Big]
\nonumber\\
&&\hspace{55mm}
\times~
\langle p|\ticalf^k\big(\beta_B+{k_\perp^2\over\sigma s}, p_\perp-k_\perp\big)
\calf^l\big(\beta_B+{k_\perp^2\over\sigma s}, p'_\perp-k_\perp\big)|p\rangle
\nonumber\\
&&\hspace{5mm}
+~{2\over k_\perp^2}\Big[
{(2k^lp'_j-k_j{p'}^l)\delta_i^k\over \sigma\beta_Bs+(p'+k)_\perp^2}
+~{(2p_ik^k-k_ip^k)\delta_j^l\over \sigma\beta_Bs+(p+k)_\perp^2}\Big]
\langle p|\ticalf_k^a(\beta_B, p_\perp)\calf_l^a(\beta_B, p'_\perp)|p\rangle
\nonumber\\
&&\hspace{5mm}
-~{4\over k_\perp^2}\langle p|
\Big[\theta\big(1-\beta_B-{k_\perp^2\over\sigma s}\big)
\ticalf_i\big(\beta_B+{k_\perp^2\over\sigma s}, p_\perp-k_\perp\big)
\calf_j\big(\beta_B+{k_\perp^2\over\sigma s}, p'_\perp-k_\perp\big)
\nonumber\\
&&\hspace{55mm}
-~{\sigma\beta_Bs\over \sigma\beta_Bs+k_\perp^2}\ticalf_i^a(\beta_B, p_\perp)\calf_j^a(\beta_B, p'_\perp)\Big]|p\rangle
\Big\}
\nonumber
\end{eqnarray}
where we performed Fourier transformation to the momentum space. Also, the  forward matrix element 
$\langle p|\ticalf_i(\beta_B, p_\perp)\calf_j(\beta_B, p'_\perp)|p\rangle$ is proportional to $\delta^{(2)}(p_\perp-p'_\perp)$. 
Eliminating this factor and rewriting in terms of $\calr_{ij}$ (see Eq. (\ref{mael})) we obtain ($\eta\equiv\ln\sigma$)
\begin{eqnarray}
&&\hspace{-1mm}
{d\over d\eta}\calr_{ij}(\beta_B,p_\perp;\eta)
\label{linear}\\
&&\hspace{-1mm}
=~-\alpha_sN_c\!\int\!\dhd^2k_\perp\Big\{
\Big[\Big({(2p-k)_k\over\sigma\beta_Bs+p_\perp^2}
{\sigma \beta_Bsg_{\mu i}-2(p-k)^\perp_{\mu}(p-k)_i\over\sigma \beta_Bs+(p-k)_\perp^2}
-~2{(p-k)^\perp_\mu g_{ik}+p_ig_{\mu k} \over \sigma\beta_Bs+p_\perp^2}\Big)
\nonumber\\
&&\hspace{22mm}
\times~
\Big({\sigma \beta_Bs\delta^\mu_j-2(p-k)_\perp^{\mu}(p-k)_j\over\sigma\beta_Bs+(p-k)_\perp^2}
{(2p-k)_l\over\sigma \beta_Bs+p_\perp^2}
-2{(p-k)_\perp^\mu g_{jl}+\delta_l^\mu p_j\over \sigma\beta_Bs+p_\perp^2}\Big)
\nonumber\\
&&\hspace{-1mm}
+~2g_{ik}
\Big({(p-k)_j(2p-k)_l-2p_j(p-k)_l\over (p-k)_\perp^2(\sigma\beta_Bs+p_\perp^2)}
+{(p-k)_j(2p-k)_l\over (\sigma\beta_Bs+(p-k)_\perp^2)(\sigma\beta_Bs+p_\perp^2)}
+~{2g_{jl}\over \sigma\beta_Bs+p_\perp^2}\Big)
\nonumber\\
&&\hspace{-1mm}
+~2g_{lj}\Big(
{(p-k)_i(2p-k)_k-2p_i(p-k)_k\over (p-k)_\perp^2(\sigma\beta_Bs+p_\perp^2)}
+{(p-k)_i(2p-k)_k\over (\sigma\beta_Bs+(p-k)_\perp^2)(\sigma\beta_Bs+p_\perp^2)}
+{2g_{ik}\over \sigma\beta_Bs+p_\perp^2}\Big)
\Big]
\nonumber\\
&&\hspace{55mm}
\times~
\theta\big(1-\beta_B-{(p-k)_\perp^2\over\sigma s}\big)\calr^{kl}\big(\beta_B+{(p-k)_\perp^2\over\sigma s},k_\perp\big)
\nonumber\\
&&\hspace{-1mm}
+~2
{\delta_i^k(k_jp^l-2k^lp_j)+\delta^l_j(k_ip^k -2p_ik^k)\over k_\perp^2[\sigma\beta_Bs+(p-k)_\perp^2]}
\calr_{kl}(\beta_B,p_\perp;\eta)
\nonumber\\
&&\hspace{-1mm}
-~
4\Big[{\theta\big(1-\beta_B-{(p-k)_\perp^2\over\sigma s}\big)\over (p-k)_\perp^2}
\calr_{ij}\big(\beta_B+{(p-k)_\perp^2\over \sigma s},k_\perp;\eta\big)
-~{\sigma\beta_Bs\over k_\perp^2(\sigma\beta_Bs+k_\perp^2)}\calr_{ij}(\beta_B,p_\perp;\eta)\Big]\Big\}
\nonumber
\end{eqnarray}

Let us demonstrate that Eq. (\ref{linear}) reduces to BFKL equation in the low-$x$ limit. Indeed, 
in this limit $\calr_{ij}$ is proportional to the WW distribution (\ref{WW}): \\
$\calr_{ij}(0,k_\perp)~\sim ~\int\! d^2xe^{i(k,x)_\perp}\langle p|{\rm tr}\{\tilU_i(x)U_j(0)\}|p\rangle$. 
In the leading-order  BFKL approximation (cf. Ref. \cite{nlobk})
\begin{eqnarray}
&&\hspace{-1mm}
\langle p|{\rm tr}\{\tilU_i(x)U_j(y)\}|p\rangle~
\label{lip1}\\
&&\hspace{-1mm}
=~{\alpha_s\over 4\pi^2}\!\int\!{d^2 q_\perp\over q_\perp^2}
q_iq_je^{i(q,x)_\perp-i(q,y)_\perp}\!\int\!{d^2 q'_\perp\over {q'}_\perp^2}\Phi_T(q')
\!\int_{a-i\infty}^{a+i\infty}\!{d\omega\over 2\pi i}\Big({s\over qq'}\Big)^\omega
G_\omega(q,q')   
\nonumber
\end{eqnarray}
Here $\Phi_T(q')$ is the target impact factor 
and  $G_\omega(q,q')$ is the partial wave of the forward reggeized gluon scattering amplitude satisfying the equation
\begin{equation}
\omega G_\omega(q,q')=\delta^{(2)}(q-q')+\int\! d^2p K_{\rm BFKL}(q,p)G_\omega(p,q') 
\label{wgw}
\end{equation}
with the forward BFKL kernel 
$$
K_{\rm BFKL}(q,p)~=~{\alpha_sN_c\over\pi^2}
\Big[{1\over (q-p)_\perp^2}-\half\delta(q_\perp-p_\perp)\!\int\!dp'_\perp{q_\perp^2\over {p'}_\perp^2(q-p')_\perp^2}\Big]
$$
Thus, in the BFKL approximation
\begin{equation}
\calr_{ij}(0,q_\perp;\ln\sigma)~=~q_iq_jR(q_\perp;\ln\sigma)
~=~{\alpha_sq_iq_j\over 2\pi^2q_\perp^2}\!\int\!
{d^2 q'\over {q'}^2}\Phi_T(q')
\!\int_{a-i\infty}^{a+i\infty}\!{d\omega\over 2\pi i}\Big({\sigma s\over qq'}\Big)^\omega
G_\omega(p,q')
\label{R}
\end{equation}
and the equation for $R$ 
\begin{eqnarray}
&&\hspace{-1mm}
{d\over d\ln\sigma}R^\sigma(q_\perp;\ln\sigma)~=~\int\! d^2p_\perp ~{p_\perp^2\over q_\perp^2}
K_{\rm BFKL}(q,p)R^\sigma(p_\perp;\ln\sigma)
\label{Rbfkl}
\end{eqnarray}
is obtained by differentiation of Eq. (\ref{R}) with respect to $\ln\sigma$ using Eq. (\ref{wgw}).

Now it is easy to see that our Eq. (\ref{linear}) reduces to Eq. (\ref{Rbfkl}) in the BFKL limit. 
As we discussed above, in this limit one may set $\beta_B=0$
and neglect ${k_\perp^2\over\sigma s}$ in the argument of $\calr_{ij}$. Substituting
$\calr_{ij}(0,k_\perp)~=~k_ik_jR(k_\perp)$ into Eq. (\ref{linear}) one obtains after some algebra
\begin{eqnarray}
&&\hspace{-1mm}
{d\over d\ln\sigma}R(p_\perp;\ln\sigma)
~=~2\alpha_sN_c\!\int\!\dhd^2k_\perp
\Big[{2k_\perp^2\over p_\perp^2(p-k)_\perp^2}
R(k_\perp;\ln\sigma)
-~{p_\perp^2\over k_\perp^2(p-k)_\perp^2}R(p_\perp;\ln\sigma)\Big]
\nonumber
\end{eqnarray}
which coincides with Eq. (\ref{Rbfkl}). We have also checked that Eq. (\ref{masterlin}) at $p_\perp\neq p'_\perp$ reduces to 
the non-forward BFKL equation in the low-$x$ limit.

Let us check now that the evolution of 
\begin{equation}
\cald(\beta_B,\ln\sigma)~=~-\half\int\! \dhd^2 p_\perp \calr_i^{~i}(\beta_B,p_\perp;\ln\sigma)
\label{rii}
\end{equation}
reduces to DGLAP equation.  
As we discussed above, in the light-cone limit one can neglect $k_\perp$ in comparison to $p_\perp$. 
Indeed, the integral over $p_\perp$ converges at $p_\perp^2\sim\sigma\beta_Bs$. On the other hand, extra $k_ik_j$ in the integral 
over $k_\perp$ leads to the operators of higher collinear twist, for example
\begin{eqnarray}
&&\hspace{0mm} 
\int\! d^2k_\perp ~k_ik_j~R_n^{~n}(\beta_B,k_\perp;\ln\sigma)
~\sim~\langle p|\partial_k\tilcaf^a_n(\beta_B, 0_\perp)\partial_j\calf^{an}(\beta_B, 0_\perp)|p\rangle^{\eta=\ln\sigma}
\nonumber\\
&&\hspace{0mm}
\sim~m^2g_{ij}\langle p|\tilcaf^a_n(\beta_B, 0_\perp)\calf^{a n}(\beta_B, 0_\perp)|p\rangle^{\ln\sigma}~\sim~m^2\cald(\beta_B,\ln\sigma)
\end{eqnarray}
(where $m$ is the mass of the target) so ${k_\perp^2\over p_\perp^2}\sim{k_\perp^2\over\sigma\beta_Bs}\sim {m^2\over\sigma s}\ll 1$.

Neglecting $k_\perp$ in comparison to $p_\perp$ and integrating over angles one obtains
\begin{eqnarray}
&&\hspace{-1mm}
{d\over d\ln\sigma}\!\int\! d^2p_\perp~\calr_i^{~i}(\beta_B,p_\perp;\ln\sigma)
\nonumber\\
&&\hspace{-1mm}
=~{\alpha_sN_c\over \pi^2}\!\int\! d^2p_\perp\Big[{1\over p_\perp^2}
-{2\over\sigma\beta_Bs+p_\perp^2}+{3p_\perp^2\over(\sigma\beta_Bs+p_\perp^2)^2}
-{2p_\perp^4\over(\sigma\beta_Bs+p_\perp^2)^3}+{p_\perp^6\over(\sigma\beta_Bs+p_\perp^2)^4}\Big]
\nonumber\\
&&\hspace{-1mm}
\times~\!\int\! d^2k_\perp
\calr_i^{~i}\big(\beta_B+{p_\perp^2\over\sigma s},k_\perp;\ln\sigma\big)
-\!\int\! d^2k_\perp{\sigma\beta_Bs\over k_\perp^2(\sigma\beta_Bs+k_\perp^2)}\!\int\! d^2p_\perp\calr_i^{~i}\big(\beta_B,p_\perp;\ln\sigma\big)
\nonumber
\end{eqnarray}
which coincides with DGLAP equation (\ref{3.27}).

 It would be interesting to compare Eq. (\ref{linear}) to CCFM equation \cite{ccfm} which also
addresses the question of interplay of BFKL and DGLAP logarithms.

\section{Rapidity evolution of fragmentation functions}

In this section we will construct the evolution equation for fragmentation function (\ref{frTMD}). 
We start from Eq. (\ref{master1alt}) which enables us to analytically continue to negative $\beta_B=-\beta_F$. 
In the operator form, the equation (\ref{master1alt}) has imaginary parts at negative $\beta_B=-\beta_F$ corresponding to poles of propagators $(\sigma\beta_Fs-p_\perp^2\pm i\epsilon)^{-1}$ 
but we will demonstrate now that for the evolution of
 a ``fragmentation matrix element'' (\ref{maelfr})\footnote{
 Again, strictly speaking we should consider 
 $\sum_X\langle 0|\tilcaf^a_i(-\beta_F, x_\perp)|p+X\rangle\langle (p+\xi p_2)+X|\calf^{a}_j(-\beta_F, y_\perp)|0\rangle$,
 see Eq. (\ref{maelfr}).
 }
\begin{equation}
\langle \tilcaf^a_i(-\beta_F, x_\perp)\calf^a_j(-\beta_F, y_\perp)\rangle_{\rm frag}
\equiv\sum_X\langle 0|\tilcaf^a_i(-\beta_F, x_\perp)|p+X\rangle\langle p+X|\calf^a_j(-\beta_F, y_\perp)|0\rangle 
\label{8.1}
\end{equation}
we have the kinematical restriction $\sigma(\beta_F-1)s>p_\perp^2$ in all the integrals
 in the production part of the kernel (\ref{master1alt}).  As to virtual part of the kernel, we will see that the imaginary parts there 
 assemble to yield the principle-value prescription for integrals over $p_\perp^2$. The ``fragmentation matrix element'' (\ref{maelfr})
 of Eq. (\ref{master2alt}) has the form
\begin{eqnarray}
&&\hspace{-1mm}
{d\over d\ln\sigma}\langle \tilcaf_i^a(-\beta_F, x_\perp) \calf_j^a(-\beta_F, y_\perp)\rangle_{\rm frag}~
\label{anni}\\
&&\hspace{-1mm}
=~-\alpha_s\langle {\rm Tr}\Big\{\!\int\!\dhd^2k_\perp\theta\big(\beta_F-1-{k_\perp^2\over\sigma s}\big)\Big[
(x_\perp|\Big(\tilU{-1\over\sigma\beta_Fs-p_\perp^2}
(\tilU^\dagger k_k+p_k\tilU^\dagger){\sigma \beta_Fsg_{\mu i}+2k^\perp_{\mu}k_i\over\sigma \beta_Fs-k_\perp^2}
\nonumber\\
&&\hspace{-1mm}
+~2k^\perp_\mu g_{ik}\tilU{1\over \sigma\beta_Fs-p_\perp^2}\tilU^\dagger
+2g_{\mu k} \tilU{p_i\over\sigma\beta_Fs-p_\perp^2}\tilU^\dagger\Big)
\ticalf^k\big(-\beta_F+{k_\perp^2\over\sigma s}\big)|k_\perp) 
\nonumber\\
&&\hspace{-1mm}
\times~(k_\perp|\calf^l\big(-\beta_F+{k_\perp^2\over\sigma s}\big)
\Big({\sigma\beta_Fs\delta^\mu_j+2k_\perp^{\mu}k_j\over\sigma\beta_Fs-k_\perp^2}(k_lU+Up_l){-1\over\sigma \beta_Fs+p_\perp^2}U^\dagger
\nonumber\\
&&\hspace{-1mm}
+~2k_\perp^\mu g_{jl}U{1\over \sigma\beta_Fs-p_\perp^2}U^\dagger
+2\delta_l^\mu U{p_j\over\sigma\beta_Fs-p_\perp^2}U^\dagger\Big)|y_\perp)+~2(x_\perp|\ticalf_i\big(-\beta_F+{k_\perp^2\over\sigma s}\big)|k_\perp)
\nonumber\\
&&\hspace{-1mm}
~\times~(k_\perp|\calf^l\big(-\beta_F+{k_\perp^2\over\sigma s}\big)
\Big(-{k_j\over k_\perp^2}{\sigma\beta_Fs-2k_\perp^2\over\sigma\beta_Fs-k_\perp^2}(k_lU+Up_l)
{1\over\sigma\beta_Fs-p_\perp^2}U^\dagger
\nonumber\\
&&\hspace{44mm}
-~2U{g_{jl}\over \sigma\beta_Fs-p_\perp^2}U^\dagger+2{k_l\over k_\perp^2}U{p_j\over \sigma\beta_Fs-p_\perp^2}U^\dagger
\Big)
|y_\perp)
\nonumber\\
&&\hspace{-1mm}
+~2(x_\perp|\Big(-\tilU{1\over\sigma\beta_Fs-p_\perp^2}
(\tilU^\dagger k_k+p_k\tilU^\dagger){k_i\over k_\perp^2}
{\sigma\beta_Fs-2k_\perp^2\over\sigma\beta_Fs-k_\perp^2}
-2\tilU{g_{ik}\over \sigma\beta_Fs-p_\perp^2}\tilU^\dagger
\nonumber\\
&&\hspace{-1mm}
+~2\tilU{p_i\over\sigma\beta_Fs-p_\perp^2}\tilU^\dagger{k_k\over k_\perp^2}\Big)
\ticalf^k\big(-\beta_F+{k_\perp^2\over\sigma s}\big)|k_\perp)(k_\perp|\calf_j\big(-\beta_F+{k_\perp^2\over\sigma s}\big)|y_\perp)\Big]
\nonumber\\
&&\hspace{-1mm}
+~2\ticalf_i(-\beta_F, x_\perp)
(y_\perp|{p^m\over p_\perp^2}\calf_k(-\beta_F)(i\!\stackrel{\leftarrow}{\partial}_l+U_l)(2\delta_m^k\delta_j^l-g_{jm}g^{kl})
U{1\over \sigma\beta_Fs-p_\perp^2+i\epsilon}U^\dagger|y_\perp)
\nonumber\\
&&\hspace{-1mm}
+~2(x_\perp|
\tilU{-1\over \sigma\beta_Fs-p_\perp^2-i\epsilon}\tilU^\dagger(2\delta_i^k\delta_m^l-g_{im}g^{kl} )(i\partial_k-\tilU_k)\ticalf_l(-\beta_F)
{p^m\over p_\perp^2}|x_\perp)\calf_j(-\beta_F, y_\perp)
\nonumber\\
&&\hspace{-1mm}
-~4\!\int\!{\dhd^2k_\perp\over k_\perp^2}\Big[\theta\big(\beta_F-1-{k_\perp^2\over\sigma s}\big)
\ticalf_i\big(-\beta_F+{k_\perp^2\over\sigma s}, x_\perp\big)
\calf_j\big(-\beta_F+{k_\perp^2\over\sigma s}, y_\perp\big)e^{i(k,x-y)_\perp}
\nonumber\\
&&\hspace{-1mm}
-~\half\big[{\sigma\beta_Fs\over \sigma\beta_Fs-k_\perp^2-i\epsilon}
+{\sigma\beta_Fs\over \sigma\beta_Fs-k_\perp^2+i\epsilon}\big]
\ticalf_i(-\beta_F, x_\perp)\calf_j(-\beta_F, y_\perp)\Big]\Big\}\rangle_{\rm frag}
~+~O(\alpha_s^2)
\nonumber
\end{eqnarray}
where we have restored $\pm i\epsilon$ in the virtual part in accordance with Feynman rules.

Let us prove that  all non-linear terms in Eq. (\ref{anni}) can be neglected with our accuracy. (Na\"ively, they were important at small $\beta_B$ but small
$\beta_F$ are not allowed due to kinematical restrictions). 
First, consider the ``light-cone'' case when the transverse momenta of fast fields $l_\perp^2$ are smaller than the characteristic transverse momenta in the gluon loop of slow fields 
$p_\perp^2\sim k_\perp^2$.
As we discussed above, in this case with the leading-twist accuracy we can commute all $U$'s  with $p_\perp$ operators until they form $UU^\dagger=1$ and disappear.
In this limit the (\ref{anni}) turns to
\begin{eqnarray}
&&\hspace{-1mm}
{d\over d\ln\sigma}\langle \tilcaf^a_i(-\beta_F, x_\perp) \calf_j^a(-\beta_F, y_\perp)\rangle_{\rm frag}~=~-4\alpha_s{\rm Tr}\Big\{   
(x_\perp|\theta(\beta_F-1-\frac{p^2_\perp}{\sigma s})
\label{masterannihi}\\
&&\hspace{-1mm}
~\times\Big[{\delta^k_ip_\perp^\mu\over p_\perp^2}
+{g^{\mu k}p_i+p_\perp^{\mu}\delta^k_i-\delta^\mu_ip^k\over \sigma\beta_F s-p_\perp^2-i\epsilon}
-{\delta^\mu_ip_\perp^2p^k+2p_ip^kp_\perp^{\mu}\over (\sigma\beta_Fs-p_\perp^2-i\epsilon)^2}
\Big]
\Big[{p^\perp_\mu \delta^l_j\over p_\perp^2}+{\delta_\mu^lp_j+\delta_j^lp^\perp_\mu-g_{\mu j}p^l\over \sigma\beta_Fs-p_\perp^2+i\epsilon}    
\nonumber\\
&&\hspace{34mm}
~-~{g_{\mu j}p_\perp^2p^l+2p^\perp_\mu p_jp^l\over(\sigma\beta_Fs-p_\perp^2+i\epsilon)^2}
\Big]|y_\perp)
\langle \tilcaf_k\big({p_\perp^2\over\sigma s}-\beta_F, x_\perp\big)  \calf_l\big({p_\perp^2\over\sigma s}-\beta_F, y_\perp\big)\rangle_{\rm frag}
\nonumber\\
&&\hspace{-1mm}
+~~\half\Big[(x_\perp|{\sigma\beta_Fs\over p_\perp^2(\sigma\beta_Fs-p_\perp^2-i\epsilon)}|x_\perp)+
(y_\perp|{\sigma\beta_Fs\over p_\perp^2(\sigma\beta_Fs-p_\perp^2+i\epsilon)}|y_\perp)\Big]
\nonumber\\
&&\hspace{77mm}
\times~\langle \ticalf_i(-\beta_F, x_\perp)\calf_j(-\beta_F, y_\perp)\rangle_{\rm frag}\Big\}
\nonumber
\end{eqnarray}

Now consider the shock-wave case when $l_\perp^2\sim p_\perp^2$. There are two ``subcases'': when 
$\beta_F\sigma_\ast\geq 1$ and when $\beta_F\sigma_\ast\ll 1$ 
(where $\sigma_\ast\sim{\sigma's\over l_\perp^2}$). In the former case we have 
$\beta_F{\sigma's\over l_\perp^2}\geq 1$ so $\sigma\beta_F s\gg p_\perp^2$ and only two last lines in Eq. (\ref{anni}) 
survive. Moreover, in this case Eq. (\ref{masterannihi}) also reduces to the last two lines so Eq. (\ref{anni}) is equivalent to Eq. (\ref{masterannihi}) in this case as well.

If $\beta_F\sigma_\ast\ll 1$, as we discussed above, one can replace $\calf_j(-\beta_F)$ (and
$\calf_j\big(-\beta_F+{p_\perp^2\over\alpha s}\big)$) by $U_j$.  We will prove now that after such replacement
the r.h.s. of Eq. (\ref{anni}) vanishes, and so does the r.h.s of Eq.  (\ref{masterannihi}),  and therefore Eqs. (\ref{anni}) 
and (\ref{masterannihi}) are equivalent in the case of large $\beta_F\sigma_\ast$ also.

 Let us now prove that if we replace all  $\calf_j(-\beta_F)$ and
$\calf_j\big(-\beta_F+{p_\perp^2\over\alpha s}\big)$ by $U_j$ the r.h.s. of Eq. (\ref{anni}) vanishes. 
Indeed, a  typical term in Feynman part of the amplitude vanishes:
\begin{equation}
\langle p+X|U_j(z)U_{z'}U_{z''}^\dagger|0\rangle ~=~0
\label{8.2}
\end{equation}
To prove this, let us consider  the shift of $U$ operator on ${2\over s}a_\ast p_1$.
 Since the shift in the $p_1$ direction does not change the infinitely long $U$ operator, we get
\begin{eqnarray}
&&\langle p+X|U_j(z)U_{z'}U_{z''}^\dagger|0\rangle ~=~
\langle p+X|e^{i{2\over s}\hat{P}_\bu a_\ast}U_j(z)U_{z'}U_{z''}^\dagger e^{-i{2\over s}\hat{P}_\bu a_\ast}|0\rangle 
\nonumber\\
&&\hspace{55mm}
~=~e^{i(\beta_p+\beta_X)a_\ast}\langle p+X|U_j(z)U_{z'}U_{z''}^\dagger|0\rangle
\nonumber
\end{eqnarray}
which can be true only if Eq. (\ref{8.2}) vanishes.  It is clear that for the same reason all terms in the r.h.s of Eq. (\ref{anni})
(and r.h.s. of Eq. (\ref{masterannihi}) as well) vanish. Summarizing, in all regimes the Eq. (\ref{anni}) can be reduced to the light-cone version (\ref{masterannihi}).

One can rewrite Eq. (\ref{masterannihi}) in the form:
\begin{eqnarray}
&&\hspace{-3mm}
{d\over d\ln\sigma}\langle\tilcaf^a_i(-\beta_F, x_\perp) \calf_j^a(-\beta_F, y_\perp)\rangle_{\rm frag}^{\eta\equiv\ln\sigma}
\label{masteranni}\\
&&\hspace{-3mm}
=~4\alpha_sN_c\!\int\!\dhd^2p_\perp\Big\{\theta\big(\beta_F-1-{p_\perp^2\over \sigma s}\big)e^{i(p,x-y)_\perp}\langle\ticalf_k^{a}\big({p_\perp^2\over\sigma s}-\beta_F,x_\perp\big)\calf_l^{a}
\big({p_\perp^2\over\sigma s}-\beta_F,y_\perp\big)\rangle_{\rm frag}^\eta
\nonumber\\
&&\hspace{-3mm}
\times~\Big[{\delta^k_i\delta^l_j\over p_\perp^2}+{2\delta^k_i\delta^l_j\over\sigma\beta_Fs-p_\perp^2}+{p^2_\perp \delta^k_i\delta^l_j+\delta^k_jp_ip^l+\delta^l_ip_jp^k-\delta^l_jp_ip^k-\delta^k_ip_jp^l-g^{kl}p_ip_j-g_{ij}p^kp^l\over (\sigma\beta_Fs-p_\perp^2)^2}
\nonumber\\
&&\hspace{20mm}
-~p^2_\perp{2g_{ij}p^kp^l+\delta^k_ip_jp^l+\delta^l_jp_ip^k-\delta^k_jp_ip^l-\delta^l_ip_jp^k\over (\sigma\beta_Fs-p_\perp^2)^3}-{p_\perp^4g_{ij}p^kp^l\over  (\sigma\beta_Fs-p_\perp^2)^4}\Big]
\nonumber\\
&&\hspace{42mm}
-~{\theta(\sigma\beta_Fs-p_\perp^2)\over p_\perp^2} 
\langle\ticalf^a_i(-\beta_F,x_\perp)\calf^a_j(-\beta_F,y_\perp)\rangle_{\rm frag}^\eta\Big\}
\nonumber
\end{eqnarray}
where we used the formula
\begin{eqnarray}
&&\hspace{-1mm}
\!\int\!{\dhd^2p_\perp\over p_\perp^2}\half\Big[{\sigma\beta_Fs\over \sigma\beta_Fs-p_\perp^2+i\epsilon}+{\sigma\beta_Fs\over \sigma\beta_Fs-p_\perp^2-i\epsilon}\Big]
 ~=~\!\int\!{\dhd^2p_\perp\over p_\perp^2}\theta(\sigma\beta_F s-p_\perp^2)
 \label{8.5}
\end{eqnarray}
The Eq. (\ref{masteranni}) is our final evolution equation for fragmentation functions valid for all $(x-y)_\perp^2$ 
(and all $\beta_F$).

 If polarizations of fragmentation hadron are not registered
we can use the parametrization (\ref{maelfr}) 
\begin{eqnarray}
&&\hspace{-1mm}
\langle \ticalf_i^a(-\beta_F,z_\perp)\calf_j^a(-\beta_F,0_\perp)\rangle_{\rm frag}^\eta~
\nonumber\\
&&\hspace{2mm}
=~
 2\pi^2\delta(\xi) \beta_Fg^2\Big[-g_{ij}\cald_{\rm f}(\beta_F,z_\perp,\eta)
-{4\over m^2}(2z_iz_j+g_{ij}z_\perp^2)\calh''_{\rm f}(\beta_F,z_\perp,\eta)\Big]
\end{eqnarray}
where 
$\calh_{\rm f}(\beta_F,z_\perp,\eta)\equiv\int\! \dhd^2k_\perp~e^{i(k,z)_\perp}\calh_{\rm f}(\beta_F,k_\perp,\eta)$ 
and\\ $\calh_{\rm f}''(\beta_F,z_\perp,\eta)\equiv\big({\partial\over\partial z^2}\big)^2\calh_{\rm f}(\beta_F,z_\perp,\eta)$,
cf. Eq. (\ref{3.22}). 
After integration over angles similar to Eq. (\ref{eveq1}) one obtains
\begin{eqnarray}
&&\hspace{-1mm}
{d\over d\eta}\Big[g_{ij}\alpha_s\cald_{\rm f}(\beta_F,z_\perp,\eta)
+{4\over m^2}(2z_iz_j+g_{ij}z_\perp^2)\alpha_s\calh_{\rm f}''(\beta_F,z_\perp,\eta)\Big]
\label{eveqfr}\\
&&\hspace{2mm}
=~{\alpha_sN_c\over \pi}\!\int_0^{\beta_F-1}\!d\beta~\Big\{
g_{ij}J_0\big(|z_\perp|\sqrt{\sigma\beta s}\big)\alpha_s\cald_{\rm f}(\beta_F-\beta,z_\perp,\eta)
\nonumber\\
&&\hspace{33mm}
\times~\Big[{\beta_F-\beta\over \beta\beta_F}+{2\over\beta_F}
+{3\beta\over \beta_F(\beta_F-\beta)}
+{2\beta^2\over \beta_F(\beta_F-\beta)^2}
+{\beta^3\over  \beta_F(\beta_F-\beta)^3}\Big]
\nonumber\\
&&\hspace{2mm}
+~J_2(|z_\perp|\sqrt{\sigma\beta s})\big(2{z_iz_j\over z_\perp^2}+g_{ij}\big)
\alpha_s\cald_{\rm f}(\beta_F-\beta,z_\perp,\eta){\beta\over \beta_F(\beta_F-\beta)}
\nonumber\\
&&\hspace{2mm}
+~
{4\over m^2}J_0\big(|z_\perp|\sqrt{\sigma\beta s}\big)(2z_iz_j+g_{ij}z_\perp^2)
\alpha_s\calh_{\rm f}''(\beta_F-\beta,z_\perp,\eta)
\Big[{\beta_F-\beta\over \beta_F\beta}+{2\over\beta_F}
+~{\beta\over \beta_F(\beta_F-\beta)}\Big]
\nonumber\\
&&\hspace{2mm}
+~{4g_{ij}\over m^2}
z_\perp^2J_2\big(|z_\perp|\sqrt{\sigma\beta s}\big)
\alpha_s\calh_{\rm f}''(\beta_F-\beta,z_\perp,\eta)
\nonumber\\
&&\hspace{33mm}
\times~\Big[
{\beta\over \beta_F(\beta_F-\beta)}
+{2\beta^2\over \beta_F(\beta_F-\beta)^2}
+{\beta^3\over  \beta_F(\beta_F-\beta)^3}\Big]\Big\}
\nonumber\\
&&\hspace{2mm}
-~ {\alpha_sN_c\over \pi}\!\int_0^{\beta_F}\!\frac{d\beta}{\beta} 
\Big[g_{ij}\alpha_s\cald_{\rm f}(\beta_F,z_\perp,\eta)
+{4\over m^2}(2z_iz_j+g_{ij}z_\perp^2)\alpha_s\calh_{\rm f}''(\beta_F,z_\perp,\eta)\Big]
\nonumber
\end{eqnarray}
where $\beta\equiv{p_\perp^2\over\sigma s}$ (and $\sigma\equiv e^\eta$ as usual).

This evolution equation can be rewritten as a system (cf. Eq. (\ref{liconefinal}) for DIS)
\begin{eqnarray}
&&\hspace{-1mm}
{d\over d\eta}\alpha_s\cald_{\rm f}({1\over z_F},z_\perp,\eta)
\label{masteranni1}\\
&&\hspace{-1mm}
=~ {\alpha_sN_c\over \pi}\!\int_{z_F}^1\!{dz'\over z'^2}\Big\{
J_0\Big(|z_\perp|\sqrt{{\sigma s\over z_F}(1-z')}\Big)
\Big[\big({1\over 1-z'}\big)_+ +{1\over z'}-2+z'(1-z')\Big]\alpha_s\cald_{\rm f}\big({z'\over z_F},z_\perp,\eta\big)   
\nonumber\\
&&\hspace{-1mm}
+~  
{4\over m^2}{1-z'\over z'}
z_\perp^2J_2\Big(|z_\perp|\sqrt{{\sigma s\over z_F}(1-z')}\Big)
\alpha_s\calh_{\rm f}''({z'\over z_F},z_\perp,\eta)\Big\},
\nonumber\\
&&\hspace{-1mm}
{d\over d\eta}\alpha_s\calh_{\rm f}''({1\over z_F},z_\perp,\eta)
\nonumber\\
&&\hspace{-1mm}
=~ {\alpha_sN_c\over \pi}\!\int_{z_F}^1\!{dz'\over z'^2}\Big\{
J_0\Big(|z_\perp|\sqrt{{\sigma s\over z_F}(1-z')}\Big)
\Big[\big({1\over 1-z'}\big)_+ -1\Big]\alpha_s\calh_{\rm f}''\big({z'\over z_F},z_\perp,\eta\big)
\nonumber\\
&&\hspace{-1mm}
+~~{m^2\over 4z_\perp^2}(1-z')z'J_2\Big(|z_\perp|\sqrt{{\sigma s\over z_F}(1-z')}\Big)
\alpha_s\cald_{\rm f}\big({z'\over z_F},z_\perp,\eta\big)\Big\}
\nonumber
\end{eqnarray}
where
$z'=1-{p_\perp^2\over\sigma\beta_F s}=1-\beta z_F$.
Here we introduced the standard notation 
$z_F\equiv{1\over\beta_F}$ for the fraction of the ``initial gluon momentum'' carried by the hadron.  
By construction, this equation
describes the evolution of fragmentation TMD at any $z_F$ and any $k_\perp\sim|z_\perp|^{-1}$. 

Let us demonstrate that Eq. (\ref{masteranni1}) agrees with  the DGLAP equation for fragmentation functions 
in the light-cone limit $x_\perp\rightarrow y_\perp$. In this limit 
\begin{eqnarray}
&&\hspace{-3mm}
{d\over d\ln\sigma}\alpha_s\cald^{\rm f}({1\over z_F},0_\perp,\ln\sigma s)~
=~{\alpha_s\over\pi}N_c\Big\{\!\int_{z_F}^1\! {dz'\over {z'}^2}~
\Big[{1\over z'(1-z')}+z'(1-z')-2\Big]
\label{8.7}\\
&&\hspace{-3mm}
\times~
\alpha_s\cald^{\rm f}({z'\over z_F},0_\perp,\ln\sigma s)
-\alpha_s\cald^{\rm f}({1\over z_F},0_\perp,\ln\sigma s)\int_0^1\! {dz'\over 1-z'}\Big\}
\nonumber
\end{eqnarray}
As explained in Eq. (\ref{3.24}),  with leading-log accuracy we can trade the cutoff in $\alpha$ for cutoff in $\mu^2$. In terms of the 
standard definition of fragmentation functions \cite{collinsbook}
\begin{eqnarray}
&&\hspace{-2mm}
d^{\rm f}_g(z_F,\ln\mu^2)~=~-{z_F^2\over 16\pi^2\alpha_sN_c(p\cdot n)}
\!\int\! du ~e^{i{u\over z_F}(pn)} \sum_X\langle 0|\tilcaf^a_\xi(un)|p+X\rangle\langle p+X|\calf^{a\xi}(0)|0\rangle^{\mu}
\nonumber\\
\label{8.8}
\end{eqnarray}
we have in the leading log approximation 
\begin{equation}
d^{\rm f}_g(z_F,\ln\mu^2)~=~{z_F\over 2N_c}\cald^{\rm f}({1\over z_F},z_\perp=0,\ln\sigma s)~+~O(\alpha_s)
\end{equation}
so we can rewrite Eq. (\ref{8.7}) in the form
\begin{eqnarray}
&&\hspace{-3mm}
{d\over d\ln\mu^2} \alpha_s(\mu)d^{\rm f}_g(z_F,\ln\mu^2)~
=~
\label{8.10}\\
&&\hspace{-3mm}
=~{\alpha_s(\mu)\over\pi}N_c\!\int_{z_F}^1\! {dz'\over z'}~\Big(\big[{1\over 1-z'}\big]_+ +{1\over z'}
+z'(1-z')-2\Big)
\alpha_s(\mu)d^{\rm f}_g\big({z_F\over z'},\ln\mu^2\big)
\nonumber
\end{eqnarray}
 easily recognizable as the DGLAP equation for fragmentation functions \cite{dglap}. (Here again the term proportional to
 $\beta$-function is absent since $\tilF^a_iF^{ai}$ is defined with an extra $\alpha_s$.)
 
 Finally, let us describe what happens if $z_F\ll 1$ and we evolve from $\sigma\sim 1$ to $\sigma\sim{z_F\over z_\perp^2s}$.
With double-log accuracy we have an equation
\begin{eqnarray}
&&\hspace{-1mm} 
{d\over d\eta}\cald^{\rm f}\big({1\over z_F},z_\perp,\eta\big)
~=~-{\alpha_sN_c\over \pi^2}\cald^{\rm f}\big({1\over z_F},z_\perp,\eta\big)
\!\int\!{d^2 p_\perp\over p_\perp^2}\big[1-e^{i(p,z)_\perp}\big]
\label{8.11}
\end{eqnarray}
(cf. Eq. (\ref{7.8})) with the solution of the Sudakov type
\begin{eqnarray}
&&\hspace{-1mm} 
\cald^{\rm f}\big({1\over z_F},z_\perp,\ln\sigma\big)
~\sim~\exp\big\{-{\alpha_sN_c\over 2\pi}\ln^2{\sigma\over z_F} sz_\perp^2 \big\}\cald^{\rm f}\big({1\over z_F},z_\perp,\ln{z_F\over z_\perp^2s}\big)
\label{8.12}
\end{eqnarray}
The evolution with the single-log accuracy should be determined from the full system (\ref{masteranni1}).

\section{Conclusions}

We have described the rapidity evolution of gluon TMD (\ref{gTMD}) with Wilson lines going to $+\infty$ in the whole range of Bjorken $x_B$ and the whole range of transverse momentum $k_\perp$. It should be emphasized that with our definition of rapidity cutoff (\ref{cutoff}) the leading-order matrix elements 
of TMD operators are UV-finite so the rapidity evolution is the only evolution and it 
describes all the dynamics of gluon TMDs (\ref{gTMD}) in the leading-log approximation.

The evolution equation  for the gluon TMD (\ref{gTMD})  with rapidity cutoff (\ref{cutoff}) is given by (\ref{masterdis}) 
and, in general, is non-linear. 
Nevertheless, for some specific cases the equation (\ref{masterdis}) linearizes. For example, let us consider the case when  $x_B\sim 1$. 
If in addition $k_\perp^2\sim s$, 
the non-linearity can be neglected for the whole range of evolution $1\gg\sigma\gg{m_N^2\over s}$ and we get the DGLAP-type system of equations (\ref{liconefinal}). 
If $k_\perp$ is small ($\sim$ few GeV) the evolution is linear
and leads to usual Sudakov factors (\ref{sudakov}). 
If we consider now the intermediate case $x_B\sim 1$ and $s\gg k_\perp^2\gg m_N^2$ the evolution
at $1\gg \sigma\gg {k_\perp^2\over s}$ will be Sudakov-type (see Eq. (\ref{7.6})) but the evolution at 
$ {k_\perp^2\over s}\gg\sigma\gg{m_N^2\over s}$ 
will be described by the full master equation (\ref{masterdis}).

For low-x region $k_\perp\sim$ few GeV and $x_B\sim{k_\perp^2\over s}$ we get the non-linear evolution described by the BK-type 
equation (\ref{7.3}).
If we now keep $k_\perp^2\sim$ few GeV and take the intermediate $1\gg x_B\equiv\beta_B\gg {k_\perp^2\over s}$ we get a mixture of 
linear and non-linear evolutions. 
If one evolves $\sigma$ ($\leftrightarrow$ rapidity) from 1 to ${k_\perp^2\over s}$ first there will be Sudakov-type  
double-log evolution (\ref{7.8}) from $\sigma=1$ to $\sigma={k_\perp^2\over\beta_Bs}$, then the transitional region
at $\sigma\sim {k_\perp^2\over\beta_Bs}$, and after that the non-linear evolution (\ref{7.3}) at 
${k_\perp^2\over\beta_Bs}\gg\sigma\gg{k_\perp^2\over s}$. The transition between the linear evolution (\ref{7.8})
and the non-linear one (\ref{7.3}) should be described by the full equation (\ref{masterdis}). 

Another interesting case is $x_B\sim{m_N^2\over s}$ and $s\gg k_\perp^2\gg m_N^2$. In this case, if we evolve $\sigma$ from 1 to  ${m_N^2\over s}$,
first we have the BK evolution (\ref{7.3}) up to $\sigma\sim{k_\perp^2\over s}$ and then for the evolution between $\sigma\sim{k_\perp^2\over s}$
and $\sigma\sim{m_N^2\over s}$ we need the Eq. (\ref{masterdis}) in full.

In conclusion, let us again emphasize that the evolution of the fragmentation TMDs (\ref{maelfr}) is always linear and the corresponding equation
(\ref{eveqfr}) describes both the DGLAP region $k_\perp^2\sim s$ and Sudakov region $k_\perp^2\sim $ few $GeV^2$.

As an outlook, it would be very interesting to obtain the NLO correction to the evolution equation (\ref{masterdis}). The NLO corrections
to the BFKL  \cite{nlobfkl} and BK \cite{nlobk,kw,nlojimwlk} equation are available but they suffer from the well-known problem that they lead to negative cross sections. This difficulty can be overcome by the ``collinear resummation'' of double-logarithmic contributions for the BFKL
\cite{resumbfkl} and BK \cite{resumbk} equations and we hope that our Eq. (\ref{masterdis}) and especially its future NLO version will 
help to solve the problem of negative cross sections of NLO amplitudes at high energies.

The authors are grateful to G.A. Chirilli, J.C. Collins, Yu. Kovchegov,  A. Prokudin,  A.V. Radyushkin, T. Rogers, and F. Yuan for valuable discussions. This work was supported by contract
 DE-AC05-06OR23177 under which the Jefferson Science Associates, LLC operate the Thomas Jefferson National Accelerator Facility, and by the grant DE-FG02-97ER41028.

\section{Appendix A: light-cone expansion of propagators}

In this section we consider the case when the transverse momenta of background fast fields $l_\perp$  are much smaller than
the characteristic transverse momenta $p_\perp$ of ``quantum'' slow gluons. As we discussed in Sect. \ref{sec2}, in this case fast fields 
do not necessarily shrink to a shock wave and one should use the light-cone expansion of propagators instead. The 
parameter of expansion is the twist of the operator and we will expand up to operators of leading collinear twist two. Such operators
are built of two gluon operators $\sim F_{\bu i} F_{\bu j}$ or quark ones $\bsi\!\!\not\! p_1\psi$ and gauge links. 
To get coefficients in front of these operators it is sufficient to consider the external gluon field of the type $A_\bu(z_\ast,z_\perp)$ and quark fields $\not\!p_1\psi(x_\ast,x_\perp)$ with all other components being zero.
\footnote{The $z_\bu$ dependence of the external fields can be omitted since due to the rapidity ordering  $\alpha$'s of the fast fields are much less than
$\alpha$'s of the slow ones.}

For simplicity, let us again start with the expansion of a scalar propagator. 

\subsection{Scalar propagator}
\subsubsection{Feynman propagator for a scalar particle in the background gluon field}
 For simplicity we will first perform the calculation  for ``scalar propagator''  $(x|{1\over P^2+i\epsilon}|y)$. As we mentioned above, we assume that the only nonzero component of the external field is
 $A_\bu$ and it does not depend on $z_\bu$ so the operator $\alpha=i{\partial\over\partial z_\bullet}$ commutes with all background fields. 
 The propagator in the external field $A_\bu(z_\ast,z_\perp)$ has the form
\begin{eqnarray}
\hspace{-1mm}
(x|{1\over P^2+i\epsilon}|y)~&=&~\Big[-i\theta(x_\ast-y_\ast)\!\int_0^\infty\!{\dhd\alpha\over 2\alpha}
+i\theta(y_\ast-x_\ast)\!\int_{-\infty}^0\!{\dhd\alpha\over 2\alpha}\Big]~
\label{A.1}\\
\hspace{-1mm}
&\times&
~e^{-i\alpha(x-y)_\bu}(x_\perp|{\rm Pexp}\big\{-i\!\int_{y_\ast}^{x_\ast}\! dz_\ast
\big[{p_\perp^2\over \alpha s}-{2g\over s}A_\bu(z_\ast)\big]\big\}|y_\perp)
\nonumber
\end{eqnarray}
 The Pexp in the r.h.s. of Eq. (\ref{A.1}) can be transformed to
\begin{eqnarray}
&&\hspace{-11mm}
(x_\perp|{\rm Pexp}\big\{-i\!\int_{y_\ast}^{x_\ast}\! dz_\ast\big[{p_\perp^2\over \alpha s}-{2g\over s}A_\bu(z_\ast)\big]\big\}|y_\perp)
\nonumber\\
&&\hspace{-11mm}
=~(x_\perp|e^{-i{p_\perp^2\over\alpha s}(x_\ast-y_\ast)}
{\rm Pexp}\Big\{{2ig\over s}\!\int_{y_\ast}^{x_\ast}\! dz_\ast
~e^{i{p_\perp^2\over\alpha s}(z_\ast-y_\ast)}A_\bu(z_\ast)e^{-i{p_\perp^2\over\alpha s}(z_\ast-y_\ast)}
\Big\}|y_\perp)
\label{A.2}
\end{eqnarray}
Since the longitudinal distances $z_\ast$ inside the shock wave are small we can expand 
\begin{eqnarray}
&&\hspace{-1mm}
e^{i{p_\perp^2\over\alpha s}(z_\ast-y_\ast)}A_\bu e^{-i{p_\perp^2\over\alpha s}(z_\ast-y_\ast)}
~=~A_\bu -{z_\ast-y_\ast\over\alpha s}\{p^i,F_{\bu i}\}
-{(z_\ast-y_\ast)^2\over 2\alpha^2 s^2}\{p^j,\{p^i,D_jF_{\bu i}\}\}+...
\nonumber\\
&&\hspace{12mm}
=~A_\bu -{z_\ast-y_\ast\over\alpha s}(2p^iF_{\bu i}-iD^iF_{\bu i})
-2{(z_\ast-y_\ast)^2\over \alpha^2 s^2}(p^ip^j-ip^jD^i)D_jF_{\bu i}
+...
\label{A.3}
\end{eqnarray}

This is effectively expansion around the light ray $y_\perp+{2\over s}y_\ast p_1$ with
the parameter of the expansion $\sim {|l_\perp|\over |p_\perp|}\ll 1$. As we mentioned, we 
will expand up to the operator(s) with twist two.

 We obtain
\begin{eqnarray}
&&\hspace{-1mm}
\calo(x_\ast,y_\ast;p_\perp)~\equiv~{\rm Pexp}\Big\{{2ig\over s}\!\int_{y_\ast}^{x_\ast}\! dz_\ast
~e^{i{p_\perp^2\over\alpha s}(z_\ast-y_\ast)}A_\bu(z_\ast)e^{-i{p_\perp^2\over\alpha s}(z_\ast-y_\ast)}\Big\}
\label{A.4}\\
&&\hspace{-1mm}
=~1+{2ig\over s}\!\int_{y_\ast}^{x_\ast}\!\!\!dz_\ast ~\big[A_\bu
-{(z-y)_\ast\over\alpha s}(2p^iF_{\bu i}-iD^iF_{\bu i})-2{(z_\ast-y_\ast)^2\over \alpha^2 s^2}(p^ip^j-ip^jD^i)D_jF_{\bu i}\big]
\nonumber\\
&&\hspace{7mm}-~{4g^2\over s^2}\!\int_{y_\ast}^{x_\ast}\!\! \!dz_\ast\!\int_{y_\ast}^{z_\ast}\!\!\!dz'_\ast
\big[A_\bu-{2(z-y)_\ast\over\alpha s}p^iF_{\bu i}\big]
\big[ A_\bu-{2(z'-y)_\ast\over\alpha s}p^jF_{\bu j}\big]~+~...
\nonumber
\end{eqnarray}
It is clear that the terms $\sim A_\bu$ will combine to form gauge links so the r.h.s. of the above equation will turn to
\begin{eqnarray}
&&\hspace{-1mm}
\calo(x_\ast,y_\ast;p_\perp)~
=~[x_\ast,y_\ast]-{2ig\over\alpha s^2}\!\int_{y_\ast}^{x_\ast}\!\!\!dz_\ast
~(z-y)_\ast\Big(2p^j[x_\ast,z_\ast]F_{\bu j}(z_\ast)-i[x_\ast,z_\ast]D^jF_{\bu j}(z_\ast)
\nonumber\\
&&\hspace{42mm}
+~2{(z-y)_\ast\over\alpha s}(p^jp^k[x_\ast,z_\ast]-ip^k[x_\ast,z_\ast]D^j)D_kF_{\bu j}\Big)[z_\ast,y_\ast]
\nonumber\\
&&\hspace{25mm}
+~{8g^2\over\alpha s^3}\!\int_{y_\ast}^{x_\ast}\!\! \!dz_\ast\!\int_{y_\ast}^{z_\ast}\!\!\!dz'_\ast~(z'-y)_\ast\Big(i
[x_\ast,z_\ast] F_{\bu j}(z_\ast) [z_\ast,z'_\ast]F_\bu^{~j}(z'_\ast)
\label{A.5}\\
&&\hspace{42mm}
-~2p^jp^k{(z-y)_\ast\over\alpha s}[x_\ast,z_\ast] F_{\bu j}(z_\ast) [z_\ast,z'_\ast]F_{\bu k}(z'_\ast)\Big)[z'_\ast,y_\ast]
+...
\nonumber
\end{eqnarray}
where dots stand for the higher twists. 

Thus, the final expansion of the propagator (\ref{A.1}) near the light cone  $y_\perp+{2\over s}y_\ast p_1$
takes the form
\begin{eqnarray}
&&\hspace{-1mm}
(x|{1\over P^2+i\epsilon}|y)~=~\Big[-i\theta(x_\ast-y_\ast)\!\int_0^\infty\!{\dhd\alpha\over 2\alpha}
+i\theta(y_\ast-x_\ast)\!\int_{-\infty}^0\!{\dhd\alpha\over 2\alpha}\Big]
\label{A.6}\\
&&\hspace{32mm}
\times~e^{-i\alpha(x-y)_\bu}(x_\perp|e^{-i{p_\perp^2\over\alpha s}(x-y)_\ast}\calo(x_\ast,y_\ast;p_\perp)|y_\perp)
\nonumber
\end{eqnarray}
Note that the transverse arguments of all background fields in Eq. (\ref{A.6}) are effectively $y_\perp$.

\subsubsection{Scalar propagator for the complex conjugate amplitude}

For calculations of the complex conjugate amplitude we need also the propagator 
\begin{eqnarray}
\hspace{-1mm}
(x|{1\over P^2-i\epsilon}|y)~&=&~
\Big[i\theta(y_\ast-x_\ast)\!\int_0^\infty\!{\dhd\alpha\over 2\alpha}
-i\theta(x_\ast-y_\ast)\!\int_{-\infty}^0\!{\dhd\alpha\over 2\alpha}\Big]
\label{A.7}\\
\hspace{-1mm}
&\times&
~e^{-i\alpha(x-y)_\bu}(x_\perp|{\rm Pexp}\big\{-i\!\int_{y_\ast}^{x_\ast}\! dz_\ast\big[{p_\perp^2\over \alpha s}-{2g\over s}\tilA_\bu(z_\ast)\big]\big\}|y_\perp)\nonumber
\end{eqnarray}
For the calculation of the square of Lipatov vertex we need to consider point $x$ inside the shock wave and point $y$ outside.  In this case
one should rewrite Eq. (\ref{A.2}) as follows
\begin{eqnarray}
&&\hspace{-11mm}
(x_\perp|{\rm Pexp}\big\{-i\!\int_{y_\ast}^{x_\ast}\! dz_\ast\big[{p_\perp^2\over \alpha s}-{2g\over s}\tilA_\bu(z_\ast)\big]\big\}|y_\perp)
\nonumber\\
&&\hspace{-11mm}
=~(x_\perp|
{\rm Pexp}\Big\{{2ig\over s}\!\int_{y_\ast}^{x_\ast}\! dz_\ast
~e^{-i{p_\perp^2\over\alpha s}(x_\ast-z_\ast)}\tilA_\bu(z_\ast)e^{i{p_\perp^2\over\alpha s}(x_\ast-z_\ast)}
\Big\}e^{-i{p_\perp^2\over\alpha s}(x_\ast-y_\ast)}|y_\perp)
\label{A.8}
\end{eqnarray}

The light-cone expansion around $x_\perp+{2\over s}x_\ast p_1$ is given by Eq. (\ref{A.3}) with $y_\ast\rightarrow x_\ast$ 
\begin{eqnarray}
&&\hspace{-1mm}
e^{i{p_\perp^2\over\alpha s}(z_\ast-x_\ast)}\tilA_\bu e^{-i{p_\perp^2\over\alpha s}(z_\ast-x_\ast)}
\label{A.9}\\
&&\hspace{12mm}
=~\tilA_\bu -{z_\ast-x_\ast\over\alpha s}(2\tilF_{\bu j}p^j+i\tilD^j\tilF_{\bu j})
-2{(z_\ast-x_\ast)^2\over \alpha^2 s^2}(\tilD_j\tilF_{\bu i}p^ip^j+i\tilD^i\tilD_j\tilF_{\bu i}p^j)
+...
\nonumber
\end{eqnarray}
(the only difference with the expansion (\ref{A.3}) is that we should put the operators $p^j$ to the right) and therefore
\begin{eqnarray}
&&\hspace{-1mm}
\ticalo(x_\ast,y_\ast;p_\perp)~\equiv~{\rm Pexp}\Big\{{2ig\over s}\!\int_{y_\ast}^{x_\ast}\! dz_\ast
~e^{-i{p_\perp^2\over\alpha s}(x_\ast-z_\ast)}\tilA_\bu(z_\ast)e^{i{p_\perp^2\over\alpha s}(x_\ast-z_\ast)}\Big\}~=~1
\nonumber\\
&&\hspace{-1mm}
+~{2ig\over s}\!\int_{y_\ast}^{x_\ast}\!\!\!dz_\ast ~\big[\tilA_\bu
-{z_\ast-x_\ast\over\alpha s}(2\tilF_{\bu j}p^j+i\tilD^j\tilF_{\bu j})-2{(z_\ast-x_\ast)^2\over \alpha^2 s^2}(\tilD_j\tilF_{\bu i}p^ip^j+i\tilD^i\tilD_j\tilF_{\bu i}p^j)\big]
\nonumber\\
&&\hspace{7mm}
-~{4g^2\over s^2}\!\int_{y_\ast}^{x_\ast}\!\! \!dz_\ast\!\int_{y_\ast}^{z_\ast}\!\!\!dz'_\ast
\big[\tilA_\bu-{2(z-x)_\ast\over\alpha s}\tilF_{\bu i}p^i\big]
\big[ \tilA_\bu-{2(z'-x)_\ast\over\alpha s}\tilF_{\bu j}p^j\big]~+~...
\label{A.10}
\end{eqnarray}
which turns to
\begin{eqnarray}
&&\hspace{-1mm}
\ticalo(x_\ast,y_\ast;p_\perp)~=~[x_\ast,y_\ast]
+{2ig\over\alpha s^2}\!\int_{x_\ast}^{y_\ast}\!\!\!dz_\ast
~[x_\ast,z_\ast]\big\{2\tilF_{\bu j}(z_\ast)[z_\ast,y_\ast]p^j
+i\tilD^j\tilF_{\bu j}(z_\ast)[z_\ast,y_\ast]
\nonumber\\
&&\hspace{23mm}
+~2{(z-x)_\ast\over\alpha s}\big(\tilD_k\tilF_{\bu j}(z_\ast)[z_\ast,y_\ast]p^jp^k
+i\tilD^j\tilD_k\tilF_{\bu j}(z_\ast)[z_\ast,y_\ast]p^k\big)\big\}(z-x)_\ast
\nonumber\\
&&\hspace{23mm}
+~{8g^2\over\alpha s^3}\!\int_{y_\ast}^{x_\ast}\!\! \!dz_\ast\!\int_{y_\ast}^{z_\ast}\!\!\!dz'_\ast~(z-x)_\ast
[x_\ast,z_\ast] \Big(
-i\tilF_{\bu j}(z_\ast) [z_\ast,z'_\ast]\tilF_\bu^{~j}(z'_\ast)[z'_\ast,y_\ast]
\nonumber\\
&&\hspace{23mm}
-~2{(z'-x)_\ast\over\alpha s}\tilF_{\bu j}(z_\ast) [z_\ast,z'_\ast]\tilF_{\bu k}(z'_\ast)[z'_\ast,y_\ast]p^jp^k\Big)
+...
\label{A.11}
\end{eqnarray}
and we get
\begin{eqnarray}
\hspace{-1mm}
(x|{1\over P^2-i\epsilon}|y)~&=&~
\Big[i\theta(y_\ast-x_\ast)\!\int_0^\infty\!{\dhd\alpha\over 2\alpha}
-i\theta(x_\ast-y_\ast)\!\int_{-\infty}^0\!{\dhd\alpha\over 2\alpha}\Big]
\label{A.12}\\
\hspace{-1mm}
&\times&
~e^{-i\alpha(x-y)_\bu}(x_\perp|\ticalo(x_\ast,y_\ast;p_\perp)e^{-i{p_\perp^2\over\alpha s}(x-y)_\ast}|y_\perp)
\nonumber
\end{eqnarray}
Here (in Eq.  (\ref{A.11})) the transverse arguments of all background fields are effectively $x_\perp$.

\subsubsection{The emission vertex \label{scalaremission}} 

For the calculation of Lipatov vertex we need the propagator in mixed representation 
$(k|{1\over P^2+i\epsilon}|z)$ in the limit $k^2\rightarrow 0$
 where $k=\alpha p_1+{k_\perp^2\over\alpha s}p_2+k_\perp$: 
\begin{eqnarray}
&&\hspace{-11mm}
k^2(k|{1\over P^2+i\epsilon}|y)~=~{2\over s}\!\int\! dx_\ast dx_\bu d^2x_\perp 
~e^{ikx}\big[-s{\partial\over\partial x_\bu} {\partial\over\partial x_\ast} +\partial_\perp^2\big](x|{1\over P^2+i\epsilon}|y)~
\label{A.13}
\end{eqnarray}
First, we perform the trivial integrations over $x_\bu$ and $x_\perp$:
\begin{eqnarray}
&&\hspace{-1mm}
\lim_{k^2\rightarrow 0}k^2(k|{1\over P^2+i\epsilon}|y)~
\label{A.14}\\
&&\hspace{-1mm} 
=\!\int\! dx_\ast d^2x_\perp 
~e^{i{k_\perp^2\over\alpha s}x_\ast-i(k,x)_\perp}\big[{\partial\over\partial x_\ast} -{i\over\alpha s}\partial_\perp^2\big]
\theta(x-y)_\ast(x_\perp|e^{-i{p_\perp^2\over\alpha s}(x-y)_\ast}\calo(x_\ast,y_\ast;p_\perp)|y_\perp)e^{i\alpha y_\bullet}
\nonumber\\
&&\hspace{-1mm} 
=~\!\int\!dx_\ast d^2x_\perp 
~e^{i{k_\perp^2\over\alpha s}x_\ast-i(k,x)_\perp}
(x_\perp|e^{-i{p_\perp^2\over\alpha s}(x-y)_\ast}{\partial\over\partial x_\ast}\theta(x-y)_\ast 
\calo(x_\ast,y_\ast;p_\perp)|y_\perp)e^{i\alpha y_\bullet}
\nonumber\\
&&\hspace{-1mm} 
=~e^{i\alpha y_\bullet}e^{i{k_\perp^2\over\alpha s}y_\ast}\!\!\int\! dx_\ast
{\partial\over\partial x_\ast} \theta(x-y)_\ast (k_\perp|\calo(x_\ast,y_\ast;k)|y_\perp)
\nonumber\\
&&\hspace{54mm}
=~e^{i\alpha y_\bullet}e^{i{k_\perp^2\over\alpha s}y_\ast-i(k,y)_\perp}\calo(\infty,y_\ast,y_\perp;k)
\nonumber
\end{eqnarray}
where $\calo(\infty,y_\ast,y_\perp;k)\equiv e^{i(k,y)_\perp}(k_\perp|\calo(\infty,y_\ast;k)|y_\perp)$. In the explicit form
\begin{eqnarray}
&&\hspace{-1mm}
\calo(\infty,y_\ast;y_\perp;k)
\nonumber\\
&&\hspace{5mm}
=~
[\infty,y_\ast]_y-{2ig\over\alpha s^2}\!\int_{y_\ast}^{\infty}\!\!\!dz_\ast
~(z-y)_\ast\Big(2k^j[\infty,z_\ast]_yF_{\bu j}(z_\ast,y_\perp)-i[\infty,z_\ast]_yD^jF_{\bu j}(z_\ast,y_\perp)
\nonumber\\
&&\hspace{15mm}
+~2{(z-y)_\ast\over\alpha s}(k^jk^l[\infty,z_\ast]_y-ik^l[\infty,z_\ast]_yD^j)D_lF_{\bu j}(z_\ast,y_\perp)\Big)[z_\ast,y_\ast]_y
\nonumber\\
&&\hspace{15mm}
+~{8g^2\over\alpha s^3}\!\int_{y_\ast}^{\infty}\!\! \!dz_\ast\!\int_{y_\ast}^{z_\ast}\!\!\!dz'_\ast~(z'-y)_\ast\Big(i
[\infty,z_\ast]_y F_{\bu j}(z_\ast,y_\perp) [z_\ast,z'_\ast]_yF_\bu^{~j}(z'_\ast,y_\perp)
\nonumber\\
&&\hspace{15mm}
-~2k^jk^l{(z-y)_\ast\over\alpha s}[\infty,z_\ast]_y F_{\bu j}(z_\ast,y_\perp) [z_\ast,z'_\ast]_yF_{\bu l}(z'_\ast,y_\perp)\Big)[z'_\ast,y_\ast]_y
\label{A.15}
\end{eqnarray}
where the transverse arguments of all fields are $y_\perp$ and $p^j$ is replaced by $k^j$.

Similarly, for the complex conjugate amplitude we get
\begin{equation}
\hspace{-1mm}
\lim_{k^2\rightarrow 0}k^2(x|{1\over P^2-i\epsilon}|k)~
=~e^{-ikx}\ticalo(x_\ast,\infty,x_\perp;k)
\label{A.16}
\end{equation}
where $\ticalo(x_\ast,\infty,x_\perp;k)\equiv (x_\perp|\ticalo(x_\ast,\infty,p_\perp)|k_\perp)e^{-i(k,x)_\perp}$, or, in the explicit form
\begin{eqnarray}
&&\hspace{-1mm}
\ticalo(x_\ast,\infty,x_\perp;k)
\nonumber\\
&&\hspace{5mm}
=~[x_\ast,\infty]_x
+{2ig\over\alpha s^2}\!\int_{x_\ast}^{\infty}\!\!\!dz_\ast
~[x_\ast,z_\ast]_x\big\{2\tilF_{\bu j}(z_\ast,x_\perp)[z_\ast,\infty]_xk^j
+i\tilD^j\tilF_{\bu j}(z_\ast,x_\perp)[z_\ast,\infty]_x
\nonumber\\
&&\hspace{15mm}
+~2{(z-x)_\ast\over\alpha s}\big(\tilD_l\tilF_{\bu j}(z_\ast,x_\perp)[z_\ast,\infty]_xk^jk^l
+i\tilD^j\tilD_l\tilF_{\bu j}(z_\ast,x_\perp)[z_\ast,\infty]_xk^l\big)\big\}(z-x)_\ast
\nonumber\\
&&\hspace{15mm}
+~{8g^2\over\alpha s^3}\!\int^{\infty}_{x_\ast}\!\! \!dz_\ast\!\int^{\infty}_{z_\ast}\!\!\!dz'_\ast~(z-x)_\ast
[x_\ast,z_\ast]_x\Big(
-i\tilF_{\bu j}(z_\ast,x_\perp) [z_\ast,z'_\ast]_x\tilF_\bu^{~j}(z'_\ast,x_\perp)[z'_\ast,\infty]_x
\nonumber\\
&&\hspace{15mm}
-~2{(z'-x)_\ast\over\alpha s}\tilF_{\bu j}(z_\ast,x_\perp) [z_\ast,z'_\ast]_x\tilF_{\bu l}(z'_\ast,x_\perp)[z'_\ast,\infty]_xk^jk^l\Big)
+...
\label{A.17}
\end{eqnarray}
In the complex conjugate amplitude we expand around the light cone $x_\perp+{2\over s}x_\ast p_1$ so 
the transverse arguments of all fields in Eq. (\ref{A.17}) are $x_\perp$.
Note that the second terms in the r.h.s. of Eqs. (\ref{A.6}) and (\ref{A.7}) (proportional to $\int_{-\infty}^0\dhd\alpha$) do not 
contribute since $\alpha>0$ for the emitted particle.

\subsection{Gluon propagator}
\subsubsection{Gluon propagator in the background gluon field}

As we saw in the previous Section, to get the  emission vertex (\ref{A.13}) it is sufficient to write down 
the propagator at $x_\ast>y_\ast$.
The gluon propagator in the bF gauge has the form
\begin{eqnarray}
&&\hspace{-1mm}
i\langle A_\mu^a(x)A_\nu^b(y)\rangle~=~(x|{1\over P^2+2igF+i\epsilon}|y)^{ab}_{\mu\nu}
\label{A.18}\\
&&\hspace{-1mm}
~\stackrel{x_\ast>y_\ast}{=}-i\int_0^\infty\!{\dhd\alpha\over 2\alpha}~e^{-i\alpha(x-y)_\bu}
~(x_\perp|{\rm Pexp}\big\{-i\!\int_{y_\ast}^{x_\ast}\! dz_\ast\big[{p_\perp^2\over \alpha s}
-{2g\over s}A_\bu(z_\ast)-{2ig\over\alpha s}F(z_\ast)\big]\big\}|y_\perp)^{ab}_{\mu\nu}
\nonumber\\
&&\hspace{-1mm}
=~
-i\!\int_0^\infty\!{\dhd\alpha\over 2\alpha}~e^{-i\alpha(x-y)_\bu}
\nonumber\\
&&\hspace{8mm}
\times~(x_\perp|e^{-i{p_\perp^2\over\alpha s}(x_\ast-y_\ast)}
{\rm Pexp}\Big\{ig\!\int_{y_\ast}^{x_\ast}\! d{2\over s}z_\ast
~e^{i{p_\perp^2\over\alpha s}(z_\ast-y_\ast)}\big(A_\bu+{i\over\alpha}F\big)(z_\ast)e^{-i{p_\perp^2\over\alpha s}(z_\ast-y_\ast)}
\Big\}^{ab}_{\mu\nu}|y_\perp)
\nonumber
\end{eqnarray}
where powers of $F$ are treated as usual, for example  $(FA_\bu FF)_{\mu\nu}~\equiv~F_\mu^{~\xi}A_\bu g_{\xi\eta}F^{\eta\lambda}F_{\lambda\nu}$.
The expansion (\ref{A.3}) now looks like 
\begin{eqnarray}
&&\hspace{-1mm}
e^{i{p_\perp^2\over\alpha s}(z_\ast-y_\ast)}\big(A_\bu g_{\mu\nu}+{i\over\alpha}F_{\mu\nu}\big)e^{-i{p_\perp^2\over\alpha s}(z_\ast-y_\ast)}
~=~A_\bu g_{\mu\nu}+{i\over\alpha}F_{\mu\nu}
\nonumber\\
&&\hspace{-1mm} 
+~i{z_\ast-y_\ast\over\alpha s}\big[p_\perp^2,A_\bu g_{\mu\nu}+{i\over\alpha}F_{\mu\nu}\big]
-~{(z-y)_\ast^2\over 2\alpha^2s^2}\big[p_\perp^2,\big[p_\perp^2,A_\bu g_{\mu\nu}+{i\over\alpha}F_{\mu\nu}\big]\big]+...   
\nonumber\\
&&\hspace{-1mm}
=~g_{\mu\nu}\Big[A_\bu-{z_\ast-y_\ast\over\alpha s}(2p^jF_{\bu j}-iD^jF_{\bu j})
-2{(z-y)_\ast^2\over \alpha^2s^2}(p^jp^kD_jF_{\bu k}-ip^kD_kD^jF_{\bu j})\Big]
\nonumber\\
&&\hspace{-1mm}
+~{i\over\alpha}F_{\mu\nu}+2i{z_\ast-y_\ast\over\alpha^2 s}p^jD_jF_{\mu\nu}+2i{(z-y)_\ast^2\over\alpha^3s^2}p^jp^kD_jD_kF_{\mu\nu}+...
\label{A.19}
\end{eqnarray}
so we get
\begin{eqnarray}
&&\hspace{-3mm}
\calg_{\mu\nu}(x_\ast,y_\ast;p_\perp)~\equiv~{\rm Pexp}\Big\{ig\!\int_{y_\ast}^{x_\ast}\! d{2\over s}z_\ast
~e^{i{p_\perp^2\over\alpha s}(z_\ast-y_\ast)}\big(A_\bu+{i\over\alpha}F\big)(z_\ast)
e^{-i{p_\perp^2\over\alpha s}(z_\ast-y_\ast)}\Big\}_{\mu\nu}
\label{A.20}\\
&&\hspace{-3mm}
=~g_{\mu\nu}+ig\!\int_{y_\ast}^{x_\ast}\! d{2\over s}z_\ast
~e^{i{p_\perp^2\over\alpha s}(z_\ast-y_\ast)}\big(A_\bu+{i\over\alpha}F\big)_{\mu\nu}(z_\ast)e^{-i{p_\perp^2\over\alpha s}(z_\ast-y_\ast)}
-~g^2\!\int_{y_\ast}^{x_\ast}\! d{2\over s}z_\ast~e^{i{p_\perp^2\over\alpha s}(z_\ast-y_\ast)}
\nonumber\\
&&\hspace{-3mm}
\times
~\big(A_\bu+{i\over\alpha}F\big)_{\mu\xi}(z_\ast)e^{-i{p_\perp^2\over\alpha s}(z_\ast-y_\ast)}
\!\int_{y_\ast}^{z_\ast}\! d{2\over s}z'_\ast
~e^{i{p_\perp^2\over\alpha s}(z'_\ast-y_\ast)}\big(A_\bu+{i\over\alpha}F\big)^\xi_{\ \nu}(z'_\ast)e^{-i{p_\perp^2\over\alpha s}(z'_\ast-y_\ast)}~
\nonumber\\
&&\hspace{-3mm}
=~g_{\mu\nu}+{2ig\over s}\!\int_{y_\ast}^{x_\ast}\!\!\!dz_\ast\Big\{g_{\mu\nu}\Big[A_\bu(z_\ast)
-{(z-y)_\ast\over\alpha s}(2p^iF_{\bu i}-iD^iF_{\bu i})-2{(z-y)_\ast^2\over \alpha^2s^2}(p^jp^kD_jF_{\bu k}
\nonumber\\
&&\hspace{9mm}
-~ip^kD_kD^jF_{\bu j})\Big]+~{i\over\alpha}F_{\mu\nu}
+2i{z_\ast-y_\ast\over\alpha^2 s}p^jD_jF_{\mu\nu}+2i{(z-y)_\ast^2\over\alpha^3s^2}p^jp^kD_jD_kF_{\mu\nu}\Big\}
\nonumber\\
&&\hspace{-3mm}
-~{4g^2\over s^2}\!\int_{y_\ast}^{x_\ast}\!\! \!dz_\ast\!\int_{y_\ast}^{z_\ast}\!\!\!dz'_\ast
\big[g_{\mu\xi}A_\bu+{i\over\alpha}F_{\mu\xi}-2g_{\mu\xi}p^j{(z-y)_\ast\over\alpha s}F_{\bu j}\big]
\nonumber\\
&&\hspace{33mm}
\times~\big[\delta^\xi_\nu A_\bu+{i\over\alpha}F^{\xi}_{~\nu}-2p^k{(z'-y)_\ast\over\alpha s}F_{\bu k}\delta^\xi_\nu\big]~+~...
\nonumber\\
&&\hspace{-3mm}
=~g_{\mu\nu}[x_\ast,y_\ast]+g\!\int_{y_\ast}^{x_\ast}\!\!\!dz_\ast~
\Big(-{2i\over\alpha s^2}(z-y)_\ast g_{\mu\nu}\big\{2p^j[x_\ast,z_\ast]F_{\bu j}(z_\ast)
-i[x_\ast,z_\ast]D^jF_{\bu j}(z_\ast)
\nonumber\\
&&\hspace{33mm}
+~2{(z-y)_\ast\over \alpha s}(p^jp^k[x_\ast,z_\ast]D_jF_{\bu k}-ip^k[x_\ast,z_\ast]D_kD^jF_{\bu j})\big\}
\nonumber\\
&&\hspace{-3mm}
+~{4\over \alpha s^2}(\delta_\mu^jp_{2\nu}-\delta_\nu^jp_{2\mu})\big\{[x_\ast,z_\ast]F_{\bu j}(z_\ast)
+{2(z-y)_\ast\over\alpha s}p^k[x_\ast,z_\ast]D_kF_{\bu j}(z_\ast)
\nonumber\\
&&\hspace{33mm}
+~2{(z-y)_\ast^2\over\alpha^2s^2}p^kp^l[x_\ast,z_\ast]D_kD_lF_{\bullet j}\big\}\Big)[z_\ast,y_\ast]
\nonumber\\
&&\hspace{-3mm}
+~{8g^2\over\alpha s^3}\!\int_{y_\ast}^{x_\ast}\!\! \!dz_\ast\!\int_{y_\ast}^{z_\ast}\!\!\!dz'_\ast\Big(
~\big[ig_{\mu\nu}(z'-y)_\ast-{2\over \alpha s}p_{2\mu}p_{2\nu}\big]
[x_\ast,z_\ast] F_{\bu j}(z_\ast) [z_\ast,z'_\ast]F_\bu^{~j}(z'_\ast)
\nonumber\\
&&\hspace{33mm}
-~2{g_{\mu\nu}\over\alpha s}p^jp^k(z-y)_\ast(z'-y)_\ast[x_\ast,z_\ast] F_{\bu j}(z_\ast) [z_\ast,z'_\ast]F_{\bu k}(z'_\ast)\Big)[z'_\ast,y_\ast]
\nonumber
\end{eqnarray}
Note that $F_{\mu\xi}F^{\xi\eta}F_{\eta\nu}$ 
and higher terms of the expansion in powers of $F_{\mu\nu}$  vanish since the only non-vanishing
field strength is $F_{\bu i}$.

Finally,
\begin{eqnarray}
\hspace{-1mm}
(x|{1\over P^2+2igF+i\epsilon}|y)^{ab}_{\mu\nu}~&=&~\Big[-i\theta(x_\ast-y_\ast)\!\int_0^\infty\!{\dhd\alpha\over 2\alpha}
+i\theta(y_\ast-x_\ast)\!\int_{-\infty}^0\!{\dhd\alpha\over 2\alpha}\Big]~
\label{A.21}\\
\hspace{-1mm}
&\times&
~e^{-i\alpha(x-y)_\bu}(x_\perp|e^{-i{p_\perp^2\over\alpha s}(x-y)_\ast}\calg^{ab}_{\mu\nu}(x_\ast,y_\ast;p_\perp)|y_\perp)
\nonumber
\end{eqnarray}

For the complex conjugate amplitude we obtain in a similar way
\begin{eqnarray}
\hspace{-1mm}
(x|{1\over P^2+2igF-i\epsilon}|y)^{ab}_{\mu\nu}~&=&~
\Big[i\theta(y_\ast-x_\ast)\!\int_0^\infty\!{\dhd\alpha\over 2\alpha}
-i\theta(x_\ast-y_\ast)\!\int_{-\infty}^0\!{\dhd\alpha\over 2\alpha}\Big]
\label{A.22}\\
\hspace{-1mm}
&\times&
~e^{-i\alpha(x-y)_\bu}(x_\perp|\ticalg_{\mu\nu}^{ab}(x_\ast,y_\ast;p_\perp)e^{-i{p_\perp^2\over\alpha s}(x-y)_\ast}|y_\perp)
\nonumber
\end{eqnarray}
where 
\begin{eqnarray}
&&\hspace{-3mm}
\ticalg_{\mu\nu}(x_\ast,y_\ast;p_\perp)~\equiv~{\rm Pexp}\Big\{ig\!\int_{y_\ast}^{x_\ast}\! d{2\over s}z_\ast
~e^{i{p_\perp^2\over\alpha s}(z_\ast-x_\ast)}\big(\tilA_\bu+{i\over\alpha}\tilF\big)(z_\ast)
e^{-i{p_\perp^2\over\alpha s}(z_\ast-x_\ast)}\Big\}_{\mu\nu}
\label{A.23}\\
&&\hspace{-3mm}
=~g_{\mu\nu}[x_\ast,y_\ast]+g\!\int_{x_\ast}^{y_\ast}\!\!\!dz_\ast~[x_\ast,z_\ast]
\Big({2i\over\alpha s^2}(z-x)_\ast g_{\mu\nu}\big\{2\tilF_{\bu j}(z_\ast)[z_\ast,y_\ast]p^j
+i\tilD^j\tilF_{\bu j}(z_\ast)[z_\ast,y_\ast]
\nonumber\\
&&\hspace{44mm}
+~2{(z-x)_\ast\over \alpha s}(\tilD_j\tilF_{\bu k}(z_\ast)[z_\ast,y_\ast]p^jp^k+i\tilD_k\tilD^j\tilF_{\bu j}(z_\ast)[z_\ast,y_\ast]p^k)\big\}
\nonumber\\
&&\hspace{-3mm}
-~{4\over \alpha s^2}(\delta_\mu^jp_{2\nu}-\delta_\nu^jp_{2\mu})\big\{\tilF_{\bu j}(z_\ast)[z_\ast,y_\ast]
+{2(z-x)_\ast\over\alpha s}\tilD_k\tilF_{\bu j}(z_\ast)[z_\ast,y_\ast]p^k
\nonumber\\
&&\hspace{44mm}
+~2{(z-x)_\ast^2\over\alpha^2s^2}\tilD_k\tilD_l\tilF_{\bullet j}(z_\ast)[z_\ast,y_\ast]p^kp^l\big\}\Big)
\nonumber\\
&&\hspace{-3mm}
+~{8g^2\over\alpha s^3}\!\int_{x_\ast}^{y_\ast}\!\! \!dz_\ast\!\int_{z_\ast}^{y_\ast}\!\!\!dz'_\ast [x_\ast,z_\ast] \Big(
\big[-ig_{\mu\nu}(z-x)_\ast
-{2\over \alpha s}p_{2\mu}p_{2\nu}\big]
\tilF_{\bu j}(z_\ast) [z_\ast,z'_\ast]\tilF_\bu^{~j}(z'_\ast)[z'_\ast,y_\ast]
\nonumber\\
&&\hspace{44mm}
-~{2g_{\mu\nu}\over\alpha s}(z-x)_\ast(z'-x)_\ast \tilF_{\bu j}(z_\ast) [z_\ast,z'_\ast]\tilF_{\bu k}(z'_\ast)[z'_\ast,y_\ast]p^jp^k\Big)
\nonumber
\end{eqnarray}

\subsubsection{Gluon propagator in the background quark field} 
\begin{figure}[htb]
\begin{center}
\includegraphics[width=88mm]{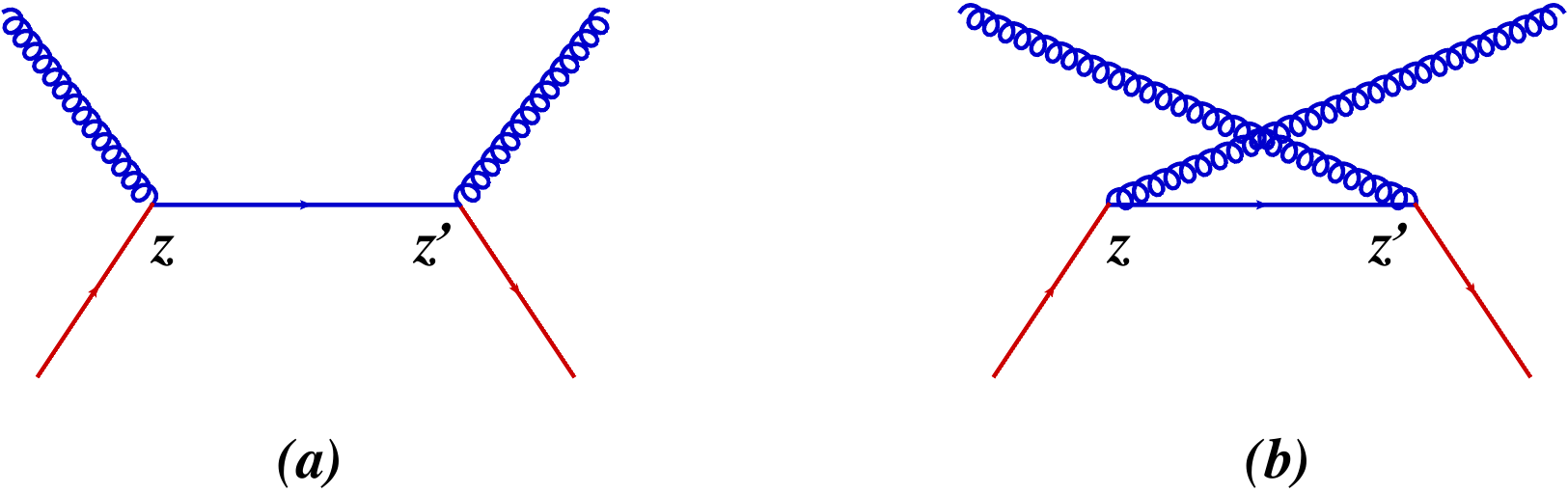}
\end{center}
\caption{Gluon propagator in an external quark field.\label{fig:7}}
\end{figure}

We do not impose the condition $D^iF_{\bu i}=0$ so our external field has quark sources $D^iF^a_{\bu i}=g\bsi t^a\!\!\!\not\!p_1\psi$ which we need to
take into consideration. The corresponding contribution  to gluon propagator comes from diagrams in Fig. \ref{fig:7}

%
\begin{eqnarray}
&&\hspace{-1mm}
\langle A_\mu^a(x) A^b_\nu(y)\rangle_{\rm Fig. ~\ref{fig:7}}~=~g^2\!\int\! d^4zd^4z'~(x|{1\over P^2+i\epsilon}|z)^{ac}(z'|{1\over P^2+i\epsilon}|y)^{db}
\label{kvark1}\\
&&\hspace{-1mm} 
\times~\big[(z|\bsi t^c\gamma_\mu\!\not\! P{i\over  \!\not\! P^2+i\epsilon}\gamma_\nu t^d\psi|z')+(z'|\bsi t^d\gamma_\nu{i\over  \!\not\! P^2+i\epsilon}\!\not\! P\gamma_\mu t^c\psi|z)\big]
\nonumber
\end{eqnarray}
As we mentioned above, we can consider quark fields with $+\half$ spin projection onto $p_1$ direction 
which corresponds to $\bsi (...)\psi$ operators of leading collinear twist. In this approximation $\!\not\!p_2\psi=0$ so
the only non-zero propagators are $\langle A_\bu(x) A_\bu(y)\rangle$, $\langle A_\bu(x) A_i(y)\rangle$ and
$\langle A_i(x) A_j(y)\rangle$. In addition, we assume that the quark fields $\psi(z)$ depend only 
on $z_\perp$ and $z_\ast$ (same as gluon fields) so the operator $\hat\alpha={2\over s}\hat{p}_\ast$ commutes with 
all background-field operators. We get
\begin{eqnarray}
&&\hspace{-1mm}
\langle A_\bu^a(x) A^b_\bu(y)\rangle_{\rm Fig. ~\ref{fig:7}}~=~2ig^2\!\int\! d^4zd^4z'~
(x|{1\over P^2+i\epsilon}|z)^{ac}(z'|{1\over P^2+i\epsilon}|y)^{db}
\label{kvark2}\\
&&\hspace{-1mm} 
\times~\big[(z|\bsi t^c\!\not\! p_1P_\bu{1\over  \!\not\! P^2+i\epsilon}\ t^d\psi|z')
+(z'|\bsi t^d{1\over  \!\not\! P^2+i\epsilon}P_\bu\!\not\! p_1 t^c\psi|z)\big]
\nonumber
\end{eqnarray}
In our gluon field $\!\not\! P=\alpha\!\!\not\! p_1+2{\not\! p_2\over s}P_\bu+\!\not\!p_\perp$ so $\!\not\! P^2=2\alpha P_\bu -p_\perp^2+{2i\over s}\!\not\! p_2\gamma^j F_{\bu j}$ and one can rewrite $P_\bu{i\over  \not\! P^2+i\epsilon}$ as
\begin{eqnarray}
\hspace{-1mm}
P_\bu{1\over 2\alpha P_\bu -p_\perp^2+i\epsilon}
~
&=&~\Big[{p_\perp^2\over 2\alpha }+{1\over 2\alpha }\big(2\alpha P_\bu -p_\perp^2\big)
\Big]{1\over 2\alpha P_\bu -p_\perp^2+i\epsilon}
\nonumber\\
&=&~{1\over 2\alpha }~+~{p_\perp^2\over 2\alpha }{1\over 2\alpha P_\bu -p_\perp^2+i\epsilon}
\end{eqnarray}
(the term ${2\over s}\!\not\! p_2\gamma^j F_{\bu j}$ does not contribute due to $\!\not\! p_2\psi=0$).
Similarly,
\begin{eqnarray}
&&\hspace{-3mm}
{1\over 2\alpha P_\bu -p_\perp^2+i\epsilon}P_\bu
=~{1\over 2\alpha }~+~{1\over 2\alpha P_\bu -p_\perp^2+i\epsilon}{p_\perp^2\over 2\alpha }
\end{eqnarray}
so one can rewrite the propagator (\ref{kvark1}) as
\begin{eqnarray}
&&\hspace{-1mm}
\langle A_\bu^a(x) A^b_\bu(y)\rangle_{\rm Fig. ~\ref{fig:7}}~=~
-ig(x|{1\over \alpha(P^2+i\epsilon)}D^jF_{\bu j}{1\over P^2+i\epsilon}|y)^{ab}
\label{kvark5}\\
&&\hspace{-1mm} 
+~ig^2(x|\big({1\over \alpha(P^2+i\epsilon)}\big)^{ac}
\big(p_\perp^2\bsi +2ip^j\partial_j\bsi-\partial_\perp^2\bsi\big)t^c{1\over P^2+i\epsilon}
\!\not\! p_1 t^d\psi\big({1\over P^2+i\epsilon}\big)^{db}|y)
\nonumber\\
&&\hspace{-1mm} 
+~ig^2(y|\big({1\over \alpha(P^2+i\epsilon)}\big)^{bd}\bsi t^d{1\over P^2+i\epsilon}t^c\!\not\! p_1
\big(\psi p_\perp^2 -2i\partial_j\psi p^j-\partial_\perp^2\psi\big)
\big({1\over P^2+i\epsilon}\big)^{ca}|x)
\nonumber
\end{eqnarray}
where in the first line we have rewritten $g^2\bsi[t^c,t^d]\!\not\! p_1\psi$ as $-g(D^jF_{\bu j})^{cd}$.

Similarly,  we get
\begin{eqnarray}
&&\hspace{-1mm}
\langle A_\bu^a(x) A^b_i(y)\rangle_{\rm Fig. ~\ref{fig:7}}
\label{kvark6}\\
&&\hspace{5mm} 
=~-ig^2(x|\big({1\over P^2+i\epsilon}\big)^{ac}(p^j\bsi -i\partial^j\bsi)\gamma_j\!\not\! p_1\gamma_it^c
{1\over P^2+i\epsilon}t^d\psi\big({1\over P^2+i\epsilon}\big)^{db}|y)
\nonumber\\
&&\hspace{11mm} 
-~ig^2(y|\big({1\over P^2+i\epsilon}\big)^{bd}\bsi t^d{1\over P^2+i\epsilon}t^c\gamma_i\!\not\! p_1
\gamma_j(\psi p^j+i\partial^j\psi)\big({1\over P^2+i\epsilon}\big)^{ca}|x),
\nonumber\\
&&\hspace{-1mm}
\langle A_i^a(x) A^b_j(y)\rangle_{\rm Fig. ~\ref{fig:7}}
~=~ig^2(x|\big({\alpha\over P^2+i\epsilon}\big)^{ac}\bsi\gamma_i\!\not\! p_1\gamma_jt^c
{1\over P^2+i\epsilon}t^d\psi\big({1\over P^2+i\epsilon}\big)^{db}|y)
\nonumber\\
&&\hspace{31mm} 
+~ig^2(y|\big({\alpha\over P^2+i\epsilon}\big)^{bd}\bsi t^d{1\over P^2+i\epsilon}t^c\gamma_j\!\not\! p_1
\gamma_i\psi\big({1\over P^2+i\epsilon}\big)^{ca}|x)
\nonumber
\end{eqnarray}
for the remaining propagators.

If now the point $y$ lies inside the shock wave we can expand the gluon and quark propagators around the 
light ray $y_\perp+{2\over s}y_\ast p_1$. It is easy to see that the expansion of the gluon fields $A_\bu$ given by Eq. (\ref{A.3}) 
exceeds our twist-two accuracy so we need only  expansion of quark fields which is
\begin{eqnarray}
&&\hspace{-1mm}
e^{i{p_\perp^2\over\alpha s}(z_\ast-y_\ast)}\psi e^{-i{p_\perp^2\over\alpha s}(z_\ast-y_\ast)}
~=~\psi +2{z_\ast-y_\ast\over\alpha s}p^jD_j\psi+...
\label{kvark7}
\end{eqnarray}
(and similarly for $\bsi$ and $D^iF_{\bu i}$). 

It is convenient to parametrize quark contribution in the same way as the gluon one (\ref{A.21})
\begin{eqnarray}
&&\hspace{-0mm}
\langle A_\mu^a(x)A_\nu^b(y)\rangle_{\rm Fig. ~\ref{fig:7}}~=~
\Big[-\theta(x_\ast-y_\ast)\!\int_0^\infty\!{\dhd\alpha\over 2\alpha}
+\theta(y_\ast-x_\ast)\!\int_{-\infty}^0\!{\dhd\alpha\over 2\alpha}\Big]~e^{-i\alpha(x-y)_\bu}
\nonumber\\
&&\hspace{-1mm}
\times~\Big[(x_\perp|e^{-i{p_\perp^2\over\alpha s}(x-y)_\ast}\calq^{ab}_{\mu\nu}(x_\ast,y_\ast;p_\perp)|y_\perp)
+(y_\perp|\bar\calq^{ab}_{\mu\nu}(x_\ast,y_\ast;p_\perp)e^{-i{p_\perp^2\over\alpha s}(x-y)_\ast}|x_\perp)\Big]
\label{kvark8}
\end{eqnarray}
In the leading order we need only the first two terms of the expansion  (\ref{kvark7}) which gives
\begin{eqnarray}
&&\hspace{-1mm}
\calq^{ab}_{\bu\bu}(x_\ast,y_\ast;p_\perp)~
\nonumber\\
&&\hspace{5mm} =~-{ig\over \alpha^2s}\!\int_{y_\ast}^{x_\ast}\! \! dz_\ast 
\big([x_\ast,z_\ast]D^jF_{\bu j}(z_\ast)[z_\ast,y_\ast]+{2(z-y)_\ast\over\alpha s}p^k[x_\ast,z_\ast]D_kD^jF_{\bu j}[z_\ast,y_\ast]\big)^{ab}
\nonumber\\
&&\hspace{5mm} 
+~{g^2\over \alpha^3s^2}\!\int_{y_\ast}^{x_\ast} \!\!dz_\ast\!\int_{y_\ast}^{z_\ast} \!\!dz'_\ast~
\Big[\big(p_\perp^2\bsi(z_\ast) +2ip^j\bsi\!\stackrel{\leftarrow}{D}_j\!(z_\ast)\big)\!\not\! p_1[z_\ast,x_\ast]t^a[x_\ast,y_\ast]t^b[y_\ast,z'_\ast]\psi(z'_\ast)
\nonumber\\
&&\hspace{26mm} 
+~2{z_\ast-y_\ast\over\alpha s} p_\perp^2p^j \bsi\!\stackrel{\leftarrow}{D}_j(z_\ast) \!\not\! p_1[z_\ast,x_\ast]t^a[x_\ast,y_\ast]t^b[y_\ast,z'_\ast]\psi(z'_\ast)
\nonumber\\
&&\hspace{26mm}
+~2{z'_\ast-y_\ast\over\alpha s}p_\perp^2p^j\bsi(z_\ast) \!\not\! p_1[z_\ast,x_\ast]t^a[x_\ast,y_\ast]t^b[y_\ast,z'_\ast]D_j\psi(z'_\ast) \Big]
\label{kvark9}
\end{eqnarray}
and
\begin{eqnarray}
&&\hspace{-1mm}
\bar\calq^{ab}_{\bu\bu}(x_\ast,y_\ast;p_\perp)~=~
\nonumber\\
&&\hspace{-1mm}   
=~-{g^2\over \alpha^3s^2}\!\int_{y_\ast}^{x_\ast} \!\!dz_\ast\!\int_{z_\ast}^{x_\ast} \!\!dz'_\ast
\Big[\bsi(z_\ast)\!\not\! p_1[z_\ast,y_\ast]t^b[y_\ast,x_\ast]t^a[x_\ast,z'_\ast]\big(\psi(z'_\ast) p_\perp^2 -2iD_j\psi (z'_\ast)p^j\big)
\nonumber\\
&&\hspace{26mm}
-~2{z_\ast-y_\ast\over\alpha s}\bsi\!\stackrel{\leftarrow}{D}_j(z_\ast) \!\not\! p_1[z_\ast,y_\ast]t^b[y_\ast,x_\ast]t^a[x_\ast,z'_\ast]\psi(z'_\ast) p_\perp^2p^j 
\nonumber\\
&&\hspace{26mm}
-~2{z'_\ast-y_\ast\over\alpha s}\bsi(z_\ast) \!\not\! p_1[z_\ast,y_\ast]t^b[y_\ast,x_\ast]t^a[x_\ast,z'_\ast]D_j\psi(z'_\ast) p_\perp^2p^j\Big]
\label{kvark10}
\end{eqnarray}
Similarly, for propagator $\langle A_\bu(x) A_i(y)\rangle$ one gets
\begin{eqnarray}
&&\hspace{-1mm}
\calq^{ab}_{\bu i}(x_\ast,y_\ast;p_\perp)~
\nonumber\\
&&\hspace{-1mm}=~-{g^2\over \alpha^2 s^2}\!\int_{y_\ast}^{x_\ast}\! \! dz_\ast\!\int_{y_\ast}^{z_\ast}\! \! dz'_\ast
\Big[
[p^j\bsi(z_\ast)-i\bsi\stackrel{\leftarrow}{D^j}(z_\ast)]\gamma_j\!\not\! p_1\gamma_i[z_\ast,x_\ast]t^a[x_\ast,y_\ast]t^b[y_\ast,z'_\ast] \psi(z'_\ast)
\nonumber\\
&&\hspace{-1mm}
+~{2(z-y)_\ast\over\alpha s}p^jp^k\bsi\!\stackrel{\leftarrow}{D}_k(z_\ast)\gamma_j\!\not\! p_1\gamma_i[z_\ast,x_\ast]t^a[x_\ast,y_\ast]t^b[y_\ast,z'_\ast] \psi(z'_\ast)
\nonumber\\
&&\hspace{-1mm}
+~{2(z'-y)_\ast\over\alpha s}p^jp^k\bsi\gamma_j\!\not\! p_1\gamma_i[z_\ast,x_\ast]t^a[x_\ast,y_\ast]t^b[y_\ast,z'_\ast] 
D_k\psi(z'_\ast)
\Big]
\label{kvark11}
\end{eqnarray}
\begin{eqnarray}
&&\hspace{-1mm} 
\bar\calq^{ab}_{\bu i}(x_\ast,y_\ast;p_\perp)~
\nonumber\\
&&\hspace{-1mm} 
=~-{g^2\over \alpha^2 s^2}\!\int_{y_\ast}^{x_\ast} \!\!dz_\ast\!\int_{z_\ast}^{x_\ast} \!\!dz'_\ast
~\Big[\bsi(z_\ast)\gamma_i\!\not\! p_1\gamma_j[z_\ast,y_\ast]t^b[y_\ast,x_\ast]t^a[x_\ast,z'_\ast](\psi p^j +iD^j\psi )(z'_\ast)
\nonumber\\
&&\hspace{-1mm}
-~2{z_\ast-y_\ast\over\alpha s}\bsi\!\stackrel{\leftarrow}{D}_k\!(z_\ast)\gamma_i\!\not\! p_1\gamma_j[z_\ast,y_\ast]t^b[y_\ast,x_\ast]t^a[x_\ast,z'_\ast]\psi(z'_\ast) p^j p^k
\nonumber\\
&&\hspace{-1mm}
-~2{z'_\ast-y_\ast\over\alpha s}\bsi(z_\ast)\gamma_i\!\not\! p_1\gamma_j[z_\ast,y_\ast]t^b[y_\ast,x_\ast]t^a[x_\ast,z'_\ast]D_k\psi(z'_\ast)p^jp^k\Big]
\label{kvark12}
\end{eqnarray}
For the propagator $\langle A_i(x) A_\bu(y)\rangle$ the corresponding expressions  $\calq^{ab}_{i \bu}(x_\ast,y_\ast;p_\perp)$ 
and\\
 $\bar\calq^{ab}_{i \bu}(x_\ast,y_\ast;p_\perp)$ are obtained from Eqs.  (\ref{kvark11}) and (\ref{kvark12}) by replacements in the r.h.s.'s\\
$\gamma_j\!\not\! p_1\gamma_i\rightarrow \gamma_i\!\not\! p_1\gamma_j$ and 
$\gamma_i\!\not\! p_1\gamma_j\rightarrow \gamma_j\!\not\! p_1\gamma_i$, respectively.

Finally, for the propagator  $\langle A_i(x) A_j (y)\rangle$ we obtain
\begin{eqnarray}
&&\hspace{-1mm}
\calq^{ab}_{i j}(x_\ast,y_\ast;p_\perp)~
=~{g^2\over \alpha s^2}\!\int_{y_\ast}^{x_\ast}\! \! dz_\ast\!\int_{y_\ast}^{z_\ast}\! \! dz'_\ast
\Big[
\bsi(z_\ast)\gamma_i\!\not\! p_1\gamma_j[z_\ast,x_\ast]t^a[x_\ast,y_\ast]t^b[y_\ast,z'_\ast] \psi(z'_\ast)
\nonumber\\
&&\hspace{25mm}
+~{2(z-y)_\ast\over\alpha s}p^k\bsi\!\stackrel{\leftarrow}{D}_k(z_\ast)\gamma_i\!\not\! p_1\gamma_j[z_\ast,x_\ast]t^a[x_\ast,y_\ast]t^b[y_\ast,z'_\ast] \psi(z'_\ast)
\nonumber\\
&&\hspace{25mm}
+~{2(z'-y)_\ast\over\alpha s}p^k\bsi\gamma_i\!\not\! p_1\gamma_j[z_\ast,x_\ast]t^a[x_\ast,y_\ast]t^b[y_\ast,z'_\ast] 
D_k\psi(z'_\ast)
\Big]
\label{kvark13}
\end{eqnarray}
\begin{eqnarray}
&&\hspace{-1mm} 
\bar\calq^{ab}_{i j}(x_\ast,y_\ast;p_\perp)~
=~-{g^2\over \alpha s^2}\!\int_{y_\ast}^{x_\ast} \!\!dz_\ast\!\int_{z_\ast}^{x_\ast} \!\!dz'_\ast
~\Big[\bsi(z_\ast)\gamma_j\!\not\! p_1\gamma_i[z_\ast,y_\ast]t^b[y_\ast,x_\ast]t^a[x_\ast,z'_\ast]\psi(z'_\ast)
\nonumber\\
&&\hspace{25mm}
-~2{z_\ast-y_\ast\over\alpha s}\bsi(z_\ast)\!\stackrel{\leftarrow}{D}_k\gamma_j\!\not\! p_1\gamma_i[z_\ast,y_\ast]t^b[y_\ast,x_\ast]t^a[x_\ast,z'_\ast]\psi(z'_\ast) p^k
\nonumber\\
&&\hspace{25mm}
-~2{z'_\ast-y_\ast\over\alpha s}\bsi(z_\ast)\gamma_j\!\not\! p_1\gamma_i[z_\ast,y_\ast]t^b[y_\ast,x_\ast]t^a[x_\ast,z'_\ast]D_k\psi(z'_\ast)p^k\Big]
\label{kvark14}
\end{eqnarray}

For the complex conjugate amplitude we get in a similar way
\begin{eqnarray}
&&\hspace{-0mm}
\langle \tilA_\mu^a(x)\tilA_\nu^b(y)\rangle_{\rm Fig. ~\ref{fig:7}}~=~
\Big[-\theta(y_\ast-x_\ast)\!\int_0^\infty\!{\dhd\alpha\over 2\alpha}
+\theta(x_\ast-y_\ast)\!\int_{-\infty}^0\!{\dhd\alpha\over 2\alpha}\Big]~e^{-i\alpha(x-y)_\bu}
\nonumber\\
&&\hspace{-1mm}
\times~\Big[(x_\perp|\ticalq^{ab}_{\mu\nu}(x_\ast,y_\ast;p_\perp)e^{-i{p_\perp^2\over\alpha s}(x-y)_\ast}|y_\perp)
+(y_\perp|e^{-i{p_\perp^2\over\alpha s}(x-y)_\ast}{\bar\ticalq}^{ab}_{\mu\nu}(x_\ast,y_\ast;p_\perp)|x_\perp)\Big]
\label{kvark15}
\end{eqnarray}
where
\begin{eqnarray}
&&\hspace{-1mm}
\ticalq^{ab}_{\bu\bu}(x_\ast,y_\ast;p_\perp)~
\nonumber\\
&&\hspace{5mm} 
=~{-ig\over \alpha^2s}\!\int_{y_\ast}^{x_\ast}\! \! dz_\ast 
\big([x_\ast,z_\ast]\tilD^j\tilF_{\bu j}(z_\ast)[z_\ast,y_\ast]-{2(x-z)_\ast\over\alpha s}[x_\ast,z_\ast]\tilD_k\tilD^j\tilF_{\bu j}[z_\ast,y_\ast]p^k\big)^{ab}
\nonumber\\
&&\hspace{5mm} 
+~{g^2\over \alpha^3s^2}\!\int_{x_\ast}^{y_\ast} \!\!dz_\ast\!\int_{z_\ast}^{y_\ast} \!\!dz'_\ast
\Big[\tilde{\bsi}(z_\ast)\!\not\! p_1[z_\ast,x_\ast]t^a[x_\ast,y_\ast]t^b[y_\ast,z'_\ast]\big(\tilde{\psi}(z'_\ast) p_\perp^2 -2i\tilD_j\tilde{\psi} (z'_\ast)p^j\big)
\nonumber\\
&&\hspace{26mm}
+~2{z_\ast-x_\ast\over\alpha s}\tilde{\bsi}\!\stackrel{\leftarrow}{\tilD}_j(z_\ast) \!\not\! p_1[z_\ast,x_\ast]t^a[x_\ast,y_\ast]t^b[y_\ast,z'_\ast]\tilde{\psi}(z'_\ast) p_\perp^2p^j 
\nonumber\\
&&\hspace{26mm}
+~2{z'_\ast-x_\ast\over\alpha s}\tilde{\bsi}(z_\ast) \!\not\! p_1[z_\ast,x_\ast]t^a[x_\ast,y_\ast]t^b[y_\ast,z'_\ast]\tilD_j\tilde{\psi}(z'_\ast) p_\perp^2p^j
\label{kvark16}
\end{eqnarray}
\begin{eqnarray}
&&\hspace{-1mm}
{\bar\ticalq}^{ab}_{\bu\bu}(x_\ast,y_\ast;p_\perp)~
\nonumber\\
&&\hspace{-1mm} 
=~-{g^2\over \alpha^3s^2}\!\int_{x_\ast}^{y_\ast} \!\!dz_\ast\!\int_{x_\ast}^{z_\ast} \!\!dz'_\ast~
\Big[\big(p_\perp^2\tilde{\bsi}(z_\ast) +2ip^j\tilde{\bsi}\!\stackrel{\leftarrow}{\tilD}_j\!(z_\ast)\big)\!\not\! p_1[z_\ast,y_\ast]t^b[y_\ast,x_\ast]t^a[x_\ast,z'_\ast]\tilde{\psi}(z'_\ast)
\nonumber\\
&&\hspace{26mm} 
-~2{z_\ast-x_\ast\over\alpha s} p_\perp^2p^j \tilde{\bsi}\!\stackrel{\leftarrow}{\tilD}_j(z_\ast) \!\not\! p_1[z_\ast,y_\ast]t^b[y_\ast,x_\ast]t^a[x_\ast,z'_\ast]\tilde{\psi}(z'_\ast)
\nonumber\\
&&\hspace{26mm}
-~2{z'_\ast-x_\ast\over\alpha s}p_\perp^2p^j\tilde{\bsi}(z_\ast) \!\not\! p_1[z_\ast,y_\ast]t^b[y_\ast,x_\ast]t^a[x_\ast,z'_\ast]\tilD_j\tilde{\psi}(z'_\ast) \Big]
\label{kvark17}
\end{eqnarray}
and
\begin{eqnarray}
&&\hspace{-1mm}
{\ticalq}^{ab}_{\bu i}(x_\ast,y_\ast;p_\perp)~
\nonumber\\
&&\hspace{5mm} 
=~
{-g^2\over \alpha^2s^2}\!\int_{x_\ast}^{y_\ast} \!\!dz_\ast\!\int_{z_\ast}^{y_\ast} \!\!dz'_\ast
\Big[\tilde{\bsi}(z_\ast)\gamma_j\!\not\! p_1\gamma_i[z_\ast,x_\ast]t^a[x_\ast,y_\ast]t^b[y_\ast,z'_\ast](\tilde{\psi} p^j+i\tilD^j\psi)
\nonumber\\
&&\hspace{24mm}
+~2{z_\ast-x_\ast\over\alpha s}\tilde{\bsi}\!\stackrel{\leftarrow}{\tilD}_k(z_\ast)
 \gamma_j\!\not\! p_1\gamma_i[z_\ast,x_\ast]t^a[x_\ast,y_\ast]t^b[y_\ast,z'_\ast]\tilde{\psi}(z'_\ast) p^j p^k
\nonumber\\
&&\hspace{24mm}
+~2{z'_\ast-x_\ast\over\alpha s}\tilde{\bsi}(z_\ast)\gamma_j\!\not\! p_1\gamma_i
[z_\ast,x_\ast]t^a[x_\ast,y_\ast]t^b[y_\ast,z'_\ast]\tilD_k\tilde{\psi}(z'_\ast) p^jp^k\Big]
\label{kvark18}
\end{eqnarray}
\begin{eqnarray}
&&\hspace{-1mm}
{\bar\ticalq}^{ab}_{\bu i}(x_\ast,y_\ast;p_\perp)~
\nonumber\\
&&\hspace{5mm} 
=~{-g^2\over \alpha^2s^2}\!\int_{x_\ast}^{y_\ast} \!\!dz_\ast\!\int_{x_\ast}^{z_\ast} \!\!dz'_\ast~
\Big[(p^j\tilde{\bsi} -i\tilD^j\tilde{\bsi})
\gamma_i\!\not\! p_1\gamma_j[z_\ast,y_\ast]t^b[y_\ast,x_\ast]t^a[x_\ast,z'_\ast]\tilde{\psi}(z'_\ast)
\nonumber\\
&&\hspace{24mm} 
-~2{z_\ast-x_\ast\over\alpha s} p^j p^k\tilde{\bsi}\!\stackrel{\leftarrow}{\tilD}_k(z_\ast) 
\gamma_i\!\not\! p_1\gamma_j[z_\ast,y_\ast]t^b[y_\ast,x_\ast]t^a[x_\ast,z'_\ast]\tilde{\psi}(z'_\ast)
\nonumber\\
&&\hspace{24mm}
-~2{z'_\ast-x_\ast\over\alpha s}p^jp^k\tilde{\bsi}(z_\ast)\gamma_i\!\not\! p_1\gamma_j
[z_\ast,y_\ast]t^b[y_\ast,x_\ast]t^a[x_\ast,z'_\ast]\tilD_k\tilde{\psi}(z'_\ast) \Big]
\label{kvark19}
\end{eqnarray}
To get ${\ticalq}^{ab}_{i\bu}$ one should again make the replacement
$\gamma_j\!\not\!p_1\gamma_i\rightarrow\gamma_i\!\not\!p_1\gamma_j$ in Eq. (\ref{kvark18}) and to get ${\bar\ticalq}^{ab}_{i\bu}$ 
the replacement $\gamma_i\!\not\!p_1\gamma_j\rightarrow\gamma_j\!\not\!p_1\gamma_i$ in 
Eq. (\ref{kvark19}).  Finally, similarly to Eq. (\ref{kvark13}) one obtains
\begin{eqnarray}
&&\hspace{-1mm}
\ticalq^{ab}_{ij}(x_\ast,y_\ast;p_\perp)~=~
{g^2\over \alpha s^2}\!\int_{x_\ast}^{y_\ast} \!\!dz_\ast\!\int_{z_\ast}^{y_\ast} \!\!dz'_\ast
\Big[\tilde{\bsi}(z_\ast)\gamma_i\!\not\! p_1\gamma_j[z_\ast,x_\ast]t^a[x_\ast,y_\ast]t^b[y_\ast,z'_\ast]\tilde{\psi}
\nonumber\\
&&\hspace{26mm}
+~2{z_\ast-x_\ast\over\alpha s}\tilde{\bsi}\!\stackrel{\leftarrow}{\tilD}_k(z_\ast)
 \gamma_i\!\not\! p_1\gamma_j[z_\ast,x_\ast]t^a[x_\ast,y_\ast]t^b[y_\ast,z'_\ast]\tilde{\psi}(z'_\ast)p^k
\nonumber\\
&&\hspace{26mm}
+~2{z'_\ast-x_\ast\over\alpha s}\tilde{\bsi}(z_\ast)\gamma_i\!\not\! p_1\gamma_j
[z_\ast,x_\ast]t^a[x_\ast,y_\ast]t^b[y_\ast,z'_\ast]\tilD_k\tilde{\psi}(z'_\ast)p^k\Big]
\label{kvark20}
\end{eqnarray}
\begin{eqnarray}
&&\hspace{-1mm}
{\bar\ticalq}^{ab}_{ij}(x_\ast,y_\ast;p_\perp)~
=~-{g^2\over \alpha s^2}\!\int_{x_\ast}^{y_\ast} \!\!dz_\ast\!\int_{x_\ast}^{z_\ast} \!\!dz'_\ast~
\Big[\tilde{\bsi} \gamma_j\!\not\! p_1\gamma_i[z_\ast,y_\ast]t^b[y_\ast,x_\ast]t^a[x_\ast,z'_\ast]\tilde{\psi}(z'_\ast)
\nonumber\\
&&\hspace{26mm} 
-~2{z_\ast-x_\ast\over\alpha s} p^k\tilde{\bsi}\!\stackrel{\leftarrow}{\tilD}_k(z_\ast) 
\gamma_j\!\not\! p_1\gamma_i[z_\ast,y_\ast]t^b[y_\ast,x_\ast]t^a[x_\ast,z'_\ast]\tilde{\psi}(z'_\ast)
\nonumber\\
&&\hspace{26mm}
-~2{z'_\ast-x_\ast\over\alpha s}p^k\tilde{\bsi}(z_\ast)\gamma_j\!\not\! p_1\gamma_i
[z_\ast,y_\ast]t^b[y_\ast,x_\ast]t^a[x_\ast,z'_\ast]\tilD_k\tilde{\psi}(z'_\ast) \Big]
\label{kvark21}
\end{eqnarray}

\subsubsection{Final form of the gluon propagator}
Assembling terms from two previous Sections we get the final result for \\
background-Feynman  gluon propagator in external field 
 in the form
\begin{eqnarray}
&&\hspace{-1mm}
\langle A_\mu^a(x)A_\nu^b(y)\rangle~
\label{A.45}\\
&&\hspace{-1mm}
=~
\Big[-\theta(x_\ast-y_\ast)\!\int_0^\infty\!{\dhd\alpha\over 2\alpha}
+\theta(y_\ast-x_\ast)\!\int_{-\infty}^0\!{\dhd\alpha\over 2\alpha}\Big]e^{-i\alpha(x-y)_\bu}
\big\{(x_\perp|e^{-i{p_\perp^2\over\alpha s}(x-y)_\ast}
\nonumber\\
&&\hspace{-1mm}
\times~\big[\calg^{ab}_{\mu\nu}(x_\ast,y_\ast;p_\perp)+\calq^{ab}_{\mu\nu}(x_\ast,y_\ast;p_\perp)\big]|y_\perp)
+(y_\perp|\bar\calq^{ab}_{\mu\nu}(x_\ast,y_\ast;p_\perp)e^{-i{p_\perp^2\over\alpha s}(x-y)_\ast}|x_\perp)\big\}
\nonumber
\end{eqnarray}
for Feynman propagator and 
\begin{eqnarray}
&&\hspace{-1mm}
\langle \tilA_\mu^a(x)\tilA_\nu^b(y)\rangle~
\label{A.46}\\
&&\hspace{5mm}
=~
\Big[-\theta(y_\ast-x_\ast)\!\int_0^\infty\!{\dhd\alpha\over 2\alpha}
+\theta(x_\ast-y_\ast)\!\int_{-\infty}^0\!{\dhd\alpha\over 2\alpha}\Big]e^{-i\alpha(x-y)_\bu}
\big\{(x_\perp|\big[\ticalg^{ab}_{\mu\nu}(x_\ast,y_\ast;p_\perp)
\nonumber\\
&&\hspace{13mm}
+~\ticalq^{ab}_{\mu\nu}(x_\ast,y_\ast;p_\perp)\big]e^{-i{p_\perp^2\over\alpha s}(x-y)_\ast}|y_\perp)
+(y_\perp|e^{-i{p_\perp^2\over\alpha s}(x-y)_\ast}{\bar\ticalq}^{ab}_{\mu\nu}(x_\ast,y_\ast;p_\perp)|x_\perp)\big\}
\nonumber
\end{eqnarray}
for the anti-Feynman propagator in the complex conjugate amplitude.

\subsection{Vertex of gluon emission}
Repeating the steps which lead us to Eq. (\ref{A.14}) we obtain 
\begin{eqnarray}
&&\hspace{-1mm}
\lim_{k^2\rightarrow 0}k^2\langle A_\mu^a(k) A^b_\nu(y)\rangle~
=~-ie^{iky}\calo^{ab}_{\mu\nu}(\infty,y_\ast,y_\perp;k),
\label{A.47}\\
&&\hspace{-1mm}
\calo^{ab}_{\mu\nu}(\infty,y_\ast,y_\perp;k)~=~\calg^{ab}_{\mu\nu}(\infty,y_\ast,y_\perp;k)
+\calq^{ab}_{\mu\nu}(\infty,y_\ast,y_\perp;k)+\bar\calq^{ab}_{\mu\nu}(\infty,y_\ast,y_\perp;k)
\nonumber
\end{eqnarray}
where
\begin{eqnarray}
&&\hspace{-1mm}
\calg^{ab}_{\mu\nu}(\infty,y_\ast,y_\perp;k)~\equiv~e^{i(k,y)_\perp}(k_\perp|\calg^{ab}_{\mu\nu}(\infty,y_\ast;p_\perp)|y_\perp),
\nonumber\\
&&\hspace{-1mm}
\calq^{ab}_{\mu\nu}(\infty,y_\ast,y_\perp;k)~\equiv~e^{i(k,y)_\perp}(k_\perp|\calq^{ab}_{\mu\nu}(\infty,y_\ast,p_\perp)|y_\perp),
\nonumber\\
&&\hspace{-1mm}
\bar\calq^{ab}_{\mu\nu}(\infty,y_\ast,y_\perp;k)~\equiv~e^{-i(k,y)_\perp}(y_\perp|\bar\calq^{ab}_{\mu\nu}(\infty,y_\ast;p_\perp)|k_\perp)
\label{A.48}
\end{eqnarray}
The explicit expressions can be read from Eqs. (\ref{A.20}) and (\ref{kvark9}) - (\ref{kvark14}) by taking  the transverse arguments of all 
fields to be $y_\perp$ and replacing the operators $p^j$ with $k^j$ similarly to Eq. (\ref{A.15}).

Similarly, for the complex conjugate amplitude the emission vertex takes the form
\begin{eqnarray}
&&\hspace{-1mm}
\lim_{k^2\rightarrow 0}k^2\langle \tilA_\mu^a(x) \tilA^b_\nu(k)\rangle~
=~ie^{-ikx}\ticalo^{ab}_{\mu\nu}(x_\ast,\infty,x_\perp;k),
\label{A.49}\\
&&\hspace{-1mm}
\ticalo^{ab}_{\mu\nu}(x_\ast,\infty,x_\perp;k)~=~\ticalg^{ab}_{\mu\nu}(x_\ast,\infty,x_\perp;k)
+\ticalq^{ab}_{\mu\nu}(x_\ast,\infty,x_\perp;k)+{\bar\ticalq}^{ab}_{\mu\nu}(x_\ast,\infty,x_\perp;k)
\nonumber
\end{eqnarray}
where
\begin{eqnarray}
&&\hspace{-1mm}
\ticalg^{ab}_{\mu\nu}(x_\ast,\infty,x_\perp;k)~\equiv~e^{-i(k,x)_\perp}(x_\perp|\ticalg^{ab}_{\mu\nu}(x_\ast,\infty;p_\perp)|k_\perp),
\nonumber\\
&&\hspace{-1mm}
\ticalq^{ab}_{\mu\nu}(x_\ast,\infty,x_\perp;k)~\equiv~e^{-i(k,x)_\perp}(x_\perp|\ticalq^{ab}_{\mu\nu}(x_\ast,\infty;p_\perp)|k_\perp),
\nonumber\\
&&\hspace{-1mm}
{\bar\ticalq}^{ab}_{\mu\nu}(x_\ast,\infty,x_\perp;k)~\equiv~e^{i(k,x)_\perp}(k_\perp|{\bar\ticalq}^{ab}_{\mu\nu}(x_\ast,\infty;p_\perp)|x_\perp)
\label{A.50}
\end{eqnarray}
Again, the explicit expressions can be read from Eqs. (\ref{A.23}) and (\ref{kvark16}) - (\ref{kvark21}) by taking  the transverse arguments of all 
fields to be $x_\perp$ (and replacing the operators $p^j$ with $k^j$) similarly to Eq. (\ref{A.17}).

\section{Appendix B: Propagators in the shock-wave background}

In this section we consider propagators of slow fields in the background of fast fields in the case when the 
characteristic transverse momenta of fast fields ($k_\perp$) and slow fields ($l_\perp$) are comparable.
In this case the usual rescaling of Ref. \cite{npb96} applies and we can again consider the external fields 
of the type $A_\bu(x_\ast,x_\perp)$ with $A_i=A_\ast=0$.

Actually, since the typical longitudinal size of fast 
fields is $\sigma_\ast\sim {\sigma' s\over l_\perp^2}$ and the  typical distances  traveled by slow gluons are 
$\sim{\sigma s\over k_\perp^2}$ our  formulas will remain correct if $l_\perp^2\gg k_\perp^2$ since the shock wave is even thinner in this case.
As we discussed above, we assume that the support of the shock wave is thin but not infinitely thin. 
For our calculations we need gluon propagators with both points outside the shock wave and propagator with one point inside and one outside.
It is convenient to start from the latter case since all the necessary formulas can be deduced from the light-cone expansion discussed in the previous Section. 
To illustrate this, let us again  for simplicity consider scalar propagator.

\subsection{Propagators with one point in the shock wave \label{secA.1}}
\subsubsection{Scalar propagator }
 For simplicity we will again perform at first the calculation  for ``scalar propagator'' \\
  $(x|{1\over P^2+i\epsilon}|y)$. 
 As usual, we assume that the only nonzero component of the external field is
 $A_\bu$ and it does not depend on $z_\bu$ so the operator $\alpha=i{\partial\over\partial z_\bullet}$ commutes with all background fields. 
 The propagator in the external field $A_\bu(z_\ast,z_\perp)$ is given by Eq. (\ref{A.1}) and (\ref{A.2}) which can be rewritten as
\begin{eqnarray}
\hspace{-1mm}
(x|{1\over P^2+i\epsilon}|y)~&=&~\Big[-i\theta(x_\ast-y_\ast)\!\int_0^\infty\!{\dhd\alpha\over 2\alpha}
+i\theta(y_\ast-x_\ast)\!\int_{-\infty}^0\!{\dhd\alpha\over 2\alpha}\Big]~e^{-i\alpha(x-y)_\bu}
\label{B.1}\\
&\times&~(x_\perp|e^{-i{p_\perp^2\over\alpha s}(x_\ast-y_\ast)}
{\rm Pexp}\Big\{{2ig\over s}\!\int_{y_\ast}^{x_\ast}\! dz_\ast
~e^{i{p_\perp^2\over\alpha s}(z_\ast-y_\ast)}A_\bu(z_\ast)e^{-i{p_\perp^2\over\alpha s}(z_\ast-y_\ast)}
\Big\}|y_\perp)
\nonumber
\end{eqnarray}
Suppose the point $y$ lies inside the shock wave  (the point $x$ may be inside or outside of the shock wave). 
Since the longitudinal distances $z_\ast$ inside the shock wave are small ($\sim {\sigma's\over l_\perp^2}$) we can use the expansion (\ref{A.3}) but the parameter of the
expansion is now ${p_\perp^2\over\alpha s}\sigma_\ast\sim{\sigma'\over\alpha}\ll 1$ rather than   twist of the operator. 
Consequently, the last term in Eq. (\ref{A.3}) can be neglected since it has an extra factor ${p_\perp^2\over\alpha s}\sigma_\ast$ in comparison to the second term:
\begin{eqnarray}
&&\hspace{-1mm}
e^{i{p_\perp^2\over\alpha s}(z_\ast-y_\ast)}A_\bu e^{-i{p_\perp^2\over\alpha s}(z_\ast-y_\ast)}
\nonumber\\
&&\hspace{15mm}
=~A_\bu -{z_\ast-y_\ast\over\alpha s}(2p^iF_{\bu i}-iD^iF_{\bu i})
-2{(z_\ast-y_\ast)^2\over \alpha^2 s^2}(p^ip^j-ip^jD^i)D_jF_{\bu i}
+...
\nonumber\\
&&\hspace{15mm}
=~A_\bu -{z_\ast-y_\ast\over\alpha s}(2p^iF_{\bu i}-iD^iF_{\bu i})+...
\label{B.2}
\end{eqnarray}
This is again the expansion around the light ray $y_\perp+{2\over s}y_\ast p_1$ but now with
the parameter of the expansion $\sim {p_\perp^2\over\alpha s}\sigma_\ast\ll 1$. 
However, we need to keep the second term of this expansion since the first term forms
gauge links (for example, it is absent in the $A_\bu=0$ gauge).

Since there are no new terms in the expansion (\ref{B.2}) in comparison to  (\ref{A.3}) we can look at the final
result (\ref{A.5}) for $\calo(x_\ast,y_\ast;p_\perp)$ and drop the terms which are small with respect to our new power counting.
This way the Eq. (\ref{A.5}) reduces to
\begin{eqnarray}
&&\hspace{-1mm}
\calo(x_\ast,y_\ast;p_\perp)~
\nonumber\\
&&\hspace{15mm}
=~[x_\ast,y_\ast]-{2ig\over\alpha s^2}\!\int_{y_\ast}^{x_\ast}\!\!\!dz_\ast
~(z-y)_\ast\big\{2p^j[x_\ast,z_\ast]F_{\bu j}(z_\ast)-i[x_\ast,z_\ast]D^jF_{\bu j}(z_\ast)\big\}[z_\ast,y_\ast]
\nonumber\\
&&\hspace{15mm}
+~{8ig^2\over\alpha s^3}\!\int_{y_\ast}^{x_\ast}\!\! \!dz_\ast\!\int_{y_\ast}^{z_\ast}\!\!\!dz'_\ast~(z'-y)_\ast
[x_\ast,z_\ast] F_{\bu j}(z_\ast) [z_\ast,z'_\ast]F_\bu^{~j}(z'_\ast)[z'_\ast,y_\ast]
+...
\label{B.3}
\end{eqnarray}
and the propagator has the form (\ref{A.6})
\begin{eqnarray}
&&\hspace{-1mm}
(x|{1\over P^2+i\epsilon}|y)~=~\Big[-i\theta(x_\ast-y_\ast)\!\int_0^\infty\!{\dhd\alpha\over 2\alpha}
+i\theta(y_\ast-x_\ast)\!\int_{-\infty}^0\!{\dhd\alpha\over 2\alpha}\Big]
\label{B.4}\\
&&\hspace{32mm}
\times~e^{-i\alpha(x-y)_\bu}(x_\perp|e^{-i{p_\perp^2\over\alpha s}(x-y)_\ast}\calo(x_\ast,y_\ast;p_\perp)|y_\perp)
\nonumber
\end{eqnarray}
As we mentioned, this formula is correct for the point $y$ inside the shock wave and the point $x$ inside or outside.

Similarly, for the complex conjugate amplitude we obtain the propagator in the form (\ref{A.12}) with
\begin{eqnarray}
&&\hspace{-1mm}
\ticalo(x_\ast,y_\ast;p_\perp)~
\nonumber\\
&&\hspace{15mm}
=~[x_\ast,y_\ast]
+{2ig\over\alpha s^2}\!\int_{x_\ast}^{y_\ast}\!\!\!dz_\ast~(z-x)_\ast
~[x_\ast,z_\ast]\big\{2\tilF_{\bu j}(z_\ast)[z_\ast,y_\ast]p^j
+i\tilD^j\tilF_{\bu j}(z_\ast)[z_\ast,y_\ast]\big\}
\nonumber\\
&&\hspace{15mm}
-~{8ig^2\over\alpha s^3}\!\int_{y_\ast}^{x_\ast}\!\! \!dz_\ast\!\int_{y_\ast}^{z_\ast}\!\!\!dz'_\ast~(z-x)_\ast
[x_\ast,z_\ast]\tilF_{\bu j}(z_\ast) [z_\ast,z'_\ast]\tilF_\bu^{~j}(z'_\ast)[z'_\ast,y_\ast]
+...
\label{B.5}
\end{eqnarray}
which is the expansion (\ref{A.11}) but with fewer number of terms.  Again, the formula (\ref{A.12}) with $\ticalo(x_\ast,y_\ast;p_\perp)$ given by the above expression 
is correct for the point $x$ inside the shock wave and 
the point $y$ inside or outside.

The expressions for particle production are the same as (\ref{A.14}) and  (\ref{A.16}) with\\
  $\calo(\infty,y_\ast,y_\perp;k)$ 
and  $\ticalo(x_\ast,\infty,x_\perp;k)$ changed to Eqs. (\ref{B.3}) and (\ref{B.5}), respectively.

\subsubsection{Gluon propagator and vertex of gluon emission}

As we saw in previous Section, the gluon propagator with one point in the shock wave can be obtained in the same way as the 
propagator near the light cone, only the parameter of the expansion is different: ${p_\perp^2\over\alpha s}\sigma_\ast$ rather than the twist of the operator. 
 Careful inspection of the expansions (\ref{A.19}) and (\ref{kvark7}) reveals that there is no leading or next-to-leading terms with twist larger than four so
 we can recycle the final formulas (\ref{A.45})  and (\ref{A.46}) for gluon propagators.  At  ${p_\perp^2\over\alpha s}\sigma_\ast\ll 1$ the expression (\ref{A.20}) for $\calg_{\mu\nu}(x_\ast,y_\ast;p_\perp)$ turns to
\begin{eqnarray}
&&\hspace{-3mm}
\calg_{\mu\nu}(x_\ast,y_\ast;p_\perp)~=
\label{B.6}\\
&&\hspace{-3mm}
=~g_{\mu\nu}[x_\ast,y_\ast]+g\!\int_{y_\ast}^{x_\ast}\!\!\!dz_\ast~
\Big(-{2i\over\alpha s^2}(z-y)_\ast g_{\mu\nu}\big\{2p^j[x_\ast,z_\ast]F_{\bu j}(z_\ast)
-i[x_\ast,z_\ast]D^jF_{\bu j}(z_\ast)\big\}
\nonumber\\
&&\hspace{-3mm}
+~{4\over \alpha s^2}(\delta_\mu^jp_{2\nu}-\delta_\nu^jp_{2\mu})[x_\ast,z_\ast]F_{\bu j}(z_\ast)
\Big)[z_\ast,y_\ast]
\nonumber\\
&&\hspace{-3mm}
+~{8g^2\over\alpha s^3}\!\int_{y_\ast}^{x_\ast}\!\! \!dz_\ast\!\int_{y_\ast}^{z_\ast}\!\!\!dz'_\ast
~\big[ig_{\mu\nu}(z'-y)_\ast-{2\over \alpha s}p_{2\mu}p_{2\nu}\big]
[x_\ast,z_\ast] F_{\bu j}(z_\ast) [z_\ast,z'_\ast]F_\bu^{~j}(z'_\ast)
[z'_\ast,y_\ast]
\nonumber
\end{eqnarray}

 Looking at quark formulas (\ref{kvark9}) - (\ref{kvark14}) we see that 
 at ${p_\perp^2\over\alpha s}\sigma_\ast\ll 1$ the only surviving  terms are the first terms in the r.h.s's  of these equations. 
 Let us compare now the size of these terms to 
the gluon contribution (\ref{B.6}).  The ``power counting'' for external
quark fields in comparison to gluon ones is 
${g^2\over s}\!\int\! dz_\ast \bsi\!\not\! p_1\psi(z_\ast)~\sim~{g\over s}\!\int\! dz_\ast D^iF_{\bu i}(z_\ast)~\sim l_\perp^2U\sim l_\perp^2$
and each extra integration inside the shock wave brings extra $\sigma_\ast$. The first lines in r.h.s.'s of Eqs. (\ref{kvark11})  and (\ref{kvark12}) are of order of 
${g^2p_i\over \alpha^2s^2}\!\int\! dz_\ast \bsi\!\!\not\! p_1\psi(z_\ast)\sim g{p_il_\perp^2\over \alpha^2 s}\sigma_\ast$ so they can
be neglected in comparison to the corresponding term ${g\over\alpha s}\!\int\! dz_\ast~F_{\bu i}(z_\ast)\sim g{l_i\over\alpha}$ in Eq. (\ref{B.6}).
As to the terms (\ref{kvark13})  and (\ref{kvark14}),  they are of the same odrer of magnitude as 
next-to-leading terms $\sim g_{\mu\nu}$ in Eq. (\ref{B.6}) so we keep them for now. With these approximations we obtain
\begin{eqnarray}
&&\hspace{-3mm}
\calq^{ab}_{\mu\nu}(x_\ast,y_\ast;p_\perp)
~=~-{4ig\over \alpha^2s^3}p_{2\mu}p_{2\nu}\!\!\int_{y_\ast}^{x_\ast}\! \! dz_\ast 
([x_\ast,z_\ast]D^jF_{\bu j}(z_\ast)[z_\ast,y_\ast])^{ab}
\label{B.7}\\
&&\hspace{24mm} 
+~{g^2\over \alpha s^2}
\!\int_{y_\ast}^{x_\ast}\! \! dz_\ast\!\int_{y_\ast}^{z_\ast}\! \! dz'_\ast
\bsi(z_\ast)[z_\ast,x_\ast]t^a[x_\ast,y_\ast]t^b
[y_\ast,z'_\ast]\gamma^\perp_\mu\!\not\! p_1\gamma^\perp_\nu\psi(z'_\ast),
\nonumber\\
&&\hspace{-3mm} 
{\bar\calq}^{ab}_{\mu\nu}(x_\ast,y_\ast;p_\perp)~=~-{g^2\over \alpha s^2}\!\int_{y_\ast}^{x_\ast}\! \! dz_\ast\!\int_{z_\ast}^{x_\ast}\! \! dz'_\ast
\bsi\gamma^\perp_\nu\!\not\! p_1\gamma^\perp_\mu[z_\ast,y_\ast] t^b[y_\ast,x_\ast]t^a[x_\ast,z'_\ast]
\psi(z'_\ast)
\nonumber
\end{eqnarray}
and the gluon propagator is given by Eq. (\ref{A.45}) with the above $\calg_{\mu\nu}$, $\calq_{\mu\nu}$, and $\bar\calq_{\mu\nu}$:
\begin{eqnarray}
&&\hspace{-1mm}
\langle A_\mu^a(x)A_\nu^b(y)\rangle~
\label{B.8}\\
&&\hspace{-1mm}
=~
\Big[-\theta(x_\ast-y_\ast)\!\int_0^\infty\!{\dhd\alpha\over 2\alpha}
+\theta(y_\ast-x_\ast)\!\int_{-\infty}^0\!{\dhd\alpha\over 2\alpha}\Big]e^{-i\alpha(x-y)_\bu}
\big\{(x_\perp|e^{-i{p_\perp^2\over\alpha s}(x-y)_\ast}
\nonumber\\
&&\hspace{-1mm}
\times~\big[\calg^{ab}_{\mu\nu}(x_\ast,y_\ast;p_\perp)+\calq^{ab}_{\mu\nu}(x_\ast,y_\ast;p_\perp)\big]|y_\perp)
+(y_\perp|\bar\calq^{ab}_{\mu\nu}(x_\ast,y_\ast;p_\perp)e^{-i{p_\perp^2\over\alpha s}(x-y)_\ast}|x_\perp)\big\}
\nonumber
\end{eqnarray}
As in the scalar case, it is easy to see that Eq. (\ref{B.8}) holds true if the point $y$ is inside the shock wave and 
the point $x$ anywhere.

Similarly, in the complex conjugate amplitude the gluon propagator is given by Eq. (\ref{A.46}) with
\begin{eqnarray}
&&\hspace{-3mm}
\ticalg_{\mu\nu}(x_\ast,y_\ast;p_\perp)
\label{B.9}\\
&&\hspace{-3mm}
=~g_{\mu\nu}[x_\ast,y_\ast]+g\!\int_{x_\ast}^{y_\ast}\!\!\!dz_\ast~[x_\ast,z_\ast]
\Big({2i\over\alpha s^2}(z-x)_\ast g_{\mu\nu}\big\{2\tilF_{\bu j}(z_\ast)[z_\ast,y_\ast]p^j
+i\tilD^j\tilF_{\bu j}(z_\ast)[z_\ast,y_\ast]\big\}
\nonumber\\
&&\hspace{-3mm}
-~{4\over \alpha s^2}(\delta_\mu^jp_{2\nu}-\delta_\nu^jp_{2\mu})\tilF_{\bu j}(z_\ast)[z_\ast,y_\ast]\Big)
\nonumber\\
&&\hspace{-3mm}
+~{8g^2\over\alpha s^3}\!\int_{x_\ast}^{y_\ast}\!\! \!dz_\ast\!\int_{z_\ast}^{y_\ast}\!\!\!dz'_\ast [x_\ast,z_\ast] \Big(
\big[-ig_{\mu\nu}(z-x)_\ast
-{2\over \alpha s}p_{2\mu}p_{2\nu}\big]
\tilF_{\bu j}(z_\ast) [z_\ast,z'_\ast]\tilF_\bu^{~j}(z'_\ast)[z'_\ast,y_\ast]\Big)
\nonumber
\end{eqnarray}
and
\begin{eqnarray}
&&\hspace{-1mm}
\ticalq^{ab}_{\mu\nu}(x_\ast,y_\ast;p_\perp)~
=~{4ig\over \alpha^2s^3}p_{2\mu}p_{2\nu}\!\int^{y_\ast}_{x_\ast}\! \! dz_\ast 
\big([x_\ast,z_\ast]\tilD^j\tilF_{\bu j}(z_\ast)[z_\ast,y_\ast]\big)^{ab}
\label{B.10}\\
&&\hspace{26mm} 
+~
{g^2\over \alpha s^2}\!\int_{x_\ast}^{y_\ast} \!\!dz_\ast\!\int_{z_\ast}^{y_\ast} \!\!dz'_\ast
\tilde{\bsi}(z_\ast)\gamma^\perp_\mu\!\not\! p_1\gamma^\perp_\nu[z_\ast,x_\ast]t^a[x_\ast,y_\ast]t^b[y_\ast,z'_\ast]\tilde{\psi}(z'_\ast),
\nonumber\\
&&\hspace{-1mm}
{\bar\ticalq}^{ab}_{\mu\nu}(x_\ast,y_\ast;p_\perp)~
=~-{g^2\over \alpha s^2}\!\int^{y_\ast}_{x_\ast} \!\!dz_\ast\!\int^{z_\ast}_{x_\ast} \!\!dz'_\ast~
\tilde{\bsi} \gamma^\perp_\nu\!\not\! p_1\gamma^\perp_\mu[z_\ast,y_\ast]t^b[y_\ast,x_\ast]t^a[x_\ast,z'_\ast]\tilde{\psi}(z'_\ast)
\nonumber
\end{eqnarray}
The expressions (\ref{B.9}) and (\ref{B.10}) are valid for point $x$ inside the shock wave (and point $y$ inside or outside).

The corresponding expressions for the Lipatov vertex of gluon production are given by Eqs. (\ref{A.47}) -(\ref{A.50}) 
with $\calg$, $\calq$, and $\bar{\calq}$ changed accordingly.

\subsection{Propagators with both points outside the shock wave \label{shockwaveout}}
In this section we will find the propagators with both points outside the shock wave.
Again, we assume that the  characteristic shock-wave transverse momenta are of order of 
transverse momenta of ``quantum'' fields with $\alpha>\sigma'$. As discussed in Sect. \ref{sec2},
we consider  the width of the shock-wave to be small but finite, consequently we can not recycle formulas from Ref. \cite{npb96} for the infinitely thin
shock-wave. 

\subsubsection{Scalar propagator}
As in the previous Section, for simplicity we start with the scalar propagator (\ref{A.1})
\begin{eqnarray}
&&\hspace{-1mm}
(x|{1\over P^2+i\epsilon}|y)~\stackrel{x_\ast>y_\ast}{=}-i\int_0^\infty\!{\dhd\alpha\over 2\alpha}~e^{-i\alpha(x-y)_\bu}
(x_\perp|{\rm Pexp}\big\{-i\!\int_{y_\ast}^{x_\ast}\! dz_\ast\big[{p_\perp^2\over \alpha s}-{2g\over s}A_\bu(z_\ast)\big]\big\}|y_\perp)
\nonumber\\
\label{B.11}
\end{eqnarray}
The Pexp in the r.h.s. of Eq. (\ref{B.11}) can be transformed to
\begin{eqnarray}
&&\hspace{-1mm}  
(x_\perp|e^{-i{p_\perp^2\over\alpha s}x_\ast}
{\rm Pexp}\Big\{ig\!\int^{x_\ast}_{y_\ast}\! d{2\over s}z_\ast
~e^{i{p_\perp^2\over\alpha s}z_\ast}A_\bu(z_\ast)e^{-i{p_\perp^2\over\alpha s}z_\ast}\Big\}e^{i{p_\perp^2\over\alpha s}y_\ast}
|y_\perp)~=~\int d^2z_\perp d^2z'_\perp
\label{B.12}
\\
&&\hspace{11mm}  
\times~(x_\perp|e^{-i{p_\perp^2\over\alpha s}x_\ast}|z_\perp)
(z_\perp|{\rm Pexp}\Big\{ig\!\int_{y_\ast}^{x_\ast}\! d{2\over s}z_\ast
~e^{i{p_\perp^2\over\alpha s}z_\ast}A_\bu(z_\ast)e^{-i{p_\perp^2\over\alpha s}z_\ast}\Big\}
|z'_\perp)(z'_\perp|e^{i{p_\perp^2\over\alpha s}y_\ast}
|y_\perp)
\nonumber
\end{eqnarray}
Next, we use the expansion (\ref{B.2}) at $y_\ast=0$
\begin{eqnarray}
&&\hspace{-11mm}
e^{i{p_\perp^2\over\alpha s}z_\ast}A_\bu e^{-i{p_\perp^2\over\alpha s}z_\ast}
~=~A_\bu-{z_\ast\over\alpha s}(2p^iF_{\bu i}-iD^iF_{\bu i})~+~...
\label{B.13}
\end{eqnarray}
This is an expansion around the light cone $z_\perp+{2\over s}z_\ast p_1$ with
the parameter of the expansion $\sim {p^2_\perp\over\alpha s}\sigma_\ast \ll 1$.
Note that similarly to Eq. (\ref{B.2})  we need to keep the second term of this expansion since the first term forms
gauge links.
 
 From Eqs.  (\ref{A.4}), and (\ref{A.5}) we obtain (cf. Eq. (\ref{A.6}))
\begin{eqnarray}
\hspace{-1mm}
(x|{1\over P^2+i\epsilon}|y)~&=&~\Big[-i\theta(x_\ast-y_\ast)\!\int_0^\infty\!{\dhd\alpha\over 2\alpha}
+i\theta(y_\ast-x_\ast)\!\int_{-\infty}^0\!{\dhd\alpha\over 2\alpha}\Big]~e^{-i\alpha(x-y)_\bu}
\label{B.14}\\
\hspace{-1mm}
&\times&
~~\int\! d^2z_\perp(x_\perp|e^{-i{p_\perp^2\over\alpha s}x_\ast}
\calo(x_\ast,y_\ast;p_\perp)|z_\perp)(z_\perp|e^{i{p_\perp^2\over\alpha s}y_\ast}|y_\perp)
\nonumber
\end{eqnarray}
where
\begin{eqnarray}
&&\hspace{-1mm}
\calo(x_\ast,y_\ast;p_\perp)~
\nonumber\\
&&\hspace{-1mm}
=~[x_\ast,y_\ast]-{2ig\over\alpha s^2}\!\int_{y_\ast}^{x_\ast}\!\!\!dz_\ast~
z_\ast\big(2p^j[x_\ast,z_\ast]F_{\bu j}(z_\ast)-i[x_\ast,z_\ast]D^jF_{\bu j}(z_\ast)\big)[z_\ast,y_\ast]
\nonumber\\
&&\hspace{-1mm}
+~{8ig^2\over\alpha s^3}\!\int_{y_\ast}^{x_\ast}\!\! \!dz_\ast\!\int_{y_\ast}^{z_\ast}\!\!\!dz'_\ast~z'_\ast
[x_\ast,z_\ast] F_{\bu j}(z_\ast) [z_\ast,z'_\ast]F_\bu^{~j}(z'_\ast)[z'_\ast,y_\ast]
+...
\label{B.15}
\end{eqnarray}
Here the transverse arguments of all fields turn effectively to $z_\perp$.
 Note that this expression is equal to  Eq. (\ref{A.5}) at $y_\ast= 0$.
 For the complex conjugate amplitude one obtains (cf. Eq. (\ref{A.12}))
\begin{eqnarray}
\hspace{-1mm}
(x|{1\over P^2-i\epsilon}|y)~&=&~
\Big[i\theta(y_\ast-x_\ast)\!\int_0^\infty\!{\dhd\alpha\over 2\alpha}
-i\theta(x_\ast-y_\ast)\!\int_{-\infty}^0\!{\dhd\alpha\over 2\alpha}\Big]~e^{-i\alpha(x-y)_\bu}
\label{B.16}\\
\hspace{-1mm}
&\times&
~\int\! d^2z_\perp(x_\perp|e^{-i{p_\perp^2\over\alpha s}x_\ast}
\ticalo(x_\ast,y_\ast;p_\perp)|z_\perp)(z_\perp|e^{i{p_\perp^2\over\alpha s}y_\ast}|y_\perp)
\nonumber
\end{eqnarray}
where
\begin{eqnarray}
&&\hspace{-3mm}
\ticalo(x_\ast,y_\ast;p_\perp)~=~[x_\ast,y_\ast]
+{2ig\over\alpha s^2}\!\int_{x_\ast}^{y_\ast}\!\!\!dz_\ast~z_\ast[x_\ast,z_\ast] (2\tilF_{\bu j}(z_\ast)[z_\ast,y_\ast]p^j
+i\tilD^j\tilF_{\bu j}(z_\ast)[z_\ast,y_\ast])
\nonumber\\
&&\hspace{-3mm}
-~{8ig^2\over\alpha s^3}\!\int_{x_\ast}^{y_\ast}\!\! \!dz_\ast\!\int_{z_\ast}^{y_\ast}\!\!\!dz'_\ast~z_\ast
[x_\ast,z_\ast] \tilF_{\bu j}(z_\ast) [z_\ast,z'_\ast]_z\tilF_\bu^{~j}(z'_\ast)[z'_\ast,y_\ast]
\label{B.17}
\end{eqnarray}
Again, this expression can be obtained from Eq. (\ref{A.11}) by taking $x_\ast\rightarrow 0$ in parentheses.

 \subsubsection{The emission vertex} 

Similarly to Eq. (\ref{A.14}) we get
\begin{eqnarray}
&&\hspace{-1mm}
\lim_{k^2\rightarrow 0}k^2(k|{1\over P^2+i\epsilon}|y)~=~\!\int\! dx_\ast d^2x_\perp 
~e^{i{k_\perp^2\over\alpha s}x_\ast-i(k,x)_\perp}
\label{B.18}\\
&&\hspace{-1mm} 
\times~\big[{\partial\over\partial x_\ast} -{i\over\alpha s}\partial_{x_\perp}^2\big]
\theta(x-y)_\ast\!\int\! d^2z_\perp~(x_\perp|e^{-i{p_\perp^2\over\alpha s}x_\ast}|z_\perp)(z_\perp|\calo(x_\ast,y_\ast;p_\perp)
e^{i{p_\perp^2\over\alpha s}y_\ast}|y_\perp)e^{i\alpha y_\bullet}
\nonumber\\
&&\hspace{35mm} 
=~\!\int\!dx_\ast 
{\partial\over\partial x_\ast}\theta(x-y)_\ast (k_\perp|\calo(x_\ast,y_\ast;p_\perp)
e^{i{p_\perp^2\over\alpha s}y_\ast}|y_\perp)e^{i\alpha y_\bullet}
\nonumber\\
&&\hspace{35mm}    
=~\!\int\! d^2z_\perp~e^{-i(k,z)_\perp}\calo(\infty,y_\ast;z_\perp;k)
(z_\perp|e^{i{p_\perp^2\over\alpha s}y_\ast}|y_\perp)e^{i\alpha y_\bullet}
\nonumber  
\end{eqnarray}
where $\calo(\infty,y_\ast,z_\perp;k)$ is obtained from Eq. (\ref{B.15}) 
\begin{eqnarray}
&&\hspace{-1mm}
\calo(\infty,y_\ast,z_\perp;k)~\equiv~e^{i(k,z)_\perp} (k_\perp|\calo(\infty,y_\ast;p_\perp)|z_\perp)
\label{B.19}\\
&&\hspace{29mm}
=~[\infty,y_\ast]_z
-{2ig\over\alpha s^2}\!\int_{y_\ast}^{\infty}\!\!\!dz_\ast~z_\ast[\infty,z_\ast]_z
(2k^jF_{\bu j}-iD^jF_{\bu j})(z_\ast,z_\perp)[z_\ast,y_\ast]_z
\nonumber\\
&&\hspace{29mm}
+~{8ig^2\over\alpha s^3}\!\int_{y_\ast}^{\infty}\!\! \!dz_\ast\!\int_{y_\ast}^{z_\ast}\!\!\!dz'_\ast~z'_\ast
[\infty,z_\ast]_z F_{\bu j}(z_\ast,z_\perp) [z_\ast,z'_\ast]_zF_\bu^{~j}(z'_\ast,z_\perp)[z'_\ast,y_\ast]_z
\nonumber
\end{eqnarray}

For the complex conjugate amplitude we get
\begin{eqnarray}
&&\hspace{-1mm}
\lim_{k^2\rightarrow 0}k^2(x|{1\over P^2-i\epsilon}|k)~=~\!\int\! dy_\ast d^2y_\perp 
~e^{-i{k_\perp^2\over\alpha s}y_\ast+i(k,y)_\perp}
\label{B.20}\\
&&\hspace{-1mm} 
\times~\big[{\partial\over\partial y_\ast} +{i\over\alpha s}\partial_{y_\perp}^2\big]
\theta(y-x)_\ast\!\int\! d^2z_\perp~(x_\perp|e^{-i{p_\perp^2\over\alpha s}x_\ast}|z_\perp)(z_\perp|\ticalo(x_\ast,y_\ast;p_\perp)
e^{i{p_\perp^2\over\alpha s}y_\ast}|y_\perp)e^{-i\alpha x_\bullet}
\nonumber\\
&&\hspace{36mm} 
=~\!\int\!dy_\ast 
{\partial\over\partial y_\ast}\theta(y-x)_\ast (x_\perp|e^{-i{p_\perp^2\over\alpha s}x_\ast}\ticalo(x_\ast,y_\ast;p_\perp)
|k_\perp)e^{-i\alpha x_\bullet}
\nonumber\\
&&\hspace{36mm}    
=~\!\int\! d^2z_\perp~e^{i(k,z)_\perp}(x_\perp|e^{-i{p_\perp^2\over\alpha s}x_\ast}|z_\perp)\ticalo(x_\ast,\infty;z_\perp;k)e^{-i\alpha x_\bullet}
\nonumber  
\end{eqnarray}
where $\ticalo(\infty,y_\ast,z_\perp;k)$ is obtained from Eq. (\ref{B.17}) in a usual way
\begin{eqnarray}
&&\hspace{-1mm}
\ticalo(x_\ast,\infty;z_\perp;k)~\equiv~e^{-i(k,z)_\perp} (z_\perp|\ticalo(x_\ast,\infty;p_\perp)|k_\perp)
\label{B.21}\\
&&\hspace{29mm}
=~[x_\ast,\infty]_z
+{2ig\over\alpha s^2}\!\int_{x_\ast}^{\infty}\!\!\!dz_\ast~z_\ast[x_\ast,z_\ast]_z
(2k^j\tilF_{\bu j}+i\tilD^j\tilF_{\bu j})(z_\ast,z_\perp)[z_\ast,\infty]_z
\nonumber\\
&&\hspace{29mm}
-~{8ig^2\over\alpha s^3}\!\int_{x_\ast}^{\infty}\!\! \!dz_\ast\!\int_{z_\ast}^{\infty}\!\!\!dz'_\ast~z_\ast
[x_\ast,z_\ast]_z \tilF_{\bu j}(z_\ast,z_\perp) [z_\ast,z'_\ast]_z\tilF_\bu^{~j}(z'_\ast,z_\perp)[z'_\ast,\infty]_z
\nonumber
\end{eqnarray}

\subsubsection{Gluon propagator in the shock-wave background}
The gluon propagator in a background gluon field (\ref{A.18}) can be rewritten as
\begin{eqnarray}
&&\hspace{-3mm}
i\langle A_\mu^a(x)A_\nu^b(y)\rangle~=~(x|{1\over P^2+2igF+i\epsilon}|y)^{ab}_{\mu\nu}
~\stackrel{x_\ast>y_\ast}{=}
-i\!\int_0^\infty\!{\dhd\alpha\over 2\alpha}~e^{-i\alpha(x-y)_\bu}
\label{B.22}\\
&&\hspace{10mm}
\times~(x_\perp|e^{-i{p_\perp^2\over\alpha s}x_\ast}
{\rm Pexp}\Big\{{2ig\over s}\!\int_{y_\ast}^{x_\ast}\! dz_\ast
~e^{i{p_\perp^2\over\alpha s}z_\ast}\big(A_\bu+{i\over\alpha}F\big)(z_\ast)
e^{-i{p_\perp^2\over\alpha s}z_\ast}\Big\}_{\mu\nu}e^{i{p_\perp^2\over\alpha s}y_\ast}
|y_\perp)^{ab}
\nonumber
\end{eqnarray}

Using the expansion (\ref{A.19}) at $y_\ast=0$ we obtain with our accuracy
\begin{eqnarray}
&&\hspace{-1mm}
e^{i{p_\perp^2\over\alpha s}z_\ast}\big(A_\bu g_{\mu\nu}+{i\over\alpha}F_{\mu\nu}\big)e^{-i{p_\perp^2\over\alpha s}z_\ast}
~\label{B.24}
\\
&&\hspace{-1mm}
=~g_{\mu\nu}\Big[A_\bu-{z_\ast\over\alpha s}(2p^jF_{\bu j}-iD^jF_{\bu j})\Big]
+~{i\over\alpha}F_{\mu\nu}+i{z_\ast\over\alpha^2 s}\Big(2p^jD_jF_{\mu\nu}-iD^jD_jF_{\mu\nu}\Big)+...
\nonumber
\end{eqnarray}

Similarly to Eq. (\ref{A.21})  we get
\begin{eqnarray}
\hspace{-3mm}
(x|{1\over P^2+2igF+i\epsilon}|y)^{ab}_{\mu\nu}~&=&~\Big[-i\theta(x_\ast-y_\ast)\!\int_0^\infty\!{\dhd\alpha\over 2\alpha}
+i\theta(y_\ast-x_\ast)\!\int_{-\infty}^0\!{\dhd\alpha\over 2\alpha}\Big]~e^{-i\alpha(x-y)_\bu}
\nonumber\\
\hspace{-3mm}
&\times&~\int\! d^2z_\perp
~(x_\perp|e^{-i{p_\perp^2\over\alpha s}x_\ast}|z_\perp)(z_\perp|
\calg^{ab}_{\mu\nu}(x_\ast,y_\ast;p_\perp)e^{i{p_\perp^2\over\alpha s}y_\ast}|y_\perp)
\label{B.25}
\end{eqnarray}
where
\begin{eqnarray}
&&\hspace{-3mm}
\calg_{\mu\nu}(x_\ast,y_\ast;p_\perp)
\label{B.26}\\
&&\hspace{-3mm}
=~g_{\mu\nu}[x_\ast,y_\ast]+~g\!\int_{y_\ast}^{x_\ast}\!\!\!dz_\ast~
\Big(-{2i\over\alpha s^2}z_\ast g_{\mu\nu}
\big(2p^j[x_\ast,z_\ast]F_{\bu j}(z_\ast)-i[x_\ast,z_\ast]D^jF_{\bu j}(z_\ast)\big)
\nonumber\\
&&\hspace{-3mm}
+~{4\over \alpha s^2}(\delta_\mu^jp_{2\nu}-\delta_\nu^jp_{2\mu})\big\{[x_\ast,z_\ast]F_{\bu j}(z_\ast)
+{2iz_\ast\over\alpha s}p^k[x_\ast,z_\ast]D_kF_{\bu j}(z_\ast)\big\}\Big)[z_\ast,y_\ast]
\nonumber\\
&&\hspace{-3mm}
+~{8g^2\over\alpha s^3}\!\int_{y_\ast}^{x_\ast}\!\! \!dz_\ast\!\int_{y_\ast}^{z_\ast}\!\!\!dz'_\ast
~\big[ig_{\mu\nu}z'_\ast-{2\over \alpha s}p_{2\mu}p_{2\nu}\big]
[x_\ast,z_\ast] F_{\bu j}(z_\ast) [z_\ast,z'_\ast]F_\bu^{~j}(z'_\ast)[z'_\ast,y_\ast]
+...
\nonumber
\end{eqnarray}

Let us consider now  quark terms coming from Fig. \ref{fig:7}. From Eq. (\ref{kvark1}) it is clear that this contribution can be parameterized similarly to Eq. (\ref{kvark8}):
\begin{eqnarray}
&&\hspace{-0mm}
\langle A_\mu^a(x)A_\nu^b(y)\rangle_{\rm Fig. ~\ref{fig:7}}~
\label{B.27}\\
&&\hspace{-1mm}
=~
\Big[-\theta(x_\ast-y_\ast)\!\int_0^\infty\!{\dhd\alpha\over 2\alpha}
+\theta(y_\ast-x_\ast)\!\int_{-\infty}^0\!{\dhd\alpha\over 2\alpha}\Big]~e^{-i\alpha(x-y)_\bu}
\!\int\! d^2 z_\perp~\Big[(x_\perp|e^{-i{p_\perp^2\over\alpha s}x_\ast}
|z_\perp)
\nonumber\\
&&\hspace{4mm}
\times~(z_\perp|\calq^{ab}_{\mu\nu}(x_\ast,y_\ast;p_\perp)e^{i{p_\perp^2\over\alpha s}y_\ast}
|y_\perp)
+(y_\perp|e^{i{p_\perp^2\over\alpha s}y_\ast}|z_\perp)
(z_\perp|\bar\calq^{ab}_{\mu\nu}(x_\ast,y_\ast;p_\perp)
e^{-i{p_\perp^2\over\alpha s}x_\ast}|x_\perp)\Big]
\nonumber
\end{eqnarray}
where $\calq^{ab}_{\mu\nu}$ and $\calq^{ab}_{\mu\nu}$ are given by expressions
(\ref{kvark9}) - (\ref{kvark14}) with $z_\ast-y_\ast\rightarrow z_\ast$ (and similarly
 $z'_\ast-y_\ast\rightarrow z'_\ast$). With our accuracy only the first terms in these expressions 
 survive so $\calq^{ab}_{\mu\nu}$ and $\bar{\calq}^{ab}_{\mu\nu}$ are given by Eq. (\ref{B.7}) from previous Section.
 
 Adding gluon contribution (\ref{B.25}) one obtains the final expression for gluon propagator in a shock-wave background:
\begin{eqnarray}
&&\hspace{-0mm}
\langle A_\mu^a(x)A_\nu^b(y)\rangle~
=~
\Big[-\theta(x_\ast-y_\ast)\!\int_0^\infty\!{\dhd\alpha\over 2\alpha}
+\theta(y_\ast-x_\ast)\!\int_{-\infty}^0\!{\dhd\alpha\over 2\alpha}\Big]~e^{-i\alpha(x-y)_\bu}
\label{B.28}\\
&&\hspace{11mm}
\times \!\int\! d^2 z_\perp~\Big[(x_\perp|e^{-i{p_\perp^2\over\alpha s}x_\ast}
|z_\perp)
(z_\perp|[\calg^{ab}_{\mu\nu}(x_\ast,y_\ast;p_\perp)
+ \calq^{ab}_{\mu\nu}(x_\ast,y_\ast;p_\perp)]e^{i{p_\perp^2\over\alpha s}y_\ast}
|y_\perp)
\nonumber\\
&&\hspace{24mm}
+~(y_\perp|e^{i{p_\perp^2\over\alpha s}y_\ast}|z_\perp)
(z_\perp|\bar\calq^{ab}_{\mu\nu}(x_\ast,y_\ast;p_\perp)
e^{-i{p_\perp^2\over\alpha s}x_\ast}|x_\perp)\Big]
\nonumber
\end{eqnarray}
 where $\calg^{ab}_{\mu\nu}$ is given by Eq. (\ref{B.26}) and 
 $\calq^{ab}_{\mu\nu}, ~\bar{\calq}^{ab}_{\mu\nu}$ are given by Eq. (\ref{B.7}).
 
Similarly, for the complex conjugate amplitude one obtains (cf. Eq. (\ref{A.46}))
\begin{eqnarray}
&&\hspace{-1mm}
\langle \tilA_\mu^a(x)\tilA_\nu^b(y)\rangle~=~
\Big[-\theta(y_\ast-x_\ast)\!\int_0^\infty\!{\dhd\alpha\over 2\alpha}
+\theta(x_\ast-y_\ast)\!\int_{-\infty}^0\!{\dhd\alpha\over 2\alpha}\Big]e^{-i\alpha(x-y)_\bu}
\label{B.29}\\
&&\hspace{11mm}
\times~
\!\int\! d^2 z_\perp~
\Big[(x_\perp|e^{-i{p_\perp^2\over\alpha s}x_\ast}
\big[\ticalg^{ab}_{\mu\nu}(x_\ast,y_\ast;p_\perp)
+\ticalq^{ab}_{\mu\nu}(x_\ast,y_\ast;p_\perp)\big]
|z_\perp)(z_\perp|e^{i{p_\perp^2\over\alpha s}y_\ast}|y_\perp)
\nonumber\\
&&\hspace{24mm}
+~(y_\perp|e^{i{p_\perp^2\over\alpha s}y_\ast}|z_\perp)
(z_\perp|{\bar\ticalq}^{ab}_{\mu\nu}(x_\ast,y_\ast;p_\perp)
e^{-i{p_\perp^2\over\alpha s}x_\ast}|x_\perp)\Big]
\nonumber
\end{eqnarray}
where 
\begin{eqnarray}
&&\hspace{-1mm}
\tilde{\calg}_{\mu\nu}(x_\ast,y_\ast;p_\perp)~
\label{B.30}\\
&&\hspace{-1mm}=~g_{\mu\nu}[x_\ast,y_\ast]
+~g\!\int_{x_\ast}^{y_\ast}\!\!\!dz_\ast~[x_\ast,z_\ast]
\Big\{{2iz_\ast \over\alpha s^2}g_{\mu\nu}(2\tilF_{\bu j}(z_\ast)[z_\ast,y_\ast]p^j
+~i\tilD^j\tilF_{\bu j}[z_\ast,y_\ast])
\nonumber\\
&&\hspace{11mm}
-~{4\over \alpha s^2}(\delta_\mu^jp_{2\nu}-\delta_\nu^jp_{2\mu})\big(\tilF_{\bu j}(z_\ast)[z_\ast,y_\ast]
+2i{z_\ast\over\alpha s}\tilD_l\tilF_{\bu j}(z_\ast)[z_\ast,y_\ast]k^l\big)\Big\}
\nonumber\\
&&\hspace{11mm}
+~{8g^2\over\alpha^2 s^3}\!\int_{x_\ast}^{y_\ast}\!\! \!dz_\ast
\!\int_{z_\ast}^{y_\ast}\!\!dz'_\ast~[x_\ast,z_\ast]\big[-i\alpha g_{\mu\nu}z_\ast-{2\over s}p_{2\mu}p_{2\nu}\big]
~\tilF_{\bu j}(z_\ast) [z_\ast,z'_\ast]\tilF_\bu^{~j}(z'_\ast)[z'_\ast,y_\ast]
\nonumber
\end{eqnarray}
and $\ticalq^{ab}_{\mu\nu}, ~{\bar\ticalq}^{ab}_{\mu\nu}$ are given by Eq. (\ref{B.10}). Note that
 the transverse coordinates of all fields are effectively $z_\perp$. 

\subsubsection{Gluon emission vertex}
Similarly to Eq. (\ref{B.18}) one obtains from Eq. (\ref{B.28})
\begin{equation}
\hspace{-0mm}
\lim_{k^2\rightarrow 0}k^2i\langle A_\mu^a(k)A_\nu^b(y)\rangle~=~
\!\int\! d^2z_\perp~e^{-i(k,z)_\perp}\calo^{ab}_{\mu\nu}(\infty,y_\ast,z_\perp;k)
(z_\perp|e^{i{p_\perp^2\over\alpha s}y_\ast}|y_\perp)e^{i\alpha y_\bullet}
\label{B.31}
\end{equation}
where $\calo^{ab}_{\mu\nu}(\infty,y_\ast,z_\perp;k)$ is given by Eqs. (\ref{A.47})-(\ref{A.48}).
With our accuracy we get
\begin{eqnarray}
&&\hspace{-3mm}
\calo^{ab}_{\mu\nu}(\infty,y_\ast,z_\perp;k)=~g_{\mu\nu}[\infty,y_\ast]^{ab}_z
\nonumber\\
&&\hspace{-3mm}
+~g\!\int_{y_\ast}^{\infty}\!\!\!dz_\ast~\Big([\infty,z_\ast]_z\big[-{2iz_\ast\over\alpha s^2} g_{\mu\nu}(2k^j-iD^j)
+{4\over \alpha s^2}(\delta_\mu^jp_{2\nu}-\delta_\nu^jp_{2\mu})\big]F_{\bu j}(z_\ast,z_\perp)[z_\ast,y_\ast]_z\Big)^{ab}
\nonumber\\
&&\hspace{-3mm}
+~{4g\over\alpha^2 s^3}\!\int_{y_\ast}^{\infty}\!\! \!dz_\ast~
\Big\{-ip_{2\mu}p_{2\nu}[\infty,z_\ast]_z D^jF_{\bu j}(z_\ast,z_\perp)[z_\ast,y_\ast]_z
\nonumber\\
&&\hspace{11mm}
+~g\!\int_{y_\ast}^{z_\ast}\!\!dz'_\ast~\big[2i\alpha g_{\mu\nu}z'_\ast-{4\over s}p_{2\mu}p_{2\nu}\big]
~[\infty,z_\ast]_z F_{\bu j}(z_\ast,z_\perp) [z_\ast,z'_\ast]_zF_\bu^{~j}(z'_\ast,z_\perp)[z'_\ast,y_\ast]_z
\Big\}^{ab}
\nonumber\\
&&\hspace{-3mm} 
+~{g^2\over \alpha s^2}
\Big\{\!\int_{y_\ast}^{\infty}\! \! dz_\ast\!\int_{y_\ast}^{z_\ast}\! \! dz'_\ast
\bsi(z_\ast,z_\perp)[z_\ast,\infty]_zt^a[\infty,y_\ast]_zt^b
[y_\ast,z'_\ast]_z\gamma^\perp_\mu\!\not\! p_1\gamma^\perp_\nu\psi(z'_\ast,z_\perp)
\nonumber\\
\hspace{-3mm} 
&&\hspace{11mm}-~\!\int_{y_\ast}^{\infty}\! \! dz_\ast\!\int_{z_\ast}^{\infty}\! \! dz'_\ast
\bsi(z_\ast,z_\perp)\gamma^\perp_\nu\!\not\! p_1\gamma^\perp_\mu[z_\ast,y_\ast]_z t^b[y_\ast,\infty]_zt^a[\infty,z'_\ast]_z
\psi(z'_\ast,z_\perp)\Big\}
\label{B.32}
\end{eqnarray}

For the gluon emission in the complex conjugate amplitude one obtains (cf. Eq. (\ref{B.20}) 
\begin{equation}
\hspace{-0mm}
-\lim_{k^2\rightarrow 0}k^2i\langle \tilA_\mu^a(x)\tilA_\nu^b(k)\rangle~
=~
\!\int\! d^2z_\perp~e^{i(k,z)_\perp}
(x_\perp|e^{-i{p_\perp^2\over\alpha s}x_\ast}|z_\perp)\ticalo^{ab}_{\mu\nu}(x_\ast,\infty,z_\perp;k)e^{-i\alpha x_\bullet}
\label{B.33}
\end{equation}
where  $\ticalo^{ab}_{\mu\nu}(\infty,y_\ast,z_\perp;k)$ is given by Eqs. (\ref{A.49})-(\ref{A.50}).
With our accuracy
\begin{eqnarray}
&&\hspace{-3mm}
\tilde{\calo}^{ab}_{\mu\nu}(x_\ast,\infty,z_\perp;k)~=~g_{\mu\nu}[x_\ast,\infty]^{ab}_z
\label{B.34}\\
&&\hspace{-3mm}
+~g\!\int_{x_\ast}^{\infty}\!\!\!dz_\ast~\Big([x_\ast,z_\ast]_z\Big\{
{2iz_\ast \over\alpha s^2}g_{\mu\nu}(2k^j+i\tilD^j)\tilF_{\bu j}(z_\ast,z_\perp)[z_\ast,\infty]_z
\nonumber\\
&&\hspace{11mm}
-~{4\over \alpha s^2}(\delta_\mu^jp_{2\nu}-\delta_\nu^jp_{2\mu})
\big(1+2ik^l{z_\ast\over\alpha s}\tilD_l\big)\tilF_{\bu j}(z_\ast,z_\perp)[z_\ast,\infty]_z\Big\}\Big)^{ab}
\nonumber\\
&&\hspace{-1mm}
+~{4g\over\alpha^2 s^3}\!\int_{x_\ast}^{\infty}\!\! \!dz_\ast~\Big([x_\ast,z_\ast]_z
\Big\{ip_{2\mu}p_{2\nu}\tilD^j\tilF_{\bu j}(z_\ast,z_\perp)[z_\ast,\infty]_z
\nonumber\\
&&\hspace{11mm}
+~g\!\int_{z_\ast}^{\infty}\!\!dz'_\ast~\big[-2i\alpha g_{\mu\nu}z_\ast-{4\over s}p_{2\mu}p_{2\nu}\big]
~\tilF_{\bu j}(z_\ast,z_\perp) [z_\ast,z'_\ast]_z\tilF_\bu^{~j}(z'_\ast,z_\perp)[z'_\ast,\infty]_z
\Big\}\Big)^{ab}
\nonumber\\
&&\hspace{-3mm} 
+~{g^2\over \alpha s^2}
\Big\{\!\int_{x_\ast}^{\infty}\! \! dz_\ast\!\int_{z_\ast}^{\infty}\! \! dz'_\ast
\tilde{\bsi}(z_\ast,z_\perp)[z_\ast,x_\ast]_zt^a[x_\ast,\infty]_zt^b
[\infty,z'_\ast]_z\gamma^\perp_\mu\!\not\! p_1\gamma^\perp_\nu\tilde{\psi}(z'_\ast,z_\perp)
\nonumber\\
\hspace{-1mm} 
&&\hspace{11mm}-~\!\int_{x_\ast}^{\infty}\! \! dz_\ast\!\int_{x_\ast}^{z_\ast}\! \! dz'_\ast
\tilde{\bsi}(z_\ast,z_\perp)\gamma^\perp_\nu\!\not\! p_1\gamma^\perp_\mu
[z_\ast,\infty]_z t^b[\infty,x_\ast]_zt^a[x_\ast,z'_\ast]_z
\tilde{\psi}(z'_\ast,z_\perp)\Big\}.
\nonumber
\end{eqnarray}
%



\vspace{5mm}
\begin{thebibliography}{99}
\bibitem{cs1}
J. C. Collins, D. E. Soper
{\it Nucl. Phys.} {\bf B194}, 445 (1982).

\bibitem{jimayuan}
X. Ji, Jian-Ping Ma, and F. Yuan, 
{\it Phys. Rev.} {\bf D71}, 034005 (2005).

\bibitem{collinsbook}
J. C. Collins, Foundations of Perturbative QCD (Cambridge University Press, Cambridge, 2011). 

\bibitem{echevidsci}
M.G. Echevarria, A. Idilbi, and I. Scimeni, 
{\it JHEP} {\bf 07}, 002 (2012).

\bibitem{domarxian}	
F. Dominguez, C. Marquet, Bo-Wen Xiao, and  F. Yuan,
{\it Phys. Rev.} {\bf D83}, 105005 (2011).


\bibitem{muldrod}
P. J. Mulders and  J. Rodrigues, 
{\it Phys. Rev.} {\bf D63}, 094021 (2001).

\bibitem{cs2}
J.C. Collins, D. E. Soper and G. Sterman,
{\it Nucl. Phys.} {\bf B250}, 199 (1985).


\bibitem{npb96}
I. Balitsky, 
{\it Nucl. Phys.}  {\bf B463}, 99 (1996);
{\it Phys. Rev.} {\bf D60}, 014020 (1999).

\bibitem{yura}
Yu.V. Kovchegov,  
{\it Phys. Rev.} {\bf D60}, 034008 (1999);
{\it Phys. Rev.} {\bf D61},074018 (2000).

\bibitem{kovsievert}
Yu.V. Kovchegov and M.D. Sievert,
{\it ``Calculating TMDs of an Unpolarized Target: Quasi-Classical Approximation and Quantum Evolution''},
e-print arXiv:1505.01176.

\bibitem{collins1}
J. C. Collins,
{\it Phys. Lett.} {\bf B536}, 43 (2002).

\bibitem{EICase}
D. Boer {\it et al},
{\it ``Gluons and the quark sea at high energies: distributions, polarization, tomography''},
arXiv:1108.1713 [nucl-th].

\bibitem{muxiyu}	
A. H. Mueller, Bo-Wen Xiao, and  F. Yuan,
{\it Phys. Rev.} {\bf D88}, 114010 (2013);
{\it Phys. Rev. Lett.} {\bf 110}, 082301 (2013).

\bibitem{nlobk}
I. Balitsky,
{\it  Phys.Rev.} {\bf D75}, 014001(2007);
I. Balitsky and G.A. Chirilli,
{\it  Phys.Rev.} {\bf D77}, 014019(2008);
{\it  Nucl. Phys.} {\bf B822}, 45 (2009),
{\it  Phys.Rev.} {\bf D88}, 111501 (2013).

\bibitem{keld}
I. Balitsky and V.M. Braun, 
{\it Phys. Lett.} {\bf B 222}, 121 (1989);
{\it Nucl. Phys.}  {\bf B361}, 93 (1991);
{\it Nucl. Phys.}  {\bf B380}, 51 (1992).

\bibitem{bejuan}
A.V. Belitsky, X. Ji, and F. Yuan,
{\it Nucl. Phys.} {\bf  B656}, 165 (2003).


\bibitem{mobzor}
I. Balitsky, {\it ``High-Energy QCD and Wilson Lines''}, 
In *Shifman, M. (ed.): At the frontier of particle 
physics, vol. 2*, p. 1237-1342  (World Scientific, Singapore, 2001)
[hep-ph/0101042]. 

\bibitem{nlolecture}
I. Balitsky, 
{\it ``High-Energy Ampltudes in the Next-to-Leading Order''},
in ``Subtleties in Quantum Field Theory'', ed D. Diakonov, 
(PNPI Publishing Dept., 2010)\\
arXiv:1004.0057 [hep-ph].

\bibitem{bbr}
I. Balitsky,
{\it Phys. Lett.} {\bf B124}, 230 (1983);
I. Balitsky and V.M. Braun,	
{\it Nucl. Phys.}  {\bf B311}, 541 (1989).

\bibitem{cgkh}
J.y. Chiu, F. Golf, R. Kelley, and A. V. Manohar,
{\it Phys. Rev.} {\bf D77}, 053004 (2008).

\bibitem{bneu}
T. Becher, M. Neubert,
{\it Eur. Phys. J.} {\bf C71}, 1665 (2011).

\bibitem{dglap}
 V.N. Gribov and L.N. Lipatov, 
{\it Sov. Journ. Nucl. Phys.} 
{\bf 15}, 438 (1972);
G.Altarelli and G. Parisi, 
{\it Nucl. Phys.}  {\bf B126}, 298 (1977);
Yu. L. Dokshitzer, 
{\it Sov. Phys. JETP} {\bf 46}, 641 (1977).

\bibitem{bal04}
I. Balitsky, 
{\it Phys. Rev.} {\bf D70}, 114030 (2004).

\bibitem{proceedings}
I. Balitsky and A. Tarasov
{\it ``Evolution of gluon TMD at low and moderate $x$''},
 Int.J.Mod.Phys.Conf.Ser. {\bf 37}, 0058 (2015);
arXiv:1411.0714[hep-ph ].

\bibitem{domumuxi}	
F. Dominguez, A.H. Mueller, S. Munier, and  Bo-Wen Xiao,
{\it Phys. Lett.} {\bf B705}, 106 (2011).

\bibitem{difope}
I. Balitsky,
{\it ``Operator expansion for diffractive high-energy scattering''},
[hep-ph/9706411].

 \bibitem{cusp}	
 G.P. Korchemsky and A.V. Radyushkin,
{\it Phys. Lett.} {\bf B171}, 459 (1986);
G.P. Korchemsky,
{\it Phys. Lett.} {\bf B217}, 330 (1989),
{\it Phys. Lett.} {\bf B220}, 629 (1989).

\bibitem{ccfm}	
M. Ciafaloni,
{\it Nucl. Phys.}  {\bf B296}, 49 (1988);
S. Catani, F. Fiorani, and G. Marchesini,
 {\it  Phys. Lett.}{\bf B234}, 339 (1990),
{\it Nucl. Phys.}  {\bf B336}, 18 (1990);

\bibitem{nlobfkl}
V.S. Fadin and L.N. Lipatov,
  {\it  Phys. Lett.} {\bf B429}, 127 (1998);
G. Camici and M. Ciafaloni,
  {\it  Phys. Lett.} {\bf B430}, 349 (1998).

\bibitem{kw}
Yu. V. Kovchegov and H. Weigert,
{\it Nucl. Phys.}  {\bf A784}, 188 (2007),
{\it Nucl.Phys.} {\bf A789}, 260(2007);

\bibitem{nlojimwlk}
A. Kovner, M. Lublinsky, and Y.  Mulian,
{\it Phys.Rev.} {\bf  D89} 061704 (2014),
{\it JHEP} {\bf 114} 1408 (2014).

\bibitem{resumbfkl}	
G.P. Salam,
{\it JHEP} {\bf 9807}  019 (1998);
M. Ciafaloni, D. Colferai, G.P. Salam,
{\it Phys.Rev.} {\bf  D60} 114036	 (1999);
 M. Ciafaloni, D. Colferai, G.P. Salam, and A.M. Stasto,
 {\it Phys.Rev.} {\bf  D68} 114003 (2003);
  A. Sabio Vera,
  {\it Nucl.Phys.} {\bf B722}, 65 (2005).

\bibitem{resumbk}
L. Motyka and A. M. Stasto
 {\it Phys.Rev.} {\bf  D79} 085016 (2009);
G. Beuf,
 {\it Phys.Rev.} {\bf  D89} 074039 (2014);
E. Iancu, J.D. Madrigal,  A.H. Mueller, G. Soyez, D.N. Triantafyllopoulos,
{\it Phys.Lett.} {\bf B744} 293 (2015).


\end{thebibliography}
\end{document}